\documentclass[3p,times,11pt]{elsarticle}

\usepackage{graphicx}
\usepackage{amssymb}

\usepackage{lineno}

\usepackage[utf8]{inputenc}
\usepackage{siunitx}
\usepackage{comment}
\usepackage{float}
\usepackage[graphicx]{realboxes}
\usepackage{bm}
\usepackage{cancel}
\usepackage{mathtools}
\usepackage{gensymb}
\usepackage{amsmath,amssymb}
\usepackage{makecell}

\usepackage{isotope}

\newcommand{\cm}{\ensuremath{\mathrm{cm}}}

\newcommand{\software}[1]{{\fontfamily{qcr}\selectfont {#1}}} 
\usepackage{xurl}

\usepackage{comment}
\usepackage{subfigure}
\usepackage{amsmath}

\usepackage{isotope}
\usepackage{longtable}
\usepackage{capt-of}
\usepackage[T1]{fontenc} 
\begin{document}

\begin{frontmatter}


\title{Particle Physics at the European Spallation Source}

\author[tuw]{H.~Abele}
\author[cern,uu]{A.~Alekou}
\author[csic]{A.~Algora}
\author[ORNL]{K.~Andersen}
\author[virg,ORNL]{S.~Bae{\ss}ler}
\author[mexc]{L.~Barron-Pálos}
\author{J.~Barrow\fnref{MIT,TAU}}
\author[iphc]{E.~Baussan}
\author[ess]{P.~Bentley}
\author[aql,infn3]{Z.~Berezhiani}

% ...
\address[MIT]{Massachusetts Institute of Technology, Dept. of Physics, Cambridge, MA 02139, USA}
\address[virg]{Department of Physics, University of Virginia, Charlottesville, VA 22904, USA}
% ...
\address[cern]{CERN, 1211 Geneva 23, Switzerland}
% ...
\address[infn3]{INFN, Laboratori Nazionali del Gran Sasso, Assergi, 67100 L'Aquila, Italy}
\address[aql]{Dipartimento di Scienze Fisiche e Chimiche, Universita' di L'Aquila, Via Vetoio, Coppito 1, 67100 L'Aquila, Italy}

% ...
\address[ess]{European Spallation Source ERIC, Box 176, SE-221 00 Lund, Sweden}
% ...
\address[mexc]{Instituto de Física, Universidad Nacional Autónoma de México, Apartado Postal 20-364, 01000, México}

% ...
\address[ORNL]{Oak Ridge National Laboratory, Oak Ridge, TN 37831, USA}
% ...
\address[iphc]{IPHC, Universit\'{e} de Strasbourg, CNRS/IN2P3, Strasbourg, France}

% ...
\address[TAU]{School of Physics and Astronomy, Tel Aviv University, Tel Aviv 69978, Israel}
% ...
\address[uu]{Uppsala University, P.O. Box 256, 751 05 Uppsala, Sweden}


\address[csic]{Instituto de Fisica Corpuscular, CSIC-Universitat de Valéncia, E-46071 Valéncia, Spain}
% ...
\address[tuw]{TU-Wien, Atominstitut, Stadionallee 2, 1020 Wien, Austria}

% ...

\fntext[ibs]{Now at The center for Exotic Nuclear Studies, Institute for Basic Science, 34126 Daejeon, Korea}

\begin{abstract}
Presently under construction in Lund, Sweden, the European Spallation Source (ESS) will be the world's brightest neutron source. As such, it has the potential for a particle physics program with a unique reach and which is complementary to that available at other facilities. This paper describes proposed particle physics activities for the ESS. These encompass the exploitation of both the neutrons and neutrinos produced at the ESS for high precision (sensitivity) measurements (searches).    
\end{abstract}

\end{frontmatter}

%

\tableofcontents


\newpage
\section{Glossary}
\begin{center}
\begin{longtable}{p{.15\textwidth}  p{.85\textwidth}}
 \hline
 {\bf Acronym/term} & {\bf Meaning}  \\
 \hline
 ANNI & Pulsed cold neutron beam facility for particle physics at the ESS \\
 aCORN & Spectrometer for measuring the electron-neutrino correlation coefficient $a$ in neutron decay \\
 $a$SPECT & Spectrometer for measuring the electron-neutrino correlation coefficient $a$ in neutron decay \\
 BD & Beam Dump \\
 Beam EDM & Experiment to measure the nEDM using a pulsed neutron beam \\
 BRAND & Spectrometer for measuring correlation coefficients in neutron decay ($BRAND$ is a subset of the targeted correlation coefficients)\\
 CAD & Computer Aided Design \\ 
 CCW & Counter Clock Wise \\
 CE$\nu$NS & Coherent Elastic Neutrino-Nucleus Scattering  \\
 CKM & Cabibbo–Kobayashi–Maskawa \\
 COHERENT & Experiment to measure CE$\nu$NS at the SNS \\
 CDR & Conceptual Design Report \\
 CP & Charge-Parity \\
 CPT & Charge-Parity-Time Reversal \\
 CRES & Cyclotron Radiation Emission Spectroscopy \\
 CW & Clock Wise \\
 DDH & Desplanques-Donoghue-Holstein \\
 DE & Dark Energy \\
 DM & Dark Matter \\
 DPA & Displacements per Atom \\ 
 DTL & Drift Tube Linac \\
 EDM & Electric Dipole Moment \\
 EDM$^n$ & Concept for a scalable multi-chamber experiment to measure the nEDM \\
 EFT & Effective Field Theory \\
 emiT & Spectrometer to search for TRIV in neutron beta decay ($D$ coefficient) \\
 ep/n separator & Electron Proton / Neutron separator (magnet separating charged neutron decay products from the neutron beam)\\
 ESS & European Spallation Source \\
 ESS$\nu$SB & European Spallation Source Neutrino Super Beam   \\
 FnPB & Research Neutron Source Heinz Maier-Leibnitz at SNS \\
 FOC & Frame Overlap Chopper \\
 FODO & FOcussing DEfocussing \\
 FRM II & Research Neutron Source Heinz Maier-Leibnitz \\
 FUNSPIN & Polarised cold neutron beam at PSI \\
 GEANT4 & A Monte Carlo simulation program for GEometry ANd Tracking \\
 GENIE & GEnerator for NeutrIno Events \\
 GRANIT & GRAvitational Neutron Induced Transitions \\
 GRS & Gravity Resonance Spectroscopy \\
 HBL & High Beta Line \\
 HIBEAM & High Intensity Baryon Extraction and Measurement - the first stage of NNBAR  \\
 HighNESS & HIGH intensity Neutron source at the ESS \\
 HWI & Hadronic Weak Interaction \\
 ILL & Institut Laue-Langevin \\
 J-PARC & Japan Proton Accelerator Research Complex \\
 LANL & Los Alamos National Laboratory \\
 LANSCE & Los Alamos Neutron Science Center \\
 LBP & ESS Large Beam Port \\
 LD$_2$ & Liquid Deuterium \\
 LEBT & Low Energy Beam Transport \\
 LINAC & LInear ACcelerator \\  
 MAC-E Filter & Magnetic Adiabatic Collimation combined with an Electrostatic Filter \\
 MCNP & Monte Carlo N-Particle program for particle transport \\
 MEBT & Media Beta Beam Transport \\
 n3He & Experiment to measure the proton asymmetry in thermal neutron capture by $^3$He, $^3{\rm He}(n,p)t$\\
 nEDM & Neutron Electric Dipole Moment  \\
 NINJA & Neutrino Interaction research with Nuclear emulsion and J-PARC Accelerator \\
 NG-C & Neutron beam line at NIST \\
 NIST & National Institute of Standards and Technology \\
 NNBAR & An experiment to search for neutrons converting to anti-neutrons at the ESS \\
 NoMoS & Neutron decay prOducts MOmentum Spectrometer \\
 NO$\nu$A & Neutrinos Off-axis $\nu_e$ Appearance \\
 NPDGamma & Experiment to measure the $\gamma$ asymmetry in thermal neutron capture by hydrogen, $p(n,\gamma)d$\\
 NDTGamma & Experiment to measure the $\gamma$ asymmetry in thermal neutron capture by deuterium, $d(n,\gamma)t$\\
 nTRV & Spectrometer to search for TRIV in neutron beta decay ($R$ coefficient) \\
 NuSTORM & Neutrino from STOred Muons \\
 PDC & Pulse-Defining Chopper \\
 PERC & Proton Electron Radiation Channel (ep/n separator at the FRM~II) \\
 Perkeo\,II, III & Spectrometer for measuring correlation coefficients in neutron decay (generation II or III)\\ 
 PF1B & Cold neutron beam facility for particle and nuclear physics at the ILL \\
 PF2 & UCN and VCN facility at the ILL \\
 PHITS & Particle and Heavy Ion Transport code System \\
 PMT & Photo-Multiplier Tube \\
 PS & Power Supply Unit \\
 PSB & Proton Synchrotron Booster \\
 PSC & Pulse-Suppressing Chopper \\
 PSI & Paul Scherrer Institut \\
 PV & Parity Violation/Violating \\
 qBOUNCE & Quantum bounce experiment with UCN \\
 QCD & Quantum Chromodynamics \\
 QNeutron & Experiment to measure the neutron electric charge \\
 RAL & Rutherford Appleton Laboratory \\
 RFQ & Radio Frequency Quadrupole \\ 
 RF  & Radio Frequency \\
 sFGD & super Fine Grained Detector \\
 SM & Standard Model \\
 SNS & Spallation Neutron Source \\ 
 T2K & Tokai to Kamiokande \\
 TPC & Time Projection Chamber \\
 TRIV & Time Reversal Invariance Violation/Violating \\
 UCN & Ultra-Cold Neutron(s) \\
 UCNA & Spectrometer for measuring the beta asymmetry coefficient $A$ with UCN \\
 VCN & Very Cold Neutrons \\
\end{longtable}
\end{center}
\newpage

\section{Introduction}
\label{sec:intro}

Progress in particle physics has traditionally been achieved by a symbiosis of experiments at the energy and intensity frontiers and model-building. This has led to the current situation of the Standard Model (SM) representing our best knowledge of particle physics but which leaves a number of open questions to be resolved. These include the composition of dark matter, the dynamic origin of the observed matter-antimatter asymmetry, and the fine-tunings needed for a light Higgs and no observable CP violation in the strong sector. Furthermore, the SM is itself known not to be complete by the existence of massive neutrinos and offers no explanation for the smallness of their masses. There thus exist many problems requiring hitherto unobserved particles and physics processes. Appropriate facilities are needed to maximise the chance of observing new physics. As the world's brightest neutron source, the European Spallation Source (ESS)~\cite{Peggs:2013sgv}  offers unique capabilities at the intensity frontier in addressing a range of the aforementioned open questions. 

Presently under construction, the ESS in Lund, Sweden, is a multi-disciplinary international laboratory with 13 European member states. At design specifications, the ESS will operate at 5~MW using a proton linac of beam energy 2~GeV, leading to spallation neutrons that are steered into a suite of instruments.  In addition to providing the most intense neutron beams, the ESS also provides a large neutrino flux.  While the ESS will start with fifteen instruments used for neutron scattering  research, a particle physics program is part of the ESS statutes~\cite{ess-statutes} and is identified as missing capability of the highest importance~\cite{ess-gap}.

Unsurprisingly, given the opportunities provided by the ESS, a number of proposed experiments and activities at the ESS are being developed by the community\footnote{Several of these, ESSnuSB~\cite{Baussan:2013zcy} and HighNESS~\cite{https://doi.org/10.48550/arxiv.2204.04051} (NNBAR~\cite{Addazi:2020nlz} together with a dedicated UCN source), are funded by Horizon 2020 actions for research leading to conceptual design reports. The construction of advanced high-pressure noble gas, cryogenic CsI, and low-noise germanium detectors for CE$\nu$NS studies at the ESS (Section ~\ref{sec:nuess}) is supported by two Horizon 2021 actions.}. It is timely to describe the various ideas put forward.  Proposed activities include the use of dedicated instruments for long-term projects such as ANNI~\cite{Soldner:2018ycf} and  HIBEAM/NNBAR~\cite{Addazi:2020nlz}. The  ANNI project can encompass precision measurements of neutron properties including decays and experiments with ultra-cold neutrons (UCN) such as a search for a non-zero electric dipole moment of the neutron.  The HIBEAM/NNBAR program provides a series of searches for baryon number transformation processes via neutron conversions, with increasing sensitivity and ultimately providing an improvement in sensitivity of three orders of magnitude compared with the earlier searches~\cite{BaldoCeolin:1994jz}. The possibility of a high-flux dedicated UCN source is also considered and it is currently under study as a part of the HighNESS project - more details are given in Section ~\ref{sec:ucn}. The nuESS experiment~\cite{Baxter:2019mcx} can exploit the ESS' ability to provide the largest pulsed neutrino flux suitable for the detection and measurement of coherent elastic neutrino-nucleus scattering. The ESSnuSB project~\cite{Baussan:2013zcy} is a large scale endeavor to use the proton linac to produce the world’s most intense neutrino beam to measure neutrino oscillation parameters and the leptonic CP phase.      

This paper is organised as follows. A section on the ESS and related infrastructure is given. The ESS configuration and specifications of the ESS for the start of operations are described. Any infrastructure changes leading to operations at full design specifications are regarded as upgrades. New infrastructures required for the various particle physics projects are briefly described in the ESS section, with further details given where needed in the following sections. After the ESS section, a theory review is given of topics in particle physics which are relevant for the proposed activities. Each activity is then described in dedicated sections. A description of the apparatus and experimental methods used in the project is given, together with a quantification of the physics potential and how it fits into the worldwide landscape of experiments (current and planned).

\section{The European Spallation Source}

Presently under construction, the European Spallation Source (ESS) in Lund, Sweden, is a multi-disciplinary international laboratory and will be one of Europe’s flagship scientific facilities. The ESS is organised as a European Research Infrastructure Consortium (ERIC) and currently has 13 member states: Czech Republic, Denmark, Estonia, France, Germany, Hungary, Italy, Norway, Poland, Spain, Sweden, Switzerland and the United Kingdom. Sweden and Denmark are the host countries. The facility's unique capabilities, when completed at full specifications, will both exceed and complement those of today's leading neutron sources, enabling new opportunities for researchers across a broad variety of scientific fields including materials, life sciences, energy, environmental technology, and particle physics. 

\begin{figure}[H]
    \centering
    \includegraphics[width=0.89\linewidth]{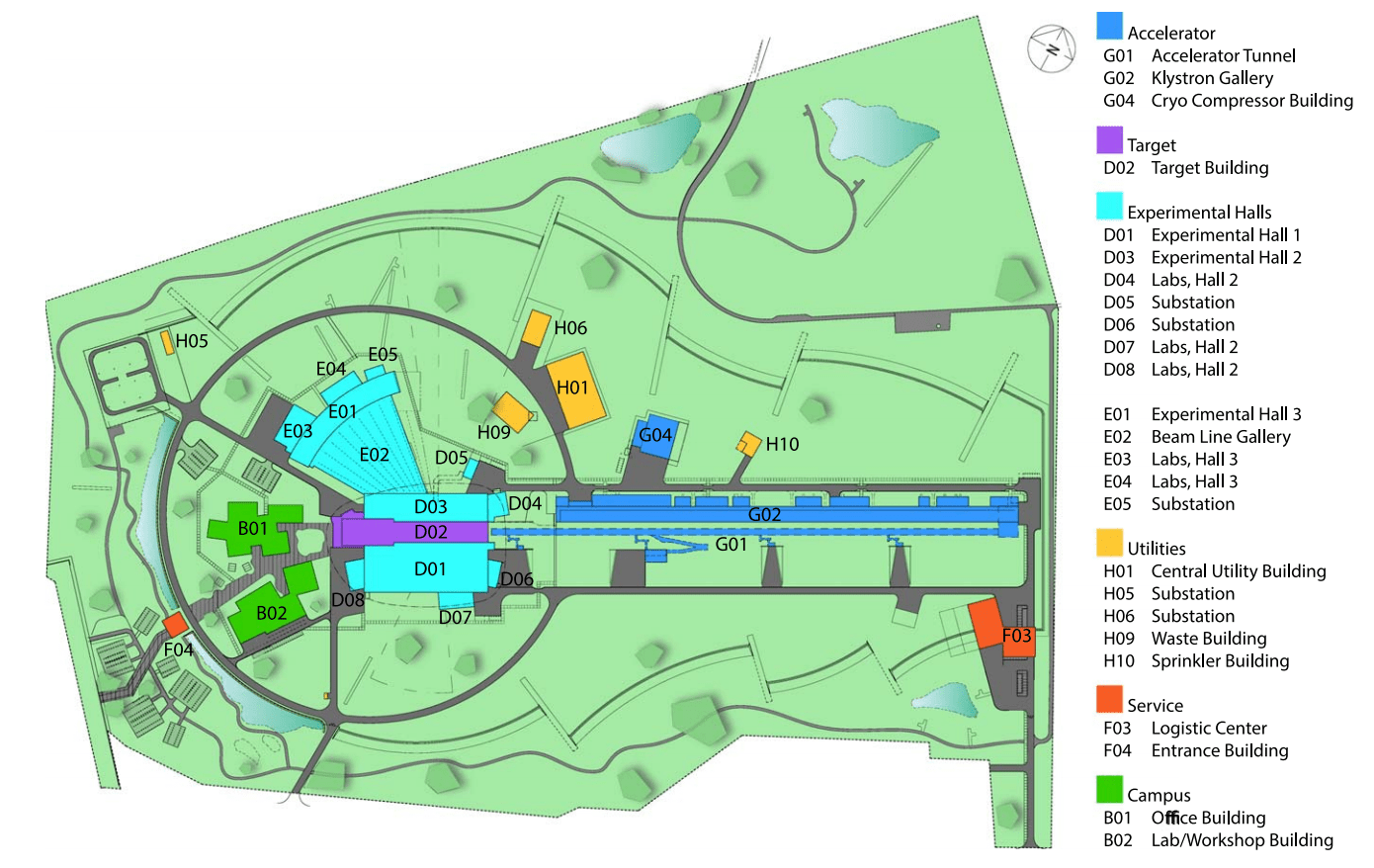}
    \caption{Layout of the ESS infrastructure.}
  \label{fig:esslayout}
  \end{figure}

An overview of the ESS facility  is shown 
in Figure~\ref{fig:esslayout} where the basic building blocks of the facility are shown: a proton linac, a target and a set of instruments located in the experimental halls. 

  \subsection{The Accelerator and the target area} \label{sec:acc}
  
  The ion source is the origin of the ESS proton beam. The beam begins as a plasma of protons, which is created by ``boiling off" the electrons from hydrogen molecules using rapidly varying electromagnetic fields. The plasma is guided into the accelerator beamline where a conducting radio frequency quadrupole (RFQ) section is followed by a Drift Tube Linac (DTL) section that accelerate the ions supplied by the electron cyclotron source up to a kinetic energy of 90~MeV. 
  \begin{figure}[H]
    \centering
    \includegraphics[width=0.89\linewidth]{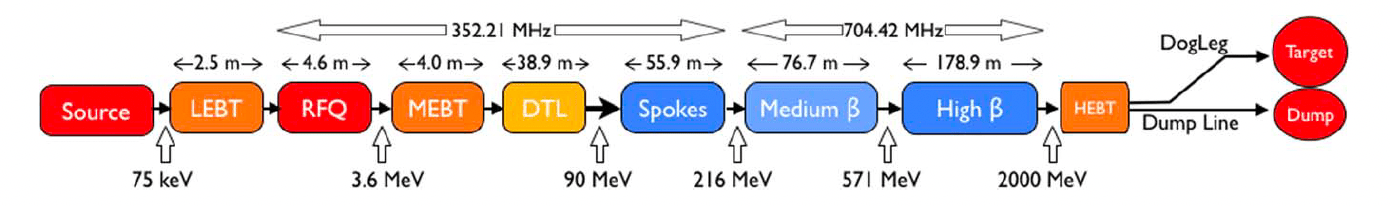}
    \caption{They layout of the ESS linac.}
  \label{fig:esslinac}
  \end{figure}

\begin{figure}[H]
    \centering
    \includegraphics[width=0.69\linewidth]{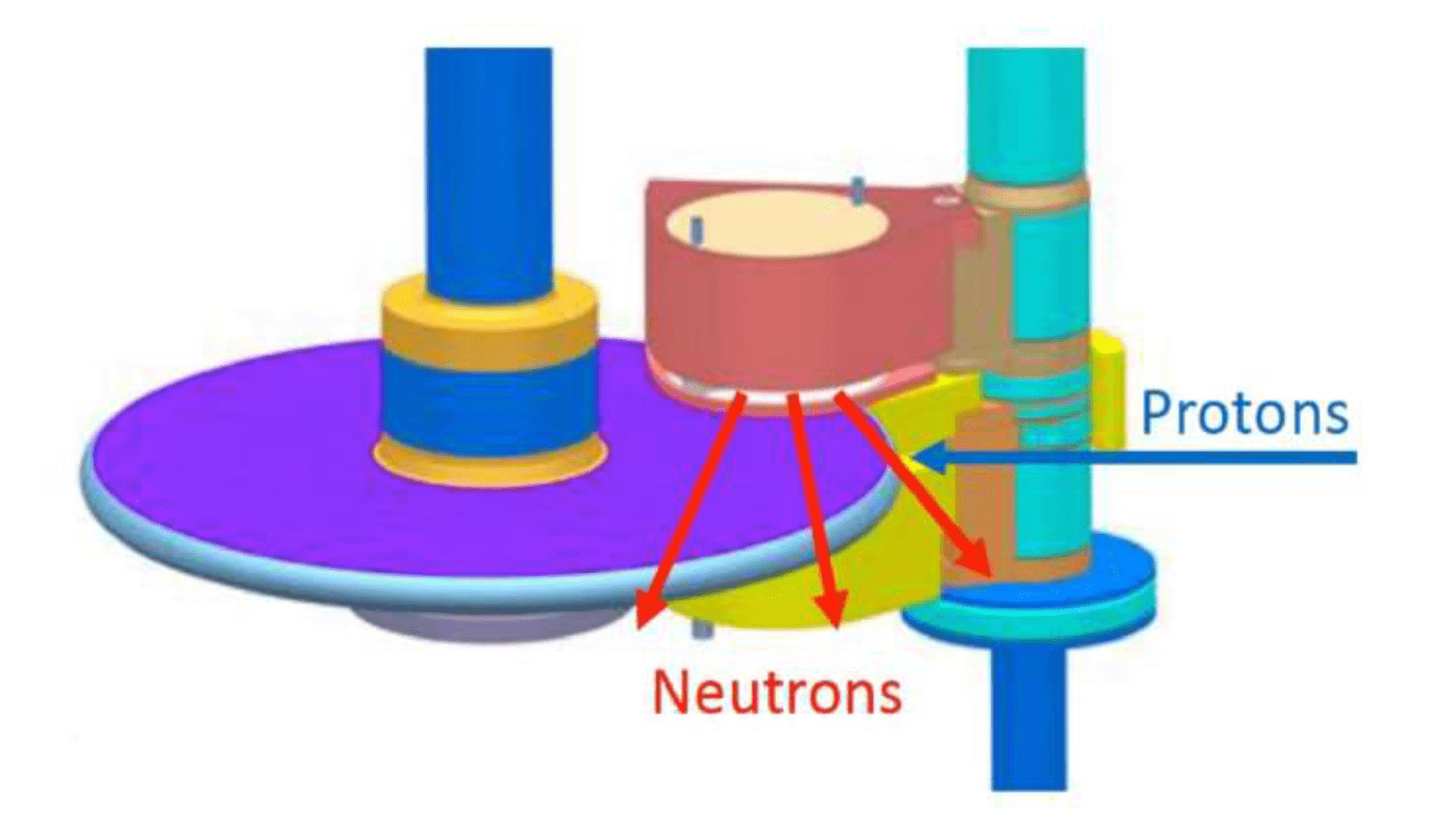}
    \caption{Overview of the ESS target area.}
  \label{fig:esstarget}
  \end{figure}

\begin{figure}[H]
    \centering
    \includegraphics[width=0.69\linewidth]{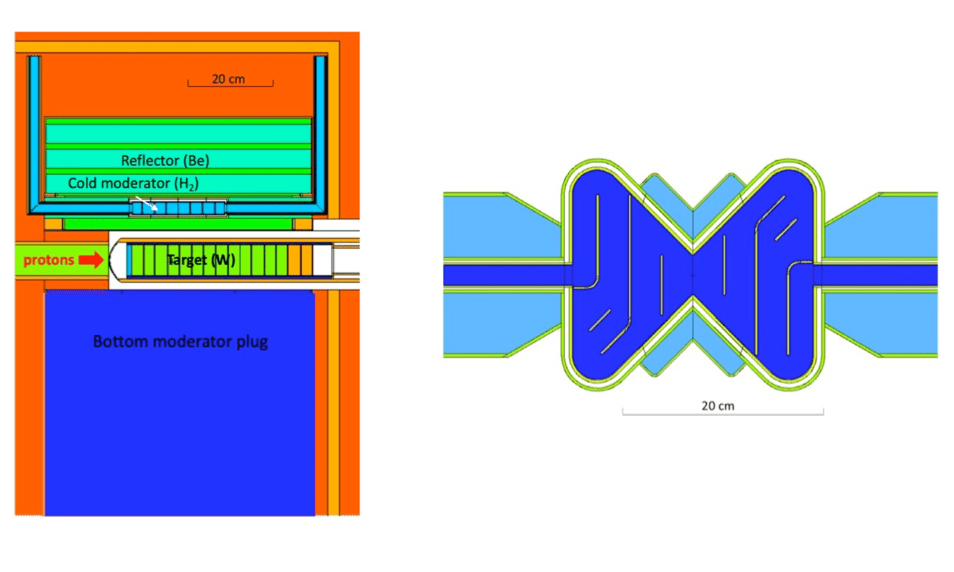}
    \caption{Overview of the ESS upper moderator. Left: MCNP model of the ESS target area in its initial configuration with only a moderator located about the spallation target. Right: Top view of the butterfly moderator. The dark blue represents para hydrogen, while the light blue is water.   }
  \label{fig:essbutterfly}
  \end{figure}

The rest of the acceleration is
achieved with superconducting accelerating structures of
three different types. ``Spoke'' cavities are used for energies up to
200~MeV, followed by medium-$\beta$ 6-cell elliptical cavities
up to 570~MeV and finally high-$\beta$ 5-cell cavities up to
2~GeV. There is a cryogenic installation that provides the cooling
power required to keep the superconducting cavities at a
temperature of 2 K and water cooling is used for the normal
conducting structures and all high power equipment
like klystrons, RF loads, etc. The ESS linac layout is shown in  Figure~\ref{fig:esslinac}. The high neutron flux from the ESS is due to the fact that the ESS, at full specifications, will possess the world's most powerful accelerator and the highest beam power on  target. The proton beam of 62.5~mA is accelerated to 2~GeV with a 14~Hz pulse structure (with each pulse  2.86~ms long). This will give 5~MW \footnote{Note that the ESS is currently committed to delivering 2~MW as accelerator power. A power of 5~MW is part of the upgrade plan (see Section~\ref{sec:timescale}). }   average power and a peak power of 125~MW. Once the proton beam has reached its final energy the beam hits a rotating tungsten target to produce neutrons by spallation (see Figure~\ref{fig:esstarget}). These are predominately evaporation neutrons at energies around 2~MeV. Tungsten blocks mounted on a wheel rotating at 23.3 revolutions per minute successively intercept the proton beam. Pressurized helium gas is used as a cooling fluid, reducing the peak temperature by 150$^{\circ}$ between two shots.
Most of the beam power is dissipated in heat in the target which is located inside a 6000 tons shielding configuration known as the monolith.
The target station converts the proton beam from the
accelerator, through the spallation process, into a number of
intense beams of slow neutrons delivered to the instruments.
The high-energy spallation neutrons are slowed down in the  neutron moderators located inside the moderator-reflector plug shown in  Figure~\ref{fig:esstarget}.  Initially, the ESS will be equipped with only a single compact low-dimensional moderator, which has been designed to deliver the brightest neutron beams for condensed matter experiments~\cite{Zanini2019}, optimized for small samples, flexibility, and parametric studies. This moderator will be located above the spallation target and due to its shape it is called the {\it butterfly} moderator (see Figure~\ref{fig:essbutterfly}). It is a cold parahydrogen moderator, 3~cm thick with a single vessel and light water moderators.  Its shape has been optimized after an intense design study optimization, allowing beam extraction in the 42 beam ports available at the ESS arranged in two 120$^\circ$ sectors. The space below the target is initially occupied by a steel plug that later can be removed and accommodate a second moderator system (see Section~\ref{sec:lowermoderator}). 


\begin{figure}[H]
    \centering
    \includegraphics[width=0.69\linewidth]{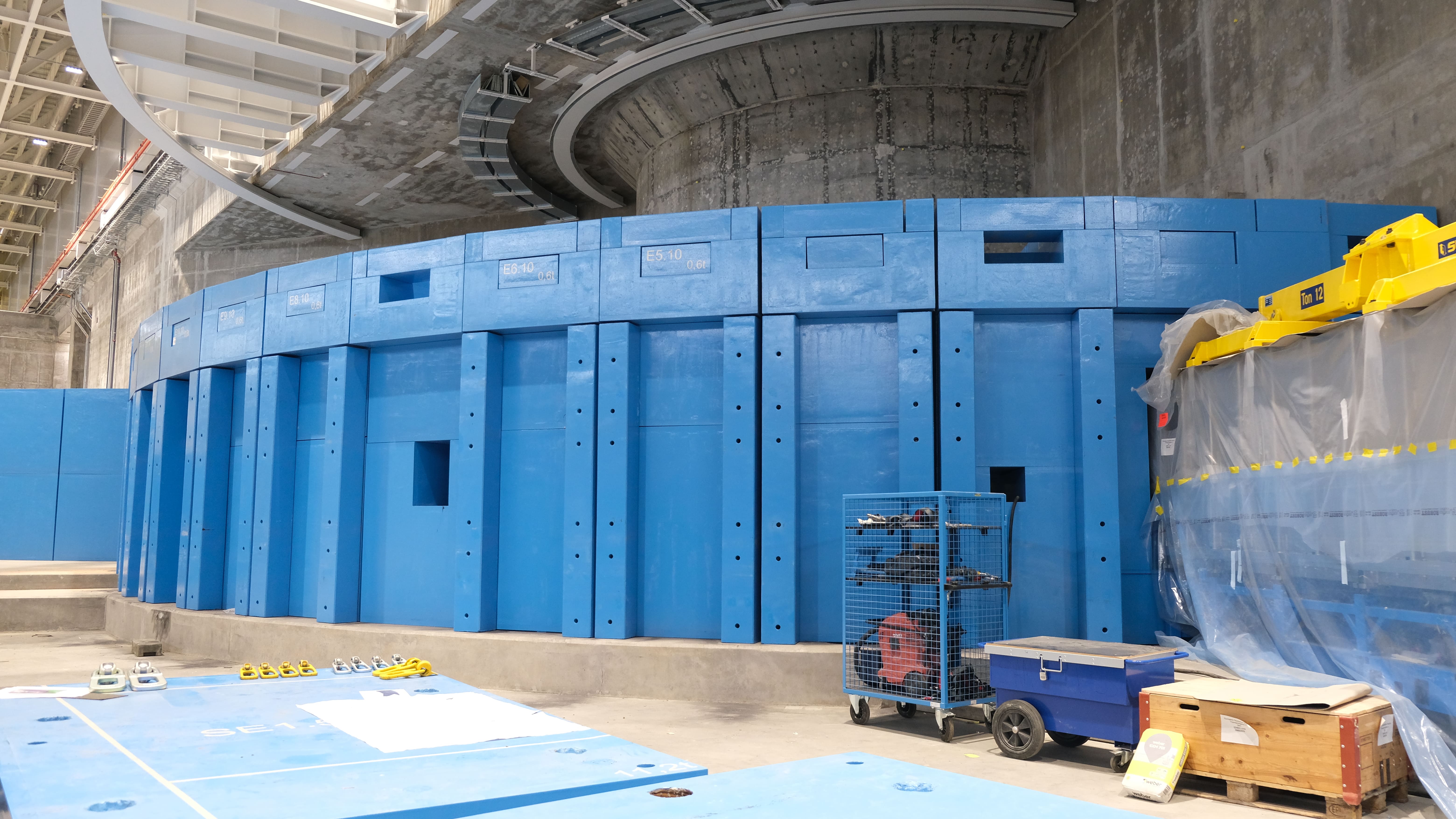}
    \caption{The ESS bunker wall.}
  \label{fig:essbunker}
  \end{figure}

The target area is surrounded by the monolith, a 3.5~m thick steel structure that also contains the neutron beamports  necessary to extract thermal and cold neutrons from the target. The beamport system of the ESS is arranged around the moderators and allows the extraction of neutrons above and below the target, a feature that is currently being investigated with the design of a lower moderator below the target. 
At the end of the monolith  the beamlines are housed in the so-called bunker (see Figure~\ref{fig:essbunker}). The bunker is a common shielding area that surrounds the ESS monolith to protect the instrument area from the high dose of ionizing radiation produced during operation. The shielding structure of the ESS bunker consists of heavy magnetite concrete walls of 3.5~m thickness, and a roof, also of heavy concrete, of variable thickness. 

Outside of the bunker, the beamlines are placed in the instrument halls.  The long instruments are located in the west sector, the short instruments in the north and east sector while the instruments with an intermediate length are located in the south sector (see Figure~\ref{fig:essinstruments}) .

\subsection{The ESS instrument suite} \label{sec:instrumentsuite}

\begin{figure}[H]
    \centering
    \includegraphics[width=0.89\linewidth]{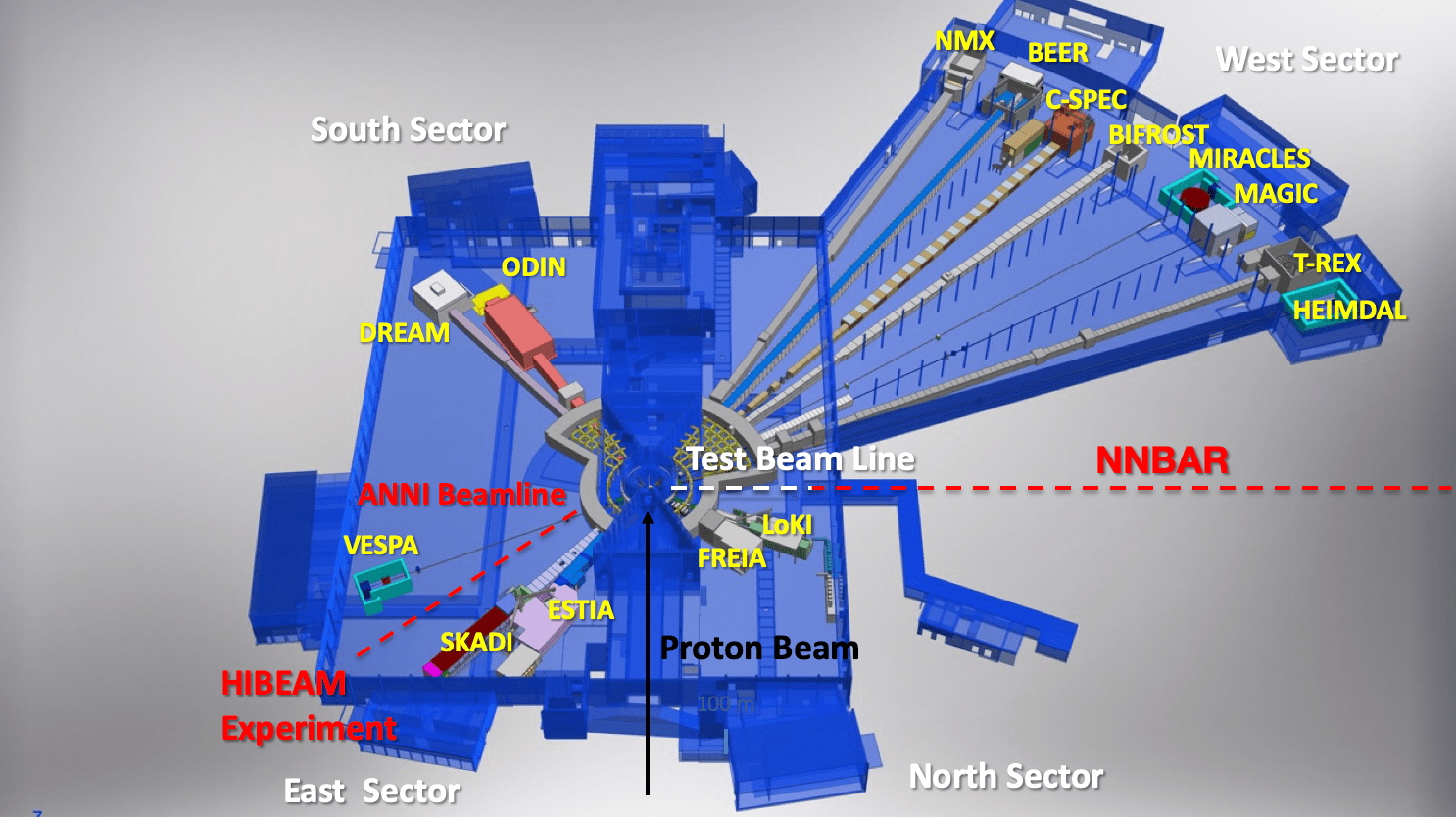}
    \caption{The ESS instrument layout.}
  \label{fig:essinstruments}
  \end{figure}

The ESS will start with fifteen instruments for neutron scattering (indicated in yellow in Figure~\ref{fig:essinstruments}). These represent only a subset of the full $22$-instrument suite required for the facility to fully realize its scientific objectives, as defined in the ESS statutes~\cite{ess-statutes}.
The ESS  mandate includes a particle physics program, and the current lack of an appropriate beamline for particle physics has been identified as one of the most important missing capabilities~\cite{ess-gap}. In Figure~\ref{fig:essinstruments} the location for the ANNI beamline is shown. This is a proposed particle physics instrument  optimised for precision measurements of neutron beta decay, hadronic weak interaction  and electromagnetic properties of the neutron (for a description of the layout of the beamline and the proposed experiments see Section~\ref{sec:ANNI}).
Figure~\ref{fig:essinstruments} also shows the position of the NNBAR beamline where the test beamline (TBL) is currently located. The TBL will be used at the start of ESS operations to characterize the target-reflector-moderator system. Once the TBL has completed its task, the NNBAR experiment could be placed at that position, at the Large Beam Port (LBP). The LBP, so named after its size compared to the standard ESS beamports, has been included to allow the extraction of a large integrated neutron flux- The LBP covers  a frame as large as three normally-sized beamports (for more details see Section ~\ref{sec:nnbar}). 
The LBP could also be used as an infrastructure for Ultra Cold Neutron production (see Section~\ref{sec:ucn}) or as a generic particle physics beamline with high integrated flux. 

\subsection{The ESS lower moderator}
\label{sec:lowermoderator}

\begin{figure}[H]
    \centering
    \includegraphics[width=0.69\linewidth]{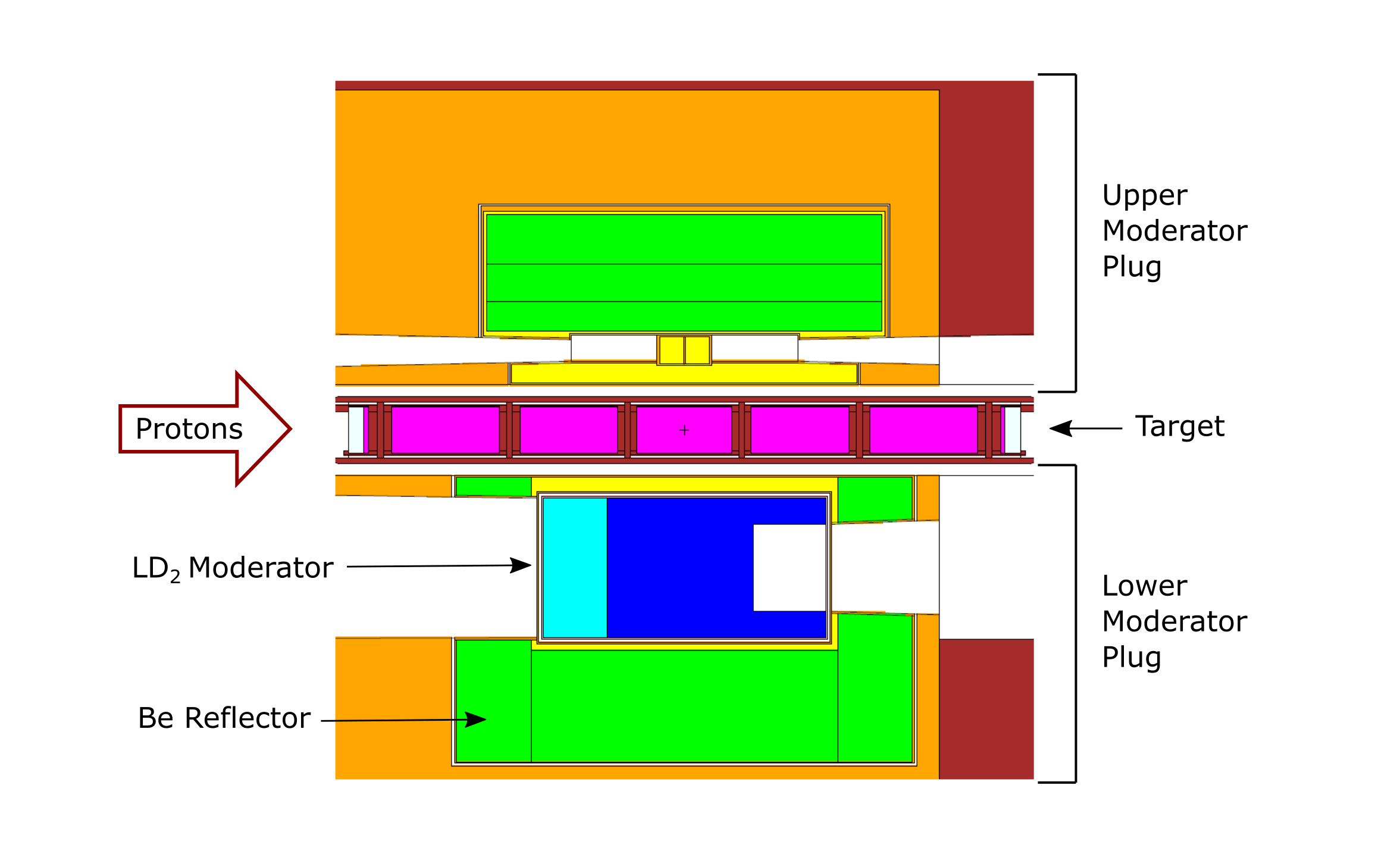}
    \caption{A lower moderator for the ESS.}
  \label{fig:esslowermoderator}
  \end{figure}

As outlined in Sections \ref{sec:acc} and \ref{sec:instrumentsuite}, the configuration of the ESS source offers the possibility of upgrades, including the design of a dedicated high intensity moderator. In fact, so far, only the upper moderator is under construction~\cite{Zanini2019} and the space below the spallation target is available for an additional moderator system. The design of the lower moderator is a part of a design study project termed HighNESS ~\cite{santoro2020development,https://doi.org/10.48550/arxiv.2204.04051}
funded as a Research and Innovation Action within the EU Horizon 2020 program. The ESS second moderator will serve two classes of applications: particle physics (the moderator is currently optimized for the NNBAR experiment but it could in principle serve every particle physics application where high integrated flux is required) and neutron scattering applications.
In the present configuration two emission windows opposite to each other are under study as can be seen in Figure~\ref{fig:esslowermoderator}. 
From a first round of optimization, the lower moderator ~\software{MCNP6} model is a ${45 \times 48.5 \times 24}$~cm$^3$ box (respectively, along and transverse to the proton beam and vertical dimension) filled with LD$_{2}$ at \SI{20}{K}. The moderator is surrounded by a light water pre-moderator (\SI{2.5}{cm} thick facing the target and \SI{1}{cm} in all the other directions) and a water-cooled room-temperature beryllium reflector. The dimensions of the openings are ${40 \times 24}$~cm$^2$  and ${15 \times 15}$~cm$^2$, for NNBAR and neutron scattering, respectively.  
Optimization of the high-intensity moderator is still on-going and the final design will be part of the Conceptual Design Report of the ESS upgrade due at the end of the HighNESS project in October 2023. In addition to the high intensity liquid deuterium (LD$_{2}$) moderator,  part of the HighNESS project is also dedicated to the design of a UCN and very cold neutrons (VCN) source that will make use of the LD$_{2}$ moderator as a primary source. Several options are currently under  study as discussed in Section~\ref{sec:ucn}. \\ 
Beyond its potential for neutron physics, the ESS will also offer exceptional opportunities for particle physics with neutrinos due to its intense neutrino flux (Section~\ref{sec:nuess}) and its powerful accelerator (Section~\ref{sec:ESSnuSB}).

\subsection{ESS timescales and accelerator power usage projections} 
\label{sec:timescale}

Owing to delays due to the Covid-19 pandemic and technical challenges, the ESS has recently conducted a rebaseline exercise to revise its construction and commissioning schedule. The newly revised baseline plan introduces a two year delay and will enable the ESS to start operation and be open for scientific users working with up to fifteen instruments in late 2027. The maximum accelerator power and proton beam energies will be 2~MW and 800~MeV, respectively.
The possibility of operating the accelerator at 2~GeV beam energy, with 5~MW power would then be part of an ESS upgrade project.
\section{Theory}


The Standard Model (SM) of particle physics \cite{Glashow:1961tr,Weinberg:1967tq,Salam:1968rm} describes a very large body of experimental results but it has a number of shortcomings and issues to which it does not have a good answer. One of the shortcomings is the large number of free parameters, especially in the so-called flavour sector where the pattern of quark and lepton masses and mixing is simply put in by hand, see e.g.~\cite{Gori:2019ybw}. A related issue is the origin of neutrino masses and mixing, see Section~\ref{sec:theoryneutrinooscillations} and e.g.~\cite{Fantini:2018itu}. These issues could be described as "close to the SM" questions. There is also the QCD-theta parameter leading to CP-violation in the purely strong sector which is unnaturally small.

The next class of issues concerns those related to the universe and the question of why matter exists. We know that a large part of the energy density of the universe is not normal matter, i.e. it cannot be explained by particles and fields present in the SM. This consists of two types, dark matter and dark energy. The former is most likely due to some particles only interacting weakly with normal matter while the nature of the latter is fully unclear, but compatible with a simple cosmological constant, see e.g. \cite{DeSimone:2019lmz}. The question of why matter and anti-matter are not created in equal amounts in the early universe also falls into this class.

The third class of issues is that where we have little experimental clue as to what to expect. These include quantum gravity, searching for violations of Lorentz and CPT symmetries and in general searching for processes and phenomena not present in the SM. 

The last class is related to studying effects of the strong interaction, especially in ways that help to answer some of the experimental uncertainties appearing for the others, but also interesting in their own right.

The resolution of the above-mentioned SM shortcomings likely requires new degrees of freedom and 
new dynamics. This strongly motivates a broad search for new physics, by either performing precision measurements of SM-allowed phenomena or by searching for phenomena that are `rare' or forbidden, because they violate approximate or exact symmetries of the SM. One should of course also be on the look-out for surprises, completely unexpected new results.

In broad brush terms one can expect that new physics has escaped detection so far because the new particles are either very heavy or very weakly coupled. Traditionally, there are two routes to search for new physics in laboratory experiments:  one is to increase the  energy of particle accelerators so as to try to  directly produce new particles -- this is known as the ``energy frontier". 
The other approach is to  perform very precise and sensitive measurements in low invariant mass systems, which allows one to indirectly access the virtual exchange of heavy degrees of freedom or directly excite light and weakly coupled particles  -- this is known as the ``precision frontier''.  
This second route  requires using ultra-sensitive detectors and/or powerful particle accelerators to generate large fluxes of particles such as neutrinos, muons, mesons, neutrons, etc. --  hence sometimes the name  ``intensity frontier''. In this language, the ESS is an intensity frontier facility. 

We emphasize that both  energy and precision/intensity frontiers are needed to infer structure, symmetries, and parameters of the underlying new physics. For example, while high-energy colliders are the most powerful probe of new particles associated to the electroweak symmetry breaking mechanism,  precision frontier experiments typically provide the strongest probes of lepton ($L$) and baryon ($B$) number violation, CP violation, flavor violation in the quark and lepton sectors, charged- and neutral-current weak interactions, neutrino properties, and dark sectors.

In the context of the intensity frontier, the ESS will provide a number of exciting opportunities, 
due to the high neutron and neutrino fluxes. First, precision measurements of neutron decay will probe non-standard interactions in the charged-current sector and help to determine the mixing element $V_{ud}$ precisely, Section~\ref{sec:theorybetadecay}. Second,  measurements of the neutron electric-dipole moment, Section~\ref{sec:theoryedm}, provide the most sensitive probe for CP-violation beyond the SM, as well as stringently bounding the QCD-theta parameter. Baryon number violation has been strongly probed in proton decay but alternatives that do not contribute to this exist and can be probed via neutron oscillations to anti-neutrons or sterile neutrons (see  Section~\ref{sec:theorynnbar}). Both of these will contribute to the understanding of net baryon number generation in the universe. Another strength of the ESS will be the searches for extra fundamental interactions and other phenomena discussed in Section~\ref{sec:theoryextra}. 
The nucleon-nucleon weak interaction is expected to have a number of nontrivial features and possible measurements are discussed in Section~\ref{sec:theoryhpv}.
Moreover, neutrino oscillation measurements will be sensitive to CP violation in the lepton sector 
and non-standard neutrino interactions with matter (see Section~\ref{sec:theoryneutrinooscillations}).
Precise measurements of coherent neutrino-nucleus scattering will test neutrino couplings to quarks and help in determining reaction rates for dark matter experiments, as outlined in Section~\ref{sec:theoryneutrinoscattering}.

In the theory part of this review we have chosen to discuss the physics relevance for the different experimental activities mainly in terms of specific models when we address sensitivities to physics beyond the Standard Model. An alternative way to do this is to use effective field theory (EFT) methods, see e.g. \cite{Penco:2020kvy} for a general introduction. We did not do this for the sake of brevity.
On the other hand, for low-energy experiments as ESS experiments typically qualify, putting the results in terms of the relevant EFT coefficients makes comparison with models easier. It also shows the importance of doing complete measurements.

Next, we discuss in greater details the opportunities discussed above.

\subsection{Neutrons}
\label{sec:theoryneutrons}

\subsubsection{Precision measurements in neutron beta decay}
\label{sec:theorybetadecay}

Beta decay has played a crucial role in the history of particle physics (neutrino hypothesis, parity violation, V-A theory, ...)~\cite{Pauli:1930pc,Fermi:1934hr,Cowan:1956rrn,Lee:1956qn,Wu:1957my,Weinberg:2009zz}. Neutron beta decay is particularly interesting for several reasons. First, it is the simplest one from a theoretical point of view (other than the pion beta decay, which has a branching ratio of $10^{-8}$), since nuclear effects are absent in this case. Second, it is sensitive to a wide variety of interactions, since both Fermi and Gamow-Teller matrix elements are non-zero. Third, significant progress is taking place on the experimental side with the arrival of powerful (ultra)cold neutron sources and new detection techniques~\cite{Abele:2008zz,Dubbers:2011ns,Baessler:2014gia,Dubbers:2021wqv}.

During the last decades there were significant tensions between some of the measurements~\cite{Abele:2008zz,Dubbers:2011ns,Wietfeldt:2011suo}, which inspired interesting theory developments that illustrate the broad discovery potential of neutron beta decay measurements, see {e.g.} Refs.~\cite{Fornal:2018eol,Berezhiani:2018eds}. Although the internal consistency of the data has greatly improved with the arrival of more precise measurements and the revision of some of the older ones, some tension remains, affecting quantities as important as the neutron lifetime~\cite{Falkowski:2020pma,Zyla:2020zbs}. It would be highly desirable to further clarify the situation. 

The exquisite theory and experimental precision achievable in several observables makes neutron beta decay sensitive to small effects generated by a wide variety of new phenomena~\cite{Herczeg:2001vk,Cirigliano:2013xha,Vos:2015eba,Gonzalez-Alonso:2018omy}. These measurements can probe heavy new particles, which can be too massive to be produced at the LHC~\cite{Bhattacharya:2011qm}. 
Such particles could play a crucial role in the generation of the matter-antimatter asymmetry or the electroweak symmetry breaking mechanism. 
Likewise, beta decay measurements are sensitive to the effects of new light particles with very small couplings to the known fields, opening an interesting connection with dark matter searches.

The physics reach of neutron beta decay goes beyond such searches of new phenomena. It gives us access to the mixing matrix element $V_{ud}$, one of the fundamental parameters of the electroweak sector in the SM. 
Recent studies have increased the accuracy of the radiative corrections to neutron beta decay, while revealing new uncertainty sources in nuclear decays, strengthening the case of the former to extract $V_{ud}$~\cite{Seng:2018yzq,Seng:2018qru,Gorchtein:2018fxl,Czarnecki:2019mwq}. 
Testing the unitarity of the CKM quark mixing matrix, whose $(1,1)$ element is $V_{ud}$, represents a crucial check of theory~\cite{Gonzalez-Alonso:2018omy}. It should be clarified whether recent tensions in this test are unaccounted SM effects or hints of new phenomena~\cite{Grossman:2019bzp,Coutinho:2019aiy,Cirigliano:2021yto}. 
Neutron decay is also of the highest importance for the study of strong interactions thanks to the precise extraction of the axial nucleon charge $g_A$ and the study of the above-mentioned radiative corrections. The comparison with QCD predictions is a challenging test~\cite{Chang:2018uxx,Gupta:2018qil} that is sensitive to nonstandard effects~\cite{Gonzalez-Alonso:2018omy}. 
Finally, neutron decay measurements are relevant as well for nuclear physics, astrophysics and cosmology, where $g_A$ and $V_{ud}$ play a crucial role~\cite{Abele:2008zz,Dubbers:2011ns,Suhonen:2017krv}. 

As explained in 
Sections~\ref{ANNIsubsec:ndecay} and~\ref{UCNsubsec:taun}, the ESS offers a unique possibility to improve our knowledge of fundamental properties of the neutron beta decay. Significant improvements are expected both on the SM parameters and the BSM searches, with the many and varied implications that were mentioned above.


\subsubsection{Searches for a non-zero neutron electric dipole moment}
\label{sec:theoryedm}

Permanent electric dipole moments (EDMs)  of non-degenerate quantum systems,  such as the neutron, 
would signal  the breakdown of time-reversal (T) and parity (P) invariance, 
or equivalently (due to the CPT theorem) of CP invariance,   where C is the charge conjugation operation that exchanges particles and antiparticles.
Given the un-observably small contribution to EDMs induced by the SM weak interactions, the current null EDM results 
and prospective sensitivities have far-reaching implications for  fundamental interactions and cosmology, as briefly summarized below 
(for more extended discussions see Refs.~\cite{Pospelov:2005pr,Engel:2013lsa,Chupp:2017rkp}).

First, the existing upper limits on neutron EDM ($\vert d_n\vert < 1.8 \times 10^{-26}$ $e$~cm at 90\% CL\cite{Abel:2020gbr})
imply that CP-violating effects in the SM strong interaction are  extremely  small 
($|\theta_{\rm QCD} |< 10^{-10}$).  This is already pointing to something very deep and historically, even before the bound 
became so stringent,  has led to the idea of a new  broken symmetry whose Goldstone boson -- the axion -- is  a viable candidate for 
the dark matter in the universe~\cite{Peccei:1977hh,Weinberg:1977ma,Wilczek:1977pj}.
Second, the current null EDM results severely constrain extensions of the SM  that include CP-violation at the TeV scale, 
and probe very strongly the CP properties of the  recently discovered Higgs boson and possible extended Higgs sectors. 
Finally, EDM searches shed light on the origin of the matter-antimatter asymmetry of the universe, 
which cannot be explained  within the SM and requires the presence of new CP violation, according to Sakharov's conditions~\cite{Sakharov:1967dj}.

To pin down the nature of new sources of CP violation,  multiple EDM searches are needed,  as EDMs of different particles species posses different sensitivity to possible contributions from quark EDMs and color-EDM,  purely gluonic CP-violating interactions, and CP-violation in the lepton sector~\cite{Chupp:2017rkp}. 
In this context the neutron stands out as the theoretically cleanest hadronic system.  In fact,  there are good prospects  to reduce the uncertainty associated with non-perturbative QCD effects in the near future with lattice QCD~\cite{Bhattacharya:2015esa,Abramczyk:2017oxr,Kim:2021qae,Dragos:2019oxn}, 
thus enhancing the model-diagnosing power of neutron EDM searches (both in case of positive and null signal).  

The gap between the existing upper limits on neutron EDM ($\vert d_n\vert < 1.8 \times 10^{-26}$ $e$~cm at 90\% CL\cite{Abel:2020gbr}) 
and the Standard Model expectation due to the Kobayashi-Maskawa phase 
($\vert d_n\vert \sim 10^{-32}$ $e$~cm~\cite{Seng:2014lea}) provides a compelling discovery window spanning six orders of magnitude. Current and next generation searches will have sensitivities a few orders of magnitude 
below the current limits. 
The major scientific merits of a one to three order of magnitude improvement in the neutron EDM sensitivity [i.e. $\vert d_n\vert \sim (10^{-29} - 10^{-27}) e~{\rm cm}$]  are:


(i) Sensitivity to a broad array of new sources of CP violation, whose origin can range from very high scale new physics to axion-like particles~\cite{Mantry:2014zsa,Dzuba:2018anu} and dark sectors~\cite{LeDall:2015ptt,Okawa:2019arp}.
In the case of new physics above the electroweak scale, existing null EDM results  already imply a lower bound of tens of TeV
on the mass scale $M$ associated with CP-violating operators in the Standard Model Effective Field Theory~\cite{Alarcon:2022ero}.  These expectations  are borne out in detailed studies of explicit models, such as the minimal supersymmetric standard model~\cite{Li:2010ax}  and the so-called split-supersymmetry
scenarios~\cite{Giudice:2004tc,ArkaniHamed:2004yi,Altmannshofer:2013lfa,McKeen:2013dma,Bhattacharya:2015esa}, where the future sensitivity reaches up to hundreds of TeV. 

(ii) Unmatched sensitivity to  CP-violating interactions of the Higgs boson~\cite{McKeen:2012av,Brod:2013cka,
Chien:2015xha} and of extended Higgs sectors~\cite{Inoue:2014nva}.
 This makes the neutron EDM a key  tool in  characterizing  the  properties  of the Higgs boson discovered at the LHC in 2012 and baryogenesis, see e.g. \cite{Fuchs:2020uoc}.

(iii)  The possibility to probe mechanisms  for the generation of the baryon asymmetry of the universe. 
 With the observation of the Higgs boson, it is particularly timely to ask whether the baryon asymmetry was produced during the electroweak symmetry-breaking era that occurred roughly 10 picoseconds after the Big Bang,  involving new particles with masses around or below the TeV scale (see Ref.~\cite{Morrissey:2012db} for a review). 
 EDM searches provide a particularly powerful probe of the associated CP violation, given their sensitivity to mass scales $M$ in the multi-TeV region.   In many scenarios 
 (with some  exceptions~\cite{Li:2008ez})   successful baryogenesis leads to strict lower bounds on the neutron and electron EDMs,  only a factor of 2 -- 3 below the current experimental limits~\cite{Cirigliano:2009yd,Morrissey:2012db}.
 The new generation of neutron EDM searches will provide very stringent tests of the  origin of matter in baryogenesis mechanisms operating near the electroweak scale. 
 
 Experimental prospects for neutron EDM searches at ESS are discussed in Sections~\ref{ANNIsubsec:EDM} and \ref{sec:UCNedm}.


\subsubsection{Neutron oscillations to anti-neutrons or to sterile neutrons}
\label{sec:theorynnbar}

The possibility of neutron--antineutron ($n-\bar{n}$) oscillation was put forward  
 in Refs. \cite{Kuzmin:1970nx,Glashow:1979nm,Mohapatra:1980qe}, 
followed by a number of papers studying different aspects 
of this phenomenon and its model realizations \cite{Mohapatra:1980de,Kuo:1980ew,Cowsik:1980np,Chetyrkin:1980ta,Barbieri:1981yr,Caswell:1982qs,Rao:1982gt,Rao:1983sd}. 
The neutron--antineutron oscillation violating baryon number $B$ by two units can have an intimate 
connection with the possibility of neutrino Majorana masses which violate lepton number $L$ 
by two units (and both violate $B-L$). 
In particular this connection can naturally emerge e.g. in a left-right symmetric model 
or its $SO(10)$ extensions,
in which $U(1)_{B-L}$ is a local symmetry and it has to be spontaneously broken at some scale 
\cite{Mohapatra:1980qe}.  

In the SM the neutron can have only Dirac mass while the neutrinos remain massless. 
But in the context of BSM theories the neutrino Majorana masses can be induced via 
the effective $\Delta L=2$ operators $C_{L=2} (\ell \phi)^2 $ of dimension 5 
involving two lepton doublets $\ell$ \cite{Weinberg:1979sa},  
where $\phi$ stands for the Higgs doublet of the SM. 
Then the neutrino mass range $m_\nu \sim 0.1$~eV implies the Wilson coefficient 
$C_{L=2} = M_{L=2}^{-1} \sim (10^{14}~\text{GeV})^{-1}$
pointing to a lepton number breaking scale close to the GUT scale.  
As for the neutron Majorana mass term $\epsilon_{n\bar{n}} n^T C n +\mathrm{h.c.}$ 
responsible for $n-\bar{n}$ mixing, 
it can be originated from the six-quark (dimension 9) operators $C_{B=2} (udd)^2$ 
involving $u$ and $d$ quarks  in combinations with different Lorentz and color structures 
\cite{Caswell:1982qs,Rao:1982gt,Rao:1983sd}.  
However, the determination of the effective scales $M_{L=2}$ and $M_{B=2}=C_{B=2}^{1/5}$ 
 is a model dependent issue, and even in the GUT context their values can be quite different. 
Namely,  $\Delta B=2$ operators induce $n-\bar{n}$ mixing with 
$\epsilon_{n\bar{n}} = \kappa(1~{\rm PeV}/M_{B=2})^5 \times 3 \cdot 10^{-25}$~eV, 
 $\kappa=O(1)$ being the operator dependent constant in the evaluation of the matrix element 
$\langle 0\vert udd \vert n\rangle$. The value of $\kappa$ is calculated by lattice QCD and models see \cite{Rinaldi:2019thf,Bijnens:2017xrz}. Indirect limits from decay of nuclei are affected by large uncertainties~\cite{Oosterhof:2019dlo,Haidenbauer:2019fyd}.
 The direct experimental bound on the $n-n'$ oscillation time 
$\tau_{n\bar{n}} = \epsilon_{n\bar{n}}^{-1} > 8.6\times 10^7$~s \cite{BaldoCeolin:1994jz}, 
implies  the upper limit $\epsilon_{n\bar{n}} < 7.7 \times 10^{-24}$~eV. 
Somewhat stronger but indirect bounds follow from the nuclear stability, namely from oxygen 
$\epsilon_{n\bar{n}} < 2.5 \times 10^{-24}$~eV,  so that present limits point  
to the effective scale of around $M_{n\bar{n}} > 0.5$~PeV. 
(For a summary of the present experimental situation, see the reviews 
\cite{Phillips:2014fgb,Babu:2013yww}.) 
Therefore, the experimental search of $n-\bar{n}$ oscillation can be meaningful if the scale $M_{6q}$ 
is below around 10~PeV. On the other hand, its discovery would be a clear evidence of $B$-violation by two units,  
and would have a great impact for our understanding of the origin of the baryon asymmetry in the universe. 
 
The last two decades showed a renaissance of the interest for new physics related to $n-\bar{n}$ mixing,  and generally for $\Delta B =2$ processes and their role in the early universe. 
In particular, new theoretical schemes involving new particle states mediating the effective six-quark 
operator were proposed and their implications for the LHC searches and for primordial 
baryogenesis were discussed 
\cite{Berezhiani:2005hv,Dutta:2005af,Babu:2006xc,Babu:2008rq,Babu:2013yca,Dev:2015uca,McKeen:2015cuz,Calibbi:2016ukt,Allahverdi:2017edd,Grojean:2018fus,Bringmann:2018sbs}.  
Mechanisms of $n-\bar{n}$ mixing were discussed also in the context of the theories with extra 
dimensions \cite{Dvali:1999gf,Nussinov:2001rb,Girmohanta:2019fsx,Girmohanta:2020qfd}. 
Different theoretical aspects of $n-\bar{n}$ oscillation were also discussed regarding the role of discrete  C, P, T symmetries
\cite{Berezhiani:2018xsx,Berezhiani:2018pcp,Tureanu:2018phm,Fujikawa:2020gsa},
as well as the possibility of the very low scale spontaneous $B-L$ violation by two units and its apparent effects \cite{Berezhiani:2015afa,Addazi:2016rgo,Babu:2016rwa}.

Another stream of research that emerged in the last two decades 
is related to a possible connection between the neutron and dark matter (DM),  
in particular in the context of the scenarios where DM is represented  by a hidden gauge sector.  
The simplest and most economic example is related to mirror particles \cite{Lee:1956qn,Kobzarev:1966qya}. 
Namely, a dark sector represented by a mirror 
SM$'$ replica of the SM,  which means that all observed particles: the electron $e$, 
proton $p$, neutron $n$ etc. must have their dark mirror twins $e'$, $p'$, $n'$ etc. which are 
sterile to our SM interactions but self-interact via the SM$'$ gauge bosons 
(for reviews see \cite{Berezhiani:2003xm,Berezhiani:2005ek}). 
Mirror matter is a viable candidate for DM with specific cosmological implications
 if, after inflation, mirror sector was reheated at lower  temperatures than the ordinary world  
\cite{Hodges:1993yb,Berezhiani:1995am,Berezhiani:1996sz,Berezhiani:2000gw,Berezhiani:2003wj}. 
During cosmological evolution, as for ordinary matter, 
the mirror nuclei and atoms, as well as stellar objects, which are dark for an ordinary observer, should form. 

Besides gravity, there can exist some feeble interactions between  particles of two sectors 
induced e.g. by kinetic mixing of the ordinary and mirror photons 
\cite{Holdom:1985ag,Glashow:1985ud} and/or  
by gauge bosons of common flavor symmetry \cite{Berezhiani:1996ii,Belfatto:2018cfo}. 
These interactions can provide a portal for the direct DM detection  
\cite{Foot:2008nw,Addazi:2015cua,Cerulli:2017jzz} and induce the mixing phenomena 
between the neutral particles of two sectors as pions and Kaons. 
Ordinary and mirror sectors can be connected also by a common axion 
\cite{Berezhiani:2000gh}. 
However, most interesting are the interactions which violate baryon and lepton numbers 
of both sectors. From the one hand, such interactions can provide a 
co-genesis mechanisms of baryon asymmetries both in ordinary and mirror worlds 
which can naturally explain the relation between the baryon and DM fractions 
in the universe \cite{Bento:2001rc,Berezhiani:2008zza,Berezhiani:2018zvs}. 
On the other hand, they can induce the mixing of the neutrinos and the neutron 
with their dark mirror partners, exactly or closely degenerate in mass. 

Namely, dimension 5 operators $C_{L=1} (\ell \phi)(\ell \phi)' $, 
which violate lepton numbers $L$ and $L'$ of both sectors by one unit,
induce active-sterile $\nu-\nu'$ mixings  
which makes  mirror neutrinos natural candidates for sterile neutrinos 
\cite{Akhmedov:1992hh,Foot:1995pa,Berezhiani:1995yi}. 

Analogously, mass mixing between the ordinary $n$ and sterile mirror $n'$ neutrons, 
$\epsilon_{nn'} n^TCn' +\text{h.c.}$, can be induced  
by dimension 9 operators $C_{B=1} (udd)(u'd'd')$ 
involving ordinary $u,d$ and mirror $u'd'$ quarks 
which violate both $B$ and $B'$ but conserve the combination $B-B'$ \cite{Berezhiani:2005hv}. 
As opposed to $n-\bar{n}$ mixing, the nuclear stability limits give no restriction 
on $n-n'$ mixing since $n\to n'$ transitions in nuclei are suppressed due to kinematic reasons 
\cite{Berezhiani:2005hv}.  
On the other hand, 
existing experimental and astrophysical limits allow $n-n'$ oscillation to be 
a rather fast process, with the characteristic oscillation time $\tau_{nn'} = 1/\epsilon_{nn'}$  
much smaller than the neutron lifetime, with interesting 
implications for cosmic rays \cite{Berezhiani:2006je, Berezhiani:2011da}  and 
neutron stars \cite{Goldman:2019dbq,Berezhiani:2020zck,Berezhiani:2021src,Goldman:2022brt}. 
Namely, the scale of underlying new physics inducing the mixed six-quark operators 
can be at TeV, $C_{B=1} \sim (1~\text{TeV})^{-5}$ 
(cosmological limits  \cite{Berezhiani:2005hv,Babu:2021mjg} 
do not exclude such a possibility),
in which case $n-n'$ mixing can be as large as $\epsilon_{nn'} \sim 1$~neV or so. 
 
 The phenomena of $n-n'$ oscillations can be experimentally searched via  neutron disappearance 
$n\to n'$ and regeneration $n\to n' \to n$ 
\cite{Berezhiani:2005hv,Pokotilovski:2006gq,Berezhiani:2017azg}. 
Such transitions can also be induced via the non-diagonal magnetic moment 
between $n$ and $n'$ states \cite{Berezhiani:2018qqw}. 
Several experiments were performed searching for $n-n'$ oscillation with the UCN   
\cite{Ban:2007tp,Serebrov:2007gw,Altarev:2009tg,Bodek:2009zz,Serebrov:2008her,Berezhiani:2017jkn,nEDM:2020ekj,Mohanmurthy:2022dbt},  
and, for a simplest scenario of the case mass degenerate $n-n'$ oscillation, 
an upper limit was set on the oscillation time, $\tau_{nn'} < 414$~s \cite{Serebrov:2007gw}. 
However, this limit becomes invalid if there is some mass splitting between $n$ and $n'$ 
or $n-n'$ oscillation is affected by the presence of mirror magnetic fields \cite{Berezhiani:2009ldq}. 
Interestingly, some of the above experiments show anomalous deviations  
\cite{Berezhiani:2012rq}  which will be critically tested in the new experiments 
\cite{Ayres:2021zbh,Addazi:2020nlz}.  

The phenomena of $n-\bar{n}$ and $n-n'$ mixings can have a common theoretical origin which unify $\Delta B=1$ and $\Delta B=2$ interactions \cite{Berezhiani:2005hv,Berezhiani:2015afa}. 
More generally, the four states $n$, $\bar n$, $n'$ and $\bar{n}'$ can be all mixed with each other.  
While the direct $n-\bar n$ mixing, i.e. the neutron Majorana mass, is strongly 
restricted by experimental limits,  $\epsilon_{n\bar n} < 10^{-24}$ eV or so, 
$n-n'$ and $n-\bar{n}'$ mixings can be both much larger. In this way,  
significant effects of $n-\bar{n}$ transition can be induced via the above mixings: 
namely, a neutron travelling to mirror world can return back as an antineutron 
\cite{Berezhiani:2020vbe}. The probability of induced  $n\to n'\bar{n}' \to \bar{n}$ 
transition can be many orders of magnitude larger than that of direct $n-\bar n$ 
oscillation, and this phenomenon can be experimentally tested
 in new experiments which can properly scan the range of the applied magnetic field 
 \cite{Yiu:2022faw}.   
 
 Experimental prospects at the ESS are the HIBEAM/NNBAR projects as discussed in Section~\ref{sec:nnbar}.

\subsubsection{Searches for extra fundamental interactions,  gravitational spectroscopy, Lorentz invariance tests}
\label{sec:theoryextra}

Fundamental interactions additional to gravitational, electromagnetic, weak and strong are assumed in many extensions of the SM. They can arise for various reasons, including the existence of additional elementary particles or dimensions of space, dark matter or energy hypotheses. The existence of additional bosons induces a spin-independent Yukawa-type interaction with a characteristic range that is inversely proportional to the boson mass. If heavy bosons are to be sought at high energies, weakly interacting light bosons would elude such detection techniques. An additional boson may even be massless \cite{Lee:1955PR}, which would violate the equivalence principle at large distances. Theories with extra large dimensions of space \cite{Rubakov:1983PLB136,Rubakov:1983PLB139,Viser:1985PLB,Arkani:1998PLB,Arkani:1999PRD,Antoniadis:1998PLB,Antoniadis:1990PLB} predict observable spin-independent interactions. The light dark matter hypothesis \cite{Fayet:2007PRD} favors them. Chameleon-type interactions \cite{Pignol:2015IJMPA} are intensively researched. The existence of axion-like particles would lead to the existence of spin-dependent extra interactions \cite{Antoniadis:2011CRP}. 

\begin{figure}
    \centering
    \includegraphics[width=\textwidth]{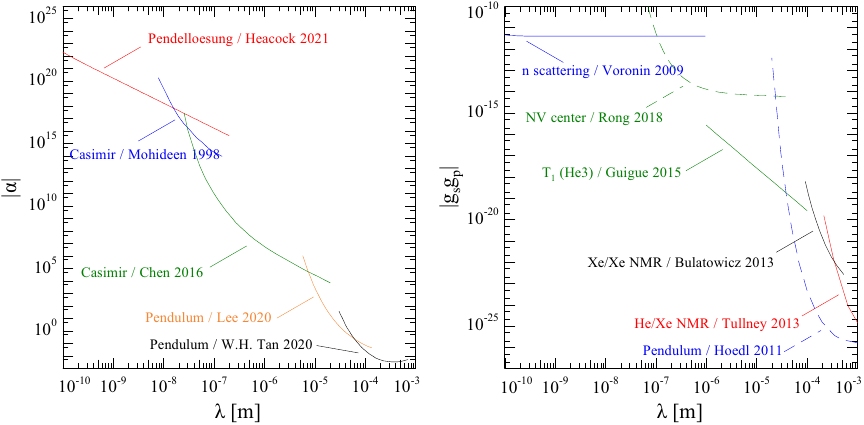}
    \caption{Left: Current most constraining limits for new spin-independent short-range forces. The first experiment (red line, \cite{Heacock:2021}) uses free neutrons. At higher $\lambda$ ranges, other techniques and probes are more sensitive \cite{Mohideen:1998,Chen:2016,Lee:2020,Tan:2020}. Right: Current most constraining limits for new spin-dependent interactions. Experiments \cite{Voronin:2009,Guigue:2015,Tullney:2013,Bulatowicz:2013} use the neutron as the polarized particle, for the first of those it is a free neutron. In \cite{Rong:2018, Hoedl:2011}, the polarized particle is an electron.}
    \label{SRfig:ShortRangeLimits}
\end{figure}

A generic functional form of a new spin-independent interactions is the one of a Yukawa-type interaction between two particles with masses $m_1$ and $m_2$ at a distance $r$: $V(r)=\alpha G m_1 m_2/r\cdot \exp(-r/\lambda)$, where $\lambda=\hbar/mc$, with the mass of the new boson $m$, is the Yukawa range of the new interaction. Here, $\alpha=1$ would mean that the new interaction is as strong as gravity for short ranges. The most simple functional form of a new spin-dependent short-range interaction between an unpolarized spin-1/2 particle (1) and a polarized spin 1/2 particle (2) at a distance $r$, mediated by a light pseudoscalar boson of mass $m$, is given by
\begin{equation}
V(r)=\frac{g_s^{(1)} g_p^{(2)} \hbar^2}{8\pi m_2} (\sigma_2\cdot \hat{r}) \left(\frac{1}{r\lambda}+\frac{1}{r^2}\right) \exp(-r/\lambda).
\end{equation}
Here, $g_s^{(1)} g_p^{(2)}$ is the product of the relevant coupling coefficients for the interaction between particle 1 (unpolarized) and 2 (polarized),  $m_2$ and $\sigma_2$ are the mass and the spin of the polarized particle, and $\lambda=\hbar/mc$ being the Yukawa range as above. The Yukawa range and the mass $m$ of the new boson are used as free parameters in the analysis. Figure~\ref{SRfig:ShortRangeLimits} shows the present most constraining limits for both of those interactions in the relevant Yukawa range. Other properties or functional forms of new short-range interactions are discussed in \cite{Dobrescu:2006,Safranova:2018,Fadeev:2019}. 

The large number of ways to search for extra interactions is reflected in the amount of data collected by the Particle Data Group \cite{Zyla:2020zbs}, e.g., in the review about ''Axions and other similar particles''. Depending on the type and range of extra interactions, neutrons play a unique or competitive role in this set of methods complementing searches at high-energy particle accelerators, in atoms, in low-background neutrino detectors, in searches for interactions on top of the van der Waals/Casimir-Polder interaction, measurements of gravity at short distances, astrophysical and cosmological constraints, and others. Several neutron experimental methods are available for such research, including neutron gravitational quantum spectroscopy \cite{Nes2002,Nesvizhevsky:2006,Jenke:2011NatP,Jenke:2014PRL,Cronenberg:NatP2018}, neutron scattering and interferometry \cite{Haddock:2018prd,Kamiya:2015, Snow:PRD,Guillaume:PLB,Nesvizhevsky:2009NatP,Brax:2018}, crystal neutron diffraction \cite{Heacock:2021,Voronin:2009,Voronin:2020}, neutron spin rotation \cite{Yan:2013,Haddock:2018plb}, and others. They allow also testing Lorentz invariance \cite{Martin:2018PRD,Escobar:2019PRD,Ivanov:2019PLB}, non-commutative quantum mechanics \cite{Bertolami:2009PRD,Saha:2007EPJC}, extensions of gravitational theories and many others. 

ESS can play a unique role among neutron sources if the experimental methods critically use the pulsed structure of ESS neutron fluxes, or unique ESS sources that can provide intense neutron fluxes with wavelengths unavailable today. In particular, this refers to the possibility of constructing an intense source of very cold neutrons. A broad window of opportunity for neutron experiments is open to constrain new interactions in the nanometer range and below, where neutrons are the best or competitive; the sensitivity of neutron experiments is rapidly improving for ranges up to about several micrometers, and is competitive for a number of scenarios. 
A distinct advantage of the neutron as a test particle is the absence of an electric charge and, compared to atoms, its small magnetic moment and the absence of van der Waals forces, 
and therefore the smallness of systematic effects. An important drawback is the low neutron flux, especially for neutrons with wavelengths optimal for these experiments. The ESS can remove precisely the most critical of the limitations of these experiments and allow a major increase in sensitivity, promising discovery potential. Some details on experimental methods can be found in Section~\ref{sec:ucn}.

\subsubsection{Studies of hadronic parity violation}
\label{sec:theoryhpv}


Nucleon nucleon (NN) weak amplitudes 
offer a unique, dynamically-rich regime in which to test the standard electroweak model.  
NN weak interactions provide a new opportunity to develop theoretical methods in low energy strong interaction theory such as effective field theory~\cite{Zhu:2004vw,deVries:2015gea} and lattice gauge theory~\cite{Wasem:2011tp}  and make predictions 
in a challenging but calculable system~\cite{Adelberger:1985ik,Desplanques:1998ak,Ramsey-Musolf:2006vfz,Haxton:2013aca,Schindler:2013yua,Gardner:2017xyl}, a recent review is \cite{deVries:2020iea}. The NN weak interaction is also a test case for our ability to trace symmetry-violating effects of a known quark-quark interaction 
in nuclei, 
which is of great interest also for 
electric dipole moments and neutrinoless double beta decay.
NN weak interactions will soon become experimentally accessible due to recent advances in atomic and molecular optics (AMO) physics through the effects of parity-odd nuclear anapole moments~\cite{Flambaum:1980sb,Wood:1997zq,Tsigutkin:2009zz,Roberts:2014bka,Safronova:2017xyt,Wieman:2019vik} on atomic and molecular structure. Advances in laser cooling also allow to detect other effects via mixing of opposite-parity states in molecules~\cite{Hutzler:2020pze,Flambaum:1984xu,DeMille:2008,Altuntas:2018ots,Norrgard:2018uba}.

The value of the ESS for this physics lies in its unparalleled combination of high intensity slow neutron beams, needed to reach the statistical accuracy to see NN weak effects, and neutron energy information using neutron time of flight from this pulsed neutron source, 
needed for the suppression of systematic errors. The experimental aspects are discussed in Section~\ref{sec:ANNIHWI}.
%

Examples of possible NN weak experiments which can take special advantage of the strengths of the ESS are: (1) neutron-proton parity-odd spin rotation, which can access the $\Delta I=2$ NN weak amplitude that can most easily be calculated in lattice gauge theory,  (2) parity-odd gamma asymmetry in $\vec{n}+D \to T+\gamma$, which is a sufficiently simple system to be treatable using effective field theory techniques in terms of two-body NN weak amplitudes, (3) a remeasurement of the parity-odd gamma asymmetry in $\vec{n}+p \to D+\gamma$ recently measured at the Spallation Neutron Source (SNS) at ORNL ~\cite{Blyth:2018aon}, which can determine the very important weak pion exchange amplitude with higher precision, and (4) a repeat of the $\vec{n}+{}^3He \to {}^3H+p$ parity violation experiment, which was also recently measured at the SNS~\cite{McCrea:2020www} and can be done with higher precision.

\subsection{Neutrinos}
The ESS will provide the largest pulsed neutrino flux, opening unprecedented opportunities to study their interaction with matter and oscillations. The main theory aspects are introduced in this section, whereas experimental details will be discussed in Sections~\ref{sec:nuess} and~\ref{sec:ESSnuSB}.

\subsubsection{Coherent neutrino-nucleus scattering}
\label{sec:theoryneutrinoscattering}

In the SM neutrinos interact with matter not only through charged current but also through neutral current 
via the tree-level exchange of a $Z$ boson. Interestingly enough at low energies the neutrino cannot resolve the nucleus, i.e., the neutrino interacts coherently with all the nucleons in the atomic nuclei, with the resulting enhancement in the cross section~\cite{Freedman:1973yd}. The challenging detection of the experimental signal of this Coherent Elastic Neutrino-Nucleus Scattering (CE$\nu$NS) was recently achieved by the COHERENT collaboration at the Oak Ridge National Laboratory~\cite{Akimov:2017ade}, opening exciting opportunities both at the technological and scientific levels.  

The ESS is particularly well suited for this kind of measurements thanks to its very high yield of low-energy neutrinos, which will allow for high-statistics measurements~\cite{Baxter:2019mcx}. It is worth noting that the use of multiple targets would allow us to disentangle various physics effects. These are discussed more in Section~\ref{sec:nuess}.

Concerning fundamental physics, CE$\nu$NS measurements offer interesting information about the nuclear structure of the target, such as the nuclear neutron distribution. Moreover, once such effects are properly parameterized, CE$\nu$NS measurements represent
an alternative
probe of the interactions between neutrinos and quarks. Some applications include the extraction of the weak mixing angle at very low momentum transfer, the study of the neutrino electromagnetic properties and the search of non-standard interactions and new light particles (sterile neutrino, dark matter, etc)~\cite{Baxter:2019mcx}. More details about the experimental opportunities are in Section~\ref{sec:nuess}.

\subsubsection{Neutrino oscillations, leptonic CP violation, non-standard neutrino interactions with matter}
\label{sec:theoryneutrinooscillations}

The quantum mechanical phenomenon of neutrino oscillations has been experimentally observed by several different so-called neutrino oscillations experiments, including the Super-Kamiokande and SNO experiments \cite{Fukuda:1998mi,Ahmad:2002jz}. In the past decades, using data from atmospheric, solar, accelerator, and reactor neutrino experiments, the various parameters that describe neutrino oscillations have been measured to excellent precision. These parameters are two mass-squared differences $|\Delta m_{31}^2|$ and $\Delta m_{21}^2$ as well as three leptonic mixing angles $\theta_{12}$, $\theta_{13}$, and $\theta_{23}$.
Nevertheless, at the moment, at least three parameters remain to be determined, i.e. the mass ordering of neutrinos, the octant of the mixing angle $\theta_{23}$, and most importantly, the leptonic CP-violating phase $\delta_{\rm CP}$. Presently, there are many dedicated proposed experiments in order to measure these parameters.

The ESSnuSB \cite{Baussan:2013zcy,Wildner:2015yaa} is a proposed experiment at the ESS. Its main aim is to discover and measure leptonic CP violation, i.e., to determine the leptonic CP-violation phase $\delta_{\rm CP}$, with unprecedented sensitivity and related to that the matter-antimatter asymmetry of the Universe. Due to the fact that the leptonic mixing angle $\theta_{13}$ is unexpectely large, the ESSnuSB is optimal for obtaining maximal sensitivity in measuring leptonic CP violation at the so-called second neutrino oscillation maximum, see for example Refs.~\cite{Huber:2005jk,Huber:2010dx,Coloma:2011pg,Coloma:2014kca}. Currently, there are two different baseline lengths that are under investigation: 360 km (Zinkgruvan, Sweden) and 540 km (Garpenberg, Sweden), which are both situated at distances from the ESS that are suitable to the second neutrino oscillation maximum. During 2018--2021, the ESSnuSB has been investigated by a European Design Study project financed by the European Commission by its Horizon 2020 Framework Programme. Comparative studies between the ESSnuSB and other proposed experimental setups in determining the leptonic CP-violating phase have been carried out \cite{Chakraborty:2017ccm,Chakraborty:2019jlv,Ghosh:2019sfi}. It has also been argued \cite{Blennow:2019bvl} that the choice of the second neutrino oscillation maximum could reduce statistics and make complementary searches of e.g. the octant of $\theta_{23}$ less fruitful, thus opting for shorter baselines (such as the baseline length of the Zinkgruvan choice).

What is the difference between the first and the second neutrino oscillation maxima? For vacuum neutrino oscillations, the first oscillation maximum is defined as $|\Delta m_{31}^2| L/(4E) = \pi/2$, where $L$ is the length of the baseline and $E$ is the average neutrino energy, whereas the second oscillation maximum is given by $| \Delta m_{31}^2 | L/(4E) = 3 \pi/2$. Using the characteristic values of the ESSnuSB ($E \simeq 0.36$~GeV), we find that the first and second oscillation maxima are situated at $\hat{L}_1 \simeq 180$~km and $\hat{L}_2 \simeq 530$~km, respectively, for $\Delta m_{31}^2 = 2.5 \cdot 10^{-3} \, {\rm eV}^2$ \cite{Esteban:2020cvm}. Indeed, the value of second oscillation maximum is close to the baseline length of the Garpenberg choice (of the detector). It should be noted that taking into consideration matter effects and three-flavor neutrino oscillations, the definition of the neutrino oscillation maxima will change quantitatively, but not qualitatively.

In addition to measuring leptonic CP-violation, the ESSnuSB is an ideal experimental setup to investigate other new physics scenarios such as non-standard neutrino interactions, sterile neutrino, and neutrino decay. For recent reviews on, for example, non-standard neutrino interactions, see Refs.~\cite{Ohlsson:2012kf,Miranda:2015dra,Farzan:2017xzy}. In phenomenological studies, so-called source and detector non-standard neutrino interactions have been explored \cite{Blennow:2015nxa}, the sensitivity to light sterile neutrinos has been studied \cite{Agarwalla:2019zex,Ghosh:2019zvl}, and the search and possibility of invisible neutrino decay has been investigated \cite{Blennow:2014fqa,Choubey:2020dhw,Chakraborty:2020cfu}. Furthermore, it has been proposed to use the ESSnuSB to test lepton flavor models \cite{Blennow:2020snb,Blennow:2020ncm}. Such models, which give rise to a non-zero leptonic Dirac CP-violating phase, could be related to the matter-antimatter asymmetry of the Universe \cite{Bertuzzo:2009im,Chen:2016ptr,Hagedorn:2016lva} and are basically testable with the ESSnuSB. In comparative studies, the ESSnuSB has also been used to put stringent constraints on fundamental symmetries such as potential CPT violation in neutrino oscillations \cite{Ohlsson:2014cha,Majhi:2021api}. The experimental aspects of ESSnuSB are discussed in Section~\ref{sec:ESSnuSB}.

\section{ANNI}\label{sec:ANNI}

\subsection{Introduction}\label{ANNIsubsec:Intro}

ANNI \cite{Theroine2015,Soldner:2018ycf} is a facility providing a cold pulsed and eventually polarised neutron beam for a multitude of experiments in particle physics with neutrons. These experiments cover a broad physics program (see also Section~\ref{sec:theoryneutrons}) and bring their own dedicated instrumentation, as outlined in Sections~\ref{ANNIsubsec:ndecay}-\ref{ANNIsubsec:AnniLast} and \ref{ANNIsubsec:hibeam}. ANNI is therefore a ``user instrument'' in the spirit of the ESS as a user facility \cite{ess-statutes}. It consists of a neutron guide transporting cold neutrons from the ESS ``butterfly'' upper moderator (see Section~\ref{sec:acc}) to the experiment while suppressing other radiation from the spallation source by its S-shaped curvature, a chopper system to tune the neutron pulses, an optional supermirror neutron polariser, and an area for the installation of experiments of up to 50~m length.

Comparable facilities exist or have been operated at continuous neutron sources\footnote{Only the latest facility at the respective neutron source is listed.}: PF1B \cite{Abele:2005xd} at the Institut Laue Langevin (ILL), FUNSPIN \cite{Zejma2005} at the Paul-Scherrer Institute (PSI), the fundamental neutron physics station NG-C \cite{Cook2009} at the National Institute of Standards NIST, Mephisto \cite{Klenke:2015hlw} at the Forschungsreaktor M{\"u}nchen II (FRM~II), as well as at pulsed spallation sources: the Fundamental Neutron Physics Beamline FnPB \cite{Fomin:2014hja} at the Spallation Neutron Source (SNS), the pulsed cold neutron beam line FP12 at LANSCE \cite{Seo2005}, and the neutron optics and physics beam line NOP \cite{Arimoto:2012zma} at J-PARC.

Many particle physics experiments with cold neutrons are limited by statistics. Beam facilities at the strongest continuous neutron sources so far provide an order of magnitude higher time-averaged neutron flux compared to beam facilities at the strongest pulsed sources. Pulsing the beam at a continuous source costs typically two orders of magnitude in intensity since both, the beam energy and the time structure, need to be restricted. In spite of the lower statistics available so far with pulsed beams, several experiments have employed them in order to overcome sources of significant systematic uncertainty, see Section~\ref{ANNIsubsubsec:benefitspulsedbeams}. ANNI promises to be the first pulsed beam facility whose time-averaged flux surpasses that of the best continuous beam facilities, providing simultaneously the advantages of pulsed beams and the world-leading neutron flux.

\subsubsection{Benefits of pulsed neutron beams for particle physics experiments}\label{ANNIsubsubsec:benefitspulsedbeams}

Pulsed neutron beams have been used at continuous sources in spite of the substantial reduction of intensity, thanks to their benefits (updated from \cite{Soldner:2018ycf}):

\begin{description}
  \item[Localisation in space:] In contrast to a continuous beam, a neutron pulse is not only limited in its transverse but also in its longitudinal direction. This allows observing neutrons in a region of well-defined response, free of edge effects. It has been used in pulsed beam measurements of the neutron lifetime \cite{Last:1988hx,Kossakowski:1989dc} where decays of cold neutrons were detected with $4\pi$ angular coverage for the neutron pulse being fully inside a strong magnetic field or a time projection chamber (TPC), respectively. At a continuous beam, edge effects need to be corrected which leads to uncertainties and usually requires dedicated measurements, for example by varying the length of the proton trap used in \cite{Yue:2013qrc}. A new project at J-PARC implements the TPC technique at a pulsed source \cite{Nagakura:2019xul,Hirota:2020mrd}. Measurements with \textsc{Perkeo\,III} \cite{Maerkisch09} have used a pulsed beam at the continuous facility PF1B in order to avoid regions of ill-defined spectrometer response, resulting in the presently most accurate measurement of the beta asymmetry parameter $A$ \cite{Markisch:2018ndu} and a limit on the Fierz interference term $b$ \cite{Saul:2019qnp}.
  \item[Localisation of neutrons in time:] If the neutron intensity arrives within a short time span, the signal-to-background ratio is increased compared to a continuous beam of the same time-averaged intensity (assuming similar levels of environmental background). The environmental background can be measured during the absence of the neutron pulse, in time intervals very close to the signal windows, allowing to subtract slowly-changing environmental background. This was exploited in \cite{Markisch:2018ndu,Saul:2019qnp}. (Note that at a pulsed source a part of the environmental background is pulsed, too. It can be measured by omitting pulses in the experiment.) Neutrons may create beam-related background with time constants large compared to the pulse duration, e.g.\ by activating windows or beta-active isotopes in a target or charging unwanted Penning traps in retardation spectrometers such as $a$SPECT \cite{Beck:2019xye}. This background is diluted in time compared to the signal, which increases the signal-to-background ratio for beam-related background. The localisation of the neutron pulse in time can also be beneficial for data acquisition and reduce the data volume, see Section~\ref{ANNIsubsubsec:CRES}.
  \item[Known velocity for each neutron:] Assuming the absence of frame overlap, the arrival time of a neutron at a given distance from a pulsed source is directly related to its velocity (or energy or wavelength). This can be exploited to calculate velocity-dependent signal or suppress systematic effects, as proposed for the Beam EDM experiment \cite{Piegsa:2013vda}, see Section~\ref{ANNIsubsubsec:beamedm}, or for accurate neutron polarimetry \cite{Penttila2005}. Employing time-dependent neutron optics, manipulation can be optimum for each individual neutron, e.g.\ by adapting the amplitude of an oscillating magnetic field of a resonance spin flipper to the neutron velocity \cite{Maruyama2003}.
 
\end{description}

\subsubsection{ANNI design guidelines and preliminary design}

ANNI was designed to provide all benefits of pulsed beams to a multitude of experiments. It was not optimised for a particular experiment but as best compromise for a reference suite of topical experiments. The reference experiments were chosen to represent all different requirements of neutron particle physics experiments with cold beams:
\begin{description}
  \item[\boldmath $a$SPECT \cite{Beck:2019xye}:] neutron decay experiment with short decay volume, potentially profiting from the localisation of the neutron pulse in time,
  \item[NPDGamma \cite{Gericke:2011zz,NPDGamma:2018vhh}:] target experiment, profiting from neutron velocity information (see Section~\ref{ANNIsubsubsec:npdgnhe3}),
  \item[PERC \cite{Dubbers:2007st,PERC:2019zjl} and \textsc{Perkeo\,III} \cite{Maerkisch09}:] neutron decay experiments with long decay volume, optimised for pulsed beams and exploiting neutron pulse localisation in space (see Section~\ref{ANNIsubsubsec:epn}),
  \item[Beam EDM \cite{Piegsa:2013vda}:] experiment featuring a very long neutron flight path, exploiting neutron velocity information and time-dependent neutron optics (see Section~\ref{ANNIsubsubsec:beamedm}).
\end{description}
The Beam EDM experiment was so far only taken into account to define the space needed for ANNI, but not in the beam line simulations. Its requirements as well as those of new proposals such as BRAND (Section~\ref{ANNIsubsubsec:BRAND}), Cyclotron Radiation Emission Spectroscopy (CRES, Section~\ref{ANNIsubsubsec:CRES}), the EDM$^n$ experiment (Section~\ref{ANNIsubsubsec:edmn}) and HIBEAM (Section~\ref{ANNIsubsec:hibeam}) will be included in the final optimisation of the beam line.

\begin{figure}
    \centering
    \includegraphics[width=\textwidth]{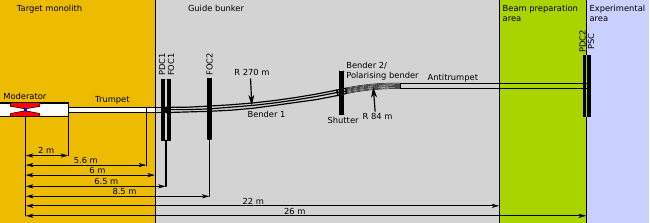}
    \includegraphics[width=\textwidth]{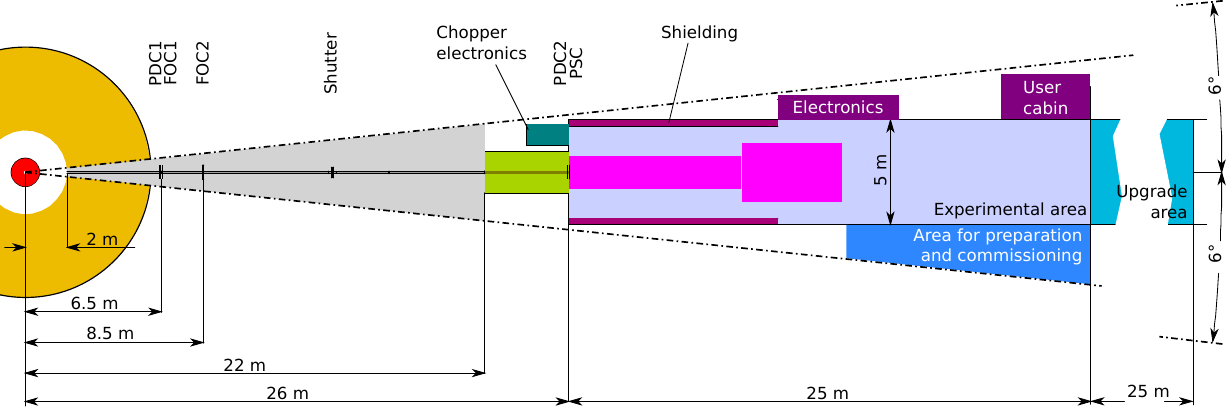}
    \caption{Top: Side view of the proposed ANNI beam line. Trumpet, Bender 1, Bender 2 and Antitrumpet assure neutron transport. For experiments with polarised neutrons, Bender 2 is replaced by a polarising bender. PDC1 and PDC2 indicate pulse-defining choppers, FOC1 and FOC2 frame overlap choppers and PSC pulse-suppressing choppers. The vertical scale is stretched by a factor of 4 for better readability. Bottom: Schematic floor plan of the ANNI facility. The magenta area indicates the footprint of PERC with a secondary spectrometer as example for an experiment. The upgrade area is needed for long experiments such as Beam EDM or HIBEAM. Taken from \cite{Soldner:2018ycf}.}
    \label{ANNIfig:scheme}
\end{figure}

We note that most of the experiments call for a rather short beam line. This is caused by two reasons: (i) The neutron pulse created by the source is polychromatic and dilutes in space and in time with increasing distance from the source. Ref.~\cite{Klauser:2014jia} discusses this effect at the example of PERC. Only experiments requiring monochromatic neutrons or a very high velocity resolution profit from large distances. This is not the case for most experiments in neutron particle physics. (ii) Although the source pulse frequency of the ESS of 14~Hz is lower than that of the SNS or the J-PARC neutron source, frame overlap between subsequent pulses limits the maximum distance of an experiment. Frame overlap for the wavelength band of 2-8~\AA{} (which includes 90\% of the cold capture flux) sets in at 34~m from the moderator and prevents unique wavelength information from the arrival time at larger distance. For these reasons the ANNI guide ends at 22~m from the ESS moderator. The experiment typically starts at 26~m from the moderator; 4~m are reserved for beam preparation or can be bridged by a straight guide section. Following ESS recommendations, the ANNI guide avoids direct view on the moderator two times, assuring that secondary showers produced by high-energy particles from the source do not reach the guide exit directly. This requires a strongly curved guide which was achieved by two benders forming an S-shaped guide in the vertical plane (vertical bending is optimum for the flat butterfly moderator). The detailed guide geometry and coatings were optimised by McStas simulations for the upper, butterfly moderator (Section~\ref{sec:acc}), see Ref.~\cite{Soldner:2018ycf}. Future optimisations utilising MCNP/PHITS simulations are required to minimise the background from the spallation source and from the guide itself. A scheme of the proposed facility is shown in Figure~\ref{ANNIfig:scheme}. 

We note that a larger, integral-flux optimised moderator as proposed for the lower moderator position (see Section~\ref{sec:lowermoderator}) increases the filling of the guide \cite{Carlile1979} and may be exploited to increase the guide cross-section (which however implies an even stronger curvature and larger neutron transport losses). However, most of the experiments at ANNI have a limited acceptance in terms of beam divergence and therefore profit from the high brightness and the directionality of the butterfly moderator, whereas unaccepted neutrons may increase backgrounds. Detailed studies based on reference experiments will finally identify the best moderator option for ANNI. 

Many experiments in particle physics require polarised neutrons. In order to offer the optimised polarised beam for each experiment, different polariser options are included in the design of the ANNI facility:
\begin{description}
  \item[Moderate polarisation at maximum flux:] The second bender of the guide is replaced by a polarising bender of the same geometry, see Figure~\ref{ANNIfig:scheme}, yielding a polarisation of $>95$\%.
  \item[Maximum polarisation with negligible angular and wavelength dependence:] The X-SM geometry of two supermirror benders \cite{Kreuz2005} can be implemented by combining the polarising bender mentioned above with a second polarising bender installed in the beam preparation area (this deflects the beam in the horizontal plane). Alternatively, an advanced compact solid-state supermirror polariser with its high magnetising field can be used, yielding a polarisation of $>99.9$\% \cite{Klauser2016,Petukhov:2016yzq}.
  \item[Polarisation with well-defined wavelength dependence:] The non-polarising guide is combined with a polarised-\isotope[3]{He} spin filter installed in the beam preparation area. This enables certain techniques for precision neutron polarimetry as described in \cite{Penttila2005}.
\end{description}

Finally, choppers are needed in order to tune the pulse emitted by the spallation source and adapt it to the requirements of the respective experiment. The chopper system consists of two frame overlap choppers (FOCs), two pulse-defining choppers (PDCs) that can run at multiples of the ESS pulse frequency of 14~Hz and a pulse-suppressing chopper (PSC), see Figure~\ref{ANNIfig:scheme}. Higher rotational frequencies of the PDCs reduce the transition times of opening and closing and thus increase the statistics since the full pulse intensity can be present for a longer period. The PSC suppresses higher-order pulses created by this operation mode of the PDCs. The following chopper modes are enabled:
\begin{description}
  \item[Full intensity:] All choppers are parked in open position. Frame overlap impairs unambiguous wavelength information.
  \item[Maximum intensity with wavelength information:] The FOCs are running, preventing frame overlap up to 130~\AA{} and simultaneously selecting a wavelength band of 6~\AA{} width.
  \item[Localisation in time:] Frame overlap is suppressed by the FOCs. The second pulse-defining chopper PDC2 is running with large opening adjusted to the highest intensity of the beam.
  \item[Monochromatic:] All choppers are phased to wavelength $\lambda_0$. The PDCs run at high frequency, in combination with the PSC, in order to maximise the resolution.
  \item[Localisation in space:] As Monochromatic mode, but the opening angles and phases of the PDCs are tuned to the dimensions of a spectrometer such as PERC or \textsc{Perkeo\,III}, see \cite{Klauser:2014jia}.
\end{description}

\begin{table}
  \centering
  \begin{tabular}{lcl}\hline
    Experiment & Gain in event rate & Comment \\\hline
    $a$SPECT     & 1.3& Full spectrum \\
               & 2.8& Localisation in time to 1/3 of ESS period\\
    NPDGamma   & 27$^{\uparrow}$ & Wavelength information\\
    PERC       & 15$^{\uparrow}$ & Localisation in space\\
    \textsc{Perkeo\,III} & 17$^{\uparrow}$ & Localisation in space\\\hline
  \end{tabular}
  \caption{Simulated gains in event rate for the reference experiments at ANNI (ESS at 5~MW) relative to the respective presently used facility. Gains marked by $^{\uparrow}$ refer to polarised beams. Taken from \cite{Soldner:2018ycf}.}
  \label{ANNItab:GainFactors}
\end{table}
Table~\ref{ANNItab:GainFactors} lists the simulated gains for the reference experiments at ANNI compared to the presently used facilities. More details on the ANNI design can be found in \cite{Soldner:2018ycf}. The following Sections discuss experiments proposed for the ANNI facility.

\subsection{Precision neutron decay}\label{ANNIsubsec:ndecay}

\subsubsection{Introduction}\label{ANNIsubsubsec:ndecayintro}

Excellent reviews of precision measurements in neutron beta decay and their analysis within and beyond the SM are available \cite{Dubbers:2021wqv,Falkowski:2020pma,Gonzalez-Alonso:2018omy,Vos:2015eba,Baessler:2014gia,Cirigliano:2013xha,Dubbers:2011ns,Abele:2008zz}. We provide an overview in the context of the experiments that have been proposed for ANNI so far (see Sections~\ref{ANNIsubsubsec:epn}-\ref{ANNIsubsubsec:CRES}).

A free neutron decays into a proton, an electron and an antineutrino, $n\rightarrow p^+ e^- \bar\nu_e$. In addition to the neutron lifetime $\tau_n$ (and decay branches involving the emission of a photon or into a hydrogen atom \cite{RDKII:2016lpd,Schott:2019ktq}), correlations between the decay products can be observed. They are defined by combining momenta or spin vectors of the involved particles into dimensionless quantities and are quantified by correlation coefficients. The orientation of the neutron spin $\bm{\sigma}_n$ can be controlled by beam polarisation and magnetic guiding fields, the electron's momentum $\bm{p}_e$ and projections of its spin $\bm{\sigma}_e$ can be measured and the antineutrino momentum $\bm{p}_{\bar{\nu}}$ can be reconstructed from observing electron and proton, whereas the spins of proton and antineutrino are difficult to measure. Combining the 4 accessible vectors and axial-vector, 6 twofold, 4 threefold, 5 fourfold and 1 fivefold scalar combinations can be defined, see \cite{Dubbers:2011ns}. Threefold and fivefold combinations test time reversal invariance. Furthermore, the electron spectrum may deviate from the pure phase space factor $\rho(E_e)$, which is described by the Fierz interference term $b$. The equations for differential decay rates of the neutron can be found in the classical papers \cite{Jackson:1957zz,Jackson:1957auh,Ebel1957} (see also \cite{Gluck:1995hs}). For illustration, we show the five ``classical'' correlations, quantified by the correlation coefficients $a$-$D$, and the terms that depend explicitly on the transverse component of the electron polarisation:
\begin{eqnarray} \label{ANNIeqn:JTWsigmaet}
\mathrm{d}\Gamma_n&\propto & \rho(E_e)\left\{1 + a\,\frac{\bm{p}_e\cdot\bm{p}_{\bar{\nu}}}{E_eE_{\bar{\nu}}} + b\,\frac{m_e}{E_e}
+ \right.\frac{\langle\bm{\sigma}_n\rangle}{\sigma_n}\cdot\left[ A\,\frac{\bm{p}_e}{E_e} + B\,\frac{\bm{p}_{\bar{\nu}}}{E_{\bar{\nu}}} + D\,\frac{\bm{p}_e\times\bm{p}_{\bar{\nu}}}{E_eE_{\bar{\nu}}} \right] \;+ \nonumber \\
&&\quad\langle\bm\sigma_{e,\perp}\rangle\cdot\left[H\,\frac{\bm{p}_{\bar{\nu}}}{E_{\bar{\nu}}} + L\,\frac{\bm{p}_e\times\bm{p}_{\bar{\nu}}}{E_eE_{\bar{\nu}}} + \right. \nonumber \\
&&\quad\qquad \left.N\,\frac{\langle\bm{\sigma}_n\rangle}{\sigma_n} + R\,\frac{\langle\bm{\sigma}_n\rangle\times\bm{p}_e}{\sigma_n\,E_e} +  S\,\frac{\langle\bm{\sigma}_n\rangle}{\sigma_n}\frac{\bm{p}_e\cdot\bm{p}_{\bar{\nu}}}{E_eE_{\bar{\nu}}} +\left.U\,\bm{p}_{\bar{\nu}}\frac{\langle\bm{\sigma}_n\rangle\cdot\bm{p}_e}{\sigma_n\,E_eE_{\bar{\nu}}} + V\,\frac{\bm{p}_{\bar{\nu}}\times\langle\bm{\sigma}_n\rangle}{\sigma_n\,E_{\bar{\nu}}} \right]\right\},
\end{eqnarray}
where $\bm{\sigma}_{e,\perp}$ represents a unit vector perpendicular to the electron momentum $\bm{p}_e$. Seven of the correlation coefficients and the kinematically dependent proton asymmetry $C = x_C (A + B)$ (with $x_C = -0.27484$) have been measured, see Table~\ref{ANNItab:summary1} in Section~\ref{ANNIsubsec:SummaryTables}.

In the SM with its $V-A$ structure, all correlation coefficients depend only on the ratio of axial-vector and vector coupling constants, $\lambda\equiv g_A/g_V$, or vanish. Here, $g_A$ accounts for the internal structure of the nucleons (compared to the quarks with axial-vector coupling 1) and can be calculated by lattice QCD, currently with precision of a few percent \cite{FlavourLatticeAveragingGroup:2019iem}. Experimental determination is the benchmark, and today is more precise by 1-2 orders of magnitude. The beta asymmetry coefficient $A=-2\lambda(\lambda+1)/(1+3\lambda^2)$ and the electron-antineutrino correlation coefficient $a=(1-\lambda^2)/(1+3\lambda^2)$ have similar intrinsic sensitivity. Since effects of CP violation on neutron decay are negligible within the SM, the time reversal violating parameters $D$ and $R$ are driven by final state effects  \cite{Ando:2009jk,Kozela:2011mc,Ivanov:2017mnz} which are one order of magnitude below the present level of experimental precision. Integrating Equation~\ref{ANNIeqn:JTWsigmaet} yields the neutron lifetime:
\begin{equation}
    \tau_n^{-1}=\frac{G_F^2\left(mc^2\right)^5}{2\pi^3\hbar(\hbar c)^6}V_{ud}^2\left(1+3\lambda^2\right)f\left(1+\delta_R'\right)\left(1+\Delta_R\right),
\end{equation}
with the Fermi coupling constant $G_\mathrm{F}$, the CKM matrix element $V_{ud}$, the phase space factor $f$ and the radiative corrections $\delta_R'$ and $\Delta_R$ (see \cite{Dubbers:2021wqv} for a recent discussion). From measuring $\tau_n$ (see Section~\ref{UCNsubsec:taun}) and $A$ (or $a$) the SM parameter $|V_{ud}|$ can be determined. The uncertainty is presently 2.5 times higher than from $0^+\rightarrow0^+$ nuclear transitions \cite{Hardy:2020qwl,Dubbers:2021wqv}, but nuclear-structure corrections are absent. In order to match experimentally the precision of today's radiative correction calculations for neutron decay, $\tau_n$ and $\lambda$ need to be measured with relative uncertainties of $1.9\times 10^{-4}$ and $1.1\times 10^{-4}$, respectively \cite{Seng:2021gmh}. Today's world averages of the experimental results have relative uncertainties of $5.7\times 10^{-4}$ and $1.0\times 10^{-3}$, respectively, where the uncertainties of the experimental inputs are scaled by factors of 1.8 and 2.7, respectively, according to the Particle Data Group procedures for averaging data \cite{PDG2022}.

Interactions beyond the SM can be expanded into $V+A$, scalar, tensor and pseudoscalar operators as corrections to the SM Hamiltonian. This modern EFT approach yields the most general Hamiltonian of beta decay already considered by Lee and Yang \cite{Lee:1956qn}. The classical papers \cite{Jackson:1957zz,Jackson:1957auh,Ebel1957} have related the coefficients of the operators -- which can be translated into the Wilson coefficients of the EFT language -- and the observable correlation coefficients. Thus, correlation coefficients can be analysed for physics beyond the SM in a model-independent way \cite{Dubbers:2021wqv,Falkowski:2020pma,Gonzalez-Alonso:2018omy,Wauters:2013loa,Severijns:2006dr}.

Whereas high-energy physics experiments are more sensitive to right-handed neutrinos \cite{Dubbers:2021wqv}, superallowed $0^+\rightarrow 0^+$ nuclear beta decays to scalar interactions \cite{Hardy:2020qwl}, and neutron beta decay is highly relevant in searches for exotic tensor interactions. The Fierz interference term describes scalar--vector and tensor--axial-vector interference and is therefore linear in the scalar and the tensor coefficients, making it particularly interesting for searching for these exotic interactions \cite{Gonzalez-Alonso:2018omy}. It can be measured from the beta spectrum or from the electron-energy dependence of beta, antineutrino or proton asymmetry. The $D$ coefficient tests time reversal violation at the TeV scale, but EDMs and other experiments provide more stringent constraints and observable deviations of $D$ from final state effects require specific fine-tuned models \cite{Ng:2011ui,El-Menoufi:2016cfo,Ramsey-Musolf:2020ndm,Falkowski:2022ynb}. Within the $V-A$ model and neglecting electromagnetic effects and recoil order corrections which can in principle be calculated, electrons emitted in beta decay are longitudinally polarised and correlation coefficients relating the transverse electron polarisation $\bm{\sigma}_{e,\perp}$ to $\bm{p}_e$, $\bm{p}_{\bar{\nu}}$ or $\bm{\sigma}_n$ vanish. Transverse electron polarisation can arise from scalar and tensor interactions beyond the SM. Said correlation coefficients have different sensitivities to the real and the imaginary parts of the scalar and the tensor couplings. Their experimental values may thus be used to establish constraints on all relevant theory parameters from neutron decay alone \cite{Bodek:2019vom}. Ref.~\cite{Dubbers:2021wqv} (Table~5) compares today's best results on the EFT Wilson coefficients from experiments at low and at high energy.

Experimentally, strong magnetic fields have been used to collect charged decay products from cold neutron beams or volumes of ultracold neutrons in measurements of spectra or asymmetries ($a$ from proton spectrum \cite{Beck:2019xye,Byrne:2002tx}, $b$ \cite{Saul:2019qnp,UCNA:2019dlk}, $A$ \cite{Markisch:2018ndu,UCNA:2017obv,Mund:2012fq}, $B$ \cite{Schumann:2007qe}, $C$ \cite{Schumann:2007hz}) because of their benefits in terms of statistics and systematics (e.g.\ separation of charged decay products according to their emission direction with respect to the neutron spin -- as already contemplated by Lee and Yang \cite{Lee:1956qn}, intrinsic determination of solid angle, substantial improvement of signal/background). At a pulsed source, experiments with the field longitudinal to the neutron beam profit in particular from spatial localisation of the neutron pulse, see Section~\ref{ANNIsubsubsec:epn} and Refs.~\cite{Saul:2019qnp,Markisch:2018ndu,Markisch:2014via}. Spatial resolution, in some cases including electron tracking, yields complementary systematics and is required in measurements of triple correlations \cite{Chupp:2012ta,Kozela:2011mc,Soldner:2004xm} or of correlations involving the transverse electron polarisation \cite{Kozela:2011mc}, see Section~\ref{ANNIsubsubsec:BRAND}. All experiments may profit from time localisation of the neutron pulse for background suppression or, in case of CRES, to reduce the data volume (Section~\ref{ANNIsubsubsec:CRES}).

The most precise published experimental results for correlation measurements in neutron beta decay are included in Table~\ref{ANNItab:summary1} of Section~\ref{ANNIsubsec:SummaryTables}. Until experiments can start at ANNI, progress is expected from several ongoing and planned activities: The \textsc{Perkeo\,III} collaboration is analysing data for the proton asymmetry $C$ \cite{RoickPhD2018} and for the Fierz interference term $b$ from measurements with polarised and unpolarised neutrons, respectively, at PF1B (ILL). The Nab experiment is setting up at the FnPB (SNS) for measurements of $a$ and $b$ with target uncertainties of $\delta a/a\sim 10^{-3}$ and $\delta b\sim 3\times 10^{-3}$, respectively \cite{Fry:2018kvq}. The uncertainty of the aCORN experiment at NG-C (NIST) of 1.7\% for $a$ \cite{Hassan:2020hrj} might be improved and, using a polarised neutron beam, the apparatus might also be used to measure $B$ with a final uncertainty of $<0.3$\% \cite{Wietfeldt:2019rhz}. PERC \cite{Dubbers:2007st,PERC:2019zjl} at Mephisto (FRM~II) aims for measurements of $a$, $b$, $A$, and $C$ with uncertainties of $\delta a/a\sim 10^{-3}$, $\delta b\sim 2\times10^{-3}$, $\delta A/A\sim 5\times10^{-4}$ and $\delta C/C\sim 10^{-3}$, respectively. BRAND plans to obtain improved results for many correlation coefficients, in  particular those involving the transverse electron polarisation, within their R\&D program at PF1B (ILL), see Section~\ref{ANNIsubsubsec:BRAND} and Ref.~\cite{Bodek:2019vom}. Concerning correlation measurements with UCN, there is an R\&D program at LANSCE to explore an upgrade to UCNA \cite{UCNA:2017obv} (UCNA+) with sensitivity for $\delta A/A \sim 2\times 10^{-3}$. 

\subsubsection{ep/n separator}\label{ANNIsubsubsec:epn}

\begin{figure}
    \centering
    \includegraphics[width=\textwidth]{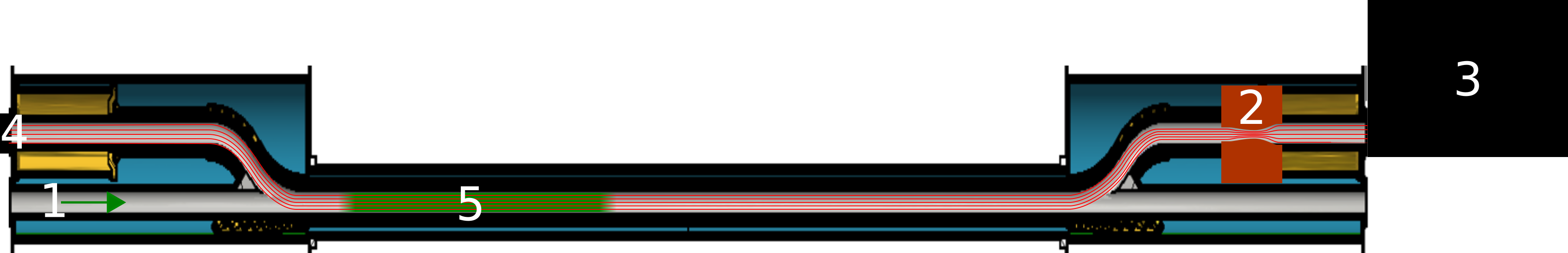}
    \caption{Sketch of the ep/n separator, an upgrade to the concept of the PERC instrument \cite{Dubbers:2007st,PERC:2019zjl} with improved backscattering detection. Neutrons enter from 1. The magnetic field (field lines indicated in red) guide charged decay products either to the tunable magnetic filter 2 and downstream detector system 3 or to the upstream detector 4. The upstream detector also detects particles reflected by the magnetic filter and from backscattering in the downstream detector system. Decay products are counted if the neutron pulse 5 (green) is fully inside the homogeneous region of the magnetic field (localisation in space). The more symmetric design of the ep/n separator compared to PERC largely improves systematics related to backscattering: the upstream detector can be shielded against background and covers 100\% of the beam.}
    \label{ANNIfig:epn}
\end{figure}

The ep/n separator, see Figure~\ref{ANNIfig:epn}, uses a high magnetic field in order to collect and magnetically guide charged decay products to both ends of a long neutron decay region and to separate them from the neutron beam. It builds on the experience with the experiments \textsc{Perkeo\,III} \cite{Maerkisch09} and PERC \cite{Dubbers:2007st,PERC:2019zjl}. The time correlation between electron and proton is washed out in the magnetic field which limits measurements at high event rates to observables that can be derived from detection of only one decay product. These are the beta and proton asymmetries (coefficients $A$ and $C$), the Fierz interference term $b$, and the electron-neutrino correlation coefficient $a$ measured via the proton spectrum. The aims of the experiments are world-leading determinations of $\lambda$, and, combined with accurate lifetime measurements, $V_{ud}$, as well as best limits on or discovery of scalar and tensor interactions.


Both, \textsc{Perkeo\,III} and PERC, are optimised to use pulsed neutron beams already at the continuous reactor sources of the ILL and the FRM~II in order to eliminate or control leading sources of systematic errors by spatial localisation of the neutron pulse, as described above and in Refs.~\cite{Markisch:2018ndu,Saul:2019qnp,Maerkisch09} for \textsc{Perkeo\,III}, and Ref.~\cite{Dubbers:2007st} for PERC.

In \textsc{Perkeo\,III}, the region with the magnetic field parallel to the neutron beam, where decay is observed under well-defined conditions, is about $\SI{2}{meters}$ long, corresponding to an observation time of about $\SI{2}{ms}$ per pulse. Its successor, the new instrument PERC, which is under construction at the FRM~II, extends the length of the active volume by a factor of $\approx 3$. A non-magnetic neutron guide preserves the high neutron density while maintaining the neutron polarisation to be stable on the $10^{-4}$ level \cite{Hollering2022,Klauser2016}. This in effect increases decay statistics by more than an order of magnitude. While \textsc{Perkeo\,III} uses a symmetric detector setup in order to minimise systematics, PERC makes use of an additional high magnetic field region (``magnetic filter'') to precisely define the solid angle of the electrons and protons detected by the downstream detector. This layout enables the use of an asymmetric setup with a single main detector system downstream, and also provides much enhanced shielding of the downstream detector (see Ref. \cite{Dubbers:2007st}). Strong magnetic fields of 1-6~T are used to reduce systematic effects and minimise the size of the detector to about $12\times\SI{12}{cm^2}$. Systematics are controlled on the $10^{-4}$ level or better \cite{Dubbers:2007st}. The magnetic field design of PERC is described in Ref.~\cite{PERC:2019zjl}. 

The magnetic filter will reflect about 95\% of all protons and electrons. While this considerably reduces the burden for the main downstream detector, all reflected particles will need to be absorbed by an upstream particle dump. In PERC, this will be an active electron detector system based on plastic scintillator and located just outside the warm bore of the PERC magnet. This detector is located close the neutron beam, though, and is hence not easy to shield. 

In Refs.~\cite{Soldner:2018ycf,Klauser:2014jia} it was demonstrated that the instruments \textsc{Perkeo\,III} and PERC massively profit from the pulse structure of the ESS, gaining one order of magnitude in event rate at ANNI, see Table~\ref{ANNItab:GainFactors}. For PERC, the upstream detector was identified as a potential limiting systematic factor. In order to match systematics and statistics, we hence proposed the improved design of the ep/n separator shown in Figure~\ref{ANNIfig:epn} which features a complete separation of the charged particles from the neutron beam also for the upstream detector \cite{Theroine2015}. A longer active volume would optimise the instrument to the low repetition rate of the ESS and hence further improve statistics.

The ep/n separator will serve as a clean, intense and versatile source of neutron decay products. Like in PERC, decay particles will be selected by a tunable magnetic filter and guided towards specialised downstream detector systems: detectors based on plastic scintillator or silicon with the possible addition of proton-to-electron conversion foils \cite{Stolarz:2002,Schumann:2007hz,Hoedl2006}, a MAC-E filter like $a$SPECT \cite{Beck:2019xye} or an advanced magnetic spectrometer like the $\bm{R}\times\bm{B}$ spectrometer NoMoS \cite{Wang:2013xjk,Moser:2020unj}.

\subsubsection{BRAND}\label{ANNIsubsubsec:BRAND}

\begin{figure}
    \centering
    \includegraphics[width=\textwidth]{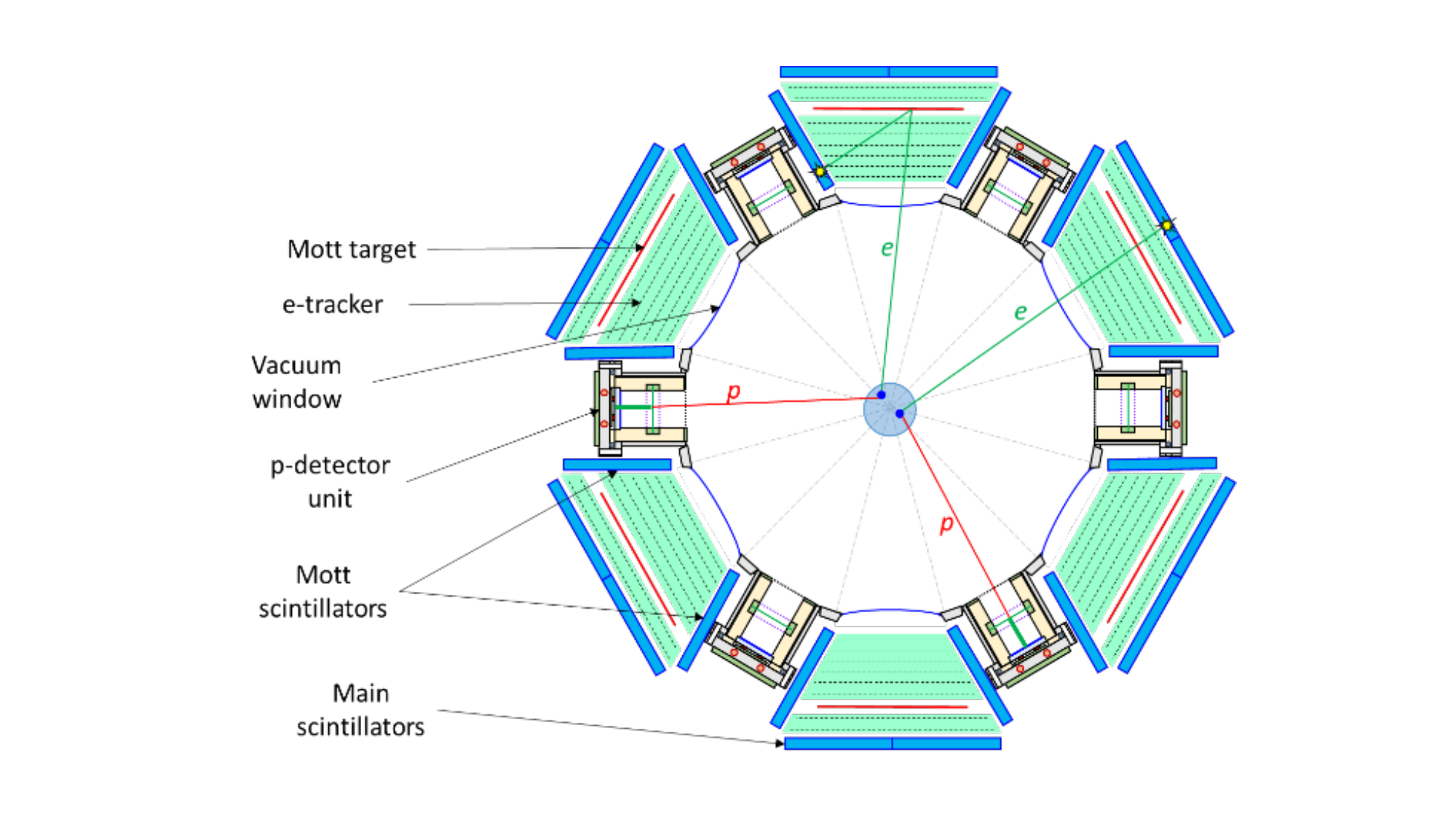}
    \caption{Cross section of the BRAND detector - conceptual design.}
    \label{ANNIfig:Brand}
\end{figure}
BRAND \cite{Bodek:2019vom} is a spectrometer with electron tracking and proton detection, see Figure~\ref{ANNIfig:Brand}. A particular feature is the observation of the transverse polarisation of the electron, giving access to the correlation coefficients $H$, $N$, $L$, $R$, $S$, $U$ and $V$ of Equation~\ref{ANNIeqn:JTWsigmaet}, where only $N$ and $R$ have been measured so far \cite{Kozela:2009ni,Kozela:2011mc}. A weak magnetic field (of about 100~$\mu$T) is used only to guide the neutron spin and does not affect trajectories or spins of the decay products significantly. Electron and proton can be measured in coincidence, giving access to the neutrino (neutrino asymmetry coefficient $B$), to correlations of two particles (electron-neutrino asymmetry coefficient $a$ from the correlation between electron and proton) and, last not least, to the time-reversal violating triple correlations (correlation coefficients $D$, $R$, $L$ and $V$) allowing to search for CP violation in neutron decay. The full approach of BRAND is complementary to the high-magnetic-field solid-angle-integrating spectrometers described in Section~\ref{ANNIsubsubsec:epn}.

The transverse polarisation of electrons is analysed in backward angle Mott scattering on high $Z$ nuclei (e.g.\ Pb or U). This method yields an exceptionally large analysing power, approaching 0.5 at backward angles. Most importantly, since Mott scattering is governed by the time reversal and parity conserving electromagnetic interaction, it is insensitive (on the level of $10^{-7}$) to false effects (e.g.\ geometry misalignment) from the longitudinal electron polarisation dominating in beta decay. This feature is unique among the techniques measuring the transverse polarisation of leptons. Mott scattering has been successfully demonstrated in neutron decay in the pioneering experiment \cite{Kozela:2009ni,Ban:2006qp,Bodek2011,Kozela:2011mc}. The achieved systematic uncertainty of $5\times 10^{-3}$ for the $R$ coefficient and $4\times 10^{-3}$ for $N$ can be further improved by at least an order of magnitude. The substantially higher angular acceptance of BRAND combined with the neutron flux at ANNI result in an expected increase in the rate of triggered events by a factor of at least 300. Further gains in precision are expected from the higher analysing power of the Mott target (in case of depleted U) and neutron polarisation at ANNI ($>99.9$\% in the maximum polarisation configuration). Thanks to its tracking capabilities, BRAND does not require the spatial localisation provided by ANNI's pulse structure. The signal-to-background ratio for coincident measurements is expected to be very high thanks to a short coincidence time window, whereas it may profit from the localisation of the neutron pulse in time for single-particle observables.

The key features of the proposed setup are: (i) efficient cylindrical detector geometry that approaches $4\pi$ acceptance, (ii) electron tracking in multi-wire drift chambers with signal readout at both wire ends, (iii) detection of both direct and Mott-scattered electrons in plastic scintillator calorimeters, (iv) conversion of protons (accelerated to 20--30 keV) into bunches of electrons ejected from a thin LiF layer \cite{Hoedl2006} followed by acceleration and subsequent detection of ejected electrons in a thin ($\sim$25 $\mu$m) plastic scintillator with position sensitivity, (v) highly symmetric detector setup (with mirror planes parallel and perpendicular to the neutron beam) for the suppression of systematic effects.

BRAND's detecting system will have a several meters long segmented structure in order to efficiently utilise the decay source. Assuming $5 \times 10^4$ decays per second in the fiducial volume expected at ANNI and two years of data taking, one can acquire $10^{12}$ direct electrons (coefficient $A$), $3 \times 10^{11}$ protons in coincidence with direct electrons (coefficients $a$, $B$, $D$), $3\times 10^8$ Mott-scattered electrons (coefficients $N$, $R$), and $10^8$ protons in coincidence with Mott-scattered electrons (coefficients $H$, $L$, $S$, $U$, $V$). These numbers are sufficient for the anticipated sensitivity of about $5 \times 10^{-4}$ for correlation coefficients involving the transverse electron polarisation. The measurements of $a$, $A$, $B$ and $D$ will not be limited by their statistical uncertainty of a few times $10^{-5}$.

BRAND's tracking capability and the ability to reconstruct the full event kinematics permit a detailed investigation of systematic effects. BRAND will be the only spectrometer capable of measuring the transverse electron polarisation. The full set of accessible correlations enables sensitive self-consistency checks as different couplings and theory corrections contribute differently to each correlation. For these reasons BRAND is complementary to all high magnetic field spectrometers. The proposed experiment is challenging and not free of risks. An experimental program has been started at the PF1B beamline of the ILL reactor in order to demonstrate feasibility and performance of the proposed components \cite{Bodek:2019vom}.

\subsubsection{Experiments with Cyclotron Radiation Emission Spectroscopy}\label{ANNIsubsubsec:CRES}

The Fierz interference term $b$ is very sensitive to exotic couplings as discussed in Section~\ref{ANNIsubsubsec:ndecayintro}. Its measurement requires accurate electron energy spectroscopy. The energy response of conventional solid state detectors for electrons is intrinsically nonlinear because of effects such as electron backscattering, energy losses in detector dead layers, quenching and depth-dependent charge- or light-collection efficiencies. Cyclotron Radiation Emission Spectroscopy (CRES) \cite{Monreal:2009za} is a novel spectroscopy technology that is not subject to these effects. As a frequency-based technology, it shows an excellent energy reconstruction \cite{Project8:2017nal} and energy resolution \cite{Project8:2022wqh}, making it particularly well suited for measurements of the Fierz interference term $b$.

CRES was proposed in Ref.~\cite{Monreal:2009za} based on the analysis of the cyclotron radiation emitted by the accelerated motion of a decay electron, spiraling in a magnetic field with the cyclotron frequency 
\begin{equation}
    \omega_\mathrm{c} =\frac{e}{m_e \gamma} B,
\end{equation}
where $e$ ($m_e$) are the charge (mass) of the electron and $B$ is the magnetic flux. The relativistic factor $\gamma=\sqrt{1+\frac{E_{e,\mathrm{kin}}}{m_e c^2}}$, where $c$ is the velocity of light in vacuum and $E_{e,\mathrm{kin}}$ is the electron's kinetic energy. In typical CRES experiments $\omega_\mathrm{c}$ is within the microwave range of the electromagnetic spectrum. CRES was first demonstrated using conversion electrons from a gaseous $\mathrm{^{83m}Kr}$ source \cite{Project8:2014ivu}.

The electron emits power into free space according to Larmor's formula \cite{Larmor1897}
\begin{equation}
    P\left(E_{e,\mathrm{kin}},\theta \right) = \frac{1}{4\pi\epsilon_\mathrm{0}} \frac{2}{3}\frac{e^4}{m_e^4 c^5}\left(E_{e,\mathrm{kin}}^2+2 E_{e,\mathrm{kin}} m_e c^2\right)\sin^2\theta,
\end{equation}
where $\theta$ is the pitch angle, defined between the electron's momentum vector and the magnetic field. The free-space equivalent radiation power for a \SI{90}{\degree}-pitch-angle, \SI{17.8}{keV} electron in a \SI{1}{T} magnetic field is \SI{1.0}{\femto\watt}.

\begin{figure}
\subfigure[]{\includegraphics[width=0.49\textwidth]{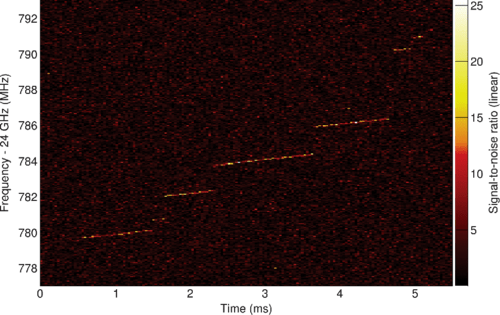}\label{ANNIfig:CRES}}
\subfigure[]{\includegraphics[width=0.45\textwidth]{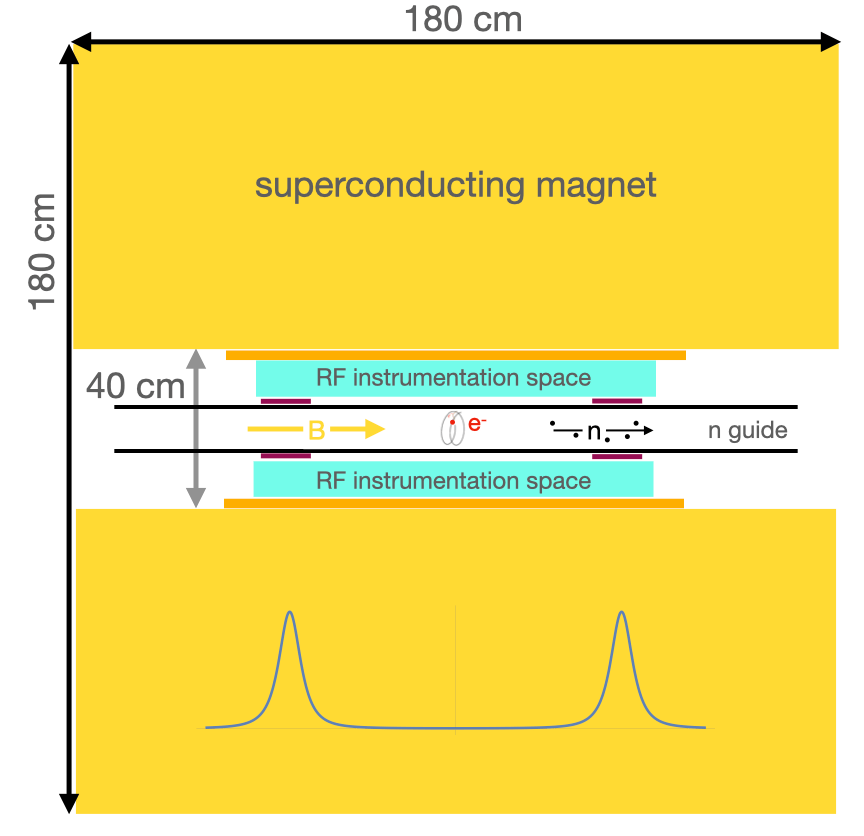}\label{ANNIfig:nCRESAtESS}}
\caption{(a) A CRES signal from a single conversion electron emitted by a $\mathrm{^{83m}Kr}$ source confined in a microwave guide. A sudden excursion of microwave power is detected when the electron is created in the decay. The signal frequency slowly increases as the electron looses energy due to the emission of cyclotron radiation. The discrete frequency jumps are related to the collision of the electron with gas molecules in the decay volume \cite{Project8:2014ivu}. (b) A conceptual layout of a CRES setup around a neutron guide with a magnetic bottle field indicated symbolically.}
\end{figure}

Most gaseous sources of decay electrons, and also neutrons, are transparent such that the microwave radiation can be received by a single antenna or antenna array surrounding the decay volume. To integrate a detectable amount of energy the electron must be confined in a magnetic bottle that sets a lower limit on the pitch angle $\theta_\mathrm{min}$. The received signal is amplified, mixed down, and bandpass-filtered before it is digitised. The data analysis can then be performed in the time or frequency domain. An individual CRES signal is shown in Figure~\ref{ANNIfig:CRES}. The energy resolution and reconstruction of CRES have been demonstrated, using $\mathrm{^{83m}Kr}$. The lines of the conversion electron doublets at \SI{30.2}{\keV} are separated by \SI{52.8}{eV} and resolved with a full width half maximum of \SI{3.3}{eV} which corresponds to a relative energy resolution of $10^{-4}$. The deviations from the predicted $1/\gamma$-scaling of the cyclotron frequency correspond to energy uncertainties of less than \SI{50}{meV} across a $>\SI{14}{keV}$ kinetic energy range. This initial demonstration opens up the opportunity for CRES applications in neutron decay spectroscopy experiments at the ANNI beamline, with fundamentally different systematic effects as compared to solid-state-based electron detectors. Dominant systematic uncertainties of these detectors like electron-surface or electron-matter interactions will be completely absent. However, new uncertainties, such as electron-gas scattering due to the electron's long travel path in the column, need to be considered.

A major challenge of CRES is the large data volume per event in sparsely occupied spectra, requiring online triggering and fast data compression. As the CRES events are distributed in frequency space, there is no detector dead time inducing event or energy pile-up. In an untriggered setup the fully differential measurement approach of CRES causes the data volume per time to be essentially independent of rate. Experiments at a pulsed source profit from pulse localisation in time since signal digitising is only needed during the presence of the neutron pulse, reducing the data volume by the ratio of the neutron pulse duration in the magnetic bottle and the pulse period. Since this ratio increases with shorter distance to the ESS moderator, a CRES experiment is best implemented in ANNI's beam preparation area (see Figure~\ref{ANNIfig:scheme}). A non-conducting supermirror neutron guide inside the magnetic bottle, instead of a gap in the guide, would allow the cyclotron radiation to be detected behind the guide walls without reducing the neutron flux. Therefore the CRES experiment could even run in parallel to another experiment downstream. A conceptual arrangement with typical dimensions is shown in Figure~\ref{ANNIfig:nCRESAtESS}.

\subsection{Electric dipole moment of the neutron}\label{ANNIsubsec:EDM}

Measurements of permanent electric dipole moments (EDMs) have a key role in fundamental particle physics, both as a precision test of the SM of particle physics, and as a sensitive probe for new physics. As first pointed out by Sakharov, the observed cosmological imbalance between matter and antimatter could only arise in the presence of significant CP-violating interactions \cite{Sakharov:1967dj}. The SM is known to violate CP symmetry in the weak interactions of quarks, and allows CP violation in the neutrino sector, but these effects are insufficient – by far – to explain the observed abundances of matter, antimatter, and photons in our universe \cite{Chupp:2017rkp}. New CP-violating physics is needed, and searches for EDMs are recognised as a ``lightning-rod'' that collects contributions from the relevant sources of CP-violation \cite{Falkowski2021b,Dekens:2013zca,Kley:2021yhn}. While a non-zero EDM measurement from any single experimental system cannot, alone, indicate the source of the CP-violation that generates it, null EDM results frequently provide the most powerful constraints on these types of new physics \cite{Cirigliano:2019vfc}.

The neutron is special in this context, due to its sensitivity to the CP-violating SM parameter $\theta_\text{QCD}$.
Although a finite value for this parameter could easily have resolved the matter-antimatter puzzle, experimental measurements of the neutron's EDM indicate that it is much too small.
Even if $\theta_{\text{QCD}}$ has a nonzero value too small to be detected in today’s experiments, this would represent a ``fine-tuning'' on the order of one part in $10^{10}$ (the so-called ``strong CP problem'').
Further details relating to theory impacts are discussed in Section~\ref{sec:theoryedm}.

The first experimental neutron EDM limit in 1957 \cite{Smith:1957ht} was already a powerful tool for testing new theories. 
The precision of neutron EDM experiments has since improved by nearly seven orders of magnitude, with the most recent result setting a limit consistent with zero at the level of $10^{-26}~e\text{ cm}$ \cite{Abel:2020gbr}. This fantastic precision, corresponding to an energy resolution of $10^{-22}\text{ eV}$, is possible because (provided adequate statistical sensitivity and control of systematic errors) frequency measurements can be improved to arbitrary precision. Yet a further improvement by a factor of approximately $10^6$ would be needed, to reach sensitivity levels where the neutron EDM that arises within the SM from CP violation in the CKM matrix can be detected \cite{Seng:2014lea}. The intervening range is considered a ``background-free'' window in which to search for new physics.

The fundamental statistical sensitivity for measuring the neutron EDM $d_{\text{n}}$, assuming a count-rate-limited frequency measurement, is

\begin{equation}
    \sigma(d_{\text{n}}) = \kappa \frac{\hbar }{\eta E T \sqrt{N}}, 
    \label{ANNIsubsubsec:edmeqn}
\end{equation}
where $E=|\bm{E}|$ is an applied electric field, and $N$ is the total number of detected neutrons. The measurement duration $T$ is the time during which an EDM-sensitive phase can be accumulated, between the two states whose frequency splitting is being measured. The dimensionless contrast parameter $\eta$ has a maximum value of 1, corresponding to 100\% spin-polarisation. The proportionality constant is $\kappa=0.5$ for (a pair of) storage measurements with ultracold neutrons, or $\kappa=2$ for the Beam EDM configuration \cite{Chanel:2018zga}.

Measurements of the neutron EDM are based on Ramsey’s Nobel prize winning technique of separated oscillating fields, which represents a very sensitive method to probe for spin-dependent interactions \cite{Ramsey:1949ts,Ramsey:1950tr}.
In this technique an effective Larmor precession frequency $\omega$ is measured by applying two consecutive phase-locked $\pi/2$ radio frequency resonance pulses to an ensemble of neutrons exposed to a magnetic field $B_0$ and an electric field $E$. By measuring $\omega$ for the two cases that the orientation of these fields is parallel ($\uparrow \uparrow$) and anti-parallel ($\uparrow \downarrow$), a value for the neutron EDM can be deduced:
\begin{equation}
  \Delta \omega = \omega_{\uparrow \uparrow} - \omega_{\uparrow \downarrow} = \frac{4 d_{\text{n}} E }{\hbar}.
\end{equation}
This formula makes the crucial assumption that the magnetic field has not changed over the duration of the two measurements.
To achieve this condition neutron EDM experiments are usually enclosed with multiple layers of active and passive magnetic shielding, and the magnetic field is monitored with various kinds of magnetic field sensors.

Early neutron EDM experiments were performed using neutron beams, while current experiments and most new projects employ stored ultracold neutrons (UCNs).
Experiments with UCNs have the advantage of much longer interaction times: on the order of 200~s, compared to the range of 10-100~ms for long neutron beam experiments.
By contrast, much larger neutron count rates can be achieved in beam experiments. Higher electric fields may also be possible for beams, since insulating wall material is not required between the high-voltage electrodes. This is needed to confine the UCNs in storage experiments, although high electric fields can also be achieved for storage experiments performed under liquid helium (see reference \cite{Phan:2020nhz} and Section~\ref{ANNIsubsubsec:edmn}).
The main limiting systematic effect in beam experiments so far has been the relativistic $\bm{v} \times \bm{E}$ effect, arising from the motion of neutrons with velocity $\bm{v}$ in the electric field \cite{Dress:1976bq}.




\subsubsection{The Beam EDM experiment}\label{ANNIsubsubsec:beamedm}

\begin{figure}
    \centering
    \includegraphics[width=\textwidth]{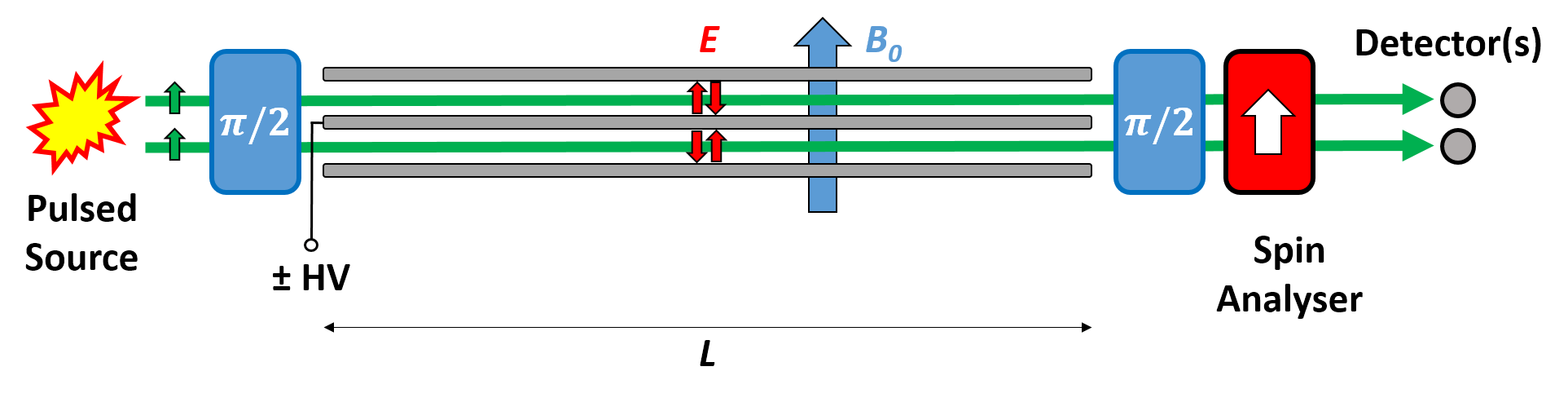}
    \caption{Schematic side-view drawing of the Beam EDM experiment. The polarised pulsed beam from the ANNI beamline is separated into two beams which pass through a neutron Ramsey spectrometer consisting of two $\pi/2$ resonance spin flip devices and a spin analyser. 
    The amplitudes of the radio frequency pulses need to be modulated in order to achieve perfect $\pi/2$ flips for neutrons of all velocities (exploiting the ESS pulse structure through use of time-dependent neutron optics).
    The two beams sense an external magnetic field $B_0$ and electric fields $E$ with opposite directions. By changing the polarity of the high voltage (HV) applied to the central electrode -- the outer two electrodes are connected to ground -- the electric field direction is inverted with respect to the direction of $B_0$. The detector(s) counting the two beam rates as a function of time-of-flight need to be capable of accepting high neutron intensities.
    The neccessary passive magnetic shielding and the vacuum beam pipe are not shown.}
    \label{ANNIfig:beamedmscheme}
\end{figure}

The Beam EDM experiment proposes to overcome the limitation of the $\bm{v} \times \bm{E}$ effect by directly measuring the shift of the Larmor frequency, i.e.\ the resulting phase-shift of the Ramsey interference pattern, as a function of the neutron time-of-flight \cite{Piegsa:2013vda}. For pulsed beams the velocity is then known for each neutron, and hence it is possible to distinguish between the velocity-independent frequency shift caused by an EDM, and the velocity-dependent false signal due to the relativistic $\bm{v} \times \bm{E}$ effect:
\begin{equation}
  \Delta \omega = \frac{4 d_{\text{n}} E }{\hbar} + \frac{2  \gamma_{\text{n}} v E \sin \alpha }{c^2 },
\end{equation}
with $v=|\bm{v}|$ the neutron velocity, $\alpha$ the misalignment angle between the electric and magnetic fields, $\gamma_{\text{n}}$ the gyromagnetic ratio of the neutron, and $c$ the speed of light in vacuum.

A scheme of the Beam EDM concept is presented in Figure~\ref{ANNIfig:beamedmscheme}.
The experiment employs two polarised neutron beams passing through a stack of electrodes and a time-of-flight Ramsey apparatus. The ``two beam method" allows to compensate for common-noise and drifts of the magnetic field, but remains sensitive to magnetic field gradient changes. With an optimised full-scale setup of length $L=50$~m (using the aforementioned upgrade area of the ANNI beamline) and an electric field $E=100$~kV/cm, it was estimated that a (one standard-deviation) statistical sensitivity of approximately $ 5 \times 10^{-26}~e\text{ cm}$ can be reached in one day of data taking  \cite{Chanel:2018zga}.
This renders this proposal complementary to approaches using UCN, and for a 100-day dataset could even be competitive with their current state-of-the-art statistical sensitivities.

A conceptually similar approach using very cold neutrons (VCN) is briefly mentioned in Section~\ref{sec:UCN_VCN_science}; in that scenario any gain in the statistical sensitivity (per detected neutron) would arise purely from longer observation times. This idea would require further study to clarify whether or not it offers any statistical or systematic advantages, in comparison to EDM experiments using cold neutrons or UCN.





\subsubsection{EDM with \emph{in-situ} UCN production}\label{ANNIsubsubsec:edmn}

\begin{figure}
    \centering
    \includegraphics[width=\textwidth]{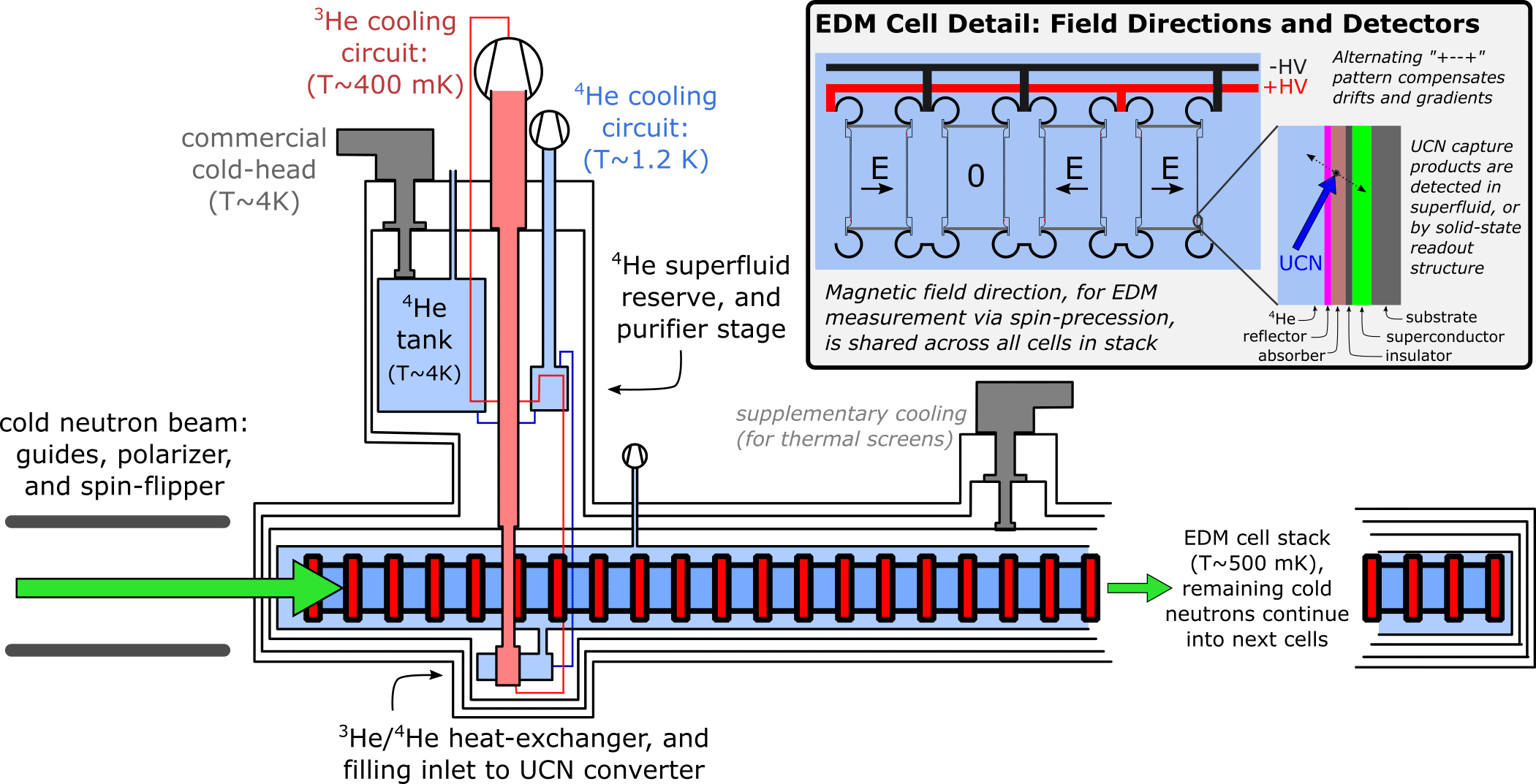}
    \caption{Schematic illustration of the EDM$^n$ experiment concept. Polarised UCNs are directly produced from cold neutrons (the production process does not alter the neutron spin), in many cells filled with isotopically-pure superfluid \isotope[4]{He}. The EDM measurement proceeds \emph{in-situ}, including both spin-precession and detection. UCN detectors in the cell walls are turned on by applying magnetic fields that partially cancel the wall potential for the high-field-seeking spin state, after a spin-precession cycle has finished. Alternative approaches could exploit absorption on $^3$He impurities, as described in reference \cite{nEDM:2019qgk}. Several readout mechanisms can be envisioned.}
    \label{ANNIfig:edmn}
\end{figure}

In recent decades neutron EDM experiments have relied on stored ultracold neutrons (UCNs), and demonstrated superb control of systematic errors despite severe statistical limitations. The advantage of UCNs mainly comes from long observation times, as discussed in Section~\ref{ANNIsubsec:EDM} above. A secondary advantage concerns practical requirements for well-controlled experimental conditions: these apply only in and around the restricted volume where UCNs are stored during a measurement. However, the low densities that can be delivered from UCN sources to external experiments represent a severe statistical limitation: the time-averaged detection rate in UCN-based EDM measurements is typically not more than 100 particles per second, whereas beam-based measurements with cold neutrons may exceed $10^8$ counts per second \cite{Chanel:2018zga}.

Significant improvements for experiments using stored UCNs will rely, to a large extent, on successfully confronting the challenge of low UCN densities. Systematic errors are also impacted, via the running time required for performing adequate supplementary studies to constrain these effects. This concern becomes more relevant with improved precision, as ever-smaller and more subtle systematic errors must be treated.

Cross-sections for converting cold neutrons to UCNs are intrinsically small, and most neutrons are lost, either by exiting the source or via other interactions. Moreover the produced in-source UCN densities suffer further reductions, typically two orders of magnitude, in extraction and delivery to experiments. Superthermal UCN sources, in which dissipative processes involving inelastic scattering are exploited to populate the phase-space more densely, provide the main avenue towards higher UCN densities. (The UCN densities produced from standard moderators are rather low.) Due to its vanishing neutron-absorption cross-section, superfluid \isotope[4]{He} is the only superthermal converter medium compatible with the long \emph{in-situ} storage times needed for sensitive EDM experiments.

The EDM$^n$ project, illustrated in Figure~\ref{ANNIfig:edmn}, explores the potential for much more sensitive neutron EDM searches \cite{in-situ-EDM}, by scalably exploiting (1) the entire moderated neutron beam available for UCN production, and (2) the full \emph{in-situ} densities already existing in UCN sources (see also Section~\ref{sec:UCNedm}). There are several further incidental advantages of the \emph{in-situ} liquid helium concept, most notably the high electric field strengths achievable in that medium. Additionally, since UCN production in superfluid $^4$He is driven by conversion of 8.9~\AA{} neutrons, the signal-to-background ratio can be increased substantially through use of a monochromatic beam -- which can be produced without loss at a pulsed source using choppers. The first point will be addressed via modular liquid-helium based sources, using small experimental cells in configurations that can be efficiently replicated many times along a neutron beam. Neutrons that pass through the first cells without interacting can still downscatter to become UCN in later cells, until the beam is depleted by other scattering processes in the helium and cell walls. Testing and development will focus on single cells, to be jointly optimised for both cold-neutron transmission and UCN storage.

\begin{figure}
    \centering
    \includegraphics[width=\textwidth]{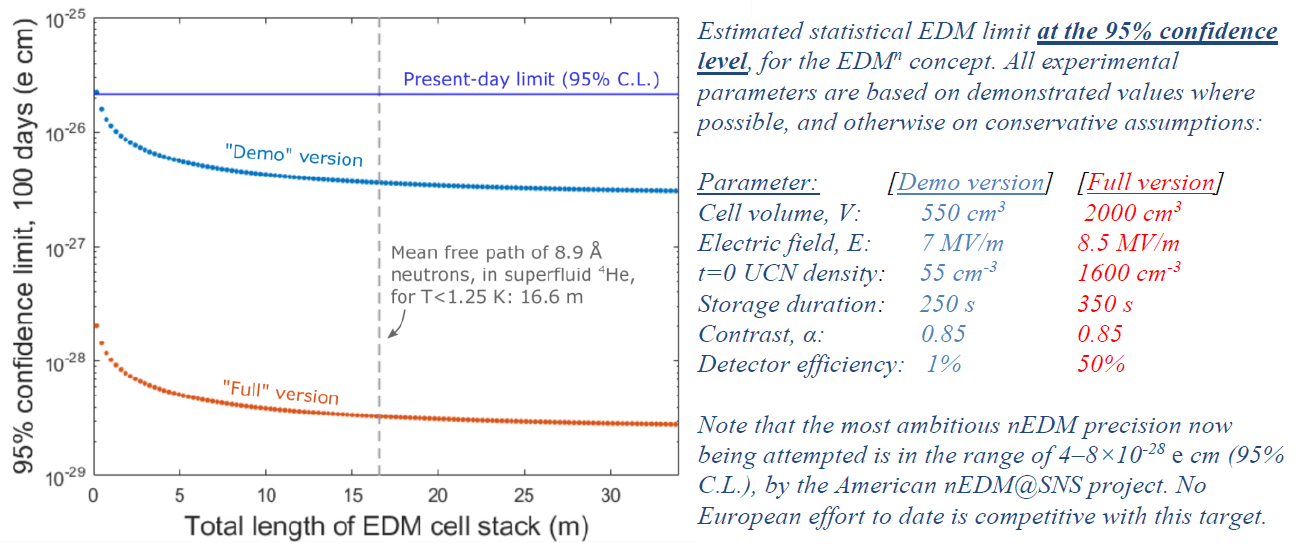}
    \caption{Statistical reach of the EDM$^n$ concept. Each data point in the left-hand plot corresponds to adding one additional cell pair (i.e., both electric field states) to the linear stack illustrated in Figure~\ref{ANNIfig:edmn}. These calculations assume that the cold neutron beam can be guided within the cell stack; the impact of beam divergence in a non-guiding scenario is discussed in \cite{in-situ-EDM}.}
    \label{ANNIfig:edmn_reach}
\end{figure}

This approach enables thorough development of the most suitable cells, independent from the challenges of scaling up to a multi-cell experiment. The multi-cell concept will also be pursued in a phased approach, with the first few-cell demonstrations already expected to be competitive in the context of today's present-generation neutron EDM experiments (see Figure~\ref{ANNIfig:edmn_reach}).

The second point will be addressed with innovative neutron detectors, based on cancelling neutron-optical potentials with magnetic fields (see Figure~\ref{ANNIfig:edmn} inset). The requirement is polarisation-sensitive UCN detection inside the cells, without negative impacts on the cells' storage properties (or unmanageable EDM systematics). Potentially low detection efficiencies are mitigated by the possibility for UCN to interact multiple times with the detector, until absorption occurs.

The developments for (1) and (2) will be combined, for single-cell demonstrations of UCN production, storage, and in-situ detection. This proof-of-principle will enable pursuing a full-scale neutron EDM experiment via a phased approach at the ESS. Such an experiment's improvements in sensitivity, while initially modest, could reach nearly three orders of magnitude beyond the present experimental limit on the neutron EDM. The only other concept so far targeting this sensitivity level would require a dedicated spallation source, where a very high UCN production rate offsets extraction inefficiencies \cite{Leung:2019pqd}.

As discussed in Section~\ref{sec:UCNedm}, the main challenge in EDM measurements is finally constraining systematic errors. The technological elements required to achieve systematic uncertainties on this level have not yet been fully demonstrated, although recent developments in magnetic shielding and magnetometry provide a foundation for focused work in this direction \cite{in-situ-EDM}. These must be pursued in parallel with the development of improved statistical reach, which as noted above is also required for studying systematic effects.


\subsection{Neutron electric charge}
\label{sec:ANNIneutroncharge}

The electric charge of the neutron is known experimentally to be zero to an extremely high precision \cite{Baumann:1988ue,Bressi:2011yfa} (see also \cite{Unnikrishnan_2004} for a review on electrical neutrality of atoms, neutrons and bulk matter).
Nevertheless, the questions of charge quantisation and the neutrality of neutrons, neutrinos, and atoms remain under debate \cite{Okun:1983vw,Holdom:1985ag,Babu:1989tq,Babu:1989ex,Foot:1992ui,Nowakowski:1992ff,Ignatiev:1993sw,Davidson:2000hf,Arvanitaki:2007gj,Cabarcas:2013hwa,VanLoi:2020kdk,Palcu:2021jst}. Moreover, there exists the possibility that the charge of a free particle might be slightly different in magnitude compared to its charge when bound in an atom \cite{Shull1967}.
Even a tiny electric charge of the neutron would forbid neutron-antineutron oscillations (see Section~\ref{sec:theorynnbar}), due to the conservation of electric charge.
Hence, measurement of the neutron electric charge could provide an important input for fundamental particle theory, as it also represents a value which is not constrained to be zero in the SM \cite{Dubbers:2011ns}.  

The first direct measurement of the neutron electric charge using a beam of free neutrons dates back to the 1950's, when Shapiro and Estulin performed a pioneering precision experiment \cite{Shapiro1956}. They employed a well collimated thermal neutron beam to search for a deflection due to a transversely applied electric field. In such an experiment, a neutron with a hypothetical non-zero electric charge $Q$ would follow a parabolic path resulting in a deflection
\begin{equation}
   y = \frac{Q E L^2}{2 m_n v^2}
\end{equation}
with $E$ the electric field applied over the length $L$, $m_n$ the neutron mass, and $v$ the neutron velocity. 
The shift is determined by measuring the neutron count rate as a function of the lateral position of a narrow aperture in front of a detector for the two cases $E=0$ and $E \neq 0$. 
In 1967 a similar experiment was performed, however, using a double-crystal spectrometer sensitive to changes of the angular orientation of a neutron beam \cite{Shull1967}.
In the 1980's Baumann, G\"ahler, Kalus and Mampe performed two experiments using a single- and a multi-slit aperture approach, respectively \cite{Gahler:1982xi,Baumann:1988ue}.
They were able to obtain a several tens of micrometer wide neutron beam using neutron optical lenses and eventually achieved the current best result for the neutron electric charge of  $(-0.4 \pm 1.1) \times 10^{-21}~e$. Improvements of this limit by 1-2 orders of magnitude have been proposed using spin interferometry with cold neutrons \cite{Voronin201325}, Ramsey spectroscopy of gravitationally bound quantum states of UCNs \cite{Durstberger-Rennhofer:2011ghz} (see also Section~\ref{UCNsubsec:GravitationalSpectroscopy}), and a deflection method with UCNs \cite{Plonka-Spehr:2009hjv,Siemensen:2018cjm}.

A novel method using a Talbot-Lau neutron grating interferometer has been proposed recently \cite{Piegsa:2018ios}.
The approach plans to employ neutron absorption gratings on silicon or quartz wafers with grating periods on the order of a few  micrometers, which are commonly used in neutron phase contrast radiography and neutron dark-field imaging in various physics and industry applications \cite{Grunzweig2006,Strobl2008,Grunzweig2013,Betz2016,Valsecchi2019}.
A symmetric Talbot-Lau setup consists of three identical absorption gratings placed at equal distances to each other and  allows to detect tiny deflections of the neutron beam \cite{Lau1948}. The first grating forms several coherent line sources. These beams are then diffracted by the second grating which produces an overlapping interference pattern in the plane of the third grating. This last grating is placed directly in front of the neutron detector and serves to analyse the microscopic pattern without the need of a high spatial-resolution detector.
Since the method relies on diffraction of the neutrons on the gratings, there exists a strong wavelength dependency which can wash out the interference pattern if a continuous white beam is employed. 
However, by combining the high-intensity pulsed neutron beam of the ANNI beamline at the ESS with a time-of-flight Talbot-Lau setup it is possible to obtain and measure interference patterns for several distinct wavelengths.
Overall, this substantially increases the sensitivity to deflection.
The electric charge of the neutron is then measured by fixing the position of the gratings such that the incident intensity on the detector is most sensitive to beam deflections, i.e.\ at the steepest slope of the interference pattern. By inverting the polarity of the applied electric field, one then searches for changes in the measured neutron count rate which would indicate a non-zero electric charge.

Another important key feature of the proposed setup is to use two separate neutron beams which are exposed to electric fields of opposite polarity -- similar to the Beam EDM experiment. This allows for a better systematic control by compensating for common noise and drifts in the individual beams. 
It was estimated that in a full-size experiment with a length $L=5$~m and an electric field $E=100$~kV/cm a sensitivity on the order $10^{-23}~e$ can be achieved within 100 days of measurement time \cite{Piegsa:2018ios}.





\subsection{Hadronic weak interaction}
\label{sec:ANNIHWI}

Due to the complex QCD structure of baryons, the Hadronic Weak Interaction (HWI) is one of the final frontiers of electroweak phenomenology.  It is customarily characterised by either 6 parity-odd meson exchange coupling constants in the Desplanques-Donoghue-Holstein (DDH) model~\cite{Desplanques:1979hn,Gardner:2022dwi} or the more modern approach of five low-energy effective field theory (EFT) couplings~\cite{Danilov:1965hyj,Zhu:2004vw,Ramsey-Musolf:2006vfz,Viviani:2014zha}.  
First attempts to compute these couplings on the lattice are made in~\cite{Wasem:2011tp}. In the limit of large number of colours $N_c$, there is a hierarchy of the expected relative magnitudes which can be tested experimentally~\cite{Phillips:2014kna,Schindler:2015nga,Gardner:2017xyl}.

The challenging experimental program of mapping out these coefficients began shortly after postulation of parity violation (PV) in 1957~\cite{Lee:1956qn} with first observation of PV in the circular polarisation of a gamma transition in \isotope[181]{Ta}~\cite{Tanner:1957zz}. PV effects in HWI are usually small but may be amplified by a factor of up to $10^6$ (in \isotope[139]{La}~\cite{Bowman:1989ci}) by the interference of closely spaced S and P resonances.  These data have confirmed statistical models~\cite{Mitchell:1999zz,Bowman:1993nu}, but the results are not directly interpretable in terms of the underlying couplings due to the complicated nuclear wave function structure.  Thus measurements in the last four decades have focused on few-body observables with exactly calculable wave functions~\cite{Adelberger:1985ik}.  Former results include the longitudinal single-spin asymmetry from polarised proton scattering from hydrogen $A^{pp}_p$~\cite{Balzer:1984qdv,Nagle:1978vn}, deuterium $A^{pd}_p$~\cite{Nagle:1978vn}, and helium~\cite{Lang:1985jv}. Measurements of $n+p\to d+\gamma$ (both gamma asymmetry with respect to the neutron spin $A^{np}_\gamma$~\cite{Cavaignac:1977uk} and gamma circular polarisation~\cite{Knyazkov:1984lzj}), $n+d\to t+\gamma$ (gamma asymmetry $A^{nd}_\gamma$)~\cite{Avenier:1984is}, and neutron spin rotation in \isotope[4]{He} ($\phi^\text{He-4}_n$)~\cite{Snow:2011zza} had uncertainties that were still too large to identify the HWI contributions to the observables.

Two recent high-precision hadronic PV inaugural experiments on the SNS FnPB of $n+p\to d+\gamma$ ($A^{np}_\gamma$) and $n+^3\mathrm{He}\to p+t$ ($A^{n\text{He-3}}_p$) have ushered a new era of precision HWI characterisation.  They offer results of 2.1~$\sigma$ \cite{NPDGamma:2018vhh} and 1.6~$\sigma$ \cite{n3He:2020zwd}, respectively, from zero, and complete the set of experiments needed to give a first experimental determination of the couplings introduced above. A three times improvement in each of these experiments at ANNI would have significant impact in constraining parameters of the HWI when taken with $A_p^{pp}$ and $A_p^{pd}$.  The next-generation measurement of $\phi_n^\text{He-4}$ is being planned at the NIST NG-C beamline~\cite{Snow:2011zza,Snow:2012vi}.  The addition of a new counting-mode measurement of $A_\gamma^{nd}$ at ANNI would over-constrain the theory, allowing for an extraction of all five pionless couplings and test the EFT formalism.  In addition, first results in a new experiment to measure spin rotation in hydrogen ($\phi_n^\text{p}$) \cite{Snow:2012vi} would help to isolate the $I=2$ coupling, which may be calculated on the lattice in the near future.  The estimated effect $\phi_n^\text{p} \approx 9 \times 10^{-7}$ rad/m, is large enough to be observed at an intense slow neutron beam such as ANNI. The scientific motivation to pursue this experiment is therefore very high. 

\subsubsection{Improvement of the SNS experiments NPDGamma and n3He}\label{ANNIsubsubsec:npdgnhe3}

The NPDGamma and n3He measurements at the SNS were both statistics limited, with a figure of merit $P^2N$, where $P$ is the polarisation and $N$ the total number of neutrons captured on target.  They both have similar requirements of a high flux neutron beamline~\cite{Fomin:2014hja}, with low background gamma rates from captures outside the target.  High polarisation helps the figure of merit, but $P\approx 100\%$ is not required to minimise the related systematic correction at the limited relative uncertainty of the $\mathcal{O}(10^{-8})$ asymmetries that can be achieved.  The beamline requires a thin neutron monitor to correct for fluctuations in the neutron flux. A polarised $^3$He spin filter was used for the preliminary measurement of NPDGamma at LANL, but it was determined that a higher figure of merit and more stable neutron polarisation could be achieved with a supermirror bender polariser.  A $\sim$1~mT holding field is needed to maintain the neutron polarisation and define the spin direction.  A relative gradient of less than $1\times 10^{-4}$~cm$^{-1}$ is required to prevent Stern-Gerlach steering~\cite{NPDGamma:2018vhh,n3He:2020zwd} and allow for efficient spin flipping.  The direction of the magnetic field must be known to $0.1^\circ$ with respect to the detector to prevent the mixing of the parity-allowed nuclear spin asymmetry associated with $\bm{k}_n \cdot \left(\bm{\sigma}_n \times \bm{k}_{\gamma,p}\right)$.  A resonant neutron spin rotator~\cite{Seo:2007aa} has been used to alternate the spin direction on a pulse-by-pulse basis to reduce systematics associated with instrumental drift.  These experiments use a novel double cos-theta RF coil with fringeless transverse fields that operates for both longitudinally and transversely polarised neutrons~\cite{Hayes2016}.

The NPDGamma experiment employs a liquid hydrogen target operated at 16~K with catalysts to thermalise into the lower energy parahydrogen state~\cite{Grammer:2014kli}, with a greatly reduced scattering cross section, and in which the spin-flip transitions are energetically forbidden, reducing beam depolarisation in the target.  A key element in the design of such a target is the balance between thin neutron windows for gamma background, and safety of the cryogenic target against explosion~\cite{Snow:2011zza}.  A new target vessel would need to be designed and fabricated for the higher flux ANNI beamline -- the old vessel was cut up to be used as a target for background asymmetry measurements~\cite{NPDGamma:2018vhh}.  The 48-segment CsI-Tl scintillator array~\cite{n3He:2020zwd} with $3\pi$ angular coverage detector used at LANL and SNS could also be used in a future measurement at the ESS.  Vacuum photodiodes were used to reduce systematic uncertainty from dependence of detection efficiency on the magnetic field from the spin rotator.

The n3He experiment uses a combined $^3$He target and ion chamber to measure the direction of the protons.  A grid of vertical wires can measure the proton asymmetry in one transverse direction as a function of depth into the target.  It was designed to measure both the longitudinal (PV) and transverse (both PV and parity conserving) asymmetries, but was run only in transverse mode, which has a figure of merit two times larger than for longitudinal mode.  This detector was run in current mode, with separate preamplifiers and 24-bit digitisers for each of the 144 sense wires. This same apparatus would be suitable for a higher-statistics measurement at the ESS.

The sensitivities of both NPDGamma and n3He were entirely determined by statistics. Thus the higher polarised neutron flux at ANNI (the simulated gain in event rate for NPDGamma is included in Table~\ref{ANNItab:GainFactors}) would allow to raise the significance of the observed asymmetries above the $5\sigma$ level with about 6 months of data per target (compare Table~\ref{ANNItab:summary1} of Section~\ref{ANNIsubsec:SummaryTables}).

\subsubsection{New experiments: NDTGamma, spin rotation}

The NDTGamma experiment presents a special challenge with the neutron capture cross section 642 times smaller than hydrogen, so that a similar design would be completely dominated by background radiation.  This was a major concern of a previous measurement of $A^{nd}_\gamma$ at the ILL~\cite{Avenier:1984is}.  In addition, the spin-1 ground state of deuterium allows for spin-flip scattering.  However, the parity odd amplitude is expected to be similar to NPDGamma, giving a 25 times larger asymmetry, which would result in similar experimental sensitivity to the PV amplitudes.  Given the lower capture rate, and the advent of new high precision digital spectroscopy~\cite{Jezghani:2020tbu}, this presents the opportunity to perform the first hadronic PV measurement of a few-body observable in counting mode with waveform digitisers and digital pulse processing.  With spectroscopic separation of backgrounds, D$_2$O could be used as the target material, which is nonflammable, and would not require a pressure vessel.
The 6.2~MeV capture gamma from deuterium is well above the gamma energy from capture of other background materials and oxygen in the target, which would only contribute 20\% of the event rate.  Preliminary measurements at LANL have shown that the average neutron polarisation at capture in D$_2$O is about 50\%.

The beamline design is similar to that of NPDGamma, but with a \isotope[3]He spin filter if the gamma background from ANNI's polarising bender should be too high in spite of its location far upstream in the guide (see Figure~\ref{ANNIfig:scheme}). The D$_2$O target would be cooled to reduce spin-flip scattering.  The entire neutron path through the detector needs to be shielded with $^6$Li to reduce the background capture gamma rate to a level comparable to that from deuterons, not to limit the total count rate.  In a highly segmented scintillator array such as the WASA detector~\cite{CELSIUSWASA:2002hcu}, the count rate would be $\sim$500~kHz in each channel.  In 6 months of beamtime on ANNI, the experiment would collect $7\times 10^{15}$ gammas, resulting in $\delta A'=1/P\sqrt{N}G=4\times 10^{-8}$ given the total neutron polarisation $P=0.6$ and geometry factor ($\cos\theta$ coverage and detection efficiency) $G=0.5$. This is 0.15 times the range of asymmetries from the extremities of the DDH reasonable range of $(-0.62\mbox{~to~+}2.1) \times 10^{-7}$ \cite{Adelberger:1985ik} .

Neutron spin rotation experiments are also well-suited for the ANNI beamline. The apparatus being developed to measure $\phi_n^\text{He-4}$ at the NIST NG-C beamline~\cite{Snow:2011zza,Snow:2012vi,Snow2015} would also be well-suited to measure $\phi_n^\text{p}$ at ANNI where the pulse structure allows separating the velocity-independent physics effect from velocity-dependent systematics such as Stern-Gerlach steering. The apparatus includes two room temperature cylindrical magnetic shields, and a superconducting, magnetic shield around the nuclear target to suppress the spin rotation due to magnetic fields.  The target is divided into two segments separated by a spin flipper, with the liquid pumped back and forth between the upstream and downstream chambers to isolate nuclear PV from magnetic effects.  In addition, the beam is segmented into two channels with the target upstream in one half and downstream in the other to suppress noise due to fluctuations in the beam intensity.  Special spin transport coils guide the spins of the polarised neutrons up to the target and then abruptly terminate the magnetic field so that neutron spins drift freely in the target chamber.  A second spin transport coil rotates the neutron spins 90$^\circ$ CW or CCW to extract the small rotation angle with a supermirror analyser followed by a 4-quadrant \isotope[3]{He} ionisation chamber.

The sensitivity of this apparatus to $\phi_n^\text{p}$ is about one third as large as that of $\phi_n^\text{He-4}$ because the optimised mean free path in hydrogen is about one third that of helium. With the flux at ANNI it is about $1.5 \times 10^{-7}$~rad/m per month, to be compared with the ``best-value'' expectation $9.1 \times 10^{-7}$~rad/m \cite{Gardner:2017xyl}.
The most technically demanding aspect of this experiment would be to engineer a cryogenic chamber which could non-magnetically transport the hydrogen between chambers and satisfy the hydrogen safety demands of the facility. The nonmagnetic bellows pump under development for the liquid helium experiment is a positive displacement pump which would work as well for liquid hydrogen in principle as it does not exploit any special properties of liquid helium.  Cryogenic chambers have been constructed and used in the past at Indiana University for the NPDGamma experiment~\cite{Blyth:2018aon} and at Thomas Jefferson National Accelerator Facility, for example, for the Qweak PV electron scattering experiment~\cite{Qweak:2014xey}, and for the PRAD experiment~\cite{Pierce:2021vkh} to measure elastic electron scattering at small angles.  These demonstrate the feasibility of running a spin rotation experiment in hydrogen at ANNI. 

\subsubsection{Hadronic TRIV measurements}

The same amplification of PV effects of up to $10^6$ in \isotope[139]{La} applies also to time reversal invariance violating (TRIV) effects, in the total absorption cross section, which is equal to the imaginary component of the forward scattering amplitude by the optical theorem:
\begin{equation}
    f(0)= A + B \bm{\sigma}_n\cdot\bm{I} + C\bm{\sigma}_n\cdot\bm{k}_n + D \bm{\sigma}_n\cdot\left(\bm{k}_n \times \bm{I}\right).
\end{equation}
The last term, which is both parity and time-reversal odd, can be extracted by passing an unpolarised beam through a \isotope[3]{He} target and a nuclear target with transverse polarisation at 90$^\circ$ with respect to each other separated by a neutron spin rotator.  Explicit time reversal symmetry without altering any of the misalignment angles can be performed by rotating the platform holding the two targets and spin flipper, but not the upstream and downstream collimators~\cite{Bowman:2014fca}.  A modular apparatus has been constructed at Los Alamos National Laboratory by the NOPTREX collaboration to perform research and development for this experiment~\cite{Schaper:2020akw}.  The collaboration is carrying out an experimental search for nuclear targets with large nuclear amplification.  Promising nuclear resonances in \isotope[139]{La} ($\Delta\sigma/\sigma=0.097$ at 0.734 eV), \isotope[131]{Xe} (0.042 at 2.2 eV), \isotope[81]{Br} (0.019 at 0.88 eV), \isotope[115]{In} (0.014 at 7 eV), and \isotope[117]{Sn} (0.008 at 1.3 eV) were measured at LANL by the TRIPLE collaboration~\cite{Mitchell:2001mvw,Mitchell:1999zz}.  These resonances are at epithermal neutron energies and would not be suitable for the ANNI beamline, but would require a thermal or hot beamline, ideally at a short-pulse station (Section~\ref{short_neutrons}) where these resonances may be resolved by neutron time-of-flight.

\label{ANNIsubsec:AnniLast}

\subsection{ANNI summary tables}\label{ANNIsubsec:SummaryTables}

Tables \ref{ANNItab:summary1} and \ref{ANNItab:summary2} summarise status, sensitivity, addressed physics and requirements with respect to the ESS for the experiments discussed in the previous sections.

\begin{table}
\begin{center}
\makebox[\linewidth]{\scriptsize
\begin{tabular}{ | c | c | c |c|c|c|}
      \hline
      \thead{Proposed\\experiment} & \thead{Measurement} & \thead{Quantity}  & \thead{Last measured} & \thead{Current value / limit} & \thead{\makecell{Statistical\\uncertainty\\ ($\bm 1\sigma$) \\ @ANNI \\ $[\text{100 days}]$}} \\
      \hline
      &\makecell{$n \to p+e+\bar{\nu}_e$}&&&&\\
      ep/n &  $A$  & \makecell{Beta asymmetry} & \makecell{\textsc{Perkeo\,III}@PF1B 
      \cite{Markisch:2018ndu}} & $-0.11985 \pm 0.00017 \pm 0.00012$  & $1\times10^{-5}$\\
      
      ep/n &  $C$  & \makecell{Proton asymmetry} & \makecell{\textsc{Perkeo\,II}@PF1B 
      \cite{Schumann:2007hz}} & $-0.2377 \pm 0.0010 \pm 0.0024$  & $1\times10^{-4}$\\
      
      ep/n &  $a$  & \makecell{$e$-$\bar\nu_e$ correlation\\ from $p$ recoil spectrum} & \makecell{$a$SPECT@PF1B 
      \cite{Beck:2019xye}} & $-0.10430 \pm 0.00084$  & $1\times 10^{-4}$\\
      
      ep/n & $b$ & \makecell{Fierz interference\\ from beta asymmetry} & \makecell{\textsc{Perkeo\,III}@PF1B 
      \cite{Saul:2019qnp}} & $0.017 \pm 0.020 \pm 0.003$  & $6\times10^{-4}$\\
      
      CRES & $b$ & \makecell{Fierz interference\\ from beta spectrum}  & \makecell{UCNA@UCN-LANL\\ 
      \cite{Hickerson:2017fzz}} & $0.067\pm0.005{}^{+0.090}_{-0.061}$  & $1\times10^{-4}$\\

      BRAND &  $a$  & \makecell{$e$-$\bar\nu_e$ correlation \\ from $e$-$p$ correlation} & \makecell{$a$CORN@NG-C 
      \cite{Hassan:2020hrj}} & $-0.10758\pm 0.00136\pm 0.00148$  & $5\times 10^{-5}$\\
      
      BRAND &  $B$  & \makecell{Neutrino asymmetry} & \makecell{\textsc{Perkeo\,II}@PF1B 
      \cite{Schumann:2007qe}} & $0.9802 \pm 0.0034 \pm 0.0036$  & $5\times 10^{-5}$\\

      BRAND &  $D$  & \makecell{Triple correlation $D$} & \makecell{emiT@NG-6 
      \cite{Chupp:2012ta}} & $(-0.94 \pm 1.89 \pm 0.97)\times 10^{-4}$  & $5\times 10^{-5}$\\
      
      BRAND &  $R$  & \makecell{Triple correlation $R$} & \makecell{nTRV@FUNSPIN \cite{Kozela:2011mc}} & $(4 \pm 12 \pm 5)\times 10^{-3}$  & $1\times 10^{-3}$\\
      
      BRAND &  $N$  & \makecell{$\sigma_n$-$\sigma_{e,\perp}$ Correlation} & \makecell{nTRV@FUNSPIN \cite{Kozela:2011mc}} & $0.067 \pm 0.011 \pm 0.004$  & $1\times 10^{-3}$\\
      
      BRAND & $H$, $L$, $S$, $U$, $V$ & \makecell{Other correlations \\with $\sigma_{e,\perp}$} & unmeasured  & unmeasured & $1\times 10^{-3}$  \\

           \hline


      Beam EDM &  $d_n$  & \makecell{Neutron electric \\ dipole moment} & \makecell{@UCN-PSI 
      \cite{Abel:2020gbr}} & $(0.0 \pm 1.1 \pm 0.2)\times 10^{-26} e\,\text{cm}$  & $5\times 10^{-27} e\,\text{cm}$ \\
      
      EDM$^n$  & $d_n$ & " & " & "  & $1.5\times 10^{-29} e\,\text{cm}$\\

      \hline
      QNeutron &  $q_n$  & \makecell{Neutron \\ electric charge} & \makecell{@ILL 
      \cite{Baumann:1988ue}} & $(-0.4\pm 1.1)\times 10^{-21} e$  & $10^{-23} e$ \\
      \hline
      NPDGamma &  $A_\gamma \left(n(p,d)\gamma\right)$  & \makecell{Gamma asymmetry} & \makecell{NPDGamma@FnPB  \cite{NPDGamma:2018vhh}} & $(-3.0 \pm 1.4 \pm 0.2)\times 10^{-8}$  & $7\times 10^{-9}$ \\
      
      n3He &  $A_p\left(n(\isotope[3]{He},t)p\right)$  & \makecell{Proton asymmetry}  & \makecell{n3He@FnPB  \cite{n3He:2020zwd}} & $(1.55 \pm 0.97 \pm 0.24)\times 10^{-8}$  & $4\times 10^{-9}$ \\

      NDTGamma &  $A_\gamma \left( n(d,t)\gamma\right)$  & \makecell{Gamma asymmetry} & unmeasured & unmeasured & $5\times 10^{-8}$ \\
      

      $\phi^\text{p}_n$ &$\phi^\text{p}_n$&\makecell{Spin rotation in\\ para-hydrogen}  & unmeasured & unmeasured & $8\times 10^{-8}$~rad/m \\

      \hline
    
\end{tabular}}
\end{center}
\caption{Summary table of observables at ANNI: Currently most precise measurement (measured by experiment@facility) and expected sensitivity of the proposed experiment at ANNI in 100 days at 5~MW ESS power. NG-6 is a cold beam line at NIST and UCN-LANL and UCN-PSI are the UCN sources at LANL and PSI, respectively. If available, current values are given with statistical and systematic error.}
\label{ANNItab:summary1}
\end{table}

\begin{table}
{\scriptsize
\begin{center}
\makebox[\linewidth]{
\begin{tabular}{ | c | c | c |c|c|c|}
      \hline
      \thead{Experiments} & \thead{New physics} & \thead{Relevant BSM}  & \thead{Addresses} & \thead{ESS edge}& \thead{Requirements}\\
      \hline
      \makecell{ep/n, \\BRAND, \\CRES} & \makecell{CKM non-unitarity, \\ Scalar, tensor couplings,\\CP violation} & \makecell{4$^\text{th}$ quark generation, \\Leptoquarks, \\ New SU(2) gauge bosons, \\New interactions} & \makecell{Fundamental symmetries,\\Baryogenesis, Dark matter} & \makecell{Neutron flux, \\Pulse structure}&\makecell{Neutron polarimetry,\\ Beam stability,\\ Background \&\\ Environment}\\
      \hline
      \makecell{Beam EDM, \\EDM$^n$} & \makecell{CP violation coupled \\to strong sector} & \makecell{CP-violating Higgs couplings, \\High-scale SUSY, \\ Baryogenesis models, \\Axion-like dark matter} &\makecell{Baryon asymmetry, \\Baryogenesis,\\Strong CP problem} & \makecell{Neutron flux,\\ Pulse structure \& \\ Neutron flight \\length (Beam EDM)}&\makecell{Background \&\\ Environment}\\
      \hline
      QNeutron & \makecell{$(\mathcal{B}-\mathcal{L})$ violation} & \makecell{Lepto-quarks \\in GUTs,\\$(\mathcal{B}-\mathcal{L})$-violating models} & \makecell{Charge quantisation,\\Neutrality of matter} & \makecell{Neutron flux, \\ Pulse structure} & \makecell{Environment}\\
      \hline
      \makecell{NPDGamma, \\n3He,\\ NDTGamma, \\ 
      $\phi^\text{p}_n$} & \makecell{nonperturbative QCD\\ quark-quark correlations} & \makecell{} & \makecell{EFTs for hadronic \\(weak) interaction} & \makecell{Neutron flux,\\ Pulse structure}& \makecell{Beam stability, \\
      Environment}\\
      
      \hline

\end{tabular}
}
\end{center}
\caption{Summary table of observables at ANNI: addressed physics, ESS edge and Requirements. ``Neutron polarimetry'' includes in particular the availability of polarised $^3$He at the ESS. ``Background'' means low background at the site of the experiment. ```Environment'' comprises, for example, stable temperature and low and stable magnetic fields.}
\label{ANNItab:summary2}
} 

\end{table}

\clearpage
\section{The HIBEAM/NNBAR program}
\label{sec:nnbar} 
\subsection{Introduction}

 Of all the empirically observed conservation laws, the conservation of baryon number is among the most fragile. Baryon number violation (BNV) is required to understand the dynamic generation of the observed matter-antimatter symmetry~\cite{Sakharov:1967dj}. It is perhaps most appropriate to discuss not {\it whether} baryon number is violated but {\it in which channels} baryon number is violated. The HIBEAM/NNBAR program\footnote{The acronym HIBEAM stands for {\bf H}igh {\bf I}ntensity {\bf B}aryon {\bf E}xtraction And {\bf M}easurement.}~\cite{Addazi:2020nlz,Backman:2022szk} takes advantage of the unique opportunity provided by the world's brightest neutron source, the ESS, for a two-stage (HIBEAM and then NNBAR) and long-term series of experiments of increasing sensitivity to search for BNV via the conversion of neutrons to antineutrons and/or sterile neutrons. The program encompasses $\Delta B=1,2$ processes and will achieve an improvement in the sensitivity for discovery of neutron conversions by three orders of magnitude compared with previous searches~\cite{Bressi:1989zd,Bressi:1990zx,Fidecaro:1985cm,BaldoCeolin:1994jz}.  
  
  Like the photon and neutrino, which may potentially mix with axion-like particles~\cite{Holdom:1985ag,Foot:1988aq} and sterile neutrinos~\cite{Pontecorvo:1967fh}, respectively, neutron mixing is an ideal portal to physics beyond the SM.  It is a quasi-stable and electrically neutral particle that can be copiously produced. The observation of neutron mixing would address open questions in modern physics. As described further in Section~\ref{sec:theorynnbar}, BNV-violating neutron conversions into antineutrons and/or sterile neutrons are generic probes of new physics, address the matter-antimatter asymmetry and could point to the existence of a hidden sector of particles, thus shedding light on dark matter\cite{Nussinov:2001rb,Dvali:1999gf,Barbier:2004ez,Dutta:2005af,Calibbi:2016ukt,Berezhiani:2015uya,Dev:2015uca,Allahverdi:2017edd,Dvali:2009ne,Berezhiani:2003xm,Berezhiani:2005ek}.   It should also be added that unlike the photon and neutrino, there have been comparatively few high sensitivity searches with neutrons, owing to infrastructure requirements, as would be satisfied at the ESS.
 
 The HIBEAM stage of the program emphasises searches for sterile neutrino processes, induced by non-zero magnetic field values, and includes a search for neutrons to antineutrons via sterile neutron states. The neutron-antineutron conversion search at HIBEAM is based on an assumption that sterile neutrons exist in a hidden sector, which would be affected by a sterile magnetic field.  The NNBAR stage is a dedicated search for free neutrons to antineutrons. This is a more generic search for neutron-antineutron conversions and is not based on any assumptions of a hidden sector. HIBEAM also includes preparations for the NNBAR stage via a low-sensitivity prototype experiment for free neutron conversion to antineutrons.
 
 The full program probes a range of possible mixing possibilities, characterised by the off-diagonal elements of the mixing matrix shown in Equation~\ref{eq:mixingnn-full}.

\begin{equation}
    \hat{\mathcal{H}}=\left(\begin{array}{cccc}
    m_n + \vec{\mu}_n \vec{B} & \varepsilon_{n\bar{n}} & \alpha_{nn'} & \alpha_{n\bar{n}'}  \\
    \varepsilon_{n\bar{n}} & m_n -\vec{\mu}_n \vec{B} & \alpha_{n\bar{n}'}  & \alpha_{nn'}  \\
    \alpha_{nn'} & \alpha_{n\bar{n}'}  & m_{n'} +\vec{\mu}_{n'} \vec{B}' & \varepsilon_{n\bar{n}}  \\
    \alpha_{n\bar{n}'}  & \alpha_{nn'} & \varepsilon_{n\bar{n}} & m_{n'} - \vec{\mu}_{n'} \vec{B}'
    \end{array}\right)\label{eq:mixingnn-full}
\end{equation}
Here, $\varepsilon_{n\bar{n}}$ is the $n \bar{n}$ Majorana 
mass mixing parameter, and  $\alpha_{nn'}$  and $\alpha_{n\bar{n}'}$
are mass mixing parameters for $n n{'}$ and for $n \bar{n}{'}$, respectively.

When discussing the mixing matrix, it is convenient to first consider the physics of NNBAR which looks for the classical $n\rightarrow \bar{n}$ ($\Delta \mathcal{B}=-2$) signature, following the quasi-free propagation of neutrons in low magnetic fields ($\lessapprox 10$~nT). This can arise due to a mixing mass amplitude $\varepsilon$ and a probability $P_{n\rightarrow \bar{n}}=\varepsilon_{n\bar{n}}^2 t^2$. More details on NNBAR are given in Section~\ref{section:nnbar}.

The conversion of neutrons to antineutrons can also arise via the second order oscillation process: $n\rightarrow n{'}\rightarrow\bar{n}$ with amplitude ($\alpha_{nn'} \alpha_{n\bar{n}'}$) or $n \rightarrow \bar{n}{'}\rightarrow \bar{n}$ with amplitude $(\alpha_{n\bar{n}'}\alpha_{nn'})$ and the interference of all three channels. If $\varepsilon_{n\bar{n}}$ is very small and $\alpha_{nn'}$ and $\alpha_{n\bar{n}'}$ are relatively large, then $n\rightarrow \bar{n}$ could be observed for a non-zero sterile magnetic field. Previous limits on free $n\rightarrow\bar n$ oscillation from experiments with a suppressed magnetic field~\cite{BaldoCeolin:1994jz} and nuclear stability limits from $n\rightarrow\bar n$ conversion in nuclei \cite{Abe:2011ky} would not strongly constrain this scenario, since a field $\vec{B}$ is necessary to compensate the magnetic field in the sterile sector in order to permit the full process $n\rightarrow n' \rightarrow \bar{n}$. 
The probability of such a process can be written as $P_{n\bar n}(t)\simeq P_{nn'}(t) P_{n\bar n'}(t)$. Existing constraints\cite{Berezhiani:2017jkn} permit $\tau_{nn'}$ 
and $\tau_{n\bar n'}$ to be as low as $1\div 10$~s

By optimising the magnetic field $\mathbf{B{}}$ to be resonantly close to $\mathbf{B{'}}$ (with a precision of several mG), the  quasi-free regime is thus reached and the probability of 
induced $n\rightarrow\bar n$ oscillation can become as large as 
\begin{equation}
\label{eq:indnnbarosc}
 P_{n\bar{n}}(t)=
 \frac14 \alpha_{n\bar{n}'}^2\alpha_{n\bar{n}'}^2t^4 \sin^2\beta
 = \frac{\sin^2\beta}{4} \left(\frac{t}{0.1~{\rm s}}\right)^4 
 \left(\frac{10^2\,{\rm s}^2}{\tau_{nn'}\tau_{n\bar n'}}\right)^2 \times 10^{-8} .
 \end{equation}  
Here, $\beta$ is the (unknown) angle between the directions of $\mathbf{B{}}$ and $\mathbf{B{'}}$ \cite{Berezhiani:2020vbe}. The probability of induced $n\rightarrow\bar n$ transition can therefore be several orders of magnitude larger than the present sensitivity in direct $n\rightarrow\bar n$ conversion~\cite{BaldoCeolin:1994jz}. As explained in Section~\ref{ANNIsubsec:hibeam} the HIBEAM part of the program looks for neutrons ''disappearing", regenerating and converting to antineutrons. 


\subsection{HIBEAM}\label{ANNIsubsec:hibeam}

\begin{figure}[!htb]
  \setlength{\unitlength}{1mm}
  \begin{center}
  \includegraphics[width=0.9\linewidth, angle=0]{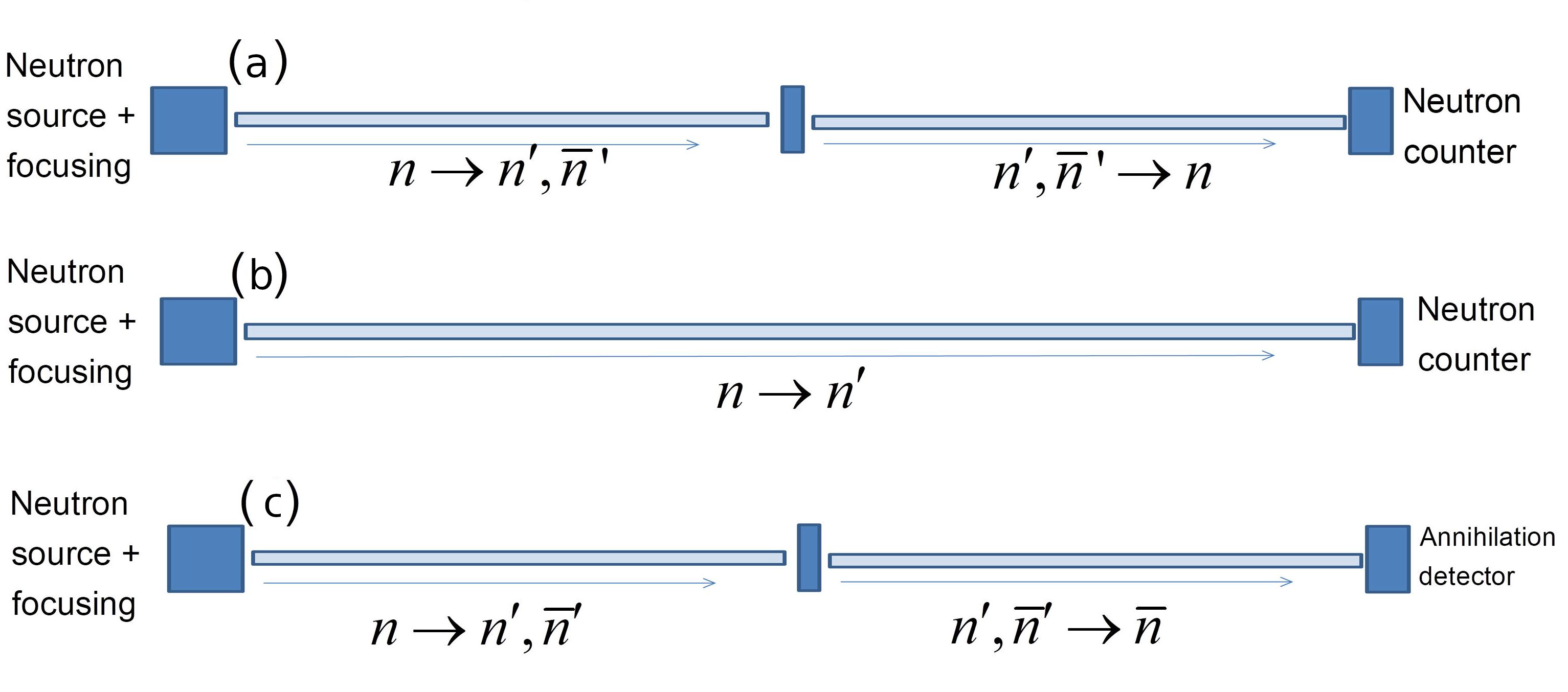}
  \end{center}
  \caption{\footnotesize Illustration of the principles of searches for sterile neutron oscillation with neutron beams. Both regeneration (a) ($n\rightarrow n'$) and disappearance (b)  ($n\rightarrow\{\bar{n},n'\}\rightarrow n$) search modes are possible. Another possibility (c) is regenerative in style, though instead leads to further mixing to an antineutron ($n\rightarrow\{\bar{n},n'\}\rightarrow \bar{n}$), requiring an annihilation detector. Each regeneration mode requires a neutron absorber to be placed at the halfway point of the beamline, preventing all ordinary neutrons from proceeding downbeam while permitting sterile species to pass unencumbered.}
  \label{fig:nnprime-cartoon}
\end{figure}

At the most basic level, HIBEAM will address transformations of $n\rightarrow \bar{n}$ and $n\rightarrow n'$. However, should a sterile neutron sector exist, processes connecting the visible and sterile sectors need not be restricted to the above processes, as proposed in Refs.~\cite{Berezhiani:2018zvs,Berezhiani:2020vbe}. As a neutron beam experiment, HIBEAM can look for the regeneration of neutrons following a beam stop (Figure~\ref{fig:nnprime-cartoon} (a)), an unexplained disappearance of neutron flux (Figure~\ref{fig:nnprime-cartoon} (b)), as well as $n\rightarrow  \bar{n}$ via a sterile neutron state (Figure~\ref{fig:nnprime-cartoon} (c)). These searches are described in further detail in the following sections. 

\subsubsection{Search for $n \rightarrow n{'}$ via disappearance}\label{sec:nnprime} In HIBEAM's $n \rightarrow \{n{'},\bar{n}'\}$ disappearance search,  neutrons would move through a 50 m long aluminum vacuum tube, with neutron rates at the start and the end of the propagation measured precisely. The presence of an unknown sterile magnetic field $\mathbf{B{'}}$ is taken into account by performing a magnetic field scan such that the sterile and visible magnetic fields are matched and a $n \rightarrow n{'}$ transition is less suppressed. The detection of a resonance would manifest itself as a reduction in the total counting rate. 


Charge-integrating counters are needed which are able to measure a charge proportional to the neutron $n$ flux with an relative accuracy up to the order $10^{-7}$. 
One possibility is the use of a \isotope[3]{He} detector ~\cite{doi:10.1063/1.4919412} in charge-integration mode.

\subsubsection{Search for the regenerative $n \rightarrow n{'},\bar{n}' \rightarrow n$ process}\label{sec:nnprimen}
The regeneration search derives from a similar theoretical basis as the disappearance search 
\cite{Berezhiani:2017azg,Broussard:2017yev}, though corresponding to a two-stage (second order) process with a consequently quadratically smaller probability. In the first stage, the $n \rightarrow n{'},\bar{n}'$ transformation takes place in an intense cold $n$ beam at the quasi-free environment limit corresponding to $|\mathbf{B}-\mathbf{B{'}}|\sim 0$. The largely untransitioned $n$ beam will be blocked by a high suppression beam trap, while the remaining sterile neutron will continue unabated through the absorber. In a second volume behind the absorber (stage two), maintaining the the same conditions of $|\mathbf{B}-\mathbf{B{'}}|\sim 0$ as the section before it, the $n{'},\bar{n}' \rightarrow n$ transformation produces detectable neutrons with momentum conserved, as though the totally-absorbing wall were not present. 


The observation of the resonance in the $\mathbf{B}$-scan would be defined by a sudden appearance of regenerated $n$'s when the $\mathbf{B}-\mathbf{B{'}} \approx 0$ condition in both volumes is met. Like with disappearance, a positive signal would be a demonstration of the $n{'},\bar{n}'\rightarrow n$ transformation as well as the existence of the sterile neutrons and photons. The requirement of matching conditions in both volumes ensures this type of measurement is significantly more robust to systematic uncertainties that could otherwise cause a false signal to be observed. 


\subsubsection{Search for $n\rightarrow \bar{n}$  by regeneration through mirror states} \label{sec:nnprimebarn}

The $n\rightarrow \bar{n}$ process can also arise due to the second order oscillation processes: $n\rightarrow n{'},\bar{n}'\rightarrow\bar{n}$.
The earlier body of  searches for free neutrons converting to antineutrons~\cite{Bressi:1989zd,Bressi:1990zx,Fidecaro:1985cm,BaldoCeolin:1994jz} were insensitive to this scenario. 




\subsubsection{Detectors for sterile neutron searches}
Discussion of antineutron detection is deferred to Section~\ref{sec:nnbardet} which describes the detector for the search for free neutrons converting to antineutrons. A smaller version of the detector described there would be deployed at HIBEAM.

The other sterile neutron searches at HIBEAM rely on measurement of the visible state of the neutrons. Detection of cold and thermal neutrons requires major technical competence for the scattering experiments of the ESS, though a standard solution for neutron detection may utilise gas detectors based on \isotope[3]{He} in a single wire proportional chamber. The detectors can be operated at low gain since the $n+\isotope[3]{He}\rightarrow t+p$ reaction produces a very large ionisation signal. 

\subsubsection{HIBEAM sensitivity}
\begin{figure}[H]
  \setlength{\unitlength}{1mm}
  \begin{center}
  \includegraphics[width=0.65\linewidth, angle=0]{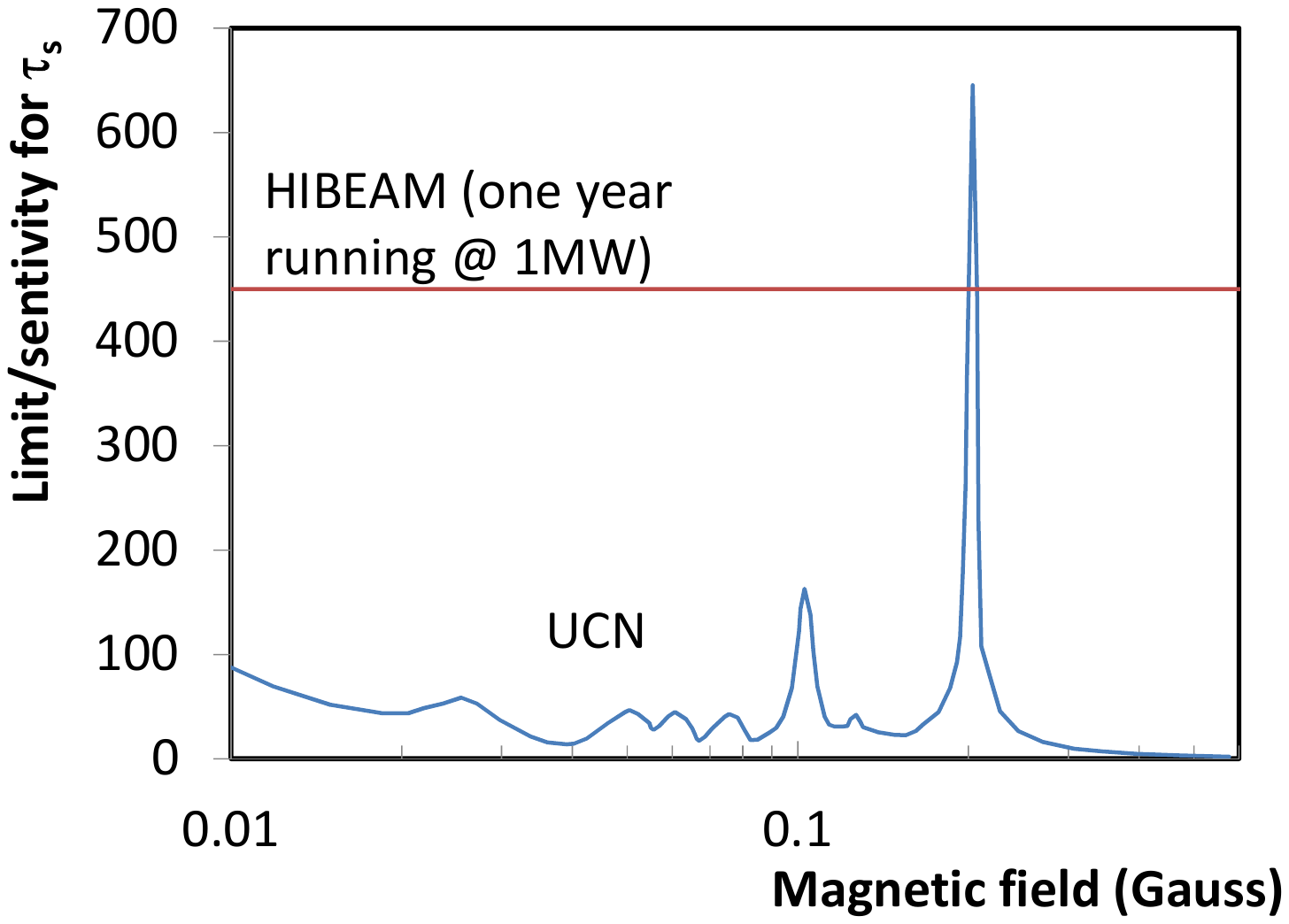}
  \end{center}
  \vspace{0.0cm}
  \caption{\footnotesize Excluded neutron oscillation times in blue for $n\rightarrow n'$ disappearance from UCN experiments~\cite{Ban:2007tp,Serebrov:2008her,Altarev:2009tg,Berezhiani:2012rq,Berezhiani:2017jkn,nEDM:2020ekj} as a function of the magnetic field $\mathbf{B{'}}$. The projected sensitivity  for HIBEAM (disappearance mode) is also shown in magenta for one year's running at the ESS assuming a power of 1MW.}
  \label{fig:nnprime-global}
\end{figure}

Figure~\ref{fig:nnprime-global} shows the current limits from trapped ultracold neutron experiments together with the expected sensitivity of the HIBEAM experiment at a beamline of the ANNI design \cite{Theroine2015,Soldner:2018ycf} (see Section~\ref{sec:ANNI}) after one year of ESS running at 1MW power (taken from Ref.~\cite{Addazi:2020nlz}).
Increases in disappearance sensitivity of more than an order of magnitude are possible depending on the value of the magnetic field used. It can be seen that HIBEAM covers a wide range of oscillation times for a given magnetic field value, many of which are unexplored by UCN-based experiments, and remain free of the model assumptions of those searches.  As a second order process, the regeneration search sensitivity is lower though can approach that of the disappearance mode for appropriate running time and a design power of 5~MW.~\cite{Addazi:2020nlz}. As an expected zero-background experiment, the search for neutrons converting to antineutrons via sterile neutrons, the sensitivity of this mode on $(\tau_{nn'}\tau_{n\bar{n'}})^{\frac{1}{2}}$ can exceed $1000$~s. Finally, the aforementioned searches will also probe the multi-Gauss region in which constraints are extremely weak. These data are also summarised at the end of this section in Table~\ref{nnbar:s1}.


\subsection{NNBAR at the ESS - measurement methodology}\label{section:nnbar}


The NNBAR experiment seeks to measure the signal of a spontaneous conversion of a neutron to an antineutron by focusing the cold neutron beam onto a thin carbon target foil. Should one of the neutrons oscillate to an antineutron on the way to the target, a striking multipion event signature, with energy typically well above existing backgrounds, could be observed. In principle, one should be able to reconstruct the 1.88 GeV two-neutron mass, although, since the annihilation is not to happen in free space but inside a nucleus, final state interaction effects and meson absorption can significantly distort the expected invariant mass resolution, even before any detector effects are taken into account \cite{Golubeva:2018mrz}.

\begin{figure}[tb]
  \setlength{\unitlength}{1mm}
  \begin{center}
\includegraphics[width=1.00\linewidth]{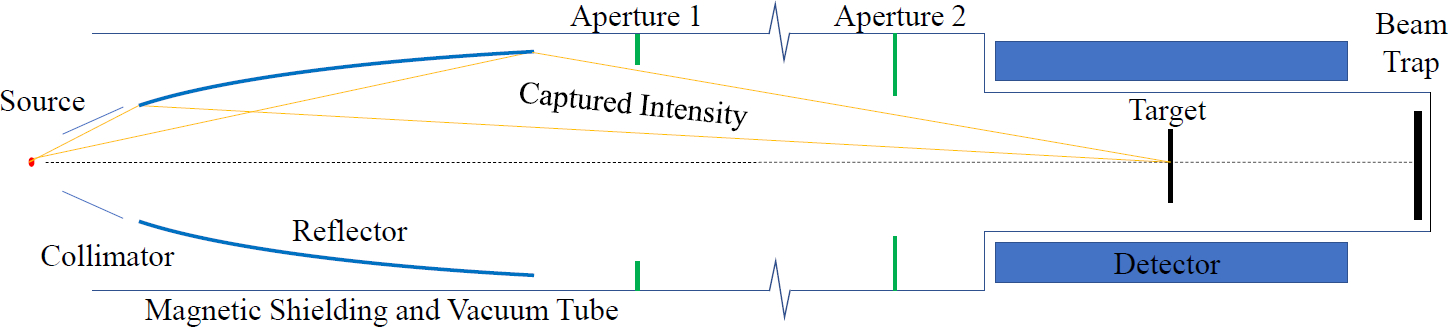}
    \end{center}
\caption{\footnotesize Baseline NNBAR experiment. Neutrons from the moderator are focused on a distant target foil surrounded by an annihilation detector.}
  \label{fig:nnbar_baseline}
\end{figure}

\begin{figure}
\begin{center}
\includegraphics[width=.666\textwidth]{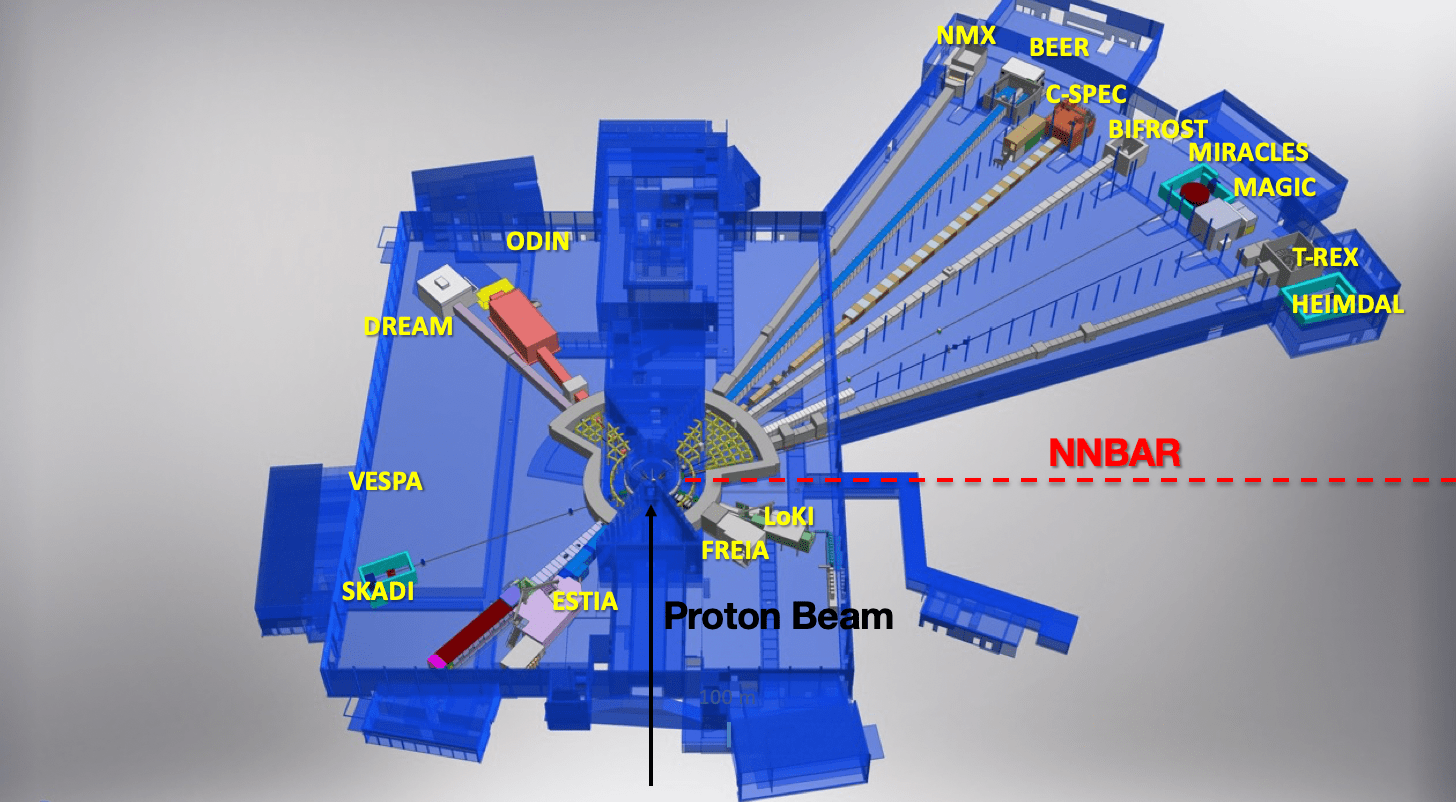}
\caption{Overview of the ESS instruments and location of the NNBAR experiment.}
\label{instruments}
\end{center}
\end{figure}

The neutrons used in the NNBAR experiment are produced by the spallation process following moderation, after which they reach typical velocities of about 1200 m/s. The beam of cold neutrons travels through a vacuum in an ultra-low magnetic field such that the quasi-free condition holds. 
The motivation for choosing an annihilation target made of carbon resides in its low cross section for neutron absorption and high cross section for annihilation. More specifically, the annihilation cross section  is proportional to $1/v$, where $v$ is the neutron velocity. For the average spectrum of neutron velocities of HIBEAM/NNBAR it is $\sim$ 4 kb ~\cite{Phillips:2014fgb} for annihilation and 4 mb for absorption i.e., a six orders of magnitude difference. The moderator-target longitudinal distance is $\sim 200 $ m for NNBAR and $\sim 50$ m for HIBEAM, with sensitivity depending quadratically on the transit time. A schematic of the experiment is shown in Figure \ref{fig:nnbar_baseline}. Neutrons passing through the foil are absorbed in a beam dump, which will possibly be made of ${}^{6}$Li. 
A second target foil can also be installed for background control purposes allowing the background estimations for NNBAR to be data$-$driven. Figure \ref{instruments} shows the future location of NNBAR within the ESS instrument suite.


\subsection{NNBAR design}
The key components of the NNBAR experiment are: 1) a cold moderator source; 2) a high precision neutron focusing system; 3) field-free propagation along a dedicated beamline and 4) detection of the annihilation signal. Each of these elements is described in the  following sections. 

\subsubsection{The NNBAR Large Beam Port } 
\begin{figure}[tb]
  \setlength{\unitlength}{1mm}
  \begin{center}
\includegraphics[width=0.75\linewidth]{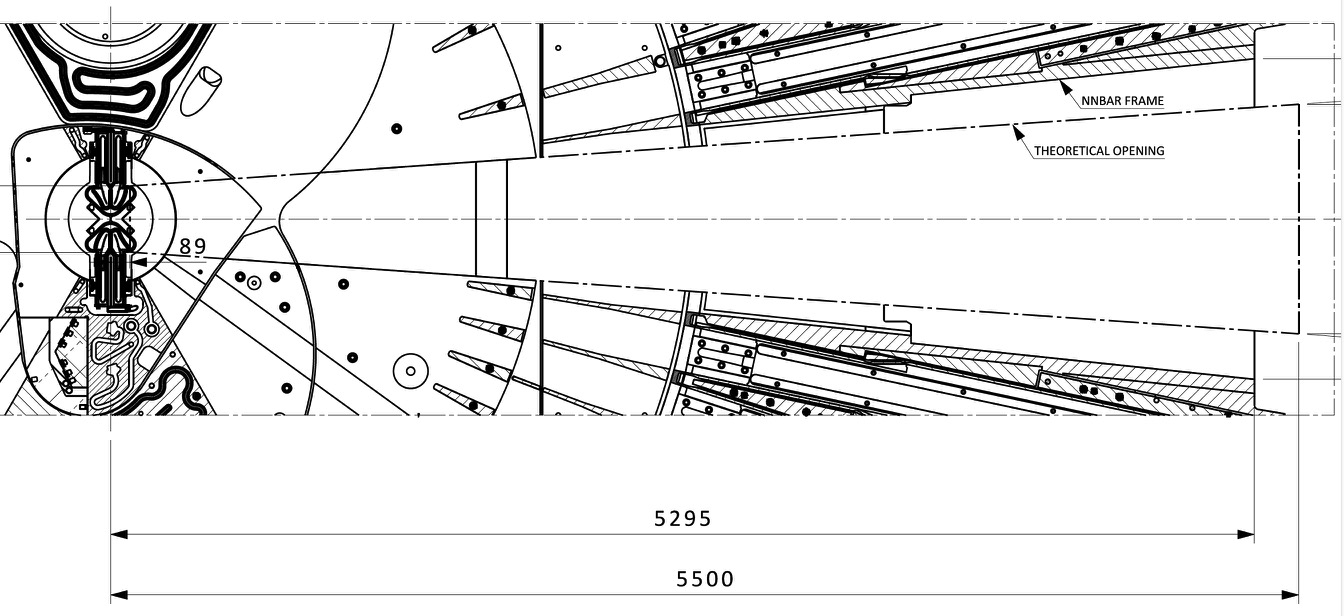}
    \end{center}
\caption{\footnotesize Top view of the NNBAR beamport. Reproduced with permission from Mats Segerup and Rickard Holmberg.}
  \label{fig:largebeamport1}
\end{figure}

\begin{figure}
\begin{center}
\includegraphics[width=.466\textwidth]{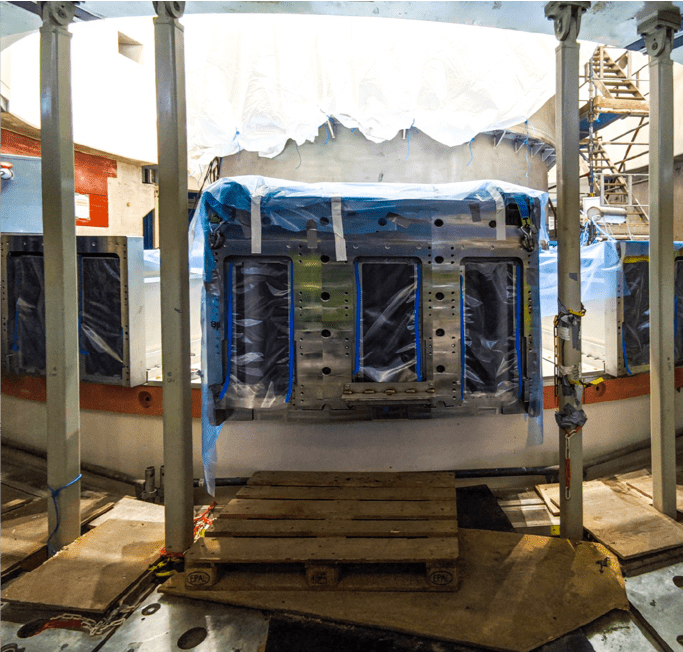}
\caption{The ESS Large Beam Port.}
\label{fig:largebeamport2}
\end{center}
\end{figure}

NNBAR will strongly profit from the large liquid deuterium moderator LD$_{2}$ described in Section~\ref{sec:lowermoderator}.
This moderator is currently under design and it will be optimized for delivering a high integrated flux to the NNBAR experiment. 
The neutrons produced by the LD$_{2}$ moderator will travel through the ESS Large Beam Port, a special beam port installed in the ESS target monolith for NNBAR to reach its goal. The Large Beam Port covers the size of three ESS standard beam ports and in the early days of ESS will be filled by three regular size beam ports. The central beam port will be used for the test beamline whose purpose is to characterize the ESS source at the beginning of the user program. 
The engineering drawing of the Large Beam Port is shown in  Figure~\ref{fig:largebeamport1} and the picture of the Large Beam Port as installed in the ESS target monolith can be seen in Figure \ref{fig:largebeamport2}. Once the Large Beam port will be opened for the operation of the NNBAR experiment it will permit the use of a large fraction of the solid angle as can be seen in Figure  \ref{fig:largebeamport1}.

With the extractions of an extremely high intensity cold beam the Large Beam Port will however transport also a very high fraction of fast neutrons and spallation background. This will require that the NNBAR beamline in the first meters close to the monolith be heavily shielded in order to be operated safely. Beamline simulations are being carried out in order to design the shielding with the required thickness~\cite{Holl:2022oif} and also to assess if any components of the ESS facility like the ESS bunker would need additional shielding due to the opening of the large beam port.  

\subsubsection{NNBAR optics}

 In order to make use of the Large Beam Port it is necessary to design special optics that allow the exploitation of such a large view of the source. The reflector needs to ensure that the large neutron flux coming from the source is directed and focused to the annihilation target $200$~m away, maximizing the NNBAR figure of merit 
  $FOM = N \cdot <t^2>$,  where $N$ is the number of free neutrons with $t$ is the transit time of a neutron in the magnetically shielded region prior to reaching the annihilation target. The flight time is measured from the last reflection that the neutron had in the optics. 
   The baseline design of the NNBAR optics is a an elliptic  guide made of reflecting material. This is  an ellipsoid with a focal distance $\SI{200}{\meter}$ and a short-axis $b$ of $\SI{2}{\meter}$. The center of the moderator is located at focal point, while the center of the detector is located at the other focal point. The reflector covers the part of the ellipse that starts $\SI{10}{\meter}$ from the source and ends at a distance of $\SI{50}{\meter}$. 
   
\begin{figure}
\begin{center}
\includegraphics[width=.666\textwidth]{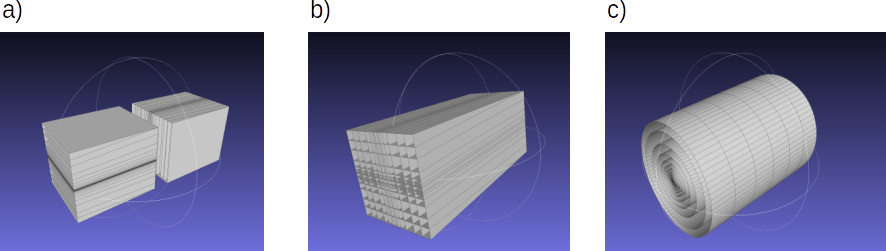}
\caption{NNBAR nested optics options a) mono planar b)double planar c) torodail (cylindrical)}
\label{opticsdesign}
\end{center}
\end{figure}

Other designs are currently under study using McStas, a neutron ray tracing code, including a differential reflector that gives a gain of 23\% compared to the baseline reflection and a "nested reflector". These reflectors are easy to assemble. Different designs are  currently being optimized within the HighNESS project and are shown in Figure \ref{opticsdesign}. So far the obtained value of the nested mirror options seems to deliver the highest $FOM$. Further studies are to be performed to allow an optimal choice of reflector for NNBAR. It is important to point it out that the reflector is being designed together with the moderator in an iterative process to enhance the sensitivity of the final experiment. 

\subsubsection{Magnetic shielding and vacuum} 

To match the quasi-free condition $\Delta E \ll  t $ \cite{Addazi:2020nlz} (where $E$ and $t$ are the energy and the propagation time of the neutron) the magnetic field should be small enough to avoid suppressing the probability of oscillations. This can be achieved by shielding it at a level below 10~nT along the 200~m neutron propagation length. This requirement can be satisfied using a two-layer mumetal shield combined with a 316L stainless steel vacuum pipe that is also part of the magnetic shielding system \cite{Backman:2022szk}. This  shielding layout with the mumetal sheets arranged in octagonal shape and clamped together  is based on the magnetic shield of the atomic fountain used for the 
Hannover Very Long Baseline Atom Interferometry facility \cite{wodey2020scalable}. A set of coils is needed around the mu-metal in order to equilibrate them.

\begin{figure}
\begin{center}
\includegraphics[width=.666\textwidth]{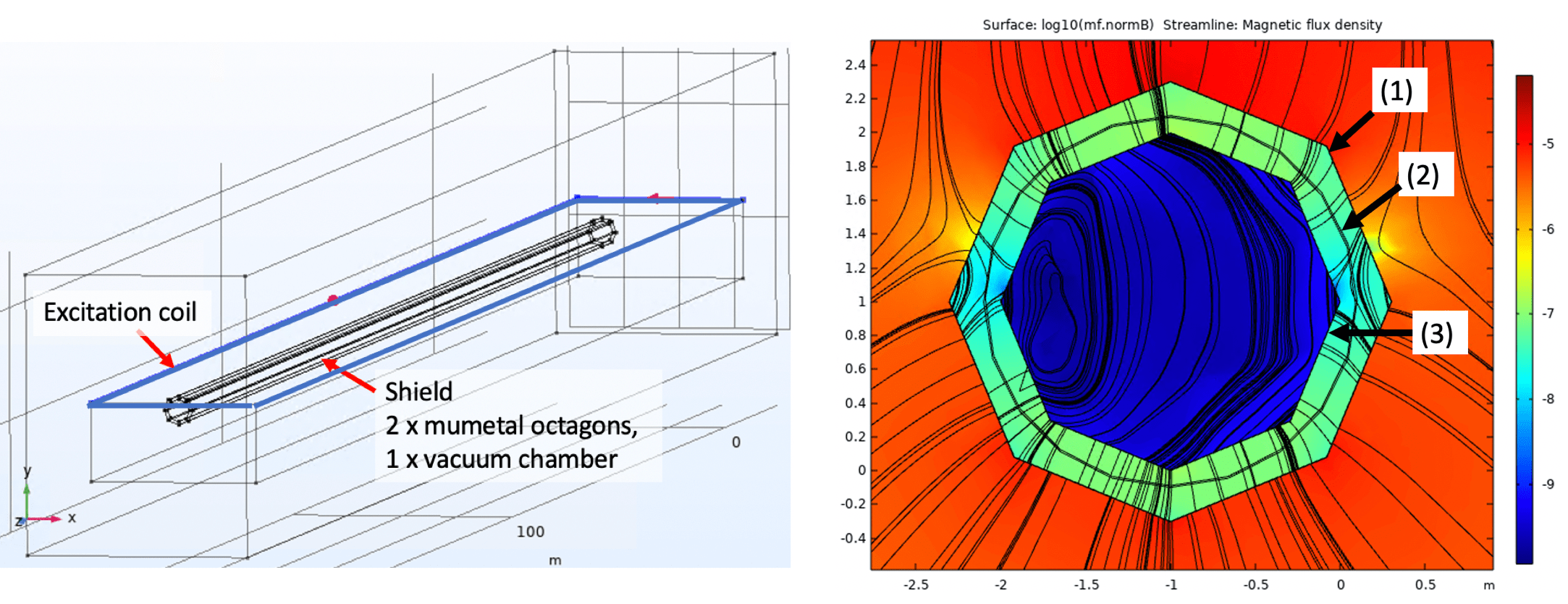}
\caption{Overview of the NNBAR magnetics shield: left the 200m magnetic shielded vacuum pipe. Right: cut through the shield from COMSOL simulations. (1) Outer mumetal layer (2) vacuum pipe (3) inner mumetal layer }
\label{magneticshield}
\end{center}
\end{figure}

Simulations performed with COMSOL \cite{multiphysics1998introduction}
a finite element calculation tool has shown that this design can satisfy the 10~nT requirement. The layout of the shielding is shown in Figure \ref{magneticshield}.

\subsubsection{Preliminary detector design}\label{sec:nnbardet} 
The NNBAR detector model~\cite{Barrow:2021deh, Yiu:2022faw,Backman:2022szk} has been implemented in \software{GEANT4}, and a range of different technologies are being considered including lead-glass, Crystal (CsI(Na)) and sampling calorimeters (liquid argon/steel). A Time projection chamber (TPC) (outer tracking) and a layer of silicon (inner tracking) are also included. The following main elements are part of the current baseline detector, starting from the center, radially outwards: a) the aforementioned annihilation target, a $100\,\mu$m thick carbon disk, with $1\,$m diameter for the initial HIBEAM stage, and $2\,$m diameter for the full NNBAR second stage; b) a charged particle tracker, necessary for the tracking and identification of pions and the annihilation vertex. The inner tracking is accomplished with a silicon layer placed within a $2\,$cm thick aluminium beampipe. The TPC surrounds the beampipe and provides particle identification through measurements of the specific continuous energy loss, $\frac{dE}{dx}$; c) a hadronic range detector for charged pion measurement composed of 10 inter-orthogonal scintillator slats and, for photon measurement (neutral pion reconstruction),  an electromagnetic calorimeter comprising lead-glass modules. Surrouding the detector d), a scintillator-based cosmic ray background veto would be deployed. A selective trigger system to be able to gather signal and signal-like background candidate events. 
Construction of a prototype ~\cite{Dunne:2021arq} detector system comprising a TPC, scintillator range and lead-glass calorimeter is underway. 

The detector is optimised so as to allow high efficiency ($>50$\%) for signal tagging and reconstruction while offering sufficient discrimination power so as to reject backgrounds. A range of background processes are being studied including cosmic rays, gamma emission due to activated detector and other material, and interactions of fast and slow neutrons from the spallation source. The aim is for a zero-background search, as achieved at ~Ref.\cite{BaldoCeolin:1994jz}. This requires precision capabilities in timing (provided by the scintillator system) and particle identification (eg via continuous ionisation energy loss in the TPC $\frac{dE}{dx}$ and $\pi^0$-tagging in the calorimeter).

\begin{figure}[H]{
  \centering
  \subfigure[]{\includegraphics[width=0.515\textwidth]{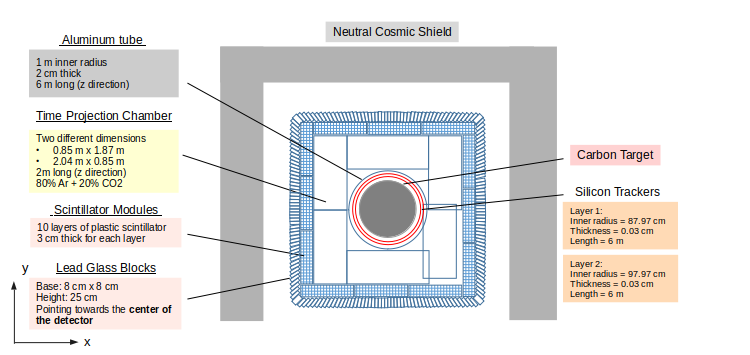}}
  \subfigure[]{\includegraphics[width=0.275\textwidth]{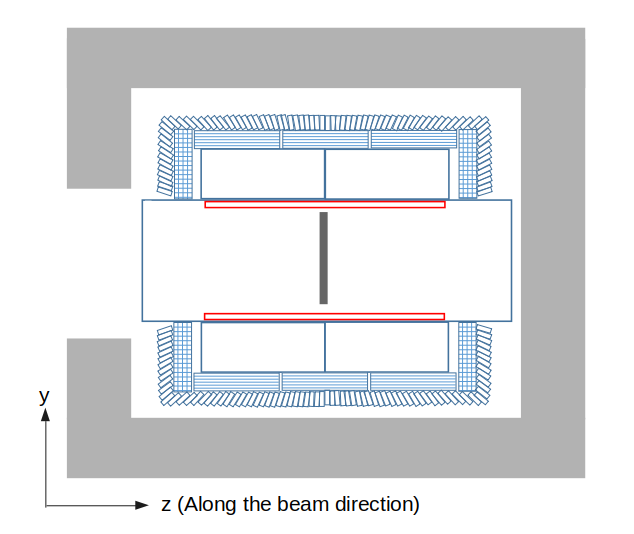}}
	\caption{Schematic overview of the NNBAR detector~design in the x-y (a) and y-z (b) views.}
	\label{box_full_det_sketch}
}
\end{figure}

Figure~\ref{box_full_det_sketch} shows an overview of the NNBAR detector. The detector is 600~cm long in the longitudinal ($z$) direction. In the transverse ($x-y$) plane, the detector has a width and height of 515~cm. The detector components and their dimensions are labelled in the figure.


\begin{figure}[H]{
  \centering
  \subfigure[ ]{\includegraphics[width=0.415\textwidth]{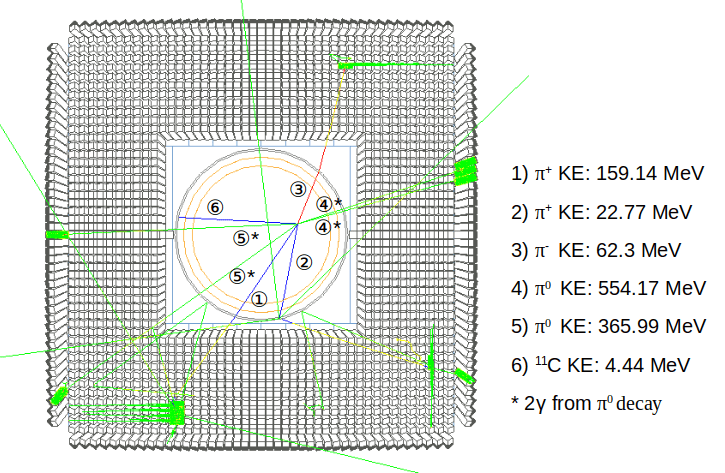}}
  \subfigure[]{\includegraphics[width=0.275\textwidth]{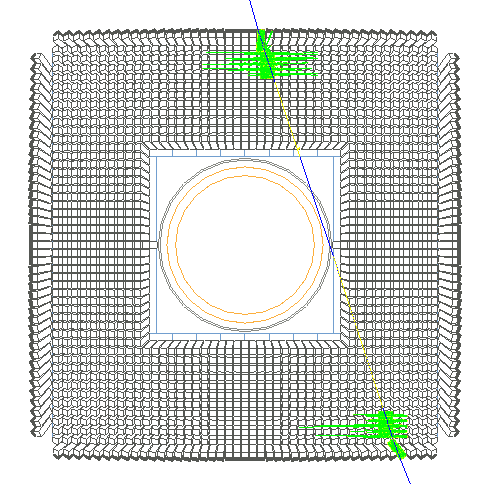}}
	\caption{Event displays with the NNBAR detector in the plane transverse to the beamline showing (\textbf{a}) a signal event with five pions; (\textbf{b}) a cosmic muon. Red (blue) color in the figure represents a negatively (positively) charged~particle.}
	\label{fig:events}
}
\end{figure}

Figure~\ref{fig:events} (a) shows a signal event with five final-state pions in the NNBAR detector with a nuclear fragment from the carbon target. Figure~\ref{fig:events} (b) depicts a cosmic ray muon traversing the NNBAR annihilation detector from top to bottom.

\subsection{Previous free neutron oscillation searches and current limits}

Previous free neutron $n\rightarrow \bar{n}$ experiments were performed at the Pavia Triga Mark II reactor~\cite{Bressi:1989zd,Bressi:1990zx} and at the Institute Lauen Langevin(ILL)~\cite{Fidecaro:1985cm,BaldoCeolin:1994jz}, the latter which~\cite{BaldoCeolin:1994jz} provides the most stringent and comprehensive limit for the free neutron oscillation time: $\sim 8.6 \times 10^7\,$s. 


\subsection{NNBAR sensitivity at the ESS}

The sensitivity of NNBAR is proportional to the number of $n\rightarrow \bar{n}$ transitions which could take place. In other words, the sensitivity of the experiment is considered in regards to the discovery reach of oscillation events. It is best quantified by the figure of merit $N_n \cdot t^2$ where $N_n$ is the free neutron flux reaching the target. 

As both neutron flux on target as well as free neutron flight time are the crucial variables influencing the figure of merit, all improvements on experimental parameters influencing either one or the other will lead to increased sensitivity. The neutron flight time is uniquely influenced by the propagation length, whereas several neutronic parameters impact $N_n$. Neutron reflector technology, in particular, is expected on its own to increase the experimental sensitivity for neutron-antineutron oscillation by a factor of around 40 when compared to the previous ILL experiment.





\subsection{The gain from using the ESS}

Table \ref{tab:gain} depicts the gain factors affecting the experiment's sensitivity. The largest gain comes from the high reflectivity ($m$) neutron mirrors. While such mirrors are now extensively available, the combination of factors makes the ESS unique in terms of pushing the limits of neutron oscillation observation. In particular, the long neutron flight path as well as a large intensity due to ESS' large beam port are unequalled at any other existing or planned neutron research facility. 

\begin{table}
\begin{center}
\begin{tabular}{| c | c |}
 \hline
 Factor & Gain wrt ILL  \\
 \hline
 Source Intensity & $\geq 2$   \\
 \hline
 Neutron Reflector & $40$  \\
\hline
 Length & $5$  \\
 \hline
 Run time & $3$ \\
 \hline
 {\bf Total gain} & $\geq 1000$ \\
 \hline
\end{tabular}
\end{center}
\caption{\footnotesize Breakdown of gain factors for NNBAR with respect to the last search for free neutron-antineutron conversions at the ILL.}
\label{tab:gain}
\end{table}

\subsection{Comparison with other future experiments}
\begin{figure}[H]
  \setlength{\unitlength}{1mm}
\centering
\hspace{-1.25cm}
  \includegraphics[width=0.9\linewidth, angle=0]{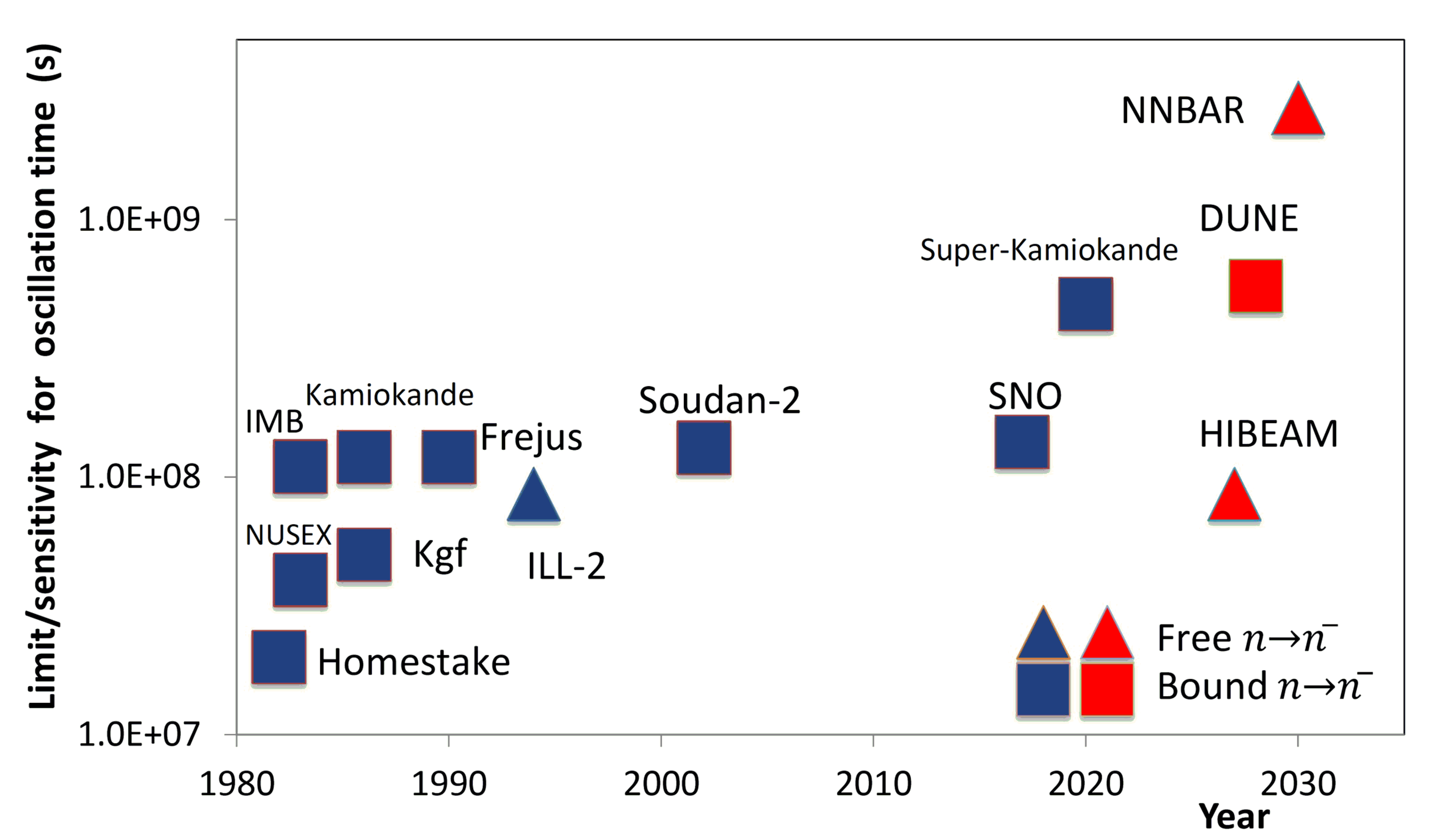}
  \caption{\footnotesize Lower limits on the free neutron oscillation time from past experiments on free and bound neutrons. Projected future sensitivities from HIBEAM and NNBAR are also shown, together with the expected sensitivity for DUNE. Searches with free neutrons include those done at the Pavia Triga Mark II reactor\cite{Bressi:1989zd,Bressi:1990zx} and the ILL~\cite{Fidecaro:1985cm,BaldoCeolin:1994jz}. The most recent and competitive result from the ILL is shown and is denoted ILL-2~\cite{BaldoCeolin:1994jz}. Limits from bound neutron searches 
   are given from Homestake~\cite{Homestake}, KGF~\cite{KGF}, NUSEX~\cite{NUSEX}, IMB~\cite{IMB}, Kamiokande~\cite{Kamiokande}, Frejus~\cite{Frejus}, Soudan-2~\cite{Soudan-2}, SNO~\cite{Aharmim:2017jna}, and Super-Kamiokande~\cite{Abe:2020ywm}. For the bound neutron experiments, various model-dependent intranuclear suppression factors are used to estimate a free neutron oscillation time lower limit.}
  \label{fig:nnbar-limits}
\end{figure}

It is instructive to compare the sensitivity of other planned $n\rightarrow \bar{n}$ experimental initiatives to NNBAR at the ESS. All other internationally competitive programs that are planned focus on searches using bound neutrons, which are complementary to searches using free neutrons. As such, NNBAR is unique, representing the most sensitive $n\rightarrow \bar{n}$ proposal using free neutrons. A comparison between sensitivities on neutron oscillations for free and bound neutron searches is difficult and model dependent \cite{Addazi:2020nlz}. In particular, such comparisons need to focus on limits that can be set on oscillation time using intra-nuclear suppression factors and not on the discovery reach itself. With that in mind, there are essentially three future experiments that include $n\rightarrow \bar{n}$ as part of their programs  a) Hyper-Kamiokande - there are no existing official estimates on the expected $n\rightarrow \bar{n}$ sensitivity from the collaboration;
b) Deep Underground Neutrino Experiment (DUNE) - the projected \textit{converted} free oscillation time lower limit for bound neutron conversions within \isotope[40]{Ar} nuclei within the future DUNE experiment is $\tau_{n\rightarrow\bar{n}}\geq5.53 \times 10^8$~s~\cite{Abi:2020evt} for an assumed exposure of 400 kt$\cdot$years ; c) NOvA - there are no estimates on future neutron oscillations time limits for NOvA. It is worth mentioning, however, that an analysis using  data from 4 months of Far Detector exposure \cite{Phan:2020urk} found the NOvA sensitivity to be below the limits set by the Super-Kamiokande experiment\cite{Super-Kamiokande:2011idx,Super-Kamiokande:2020bov}. A summary, taken from Ref.~\cite{Addazi:2020nlz} for projected future sensitivities, as well as past limits obtained by other experiments can be see in Figure \ref{fig:nnbar-limits}.



Sensitivities for the HIBEAM/NNBAR program on oscillation times are summarised in Table \ref{nnbar:s1}. Results from earlier experiments and the gain from making searches at the ESS are also shown. Note that the increase of a factor $\sim 10^{3}$ for NNBAR corresponds to the rise in the discovery potential due to the rise in the figure of merit at the ESS. The limit on the oscillation time that can be reached with respect to the ILL experiment sensitivity increases by $\sqrt{10^{3}}\sim 30$.


{\scriptsize
\begin{center}
\captionof{table}{Summary table of NNBAR and HIBEAM sensitivities. Limits from previous searches are also given.}
\makebox[\linewidth]{
\begin{tabular}{ | c | c | c |c|c|c|c|}
      \hline
      \thead{Experiment} & \thead{Measurement} & \thead{Quantity}  & \thead{Recent\\ measurements} & \thead{Current limit} & \thead{Gain (ESS)} & \thead{Experimental conditions} \\
      \hline
      NNBAR &  \makecell{$n\rightarrow\bar{n}$}  & \makecell{Free neutron oscillation \\ time ($\tau$)} & \makecell{ \cite{BaldoCeolin:1994jz}} & $\tau$ $>$ $8.7 \times10^{7}$~s  & \makecell{$\sim$ $10^{3}$ (discovery) \\  $\sim$ 30 ($\tau$)} &  \makecell{beam energy = 800 MeV; \\ power = 2 MW} \\
      \hline
     HIBEAM &  \makecell{$n\rightarrow [n',\bar{n}']$ \\ $n\rightarrow [n',\bar{n}'] \rightarrow n$ \\ $n\rightarrow [n',\bar{n}'] \rightarrow \bar{n}$}  &
     \makecell{neutron disappearance,  \\  regeneration and \\ antineutron conversion \\ } & \makecell{\cite{Serebrov:2007gw,Berezhiani:2017jkn,nEDM:2020ekj}  \\  \cite{Berezhiani:2017jkn,nEDM:2020ekj}} & 
     \makecell{ $\tau_{nn',n\bar{n'}} > ~50$~s   \\  $(\tau_{nn'}\tau_{n\bar{n}}')^{\frac{1}{2}}  > ~1000$~s \\ 
     $B$-field dpendent} & 
       \makecell{$\sim 10$ ($B$-field dependent) \\} &
     \makecell{beam energy = 800 MeV; \\ power = 2 MW}
      \\ 
   
      \hline

\end{tabular}
}
\label{nnbar:s1}
 \end{center}
  }

\subsection{An option to use the coherent reflection of antineutrons and neutrons}

The interesting option to simultaneously increase the sensitivity of NNBAR and decrease its cost is to implement the new idea of a guide both for neutrons and antineutrons \cite{Nes:2019PRL,Nes:2019PLB,Pro:2020PRD}. A typical effective critical velocity of the guide material for antineutrons is $\sim 4$~m/s. A typical lifetime of antineutrons in the guide is $\sim 2~$s; it is defined both by antineutron annihilation and dephasing between neutrons and antineutrons. With these values and a specially designed guide, one could probably deliver a large fraction of the total neutron/antineutron intensity to the annihilation detector while the systematic uncertainties associated with the interaction of antineutrons with the guide walls are quite small.

This approach has been applied in the analysis of a possible (quite short) experiment at the PF1B facility at ILL \cite{Abele:2005xd}. It was found that an increase of $3-10$ times in sensitivity over the best existing constraint is feasible \cite{Gud:2021As}. An analogous study for the NNBAR experiment is planned in the near future. Potential advantages of the new experimental design, to be verified by a detailed analysis, include a larger sensitivity in a longer setup, a lower radiation around the exit from the ESS radiation protection, a smaller detector background, a significantly lower cost due to a more compact design. An additional order of magnitude gain in sensitivity might be possible if a dedicated VCN source is built at ESS.

\clearpage
\section{UCN/VCN}\label{sec:ucn}


The energy of ultracold neutrons (UCNs) is so low ($\sim10^{-7}~\text{eV}$) that they can be reflected from many materials at all angles of incidence (different energy ranges correspond to different materials). 
UCNs can be stored in material traps \cite{Zeldovich1959}, and such low energies also make it possible to store them in magnetic traps \cite{Vladimirskii1962}. The possibility of long holding times (several hundred seconds) for UCNs in experimental setups makes these neutrons extremely sensitive to small effects. In some cases, UCNs turn out to be a more sensitive probe in comparison with thermal and cold neutrons, despite significantly lower fluxes. In fact they are the only means by which some topics, such as gravitationally bound states of the neutron, can be experimentally addressed. 
The work of F.L. Shapiro \cite{Shapiro1968} first drew attention to this possibility for precision measurements using UCNs, and the first experimental observations of UCNs \cite{Luschikov1969,Steyerl:1969csa} led to a rapid development in this area of research beginning in the 1970s \cite{ignatovich1990physics}. This in turn led to significant improvements in accuracy for measurements of the neutron lifetime \cite{PDG2020}, constraints on the size of a neutron electric dipole moment (EDM) \cite{Abel:2020gbr}, and the discovery and study of UCN quantum states above a flat mirror in a gravitational field \cite{Nes2002}. Since the first detection of UCNs, a great deal of methodological experience has been accumulated in the production, detection, spectrometry, and storage of UCNs. UCN losses on the walls of the storage volumes can be lower than a few percent of the losses associated with neutron decay. Processes involving small energy transfers were also discovered in the interaction of UCNs with material surfaces \cite{Nes1999}, which had not been taken into account in early experiments. These processes can be a source of systematic errors in experiments measuring the neutron lifetime, on the one hand, and on the other hand, they open up prospects for studying dynamics on the surfaces of condensed matter. It remains a challenge for experimentalists to obtain the theoretically predicted loss factors on weakly-absorbing materials such as solid oxygen and beryllium \cite{Alf1992}. 

Throughout this time many advances in experimental accuracy were obtained by using more intense UCN sources, while others -- especially for neutron lifetime measurements -- were driven by improvements of experimental apparatus. 
The program for the use of UCN to study the surface of condensed matter was proposed some time ago \cite{Golub1996}, but even today the intensity of existing sources is not enough to implement this program. The active development of other, established laboratory methods for studying surfaces, and the limited coverage of $(|\bm{q}|,\omega)$ for UCN, seems to indicate that UCN science is not yet competitive in this area. Today, the potential of very cold neutron (VCNs, energies $< 10^{-4}~$eV) experiments is not fully exploited. Despite the periodic discussion of the advantages of their use \cite{ANL2005,ORNL2016,ESS2022}, the development of this field is hampered both by the lack of intense sources of these neutrons (the only available source of VCN is PF2 at the ILL \cite{pf2_VCN}) and by the need to adapt experimental setups and existing techniques. The development of much more intense UCN and VCN sources could possibly enlarge the field of UCN applications, and create whole new fields of application for VCNs in particle physics -- as well as in neutron scattering. 
 
The concrete implementation of UCN and VCN sources for the ESS is presently under study. Because experiments exploiting such sources are strongly affected by specific source parameters -- especially the energy spectrum and phase-space density -- the experimental concepts at the present time can only be discussed at a schematic level. The remainder of this section surveys a range of science experiments in these fields that could potentially be pursued at the ESS, provided a competitive UCN density is produced, and comments on certain aspects of the source design that could impact their practical implementation or sensitivity.

\subsection{Implementing a UCN source at the ESS}
Delivering higher numbers of UCN into experiments remains an important possibility for improving experimental accuracy. For storage experiments in which the free neutron lifetime is not the limiting loss mechanism, such as measurements of neutron EDM, significant gains may also be made by increasing the holding time. One should also note that some experiments rely on a high transmitted flux of UCNs, rather than a high stored density. In all cases, however, present-day experiments with UCNs are strongly limited by counting statistics.

The longest-running and most reliable user-mode UCN source is PF2 at the Institut Laue-Langevin (ILL), based on the concept of a Doppler-shifting neutron turbine \cite{pf2_VCN}. This source has a total UCN current density of about $2.6\times 10^4 $~cm$^{-2}$ s$^{-1}$ up to $v_ z = 6.2$ m/s, and $3.3\times 10^4$ cm$^{-2}$ s$^{-1}$ up to $v_z =$ 7 m/s, and corresponding stored UCN densities up to about 40 neutrons per cm$^3$ (or $20~\text{cm}^{-3}$ with an aluminum safety foil installed, in the usual configuration). All  UCN sources implemented after PF2 use solid deuterium or superfluid helium as superthermal converters for cold neutrons \cite{Golub1975}. In contrast to moderation, superthermal conversion allows to accumulate phase-space density of UCN over the equilibrium value, via inelastic scattering where a Boltzmann factor suppresses those interactions which lead to an increase of the neutron energy. Both of these converter materials are well studied \cite{Kasprzak-thesis,Doge-thesis,BAKER200367,Schmidt-Wellenburg:2015ema,Zimmer2014}.


Solid deuterium provides a higher conversion rate, where the optimal spectrum for incident neutrons has an effective temperature of $\sim30\,$K, and the optimal converter temperature is $\sim5\,$K. The high UCN production rate makes this an attractive option for pulsed neutron sources, but absorption losses limit the lifetime of UCNs inside the source. 
While it would be desirable for a UCN source at the ESS to deliver UCN densities limited only by the pulsed neutron flux density, no experimentally-validated implementation of such a concept is presently available. The theoretical approaches to such a concept \cite{Shapiro1970} would require placing the UCN source very close to the spallation target, since velocity dispersion leads to pulse spreading after neutrons have left the source. 
The prospective gain factor relative to the average UCN density  is ultimately limited by the ratio of the time interval between pulses to the duration of a single pulse. For the ESS (with a pulse duration of 2.86~ms and a pulse frequency of 14~Hz) it would be $\sim25$ times, neglecting any additional losses. A further theoretical proposal building on this concept is the ``time-focusing'' of neutrons, to re-compress pulses that have undergone some spreading outside the source \cite{Frank1996}. Non-adiabatic spin-flips in a variable magnetic field were later considered as a means to implement time-focusing; this idea has been tested at some level \cite{Arimoto2012,Imajo2021}, but further developments are required for practical use. Recently, two other mechanisms have also been proposed for time-focusing of neutron pulses: non-stationary diffraction by a moving grating \cite{AFrank_TFgrid}, and a homogeneous time-varying magnetic field accompanying deceleration of very cold neutrons to UCN energies \cite{Nes2022_RT}. The practical implementation of these ideas requires further investigation of the proposed techniques.



Superfluid helium has a specific rate of UCN production almost 10 times lower than solid deuterium (assuming a neutron spectrum with effective temperature of around $30$~K), and the optimal spectrum for incident neutrons has an effective temperature of $\sim6$~K. 
At converter temperatures below $\sim0.8$~K, superfluid helium permits very long UCN lifetimes in the converter (finally limited by beta decay below $\sim0.6$~K), making it possible to accumulate UCNs over long times and obtain high ``\emph{in-situ}'' densities within the converter volume. The \emph{in-situ} UCN density $\rho_\text{source}$ will be reduced by dilution and transfer losses, for any extraction-based experiment in which the UCNs are transferred to an \emph{ex-situ} storage volume for measurement. For a perfect guide system in which dilution losses dominate, the density $\rho_\text{expt}$ of UCNs delivered into such an experiment is determined by the volumes of the source ($V_\text{source}$), guide system ($V_\text{guide}$), and experiment ($V_\text{expt}$):

\begin{equation}
    \rho_\text{expt} = \frac{V_\text{source}\rho_\text{source}}{V_\text{source} + V_\text{guide} + V_\text{expt}} .
\end{equation}
The long times required for obtaining the saturated UCN density in a helium converter make it unrealistic to exploit the time-structure of a pulsed neutron source for further gains in UCN density. The time-structure associated with neutron delivery can, however, be experimentally useful for disentangling beam-related backgrounds. 
In addition, since UCN production in liquid helium is dominated by neutrons in a narrow wavelength range around 8.9 \AA{}, the time structure allows for significant suppression of the thermal load, activation, and requirements for biological shielding. This can be achieved by using choppers to remove the other wavelengths without attenuating the time-averaged flux at 8.9 \AA{} (see also the discussion in Section~\ref{ANNIsubsubsec:edmn}). A similar effect can be achieved using monochromators at a steady state neutron source, but only with significant loss of 8.9 \AA{} neutrons. 
There are also possibilities to achieve significant gains in UCN statistics at the ESS via delivery of large-area neutron beams, or through use of novel imaging concepts \cite{zimmer2018imaging} which are not practical to implement at completed, running facilities.


The ideas for intense UCN sources proposed in relation to the WWR reactor \cite{Serebrov2015} or the PIK reactor \cite{Lych2016} cannot be implemented directly in the ESS, due to design features of the spallation source: a very compact and bright zone for neutron production, and a fast decrease of neutron density away from this zone accompanied by intense gamma-ray loads.

\begin{figure}[H]
    \centering
    \includegraphics[width=0.89\linewidth]{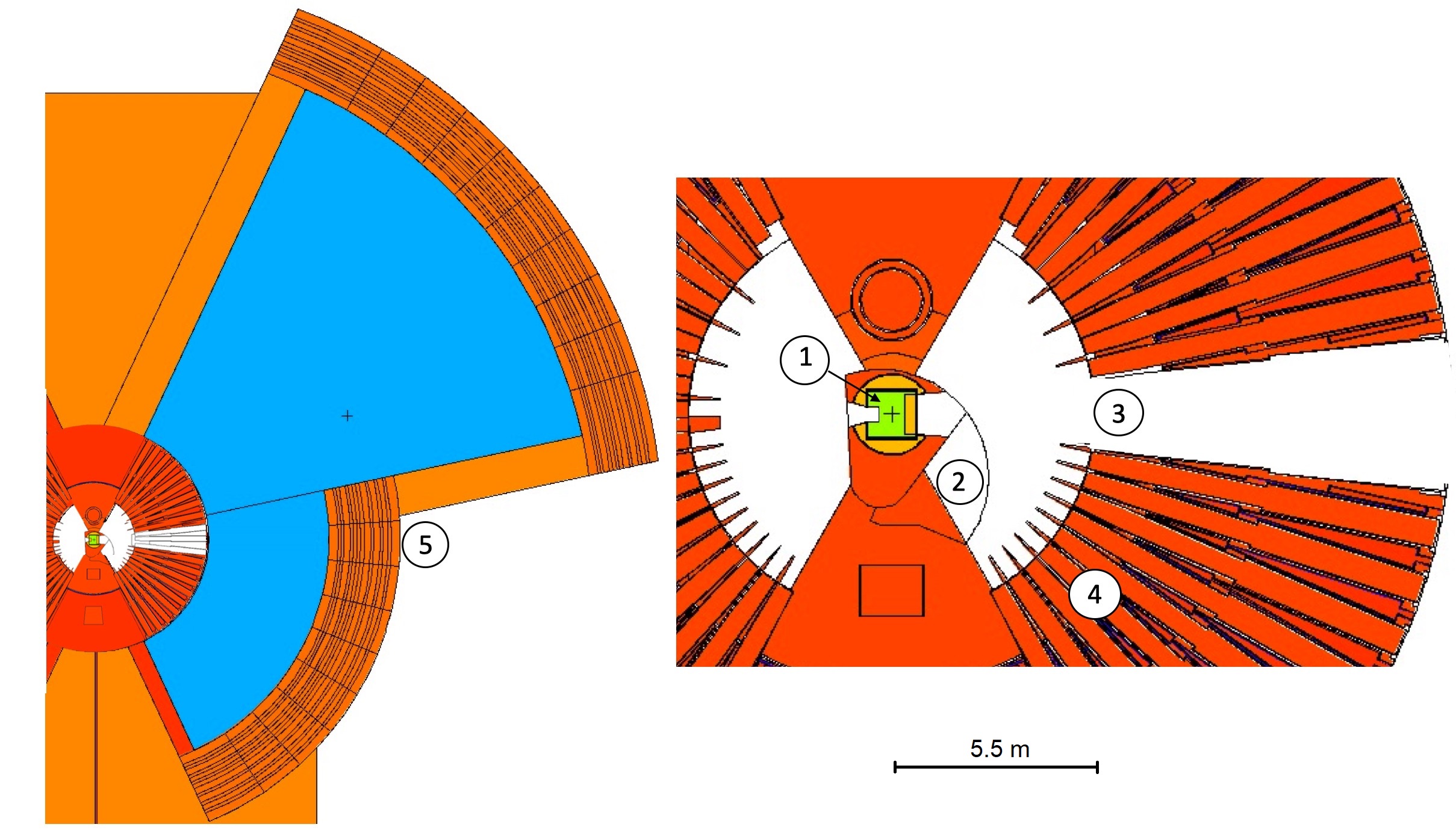}
    \caption{Horizontal cut through the target and bunker region (cf. Figure~\ref{fig:essbunker}), at the height of the liquid-deuterium moderator (shown in green) situated below the spallation target (not visible). The drawings are obtained from a detailed MCNP \cite{UCNmcnp} geometry. The cylindrical region of radius of 5.5 m around the center represents the shielding monolith (shown in red). About half of its 42 standard beamports are visible in the cut plane. Possible locations of UCN sources, as studied within the HighNESS project are: (1) inside the ``twister'', a structure which contains the upper (liquid-parahydrogen, high-brightness) and lower (liquid-deuterium, high-intensity) moderators; in this position the UCN source is closest to the spallation target and would either be placed below the lower moderator or replace it, (2) inside the moderator cooling block, (3) in the large beamport (shown as a white segment in the monolith) that is initially foreseen to be used for the NNBAR experiment, (4) in a standard beamport, (5) outside the ``bunker'', a heavy concrete shielding structure (shown in orange) placed around the monolith; the minimum distance of this location from the moderator is 15 m, the large beamport being used to image the neutron emission surface of the lower moderator onto a volume of superfluid helium for down-conversion of cold neutrons in a UCN source or \emph{in-situ} in an experiment.}
  \label{fig:essucn}
  \end{figure}

 
Various ideas and possibilities for implementing UCN sources at the ESS are presently under discussion \cite{ESS2022} and require further evaluation. Many are linked to the availability of a high-intensity lower moderator, described in Section~\ref{sec:lowermoderator}. The option of \emph{in-situ} UCN production at the ANNI facility is discussed more thoroughly in Section~\ref{ANNIsubsubsec:edmn}. 
Possible source locations close to the target are outlined in Figure~\ref{fig:essucn} \cite{zanini}.
UCN sources can be placed in several locations inside the target monolith structure (in red in Figure~\ref{fig:essucn}, right) which has a radius of 5.5 m. Location ``1'' is inside the ``twister'', 
a structure that contains the upper and lower moderators.
In this position the UCN source is closest to the spallation target, and would either be placed below the liquid deuterium moderator or replace it.
Location ``2'' is inside a shielding plug (called Moderator Cooling Block) which is originally intended merely for shielding, but can be adapted to accommodate a secondary source. This position is at a further distance from the target,  facilitating the cooling of the UCN converter, but still receives a large flux from the high-intensity moderator.  
The use of the Large Beam Port (LBP) is also very attractive due to a large solid angle viewed from the moderator and a high calculated brightness for 8.9~\AA{} neutrons of $3.4\times10^{11}~\text{n cm}^{-2}\text{s}^{-1}\text{sr}^{-1}\text{\AA{}}^{-1}$, averaged from a large emission surface of $40 \times 24~\text{cm}^2$. This brightness is comparable to that of the ILL's cold source at 8.9~\AA{}, $5\times10^{11}~\text{n cm}^{-2}\text{s}^{-1}\text{sr}^{-1}\text{\AA{}}^{-1}$ \cite{Ageron1989}.
A large He-II vessel placed in location ``3'' is also a possibility \cite{serebrov}, while due to constraints of space the use of a standard beamport at location ``4" presently appears less favorable. Another option could be to use innovative reflector optics \cite{Zimmer:2016tyc} to maximise the neutron flux at position ``5''. Based on the same idea, a neutron-optical device is being developed to feed a He-II vessel located a greater distance away \cite{zimmer}.
Position ``5'' represents also the in-beam option of a UCN source, placed in a beamline pointing at the cold moderator.

All potential source positions should be considered in more detail in the future. Both helium and deuterium converters present a problem for cooling when placed in intense radiation fields. Recall that the optimum temperature for helium is below 0.8 K, and for deuterium is about 5 K.
A liquid deuterium source of cold neutrons maintained at 20 K is much easier to cool, and is a good spectrum shaper for both converters. The thermal load is mainly provided by gamma radiation, and to a first approximation depends only on the total converter mass. Because the required temperature is lower, and the required mass higher, in the case of superfluid helium for a similar complexity of cooling apparatus the distance to the target should be increased relative to solid deuterium. Thermal conductivity of the converter medium is an additional concern, and cooling technologies in the ranges above and below 1.2 K are fundamentally different. The optimal placement for ``in-pile'' UCN production (i.e., very close to the primary neutron source) thus depends strongly on the source configuration.

The reasonable thickness of the converter is limited by the depth of UCN escape from its material. Thus, it is practically unlimited for helium and does not exceed a few centimeters for deuterium due to elastic scattering by crystal inhomogeneities at the indicated temperature. Technologies for creating a perfect deuterium crystal are a subject of ongoing research, and there are promising recent results in the field \cite{Korobkina2022_ESS}. When comparing sources based on superfluid helium and solid deuterium for the ESS, with horizontal beam lines, it should be kept in mind that these converters have quite different neutron-optical potentials (18 neV for helium versus 105 neV for deuterium). Accordingly, the extraction system must be optimised for the respective UCN spectrum. In the case of a deuterium source, for certain specific experimental configurations the experimental apparatus can be raised to shift the spectrum into the energy region that provides long storage times. This spectrum transformation could be a source of additional losses and should be taken into account in a detailed comparison. 

Solid deuterium also offers a possibility to produce Very Cold Neutrons (VCN) from the same converter (see Section~\ref{UCNsubsec:VCN}).

 
An alternative approach for some applications is to set up experiments within the converter of a superfluid helium source. This approach makes it possible to use the full accumulated \emph{in-situ} UCN density, which at $220~\text{cm}^{-3}$ \cite{Wurm:2019yfj} for existing prototype sources already exceeds the values anticipated for next-generation storage experiments. (Only much lower densities can presently be delivered to \emph{ex-situ} experiments, due to extraction and transport losses.) A further advantage is that by eliminating transport losses, one automatically uses the low-energy part of the UCN spectrum which can be most efficiently stored. Extraction-based measurements at any UCN source suffer from the practical compromise that faster UCNs are more easily transported, whereas slower UCNs are more easily stored.

The clear advantages of the \emph{in-situ} approach have been recognized in the past, leading to at least three distinct experimental efforts in this direction \cite{nEDM:2019qgk,Baker:2010zza,Huffman:2000fh,HUFFMAN201440,Young:2014mxa}. All have confronted significant challenges in the technical implementation of the \emph{in-situ} measurement concept, due to challenges of scale and the challenge of jointly satisfying the stringent requirements for UCN production and for precision measurements in the same apparatus. 
\emph{In-situ} experiments may provide the most promising avenue to increase the storable density of UCNs with velocities up to 4 m/s, since the required technology for UCN production has already been demonstrated. 

The phase-space density in the experimental apparatus is most important: this results from not only the available UCN density at storable velocities, but also extraction efficiency and all loss processes that must be taken into account for UCN delivery to the experiment. New UCN sources delivering higher phase-space densities in experiments (e.g., storage experiments with number density exceeding $10^2~\text{cm}^{-3}$ for UCN velocities $<4$ m/s) would enable significantly improved experimental sensitivities. In particular, the following cases would profit from increases of flux or stored density:
\begin{itemize}
   \item [--] improving the accuracy of measuring the neutron lifetime;
  \item [--] precision gravitational neutron spectroscopy;
  \item [--] improving the sensitivity to the electric dipole moment of the neutron;
  \item [--] experiments to study non-stationary quantum effects \cite{Felber1996,AFrank2016,AFrank2020,AFrank2021}.
\end{itemize}
In the following sections, these experiments are briefly discussed.


\subsection{Measurements of the neutron lifetime}\label{UCNsubsec:taun}

As mentioned in Section~\ref{sec:theorybetadecay} precision measurements of neutron beta decay, including the neutron lifetime, are of key theoretical importance. The neutron lifetime is used to determine coupling constants of the electroweak interaction, and to check the unitarity of the CKM quark mixing matrix. The neutron lifetime is also an important parameter for astrophysics and cosmology.

One may conceptually distinguish two different experimental approaches for measuring the neutron lifetime: (1) those which directly measure the total decay rate, and (2) those which are sensitive only to a specific decay branch. The first approach describes all UCN-based approaches of which we are aware, including those detecting decay products: time-dependence in the detected rate of decay products is driven by the surviving number of UCN. The proposal to measure the neutron lifetime in a beam-based experiment, by observing transformation of the neutron time-of-flight spectrum \cite{Kuznetsov:2018aby}, also belongs to the first approach. Experiments following the second approach include the remaining beam-based lifetime measurements, which in addition to measuring the absolute rate of decay products for a specific branch, rely on absolute measurements of the total neutron flux in order to determine the specific activity for producing the observed decays. This second approach thus enables checking if the detected branch fully explains the decay rate measured via the first approach.

In particular the second approach is insensitive to any decay branches that are not directly detected, and thus comparing determinations of the neutron lifetime between these two approaches provides a conceptual method to constrain invisible decay channels, such as the decay into a hydrogen atom and a neutrino \cite{Green:1990zz}. One or more ``dark'' decay channels could hypothetically explain the observed discrepancy between beam-based and storage-based measurements of the neutron lifetime \cite{Fornal:2018eol}. Subsequent evaluations of this hypothesis have indicated that it is not well-supported by available data \cite{Czarnecki:2019mwq,Dubbers:2018kgh}.

Experiments based on UCN storage in traps measure the neutron lifetime by observing the time-evolution of the number of UCNs in the trap. In the case of magnetic storage, it is assumed that when the UCN spectrum is formed below the magnetic potential, in the absence of depolarisation or inelastic interactions, the storage time is equal to the neutron lifetime. In experiments, researchers try to control the process of depolarisation or to detect depolarised UCN and try to reduce the influence of marginally-trapped neutrons. Marginally-trapped neutrons have energies above the value of the trap potential, but could be stored for a long time in trajectories that provide a small velocity component parallel to the magnetic field gradient (or normal to the trap walls in the case of material traps). In the case of storage in a material trap, one must take into account UCN losses due to the interaction with the walls by measuring the dependence of the storage time on the frequency of collisions of neutrons with the walls and extrapolating this dependence to zero frequency \cite{Kosv1986}. Changing the frequency is achieved either by changing the spectrum of stored neutrons (energy calibration), or by adding additional surfaces to the storage volume, or by changing the volume of the trap (geometric calibration) \cite{Alfimenkov1990}. The losses can be controlled by detecting neutrons escaping the trap via the inelastic scattering channel \cite{Morozov2015}. It is postulated that additional surfaces or surfaces of different traps provide the same loss coefficients. It is possible to calibrate the loss at a fixed collision frequency and change the loss factor, allowing one to get away from this postulate \cite{Alfimenkov1993}.  

Possibilities for improving the statistical uncertainty in this approach, by increasing the phase-space volume of the experimental setup, are significantly limited. A significant increase in the intensity of permanent magnetic fields is technically unrealistic to implement in such experiments. Increasing the geometric volume is possible as long as the characteristic filling times remain shorter than the neutron lifetime, but this can lead to an increase in systematic effects: in particular, it would be necessary to increase the spectral ``cleaning time'' for removing marginally-trapped neutrons that can escape the trap before decaying, and thus shorten the apparent neutron lifetime. The best measured statistical uncertainty to date for the neutron lifetime is 0.28~s \cite{UCNt:2021pcg}, measured with magnetic UCN storage.

Similarly, in experiments with material traps, the best statistical uncertainty to date is 0.7~s \cite{Ser2017}, obtained in a large setup. Further increases in the geometric dimensions of the trap are not expected to lead to improvements, since the characteristic time for filling the trap will exceed its storage time.

The broadening of the spectrum of stored neutrons will automatically lead to an increase in systematic errors (due to the energy dependence of all characteristic parameters describing the storage and detection of UCNs). It should be noted that in experiments with material traps, it is still tempting to use materials with theoretically low loss factors (such as beryllium and solid oxygen), which have not yet been observed in practice \cite{Alf1992}. It follows from the discussion above that improving the accuracy in any of these approaches requires an increase in the UCN density delivered to an experimental setup. Provided high densities can be reached, it is interesting to consider storage of UCNs on the surface of superfluid helium \cite{Alfim2009} and experiments on transformations of the UCN phase-space volume \cite{Andreev2013}. Such transformations could make it possible to efficiently fill the traps with the soft part of the UCN spectrum.

To date, UCN measurements detecting charged decay products \cite{Dzhosyuk2005,Picker2005} have not reached a competitive level of accuracy in comparison to those experiments discussed above that count stored UCN directly. 

For measuring the neutron lifetime at the ESS, one can draw attention to the proposal to determine it from the transformation of the neutron time-of-flight spectrum \cite{Kuznetsov:2018aby}. This proposal has nothing to do with UCN directly, but the observed effect in the proposed experimental design becomes larger for slower neutrons. It is interesting to consider the possibility of implementing such an approach with UCN, although a special implementation involving vertical extraction may be required to account for effects due to gravity.

One may also consider implementing experiments at the ESS that follow the second approach of detecting a specific decay branch, either UCN-based or otherwise. Such implementations require very good knowledge of the detection efficiency for neutron decay products in order to determine the absolute rate for the observed decay branch, as well as absolute measurement of the neutron flux or stored neutron number. Beam-based neutron lifetime measurements already use this approach \cite{Last:1988hx,Kossakowski:1989dc,Yue:2013qrc,Wietfeldt:2011suo}, which is further discussed in Section~\ref{ANNIsubsubsec:benefitspulsedbeams}. We are not aware of any measurement approaches based on UCN storage that follow this concept; in principle it could be implemented via substantially-improved knowledge of both the absolute UCN number and the detection efficiency for decay products.

\subsection{Gravitational spectroscopy with neutrons}\label{UCNsubsec:GravitationalSpectroscopy}
\begin{figure}[H]
  \setlength{\unitlength}{1mm}
\centering
\hspace{-1.25cm}
  \includegraphics[width=0.61\linewidth, angle=0]{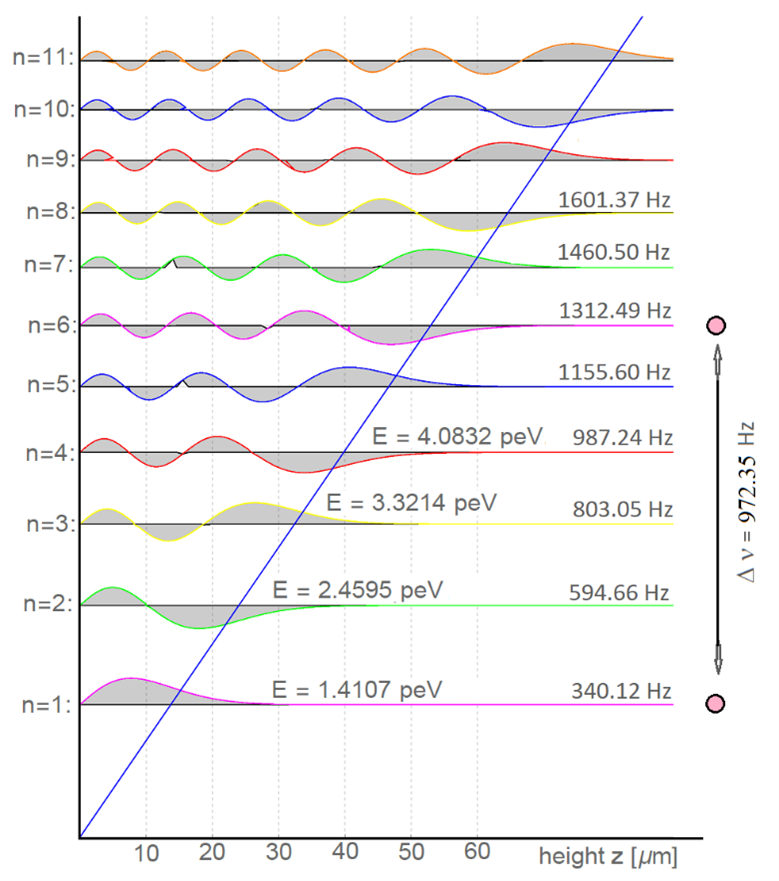}
  \caption{\footnotesize Energy eigenvalues and eigenfunctions of a neutron bound in the gravity potential of the Earth, and corresponding equivalent frequencies. By oscillating the mirror at a frequency that corresponds to the energy difference between two quantum states, transitions are introduced. Shown is a resonant transition between $|1\rangle \leftrightarrow |6\rangle$ at a frequency of 972.35~Hz. The oscillation frequencies are in the acoustic frequency range.}
  \label{fig:GravLev}
\end{figure}

This section is devoted to the control and understanding of the behavior of an elementary quantum system, such as a neutron with extremely low vertical velocity component, in a gravitational field above a mirror \cite{Lusch1978,Nes2002,Abele2010} using resonance spectroscopy methods \cite{Nes2006_book}. It offers a way, based on quantum interference and spectroscopy, to search for new  hypothetical interactions beyond the standard model and modifications of gravity at short distances. The neutron gives access to parameters that characterize gravity in the standard framework of general relativity, as well as more speculative extensions such as torsion~\cite{Ivanov:2016eug,Abele:2015uua}.
This is possible because the neutron is bound between a flat lying mirror and the rising potential of gravity above, $V=mgz$, where $m$ denotes the neutron mass, $g$ the local gravitational acceleration, and $z$ the distance above the mirror. Every bound system in quantum mechanics has discrete energy levels. In our case the lowest energy eigenvalues $E_n$ ($n = 1, 2, 3, 4, 5$) are 1.41~peV, 2.46~peV, 3.32~peV, 4.08~peV, and 4.78~peV. In Figure~\ref{fig:GravLev} they are shown together with the corresponding neutron wavefunctions, given by the well-known mathematical Airy functions that solve Airy’s differential equation. 

In 2002 the lowest stationary quantum state of neutrons in the gravitational field was clearly identified \cite{Nes2002,Nes2003,Nes2005,West2007}. 
The proof of principle triggered new experiments and activity in this direction, namely the setting up of the GRANIT \cite{Granit}, the qBOUNCE \cite{qB} collaborations, and teams from Tokyo \cite{Ich2014} and Los Alamos National Laboratory. Spectroscopy is a technique used to assess -- in most cases -- an unknown quantity of energy by means of a frequency measurement. Basic quantities like atomic masses, particle energies, momenta, and magnetic moments have been determined in this way. Transitions can be driven between the gravitational energy eigenstates, by a time-dependent perturbation provided by an external oscillator. These transitions occur with a characteristic energy transfer, and the resonance condition is observed when the oscillator frequency matches the difference between two eigenstates. The transition is observed via the change in population of the various eigenstates, after the time-dependent perturbation has been applied.

In this Gravity Resonance Spectroscopy (GRS) technique \cite{Jenke2011}, the energy difference between these states has a one-to-one correspondence to the frequency of the modulator, in analogy to the Nuclear Magnetic Resonance technique, where the energy splitting of a magnetic moment in an outer magnetic field is related to the frequency of a radio-frequency field. 
The linear gravity potential leads to discrete non-equidistant energy eigenstates $|n\rangle$. A combination of any two states can therefore be used, as each transition can be addressed by its unique energy splitting, or, by vibrating the mirror mechanically with the appropriate frequency. In that way GRS is becoming a tool, which combines the virtues of UCNs
-- namely long observation time together with insensitivity to electrostatic and van der Waals interactions
-- with spectroscopy.
Highest sensitivity can be reached by applying Ramsey’s method of separated oscillating fields \cite{Abele2010} to GRS.  This method will allow a precise measurement of energy differences with a precision similar to the magnetic resonance technique. In a beam experiment, this precision is determined by the level's width according to the ratio: $\Delta E/E\sim \hbar v/LE$, where $\Delta E$ is the level's energy width, $E$ is an energy  difference between gravitational levels, $v$  is the neutron velocity, and $L$ is the mirror length defining the duration of the interaction; for qBOUNCE, $\Delta E/E\sim 3\%$.  As a consequence, the coupling of residual fluctuations of the magnetic field to the magnetic moment of the neutron must be highly suppressed. 
The search for generalized gravitational theories is highly topical now. Observational cosmology led to the introduction of the concepts of dark matter (DM) and dark energy (DE). 
At the level of precision, GRS provides constraints on some possible gravity-like interactions. If some as yet undiscovered particles interact with neutrons, this should result in a measurable energy shift of the observed quantum states shown in Figure~\ref{fig:GravLev}  \cite{Jenke2011,Jenke2014,Cron2018}. 

The accuracy of measuring the gravitational quantum states of neutrons depends on two factors: a) the UCN density in phase space, and b) the observation time for UCN in quantum states. Therefore an \emph{in-situ} experiment at the ESS would be of interest, using the highest UCN densities inside a helium converter (without extracting the UCNs). The observation time for quantum states could ultimately be comparable to the lifetime of the neutron, as proposed in \cite{Nes2020}. The practical implementation of such an experiment, however, is challenging. A convenient method of analysis of gravitational quantum states in such an experiment is time-resolved spectroscopy as proposed in \cite{Nes2021PRL}.

\subsection{Measurements of the neutron's permanent electric dipole moment (EDM)}
\label{sec:UCNedm}

Ramsey's method of separated oscillatory fields is the foundational technique used for measuring the neutron EDM, whether using cold neutrons or UCNs -- see also Section~\ref{ANNIsubsec:EDM}. (Other approaches have been proposed, e.g. continuous measurements exploiting spin-dependent absorption of neutrons on $^3$He nuclei \cite{nEDM:2019qgk}, or beam experiments exploiting neutron spin-rotation in crystals with very high internal electric fields \cite{Fedorov:2010sj}. To date these approaches have not delivered results at a competitive sensitivity level.) The Ramsey method was initially applied in beam experiments \cite{Smith:1957ht}, until that approach reached systematic limitations due to beam divergence and motional magnetic fields \cite{Dress:1976bq}; see Section~\ref{ANNIsubsubsec:beamedm} for a discussion of how these effects might be mitigated in modern experiments at the ESS. At that juncture, UCNs offered a viable way forward both by mitigating these particular systematic effects and by opening the door to much longer observation times \cite{Altarev:1980rgh}. Since then, all subsequent advances in the precision and accuracy of neutron EDM measurements have been based on storage experiments with UCNs in a measurement volume separate from the UCN source.

A key subsequent development was the concept of ``comagnetometry'' \cite{Lamoreaux:1989gy}, in which an atomic species whose intrinsic sensitivity to magnetic fields is similar -- but whose EDM is expected, or has been measured, to be small -- occupies the measurement volume together with stored UCNs. This enables a simultaneous differential measurement, in which all effects seen by the comagnetometer are subtracted in common-mode. This approach greatly improves the sensitivity of such measurements, at the expense of experimental complexity and certain limitations on the experimental conditions (e.g., electric field strength) that are required in order for the comagnetometer to function. We note, however, that the leading systematic errors are quite different for atomic comagnetometers and UCNs. In fact, the largest source of error in the most recent neutron EDM measurement was associated with the comagnetometer correction, arising from differences in how spatially inhomogeneous fields within a measurement cell are probed by comagnetometer atoms and UCNs. This leads to the possibility that high-order gradients are sampled differently by the two ensembles \cite{Abel:2020gbr}. Recent proposals for new measurements of the neutron EDM (e.g., reference \cite{Wurm:2019yfj} as well as the proposals discussed in Section~\ref{ANNIsubsec:EDM}) in some cases favor other approaches to achieve stringent control of experimental conditions, rather than contending with the increasing challenges of comagnetometry at higher levels of sensitivity.

A further conceptual advance, now universally adopted in the present generation of neutron EDM experiments, was to perform two measurements simultaneously with a common magnetic field and oppositely-directed electric fields \cite{Altarev:1996xs}. This additional common-mode correction effectively removes offset field drifts within the pairs of measurements that must be combined in order to determine the EDM. When in addition a comagnetometer is present for both measurements, one can also remove in this way the leading effects connected with magnetic field gradients. A generalization of this concept to many cells has been proposed \cite{serebrov2007multi}, which in suitably stable conditions would permit controlling this type of systematic effect at arbitrary order. A conceptually similar approach employing four cells (and no comagnetometer) is used in measurements of the $^{199}$Hg atomic EDM \cite{Graner:2016ses}; the multicell concept is also central to the implementation of the EDM$^n$ project that has been proposed for the ESS, in which cells with zero electric field are intended to be used for averaging and correction of systematic errors (see Section~\ref{ANNIsubsubsec:edmn}).

As for other precision physics experiments, the major effort in EDM measurements is spent in connection with constraining systematic errors. Important systematic effects include (1) those arising from spatial and temporal variation of magnetic fields (and in particular, of magnetic field \emph{gradients}), (2) nondynamical phases arising from a combination of motional magnetic fields and gradients, and (3) direct false-EDM effects that correlate with the applied electric field, e.g. those arising from leakage currents or induced magnetization of inner components. Achieving higher densities of slower UCNs in smaller storage chambers would directly reduce the impact of (1) and (2), while also increasing the rapidity with which systematic studies can be conducted by intentionally varying or amplifying particular systematic errors: the analysing power available from UCN statistics is frequently the rate-limiting consideration. While these classes of effects have been extensively studied at present-day sensitivities, constraining them in future experiments will rely on dedicated research-and-development efforts. It is also to be expected that new systematic effects will become relevant as experimental precision increases.

While increasing the number of stored UCNs may be the main route to long-term improvements in measurements of the neutron EDM, one should also note that gains can still be had from increasing the observation time and electric field strength -- parameters for which the scaling of sensitivity for a single measurement is actually more favorable, see Equation~\ref{ANNIsubsubsec:edmeqn}. Unlike experiments measuring the neutron lifetime, the longest holding times achieved so far in neutron EDM measurements are on the order of $200~\text{s}$ -- significantly shorter than the ultimate limit imposed by the neutron lifetime. This holding time must be chosen as a compromise among the cycle time for re-filling the measurement cells, the storage time constant of the measurement cells, and numerous constraints including the performance of other key components such as a comagnetometer (if any). Recent demonstrations of long UCN storage lifetimes in small bottles \cite{neulinger-thesis,Neulinger:2022oue}, with material coatings intended for EDM applications and using soft UCN spectra extracted from a superfluid-helium source, have given rise to hopes that the storage time in particular may provide a route toward improvements in the next generation of experiments. The electric field strength is typically limited to the order of $2~\text{MV/m}$ (or less, in the presence of a comagnetometer) by the surface conductivity of the insulating ring which forms one wall of a UCN storage cell. However, recent advances in cryogenic technology developed for an \emph{in-situ} neutron EDM measurement by the American collaboration nEDM@SNS have indicated that much higher fields, on the order of $10~\text{MV/m}$, can be sustained when the cell is filled with superfluid helium \cite{Phan:2020nhz}. This brings the electric field strength into the same range as for the Beam EDM experiment (see Section~\ref{ANNIsubsubsec:beamedm}).

In contrast to the scenario described in Section~\ref{ANNIsubsubsec:edmn}, one may also envision implementing a multicell \emph{in-situ} EDM measurement with a smaller number of cells at a dedicated UCN facility. Through use of multimirror imaging optics \cite{zimmer2018imaging,Herb:2022owh} it may be possible to deliver very high neutron flux in the 8.9~\AA{} wavelength range, from a large area of the lower moderator surface at the ESS, albeit only over a short axial range due to the beam divergence. This might open the possibility of achieving UCN densities in the range of $10^4~\text{cm}^{-3}$ -- if such cold-neutron-extraction concepts can be efficiently realized, without excessive backgrounds or unanticipated losses. While the 8.9~\AA{} component of the feeding beam cannot be fully exhausted within a few-meter installation, such an approach has the advantage that any requirements for guiding cold neutrons within the multicell stack could be substantially relaxed. This concept requires further study. 

\subsection{Study of non-stationary quantum effects}
The development of ideas about the optical phenomenon called the accelerating matter effect \cite{AFrank2008,Braginetz2017,AFrank2016_PPN}, which consists in a shift in the frequency of a wave when passing through an accelerating refracting sample, led to the hypothesis of the existence of a very general acceleration effect \cite{AFrank2020_PU}. As applied to the physics of the microscopic world, its formulation is that the result of a particle interacting with any accelerating object should be a change of its energy and frequency determined by the ratio $\Delta \omega = ka\tau $, where $k$ is the wave number, $a$ is the acceleration of the object, and $\tau$ is the interaction time. 

It should be expected that any elementary scatterer moving with acceleration should change the frequency of the wave, and this means an effect that complements the usual Doppler shift but is proportional not to speed, but to acceleration. The validity of the acceleration effect hypothesis in quantum mechanics has recently been confirmed by numerically solving a number of problems related to the interaction of a wave packet with potential structures moving with acceleration \cite{Zakharov2021}. If the hypothesis on universality of the acceleration effect is true, it applies to the case of neutron scattering on the atomic nuclei of accelerating matter. Assuming that this elementary process can change the frequency of the neutron wave function, we will be forced to take a new look at the problem of neutron optics of an accelerating medium.

Different aspects of this hypothesis must be proved experimentally. The use of UCN for this aim is highly preferable, because the interaction time of neutrons with nuclei is inversely proportional to the neutron velocity. In addition, the study of non-stationary quantum effects requires either monochromatisation or interruption of the neutron beam. Thus for these experiments, the high phase-space density that could potentially be achieved at the ESS is advantageous. 

\subsection{Implementation of a VCN Source at the ESS}
\label{UCNsubsec:VCN}
Intense fluxes of neutrons in the range between 10~\AA{} to about 120~\AA{} are of interest for a variety of applications, some of which are listed in Section~\ref{sec:UCN_VCN_science}; however, so far the development of applications for neutrons in this wavelength range has been hindered by the challenging task of realizing VCN sources. At present the only existing VCN beamline is PF2 at the ILL, where slow neutrons are extracted vertically from one of the cold moderators, but this beamline has low intensity for realization of most potential applications. Its neutron flux at $v=40~\textrm{m}/\textrm{s}~(100~\textrm{\AA})$ is about $10^5~\textrm{cm}^{-2}\textrm{s}^{-1}(\textrm{m}/\textrm{s})^{-1}~ (=0.4\times 10^5~\textrm{cm}^{-2}\textrm{s}^{-1}\textrm{\AA}^{-1})$.
In recent years however there have been promising developments that hint at the possibility of realizing new sources. On one hand research has progressed in identifying materials for production of intense VCN fluxes: at least an order of magnitude higher, relative to PF2.
The innovative use of cascaded cooling in paramagnetic systems such as deuterated clathrate hydrates has been suggested \cite{Zimmer:2016qai}, and experimental investigations of these materials are ongoing within the HighNESS project. Solid deuterium is another candidate material for VCN production. 

One of the most promising developments related to VCN sources concerns the use of effective reflectors which allow to direct neutrons to experimental set-ups. In recent years it was shown that diamond nanoparticle powders can be used as such reflectors \cite{Nesvizhevsky:2008rk}, and VCN storage in a  cavity made of the powder was realised  \cite{Lych2009_store}. The influence of hydrogen impurities \cite{Kry2011}, powder structure \cite{Aleks2021_clast} and particle size  \cite{Aleks2021_ps} on reflection was determined and a method was found to remove hydrogen from the powder \cite{Nes2018_Car,Bosak2020}. The interaction of neutrons with powders of nanostructures,  
possible applications, 
and neutron transport simulation inside, 
are widely studied in various scientific centers both theoretically and experimentally. The possibility to use these reflectors for cold neutrons was also shown \cite{Cubitt2010,Nes2018PRA}. 


The neutronic study of VCN sources  within the HighNESS project has started, considering either the use of dedicated VCN converter materials, or the design of an optimized geometry based on reflectors such as nanodiamonds \cite{zanini}, which can increase the reflection of the longer wavelength neutrons from  a cold moderator \cite{jamali}.

\begin{figure}[H]
    \centering
    \includegraphics[width=0.86\linewidth]{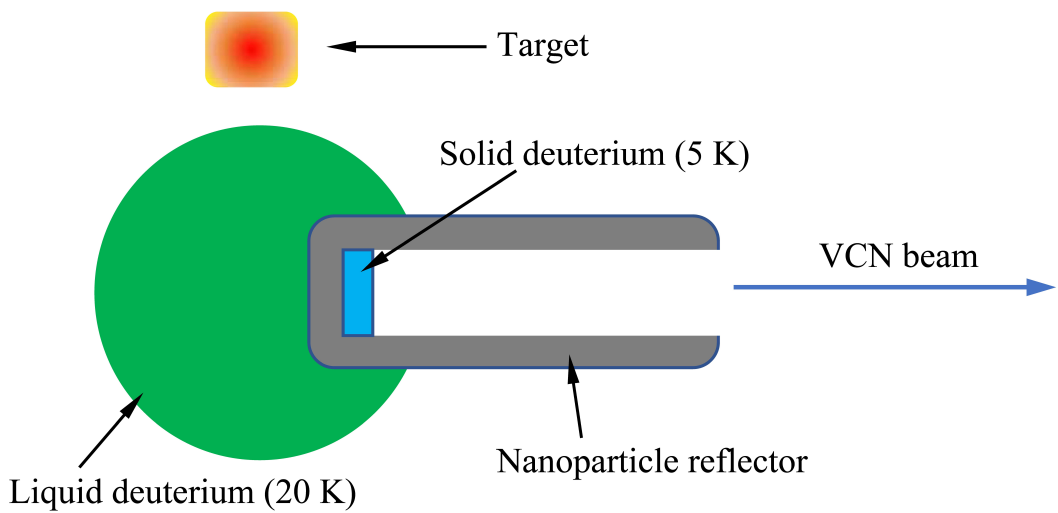}
    \caption{A conceptual diagram of the VCN source lay-out.}
  \label{fig:VCNsource}
  \end{figure}

It is possible to combine the use of a dedicated VCN converter material with advanced reflectors for a more efficient VCN source.
A possible configuration of the VCN source at ESS (or another intense neutron facility) could look like this: a large liquid deuterium moderator is located under the ESS target. A capsule of a nanoparticle reflector, open on one side, is buried with its closed end into this source. At the bottom of the capsule is a relatively small solid deuterium converter (this principal is illustrated at Figure~\ref{fig:VCNsource}). Solid deuterium as VCN moderator is  attractive because it will produce a large amount of VCN and it is possible to combine a source of UCN/VCN. The use of reflectors made of nanoparticles  will significantly increase the direction flux  of extracted VCN \cite{Chernyavsky2022} and allow reducing the thickness of the solid deuterium by a factor of two at fixed VCN flux. Calculations for realistic geometries, and optimization, are planned to be carried out in the near future.

\subsection{Scientific Prospects for Use of VCN}
\label{sec:UCN_VCN_science}

We are aware of two ideas where large potential gains from the use of VCN have been investigated in detail:
\begin{itemize}
    \item Search for neutron-antineutron oscillations \cite{VCN_n_nbar}. The idea of an experiment with cold neutrons was discussed in Section~\ref{sec:nnbar}. The sensitivity of the experiment is proportional to the product of the observation time of oscillations and the square root of the neutron flux. The switch to slower neutrons with the same experimental setup does not  give a gain at first glance. The observation time is inversely proportional to the neutron longitudinal velocity, and the neutron flux in the neutron guide is proportional to the square of the velocity (under the natural assumption that the neutron phase-space density in the source is constant). However, the use of the above-mentioned reflectors can in principle make it possible to significantly increase the phase-space density of the VCN \cite{Nes2022_VCN}.
    \item Search for additional types of interactions \cite{Nes2008_ESR_inter,Snow2022} by studying the angular distribution of neutrons scattered by atoms of noble gas. Full details of the method are presented in reference \cite{Nes2008_ESR_inter}.
\end{itemize}

In contrast to the above ideas which are relatively well-developed in terms of estimated gain factors, the following points can also be considered but remain at an earlier stage of development where the balance of potential gains (if any) against various disadvantages are less clear at this point:
\begin{itemize}
    \item Search for the electric dipole moment (EDM) of the neutron \cite{Piegsa2013}. In comparison to the cold-neutron beam experiment discussed in Section~\ref{ANNIsubsubsec:beamedm}, the use of VCN for an apparatus with the same length could deliver a gain in statistical sensitivity due to longer observation times. However at the present level of analysis it is unclear if this approach could be competitive with experiments using cold neutrons or UCN. 
    A detailed analysis is also needed for systematic effects, which are different for different neutron velocities.

    \item The sensitivity of certain interferometers (for instance based on diffraction gratings) depends on both the intensity and the parameters of the optical system. The deviation of the refractive index from
unity is greater, the slower the neutrons.  As result the the optical effects are more pronounced and requirements for the optical system become not so strong. 
    Interferometers can be used for precision measurements of the coherence length of neutron scattering by nuclei, or searches for new types of interactions. 
    \item The use of VCN in a beam experiment to measure the neutron lifetime, similarly to \cite{Nico2005}, where cold neutrons were used, will lead to an increase in the probability per neutron to observe a decay, that will affect the statistical accuracy of the result. It should be noted that the systematic error in this experiment, associated with the determination of the absolute neutron flux, still exceeds the statistical one, although some progress has been made towards its reduction \cite{Nico2013}.
    
    \item Increasing the sensitivity of commonly used neutron scattering techniques such as spin echo \cite{Mezei1972,Hino2003,Oda2017} (by increasing the observation/interaction time) and reflectometry (by expanding the dynamic range of the setup) \cite{Baes2011}.
    \item Use of tomography with VCN for solving specific problems \cite{Maru2005}.
\end{itemize}

A detailed analysis of the proposed UCN/VCN sources should be performed in order to reliably evaluate their performance. The types of chosen sources would determine the list of scientific problems that could be solved using them. An assessment of the accuracy and sensitivity of prospective research could only be made after evaluating the actual source parameters.
\section{CE$\nu$NS at the ESS} \label{sec:nuess}
\subsection{Introduction}

Coherent elastic neutrino-nucleus scattering (CE$\nu$NS) is the most probable mechanism for low-energy neutrino interaction. Nucleons participate coherently in this neutral-current scattering process, resulting in a coupling effectively proportional to the square of the number of neutrons in the target nucleus \cite{Freedman:1973yd,PhysRevD.30.2295}. This large enhancement to the scattering cross section facilitates the detection of these elusive particles while considerably reducing the target mass involved. Nevertheless, this process remained undetected until recently \cite{Akimov:2017ade,bjornthesis,nicolethesis}, more than four decades following its theoretical description. This somewhat puzzling delay was  due to a combination of factors: the modest energy of the nuclear recoils induced -the single observable from this process-, and  the limited intensity of available neutrino sources in the favorable energy range -coherence is lost above few tens of MeV-. 

Coherent nuclear scattering mediates the interactions of popular dark matter candidates (Weakly Interacting Massive Particles), and dominates neutrino transport within supernovae and neutron stars \cite{PhysRevLett.32.849,doi:10.1146/annurev.ns.27.120177.001123}. These aspects added to a broad interest in obtaining an experimental verification for CE$\nu$NS. Most importantly, the observation of this process unlocked four decades of phenomenological proposals aiming to exploit this new mechanism of neutrino interaction. CE$\nu$NS measurements promise to improve our knowledge of both neutrino properties and nuclear structure, each providing new opportunities for revealing  physics beyond the Standard Model. The fact that evidence for the incompleteness of the Standard Model comes from the neutrino sector, an area of still incipient knowledge, adds to the appeal of these investigations. For instance, a neutral-current detector responds almost identically to all known neutrino types \cite{SEHGAL1985370,Tomalak:2020zfh}. Therefore, observation of neutrino oscillations in such a device would provide unequivocal evidence for sterile neutrinos \cite{PhysRevD.30.2295,PhysRevD.96.063013,PhysRevD.101.075051,PhysRevD.102.113014}. In addition to this, the differential cross section for this process is strongly dependent on  values of the neutrino magnetic moment otherwise beyond reach \cite{DODD1991434,PhysRevD.97.033003,Billard:2018jnl,em4}. Numerous recent studies have described the sensitivity of CE$\nu$NS to non-standard neutrino interactions with quarks \cite{Barranco_2005,PhysRevD.76.073008,PhysRevD.96.095007,LIAO201754,PhysRevD.97.035009,Farzan:2018gtr,PhysRevD.98.015005,Esteban:2018ppq,PhysRevD.98.075018,PhysRevD.95.115028,PhysRevD.101.035039,Denton:2018xmq,ARCADI2020115158}, to the effective neutrino charge radius \cite{Papavassiliou:2005cs,PhysRevD.98.113010},  to neutron density distributions  \cite{Amanik_2008,PhysRevC.86.024612,PhysRevLett.120.072501,PhysRevD.97.113003,PAPOULIAS2020135133,PhysRevD.101.033004}, and to accelerator-produced dark matter \cite{PhysRevD.102.052007,Ge:2017mcq,Brdar:2018qqj,PhysRevLett.124.121802}. CE$\nu$NS data represent a valuable ingredient in the electroweak precision program, clearly superior to previous neutrino scattering experiments \cite{Breso-Pla:2023tnz}. A precise measurement of the CE$\nu$NS cross section would also provide a sensitive appraisal of the weak nuclear charge \cite{KRAUSS1991407,Canas:2018rng,PhysRevD.99.033010,PhysRevD.100.071301}. There is a crucial interplay between nuclear structure and CE$\nu$NS: the low energy scattering explores the neutron distribution of ground state and low--energy excitations. Therefore, nuclear structure provides key information to correctly interpret the measured cross section, and the cross section probes the weak sector of the nucleus \cite{Almosly19,Hoferichter20}. In turn, this input will contribute to calibrate other standard model tests in nuclei based on parity violation \cite{Dobaczewski05}. A complete review of the vibrant field of CE$\nu$NS phenomenology is beyond the scope of this paper: a more extensive compilation can be found in \cite{Baxter:2019mcx}.

In a seminal paper \cite{PhysRevD.30.2295}, Drukier and Stodolsky described the prospects for CE$\nu$NS detection from a variety of low-energy neutrino sources (solar, terrestrial, supernova, reactor, and spallation source). Of these, spallation sources were highlighted as  the most convenient, due to higher recoil energies -easier to detect- and pulsed operation -leading to a  favorable steady-state environmental background reduction-. While the main use for these facilities is as intense neutron sources, their protons-on-target (POT) produce positive pions, which decay at rest generating a monochromatic flash of 30 MeV prompt $\nu_{\mu}$. This is followed by delayed  $\nu_{e}$ and $\bar{\nu}_{\mu}$ emissions with a broad energy (Michel spectrum, up to few tens of MeV), over the $\sim$2.2 $\mu$s timescale characteristic of $\mu$ decay. Specifically, the neutrino flux 20 m away from target at a 5 MW ESS is expected to be $\sim 1.6\times10^{8}$ cm$^{-2}$ s$^{-1}$ \cite{Baxter:2019mcx}.

\begin{figure}[!htbp]
    \centering
    \includegraphics[width=1\linewidth]{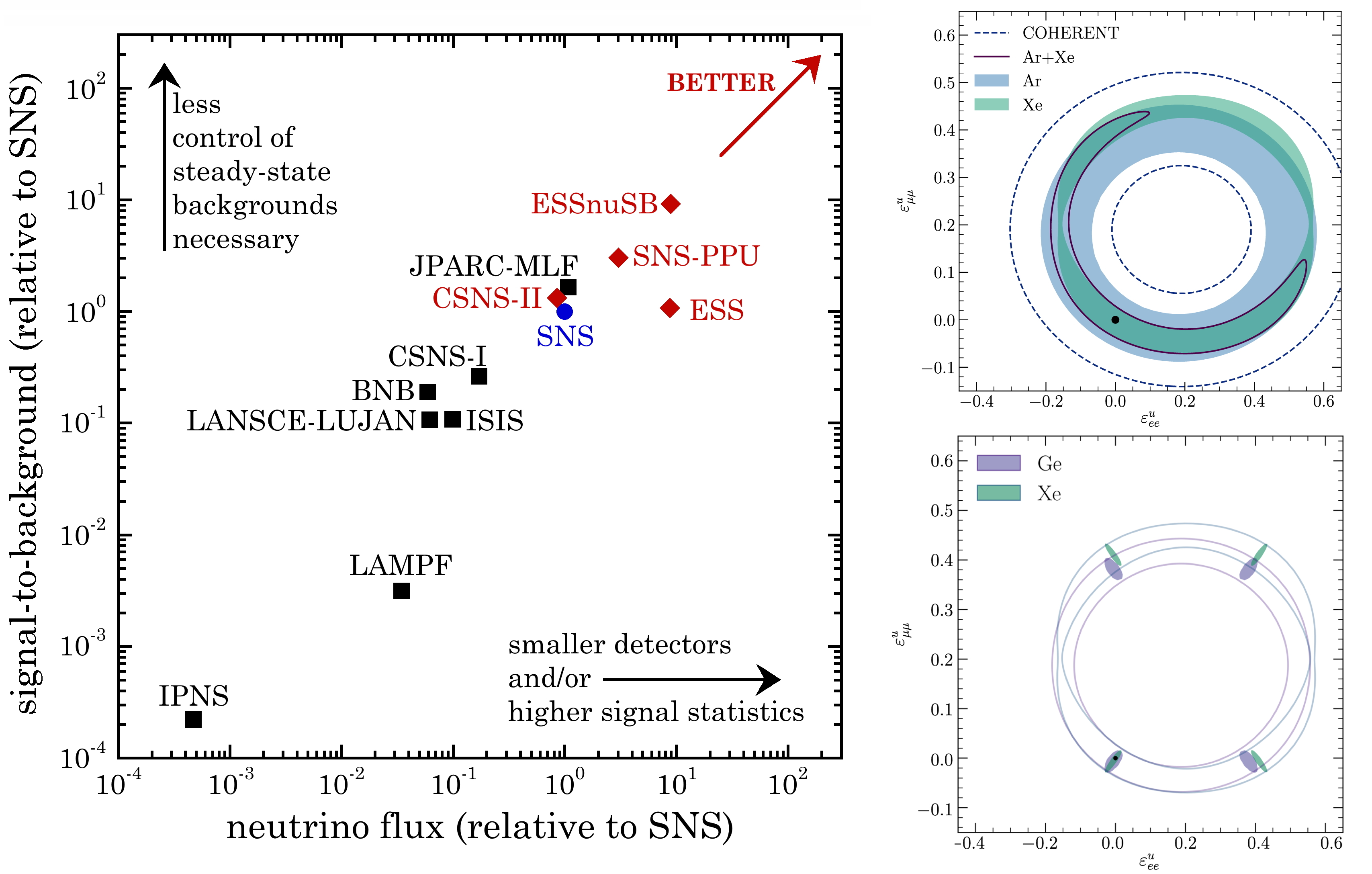}
    \caption{\footnotesize 
{\it Left:} Balanced comparison of pulsed spallation facilities as CE$\nu$NS sources. Future  (red diamond) and past (black square) sources are shown relative to the SNS at the time of the first CE$\nu$NS measurement (blue circle). The separate impact of proton current and proton energy on neutrino flux are correctly included as in \cite{Baxter:2019mcx}. Similarly, the effect of beam pulse time-profile on  signal-to-background ratio is properly computed assuming 10 $\mu$s windows for full detection of the delayed $\nu_{e}$,$\bar{\nu}_{\mu}$ component, and accounting for the ability to subtract steady-state backgrounds when characterized in anti-coincidence \cite{Akimov:2017ade,Baxter:2019mcx}. This plot is in contrast to more simplistic comparisons \cite{scholberg,Kelly:2021jgj}.  {\it Top right:} (adapted from \cite{Baxter:2019mcx}) Improvement in sensitivity to a parametrization of non-standard neutrino interactions, expected  from use of high-pressure Xe and Ar TPCs at a 5 MW, 2 GeV ESS (see text). Parameter space outside of the tori is excluded (``COHERENT" indicates present SNS limits). The advantage of combining information from different targets (solid line contour)  is evident. {\it Bottom right:} Further improvements expected from the squeezed beam bunching in a ESSnuSB configuration, leading to a  steady-state background reduction \cite{ivan}. Tori are for ESS sensitivity, colored regions for ESSnuSB. }
  \label{fig:nuESS.jpg} \end{figure}

The physics potential for CE$\nu$NS at the ESS has been recently described in \cite{Baxter:2019mcx}. Additional studies have further examined it \cite{PhysRevD.102.113014,Coloma:2020nhf,Esteban:2021axt,Chatterjee:2022mmu}. A 5 MW, 2 GeV ESS is expected to provide an order of magnitude increase in neutrino flux with respect to the Spallation Neutron Source (SNS) at ORNL, the site of the first CE$\nu$NS observation.  Compact, low-cost detectors (few kg to few tens of kg) operated at this new European facility, will produce CE$\nu$NS measurements not limited by statistics of the signal, but instead by systematics of order few percent, such as the specific response of different detector materials to low-energy nuclear recoils \cite{PhysRevD.100.033003}. As a result, the ESS will provide a definitive sensitivity to the broad body of new physics reachable via CE$\nu$NS, unsurpassed for the foreseeable future \cite{Baxter:2019mcx,Coloma:2020nhf,PhysRevD.102.113014} (Figure~\ref{fig:nuESS.jpg}).

\subsection{CE$\nu$NS at the ESS}

A point emphasized in \cite{Baxter:2019mcx} is that the best foreseeable neutrino source deserves the most-advanced  next-generation nuclear recoil detectors, so as to obtain the best possible results from the extraordinary opportunity the ESS provides. Several promising new technologies were described in that publication. In the interest of brevity, only two examples are  highlighted here. The first is the proposed use of undoped cryogenic (80 K) CsI as a CE$\nu$NS detecting medium. This material holds vast potential for this application \cite{PhysRevC.104.014612} when operated in combination with state-of-the-art light sensors and novel waveshifters \cite{Baxter:2019mcx}. To be specific, for comparable neutrino fluxes (i.e., before accounting for the $\times$10 increase at the ESS), the CE$\nu$NS rate per kg of scintillator can be increased by a factor of eight with respect to the room-temperature CsI[Na] employed for the first CE$\nu$NS detection at the SNS,  reaching in the process a sensitivity to nuclear recoils down to a $\sim$1 keVnr  threshold  \cite{Baxter:2019mcx,PhysRevC.104.014612} (the suffix ``nr" makes reference to nuclear recoil energy, as opposed to ``ee" for electron-equivalent, i.e., observable energy). As a reference, the CsI[Na] threshold in \cite{Akimov:2017ade} was 5 keVnr, with a significant signal acceptance  achieved only above 10 keVnr. Besides increasing the signal rate, a reduced energy threshold maximizes the sensitivity to new neutrino physics, which arises preferentially at the lowest recoil energies \cite{Baxter:2019mcx}.  For perspective, a compact 31.5 kg cryogenic CsI detector will register $\sim$12,000 CE$\nu$NS events per year at the ESS, with a superior energy resolution derived from the production of $\sim$47 primary electron-hole pairs per keVee at a silicon light sensor \cite{Baxter:2019mcx}. As a measure of   potential for CE$\nu$NS at the ESS, these figures can be contrasted with the $\sim$3,000 events/yr, $\sim$ 4.2 photoelectrons per keVee, and $\sim$20 keVnr threshold that a  massive 750 kg liquid argon detector would achieve at the SNS \cite{PhysRevD.102.052007,tayloe2}.

Innovative high-pressure gaseous xenon TPC detectors have been recently developed for neutrinoless double-beta decay searches \cite{Renner_2018}. These devices are sensitive to both primary scintillation (S1) and electroluminescent amplification of charge ionization (S2) signals, a property that can be exploited to separate nuclear recoil from electron recoil signals \cite{RENNER201562}. They are also insensitive to charge-trapping effects at low-energy that can limit the application of two-phase noble liquid detectors to CE$\nu$NS experimentation \cite{Baxter:2019mcx}. Additionally, they feature an outstanding energy resolution \cite{Mart_nez_Lema_2018}, and the ability to easily swap the active gas, which can help enhance the sensitivity to some aspects of CE$\nu$NS  phenomenology (Figure~\ref{fig:nuESS.jpg}). For all these reasons, they are prime candidates to be adopted as CE$\nu$NS detectors at the ESS. Their response to low-energy nuclear recoils is an incipient area of knowledge \cite{Baxter:2019mcx}, to be studied in the interim until ESS operation.

A curious coincidence increases the interest in simultaneously using these two technologies for this application: due to their similar number of neutrons per nucleus, with cesium and iodine immediately surrounding xenon in the table of elements, CsI and Xe detectors are essentially indistinguishable in their expected response to CE$\nu$NS \cite{Baxter:2019mcx}. In contrast to their identical CE$\nu$NS response, these two detector technologies, and in turn their expected systematics, are fundamentally different.  Their side-to-side use at the ESS would bring about a unique situation in experimental particle physics: two highly-complementary and yet distinct technologies, able to provide robust independent confirmation for any subtle signatures of new physics that might be found in one of them.  

\subsection{Timeline and expected sensitivity}

The presently envisioned slow ramp-up leading to a 5 MW, 2 GeV ESS performance (Section~\ref{sec:timescale}) is an excellent match to the R\&D necessary to develop next-generation nuclear recoil detectors able to fully profit from this unique neutrino source. The  construction and characterization of advanced high-pressure noble gas, cryogenic CsI, and low-noise germanium detectors for this application \cite{Baxter:2019mcx} has been recently supported by two ERC Horizon 2021 actions. This timeline provides room for the development of the novel detector technologies envisioned, as well as for the required  studies of detector response to low-energy nuclear recoils. The latter are of crucial importance in order to correctly identify indications of new physics via  CE$\nu$NS \cite{PhysRevD.100.033003,PhysRevC.104.014612,PhysRevD.103.122003}. 

It is worth emphasizing at this point that CE$\nu$NS detectors are devoid of a significant burden on the ESS, being nonintrusive to its established neutron science goals. Due to their compact design they require only the allocation of a modest floor area for their installation. As a result, in case of unanticipated ESS delays in achieving design power, these technologies can continue to be improved {\it in situ} for a period of time (e.g., by characterizing and abating backgrounds) up until maximum ESS power is achieved.  New physics results can be generated in the interim by virtue of the mentioned improved signal rate per unit detector mass that is expected from advanced detectors. Table III in \cite{Baxter:2019mcx} provides detailed information on the increase in sensitivity to several areas of neutrino phenomenology expected from next-generation  CE$\nu$NS detectors at the ESS. Further significant improvements can be generated by an eventual upgrade of the facility to neutrino Super Beam specifications (Figure~\ref{fig:nuESS.jpg}, Section~\ref{sec:ESSnuSB}). 

Suitable potential locations for CE$\nu$NS experimentation within the ESS facility have already been identified. Besides a proximity to the ESS target able to maximize neutrino flux, such sites must guarantee that backgrounds from prompt neutrons able to escape  the  target monolith do not compete with CE$\nu$NS signals. Profiting from previous experience in site selection at the SNS, two promising locations are being investigated: a utility room \mbox{15 m} from target, with 5.5 meters of steel monolith and a minimum of 6 meters of magnetite-loaded  heavy  concrete  (3.8 g/cm$^{3}$) separating them, and additional underground galleries as close as 23 m to target, with soil and concrete pylons as intervening shielding. Full Monte Carlo simulations of neutron generation and transport in the ESS building are under way: first results confirm the suitability of these locations. These use MCNPX and GEANT4 geometries of the target monolith, and detailed NAVISWORKS 3-D building layouts.  Once neutron production has commenced at the ESS, neutron background measurements will be performed to confirm these predictions, a process similar to that followed at the SNS previous to the first CE$\nu$NS measurement \cite{Akimov:2017ade,bjornthesis}.

The advent of CE$\nu$NS has opened  new exciting avenues in the search for physics beyond the Standard Model, spanning both particle and nuclear realms.  CE$\nu$NS studies will expand our incipient knowledge of neutrino properties through multiple new paths, each providing an opportunity for discovery. The ESS will soon become the most intense pulsed neutrino source suitable for CE$\nu$NS experimentation, remaining  the leader for the foreseeable future. Advanced next-generation CE$\nu$NS detector technologies are set to profit from the unique opportunity that the ESS provides.

\section{The ESS neutrino Super Beam (ESS$\nu$SB)} \label{sec:ESSnuSB} \label{ESSnuSB_Intr}

\begin{figure}[H]
\centering
\includegraphics[width=5.cm]{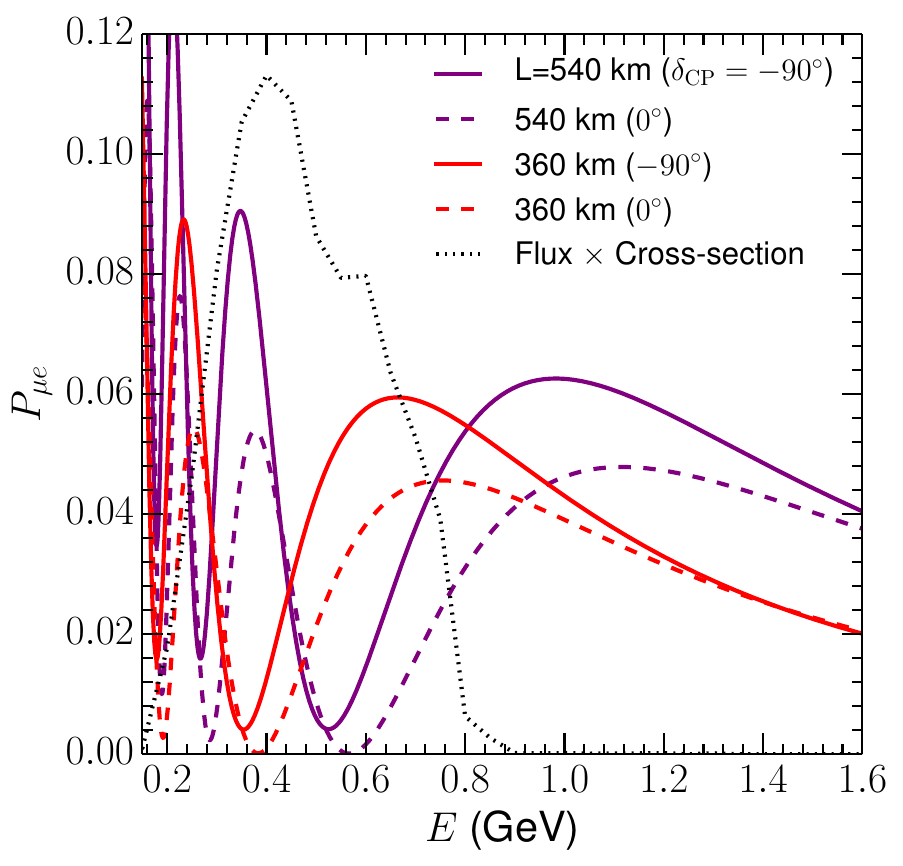}
\includegraphics[width=5.cm]{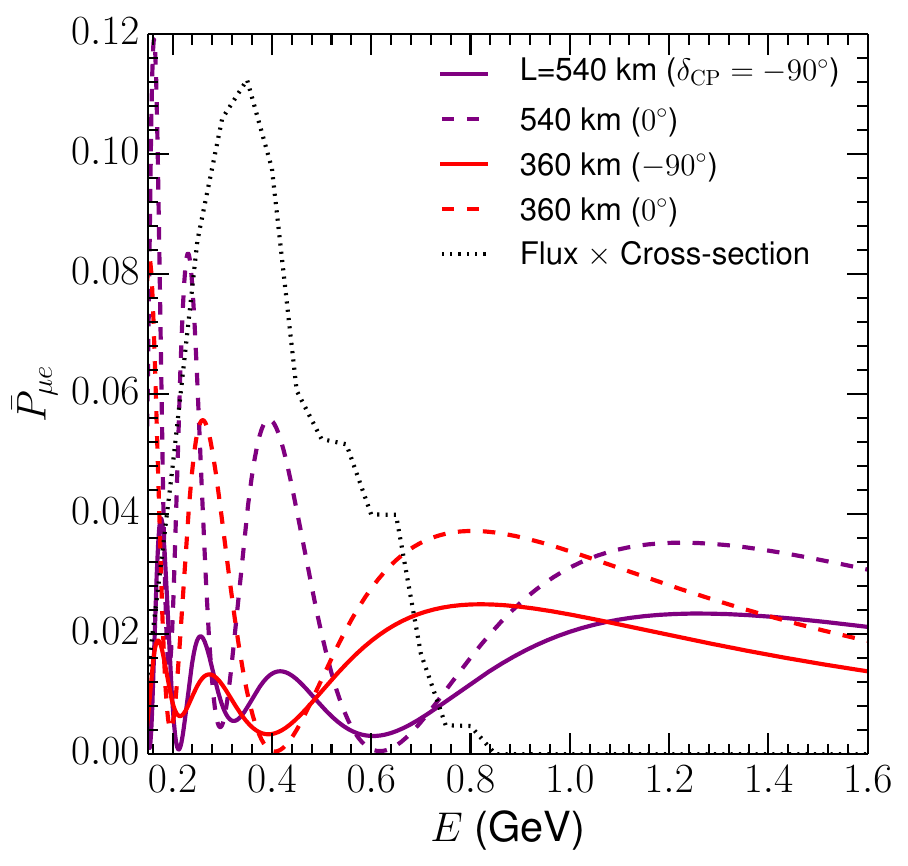}
\caption{Oscillation probabilities for the \SI{360}{km} (red curves) and \SI{540}{km} (purple curves) baselines as a function of the energy for neutrinos (left panel) and antineutrinos (right panel). The solid (dotted) lines are for $\delta = -90^\circ$ ($\delta =0$). The grey dotted lines show the convolution of the signal component of the neutrino flux with the detection cross section. Thus, they serve as a guide of what energies of the oscillation probability would be well-sampled by the ESS$\nu$SB setup.}
\label{fig:probs}
\end{figure}

In the standard three flavour scenario, the phenomenon of neutrino oscillation can be described by three mixing angles: $\theta_{12}$, $\theta_{13}$ and $\theta_{23}$, two mass squared differences $\Delta$m$_{21}^{2}$ (= m$_{2}^{2}$ - m$_{1}^{2}$) and $\Delta$m$_{31}^{2}$ (= m$_{3}^{2}$ - m$_{1}^{2}$) and one Dirac type phase $\delta_{CP}$. During the past few decades, some of these parameters were measured with good precision. At the moment, the unknown parameters are: (i) the mass hierarchy of the neutrinos, which can be either normal, i.e. $\Delta$m$_{31}^{2}$ $>$ 0, or inverted, i.e. $\Delta$m$_{31}^{2}$ $<$ 0, (ii) the octant of the mixing angle $\theta_{23}$, which can be either at the lower octant, i.e. $\theta_{23}$ $<$ 45$\degree$, or at the higher octant, i.e. $\theta_{23}$ $>$ 45$\degree$ and (iii) the Dirac violating phase, $\delta_{CP}$. In the search for the CP-violation in the leptonic sector, crucial information has been obtained from reactor and accelerator experiments \cite{An:2012eh,Abe:2011sj}. The discovery and measurement of the third neutrino mixing angle, $\theta_{13}$, with a value $\sim$~\SI{9}{\degree}, corresponding to $\sin^{2}$2$\theta_{13}$~$\sim$~0.095, confirmed the possibility of discovering and measuring a non-zero value of the Dirac leptonic CP violating angle, $\delta_{CP}$. Before this measurement, a significantly smaller value of $\theta_{13}$ was assumed, with a range of values of $\sin^{2}$2$\theta_{13}$~$\sim$~ 0.01 and 0.09, with 0.04 as a standard value. In the light of this new finding, the sensitivity to CP violation observation and measurement, with precision, of $\delta_{CP}$ has shown a strong enhancement at the second oscillation maximum compared to that at the first oscillation maximum~\cite{Nunokawa:2007qh, Coloma:2011pg, Parke:2013pna}. This can be seen in Figure~\ref{fig:probs} where the change in the probability, upon changing the values of $\delta_{CP}$, is much more significant at the second peak maximum, as counted from the right side in the figure, as compared to the first maximum. Moreover, by placing the far detector at the second oscillation maximum, the experiment is significantly less affected by, and hence more robust against, systematic uncertainties. This is particularly important since the improvement of the present systematic errors is known to be very hard. However, placing the far detector at the 2nd oscillation maximum implies the need to use very high intensity "super" neutrino beams to compensate for the longer baseline, hence lower statistics.

The long-baseline experiments which are currently running to measure these unknowns are T2K \cite{Abe:2020vii} in Japan and NO$\nu$A \cite{NOvA:2019cyt} in the USA. Regarding the true hierarchy of the neutrino mass, the results of both T2K and NO$\nu$A favour normal hierarchy over inverted hierarchy. Regarding the true nature of the octant of $\theta_{23}$, both these experiments support a higher octant, however the maximal value, i.e. $\theta_{23}$ = 45$\degree$, is also allowed within 1$\sigma$. Regarding the value of $\delta_{CP}$, there is a mismatch between T2K and NO$\nu$A. Considering the branch for $\delta_{CP}$ as -180$\degree$ $\leq$ $\delta_{CP}$ $\leq$ 180$\degree$, T2K supports the best-fit value of $\delta_{CP}$ around -90$\degree$, i.e. maximal CP violation, and the best-fit value measured by NO$\nu$A is around 0$\degree$, i.e. CP conservation. However, it is important to note that both the values of $\delta_{CP}$ = 0$\degree$ and -90$\degree$ are allowed at 3$\sigma$ and it requires more data to establish the true value of $\delta_{CP}$. It is believed that these two experiments will give a hint towards the true nature of these unknown oscillation parameters, while the future generation of the long-baseline experiments will establish these facts with a significant confidence level.

The European Spallation Source neutrino Super-Beam (ESS$\nu$SB) is a proposed accelerator-based long-baseline neutrino experiment in Sweden \cite{Baussan:2013zcy}. In this project, a high intensity neutrino beam will be produced at the upgraded ESS facility in Lund. The neutrinos will be detected in a large water Cherenkov detector located at a distance of \SI{360}{\kilo\meter} from the ESS site, which is the baseline distance, or at a distance of \SI{540}{\kilo\meter}, which is a possible back-up option. The primary goal of this experiment is to measure the leptonic CP phase, $\delta_{CP}$, by probing the second oscillation maximum. As the variation of neutrino oscillation probability with respect to $\delta_{CP}$ is much higher in the second oscillation maximum, as compared to the first oscillation maximum \cite{Nunokawa:2007qh,Coloma:2011pg, Parke:2013pna}, ESS$\nu$SB, as a second generation super-beam experiment has the potential of measuring $\delta_{CP}$ with unprecedented precision compared to the first generation long-baseline experiments.

An EU supported design study of ESS$\nu$SB has been conducted by 17 laboratory and university groups in 11 European countries during the period January 2018 till April 2022. The detailed results of this study are described in the ESS$\nu$SB Conceptual Design Report to appear in The European Physical Journal Special Topics \cite{ESSnuSBCDR}. In the following sections \ref{Inst_Intr} to \ref{ESSnuSBPhysics} the most important results of this study are outlined. In July 2022 EU decided to support the continuation during the period 2023-2026 of this design study under the project name ESSnuSBplus, which shall result in a Technical Design Report to be published at the end of this period. During this continued design study the details of the civil engineering required for the realization of the project shall be studied as well as the further plans outlined in the following section~{\ref{essnusb_syn}}.

\subsection{Instrumentation}
\label{Inst_Intr}

\begin{figure}
    \centering
    \includegraphics[width=0.55\linewidth]{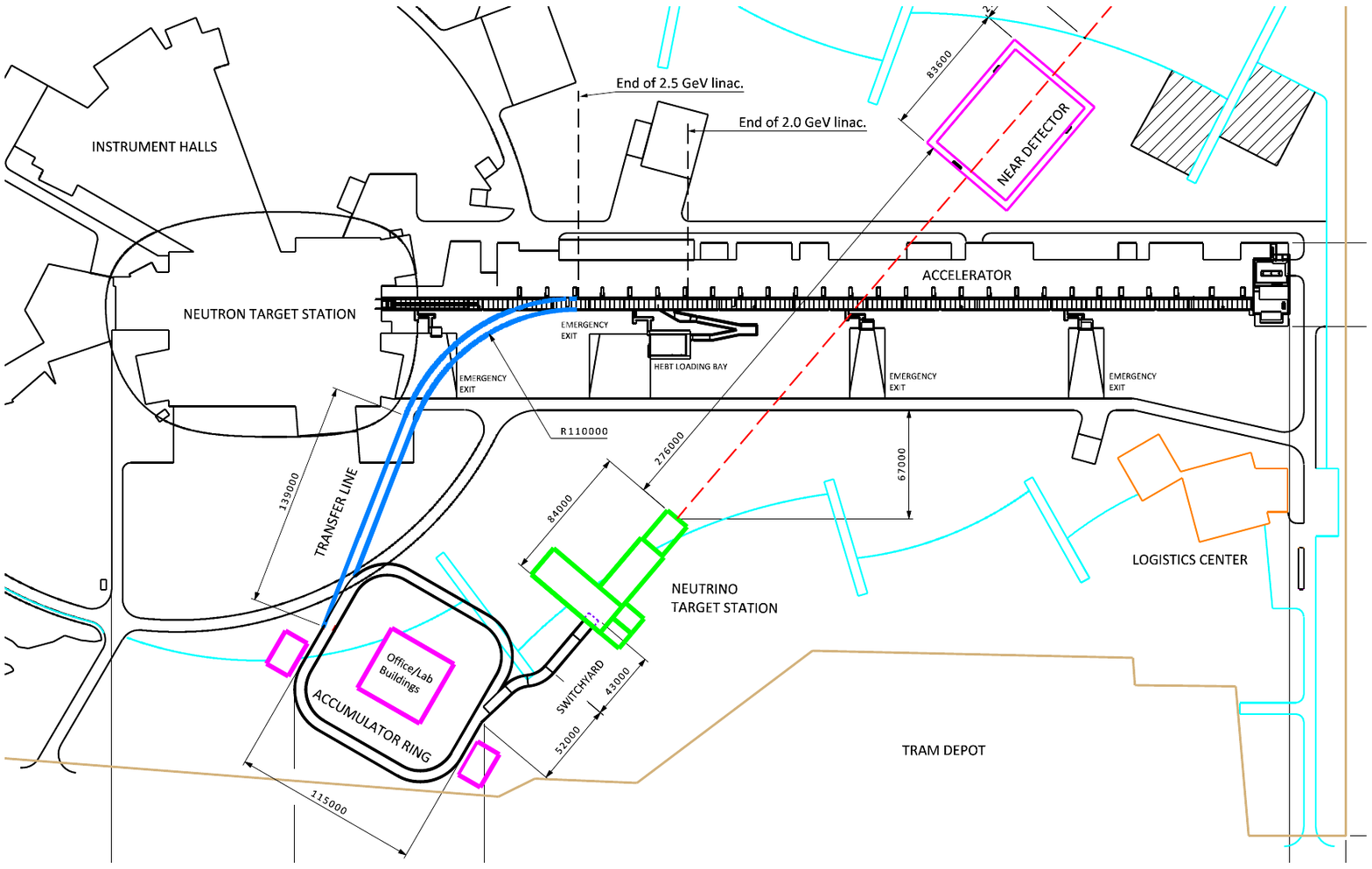}
    \caption{Schematic layout of the required ESS upgrades. The ESS linear accelerator is shown in black, the transfer line from the LINAC to the accumulator ring is shown in blue, the accumulator ring and the switchyard are shown in black, the target station is shown in green, and the near detector site is shown in magenta. (Units in mm)}
    \label{fig:ess-upgrade}
\end{figure}

An initially pure $\nu_{\mu}$ beam is obtained from the decays of charged pions produced in the colliosins of high energy protons with a target. The determination of the leptonic CP violating phase angle $\delta_{CP}$ from the oscillations of such a $\nu_{\mu}$  beam is made by measuring at the baseline distance the absolute number of the appearing $\nu_{e}$ and the shape of  the spectrum of their relative momentum as well as from the comparison of these quantities for a neutrino and an antineutrino beam, respectively. To be able to produce a neutrino beam sufficiently intense for such measurements, concurrently with the intense spallation neutron beam of the ESS, it is necessary to apply a number of upgrades to the ESS facility. The proposed upgrades are schematically presented in Figure~\ref{fig:ess-upgrade}. The pulse frequency of the ESS LINAC (the proton driver) must be increased from \SI{14}{\hertz} to \SI{28}{\hertz} to obtain additional acceleration cycles that will be used for neutrino production, without affecting the neutron programme. Moreover, during neutrino cycles, H$^{-}$ ions instead of protons need to be accelerated in order to ease the filling of the accumulator ring. An accumulator ring will be built to compress the ESS pulse to about \SI{1.2}{\micro\second}. A neutrino production target station, composed of four identical targets enveloped by four magnetic focusing devices (horns), will be built. The horns will be used to sign select and focus the pions, and thus also the neutrinos resulting from their decay, toward the near and far detectors. A near detector will be used to monitor the neutrino beam and to measure neutrino interaction cross-sections, especially electron neutrino cross-sections, at a short baseline from the neutrino source, while the far detector will be used to detect the oscillated neutrino beam at the long baseline distance. In the following sections, these parts are discussed together with the physics potential of the experiment.

\subsubsection{Proton Driver}
\label{protondriver}

\begin{figure*}[hbt]
\begin{center}
\includegraphics[width=1\textwidth]{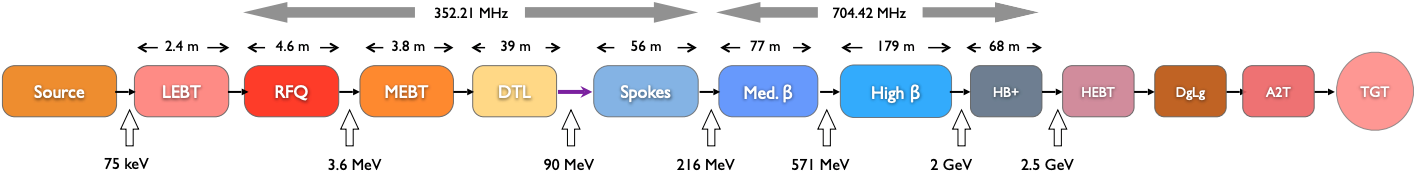}
\caption{\small Proton driver layout.}
\label{fig:protondriver}
\end{center}
\end{figure*}

The ESS LINAC, currently being constructed, will accelerate 14 proton pulses of \SI{2.86}{\milli\second} length per second and \SI{62}{\milli\ampere} current to \SI{2}{\giga\electronvolt} energy, implying the production of a \SI{5}{\mega\watt} proton beam. Figure~\ref{fig:protondriver} shows the layout of the ESS proton driver. The proposed power increase of the LINAC from \SI{5}{\mega\watt} to \SI{10}{\mega\watt} will be realised by increasing the pulse frequency from \SI{14}{\hertz} to \SI{28}{\hertz}, adding 14 more H$^{-}$ pulses of \SI{50}{\milli\ampere} to be accelerated to \SI{2.5}{\giga\electronvolt}, interleaved with the proton pulses. The reason for accelerating H$^{-}$ ions is that they can be stripped of their electrons to inject protons in a manner that increases the phase space density for an existing proton beam, otherwise limited by Liouville's theorem. Each of the H$^{-}$ pulses will be chopped into four sub-pulses separated by gaps of $\sim$\SI{100}{\micro\second} length. In addition, there are \SI{133}{\nano\second} gaps every \SI{1.33}{\micro\second}, corresponding to the revolution period in the ring. The need for these gaps is dictated by the rise-time requirement of the extraction kicker magnets in the accumulator ring and the time needed to reconfigure the ring for the injection of the next sub-pulse.

The total number of particles delivered to the accumulator ring will be $8.9\times 10^{14}$ per pulse cycle (macro-pulse), divided into four batches of $2.2 \times 10^{14}$. Each batch is stacked in the accumulator ring over about 600 turns, compressing the \SI{3.5}{\milli\second}, including gaps, to four \SI{1.2}{\micro\second} pulses, which are subsequently extracted to the target. The acceleration of H$^{-}$ ions requires the addition of a H$^{-}$ ion source to the side of the LINAC proton source and a doubling of the front-end accelerator elements up to the point where the proton and H$^{-}$ beam-lines are merged. For the H$^-$ production, the Penning source~\cite{dudnikov:2012} at ISIS-RAL~\cite{isis} and the RF source at SNS~\cite{shishlo_2012,folsom:2021strp} were identified as the most promising ion sources to meet the ESS$\nu$SB requirements.

A LINAC-to-ring (L2R) transfer line has been designed to transport the \SI{2.5}{\giga\electronvolt} H$^-$ beam output from the upgraded high-$\beta$ line (HBL) at the end of the LINAC to the Accumulator Ring \cite{alekou:2020_D3.2}. Several mechanisms of unwanted H$^-$ stripping mechanisms, that would lead to large beam losses, have been considered in the design~\cite{folsom:2021_hmin_strp}. At beam energies greater than ${\sim}$\SI{100}{\mega\electronvolt}, activation of machine components could become a concern if the loss values would exceed acceptable limits. The beam-loss will be kept below \SI{1}{\watt}/m, which ensures a maximum value of 1~mSv/h ambient dose rate due to activation at \SI{30}{\cm} from a surface of any given accelerator component, after 100~days of irradiation and \SI{4}{\hour} of cool-down~\cite{Tchelidze2019}. Moreover, as mentioned before, several hardware modifications will need to be applied on the ESS LINAC in order to make it able to produce the intense neutrino beam. The modification programme includes, but is not limited to: the upgrade of the low energy beam transport (LEBT), the medium energy beam transport (MEBT), the radiofrequency quadrupole (RFQ), the drift-tube linac (DTL) tank, the Modulator Capacitor and the Cooling System.

\subsubsection{Accumulator Ring}
\label{accumulator}

There are several reasons for having an Accumulator ring to compress the \SI{2.86}{\milli\second} (\SI{3.5}{\milli\second} including beam-free gaps) long ESS linac pulse to \SI{1.2}{\micro\second}. One is that the background in the large Far Detectors caused by neutrinos from cosmic rays would be overwhelming even at \SI{1000}{\meter} below ground level with a data-taking gate of \SI{3}{\milli\second}, whereas this background becomes negligible with a \SI{1.2}{\micro\second} gate. Another is that it is not possible to maintain a flat top longer than a few microseconds of the \SI{350}{\kilo\ampere} pulse that is fed to the neutrino horns 14 times a second because of the very large Joule heating caused by the high current.

The underground accumulator ring, which has a circumference of \SI{384}{\meter}, will receive the four \SI{0.79}{\milli\second} long sub-pulses separated by \SI{100}{\micro\second}, which is the time needed to prepare the injection kicker magnets for the injection of the next sub-pulse. Each sub-pulse will be injected during ca 600 turns and then extracted in one turn, thus producing four ca \SI{1.2}{\micro\second} long proton pulses separated by almost 0.9 ms that will each be sent to one of the four separate targets. The H$^{-}$ pulse, which will be injected into the accumulator ring using so-called phase-space painting, will be stripped at the entrance of the accumulator ring using thin carbon foils. The temperature to which these foils are heated must be kept below \SI{2000}{\kelvin} above which the foil sublimation rate will be too high. To achieve this, the phase-space painting procedure is optimized so that the number of foil crossings from already stripped particles is minimized. This also reduces detrimental effects from space charge in the ring. As a round beam cross-section is desirable when the beam hits the target, anti-correlated painting of the beam has been opted for. The peak temperature in the stripper foil is further reduced by injecting a mismatched beam, to dilute the thermal load from the injected beam, and, by using multiple thin foils instead of a single thicker foil, in order to increase the effective foil surface area. It is planned to investigate, as an alternative to foil stripping, the use of laser-assisted stripping of the incoming H$^{-}$ ions. This method is currently being developed at the SNS in the US~\cite{Gorlov:2019:laserstripping}.

\begin{figure}[h]
  \begin{center}
    \includegraphics[width=0.50\linewidth]{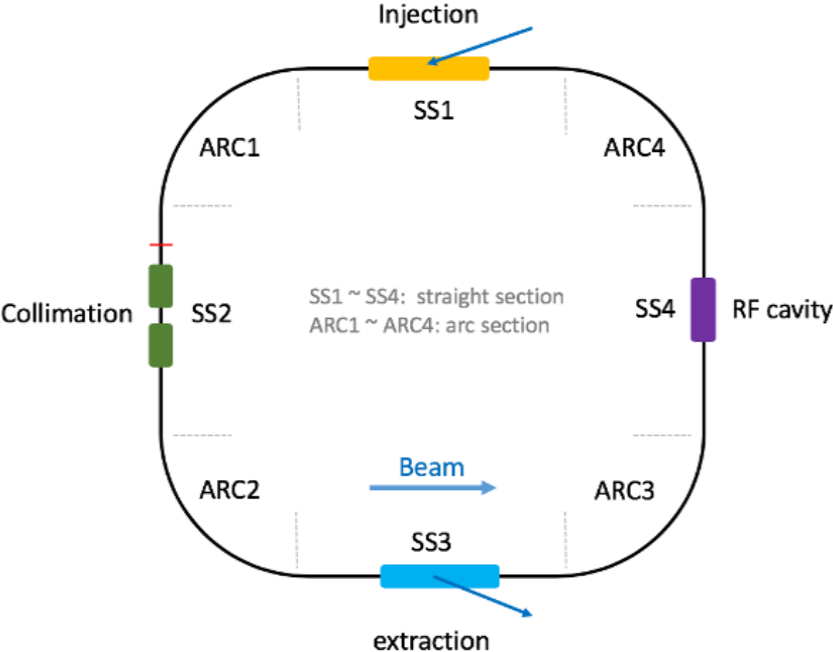}
\caption{The accumulator ring layout.}
    \label{fig:accumulator_layout}
  \end{center}
\end{figure}

The performance of the accumulator ring design has been verified through multi-particle simulations including direct and indirect space charge. The simulations show that a geometric 100$\%$ emittance as low as 60\,$\pi$ mm mrad is achievable. The total tune spread expected is around 0.05, which implies that space charge is not a limiting factor for the accumulator ring design, in the operating scheme that we envision. We chose an accumulator layout that has four rather short arcs connected by relatively long straight sections, see Figure~\ref{fig:accumulator_layout}. The arcs contain four Focusing-Defocusing (FODO) cells, each with two dipole magnets, with the dipole magnet centered between two quadrupole magnets. The number and length of the dipole magnets have been chosen to reach the desired bending radius using a moderate magnetic field strength of \SI{1.3}{\tesla}. The main challenges of the design are to control the beam loss and to find a H$^{-}$ stripping scheme that is reliable over time. The design of a two-stage collimation system has been made to meet these challenges. A barrier RF cavity will be used to contain the beam pulses longitudinally and preserve the \SI{100}{\nano\second} particle-free gap required for extraction. The beam will be extracted from the ring using a set of vertical kicker magnets and a horizontal septum and the four sub-pulses will be guided by a \SI{72}{\meter} transfer line up to the beam switchyard.

\begin{figure}[h]
\centering
\includegraphics[width=0.60\linewidth]{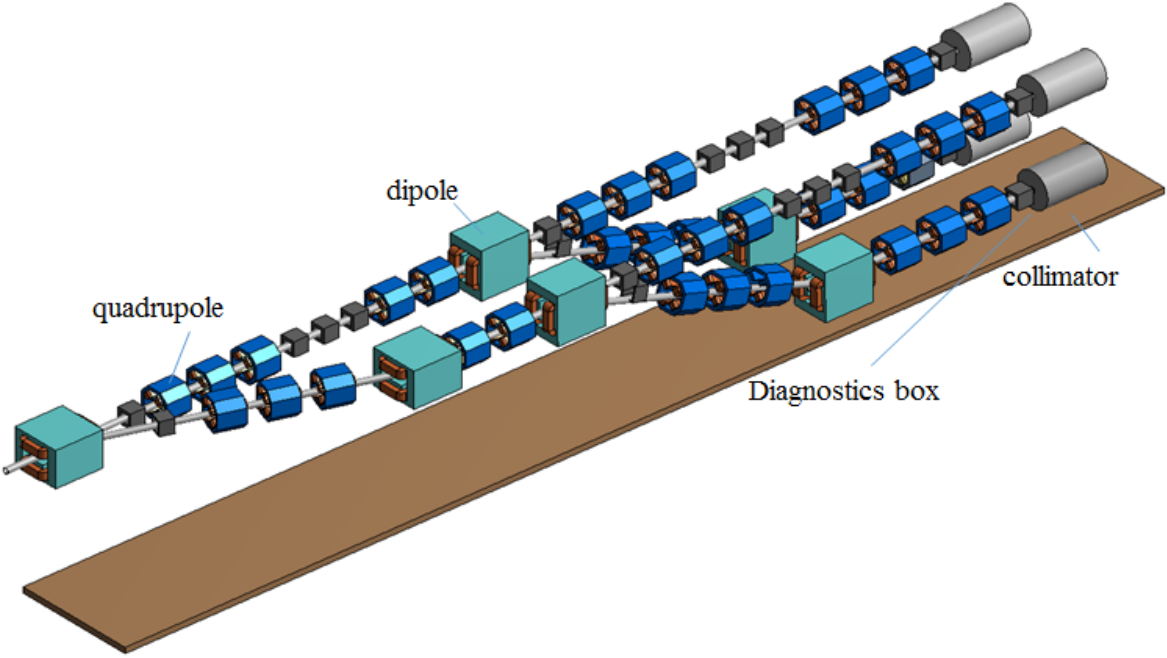}
\caption{\small A 3D isometric view of the beam switchyard.}
\label{fig5.9}
\end{figure}

The beam switchyard directs each compressed sub-pulse to one of the four targets by fast switching of dipole magnets. Figure~\ref{fig5.9} shows a 3D CAD drawing of the \SI{45}{\meter} long switchyard, where the first switching occurs in the horizontal plan and the second level switching is in the vertical plane. Switching during less than \SI{0.9}{\milli\second} between the sub-pulses requires dipole magnets with low inductance. As an alternative, a kicker in combination with a septum dipole will be considered. Four collimators are included at the end of the line to perform a final beam cleaning before it hits the target. Both the transfer line and the switchyard have been designed so as to produce a round beam suitable for the target and for minimal beam loss.

\subsubsection{Short pulses for neutron production}
\label{short_neutrons}
Unlike ESS, the SNS is a short-pulse spallation facility where an accumulator ring is used to produce proton pulses of \SI{1}{\micro\second} duration, which, in turn, are used for generating short pulses containing a large number of spallation neutrons in a liquid mercury target. Such a short proton pulse would induce very high thermomechanical stress on a solid target. In Sweden, where a mercury target is not allowed, a liquid lead target could be considered for a similar scheme. On the other hand, the neutron moderation process will later stretch the initially short pulse to about \SI{100}{\micro\second}. Therefore, as an alternative, the ESSnuSB accumulator could be used in a resonant slow-extraction mode~\cite{McGinnis:2013sza}, where the compressed sub-pulse is extracted over approximately 75 turns. This would produce a \SI{100}{\micro\second} proton pulse, which is then matched to the neutron moderation time, thus substantially reducing the thermomechanical stress on the target. It may be possible to find a solid-target design that can withstand such a pulse. However, resonant slow extraction is an inherently lossy process where high irradiation of the extraction zone is inevitable. The accumulator ring design would have to be carefully studied if single and multi-turn extraction were to be used simultaneously.

\subsubsection{Target Station}
\label{targetstation}

\begin{figure}[h!]
\begin{center}
\includegraphics[width=0.75\linewidth]{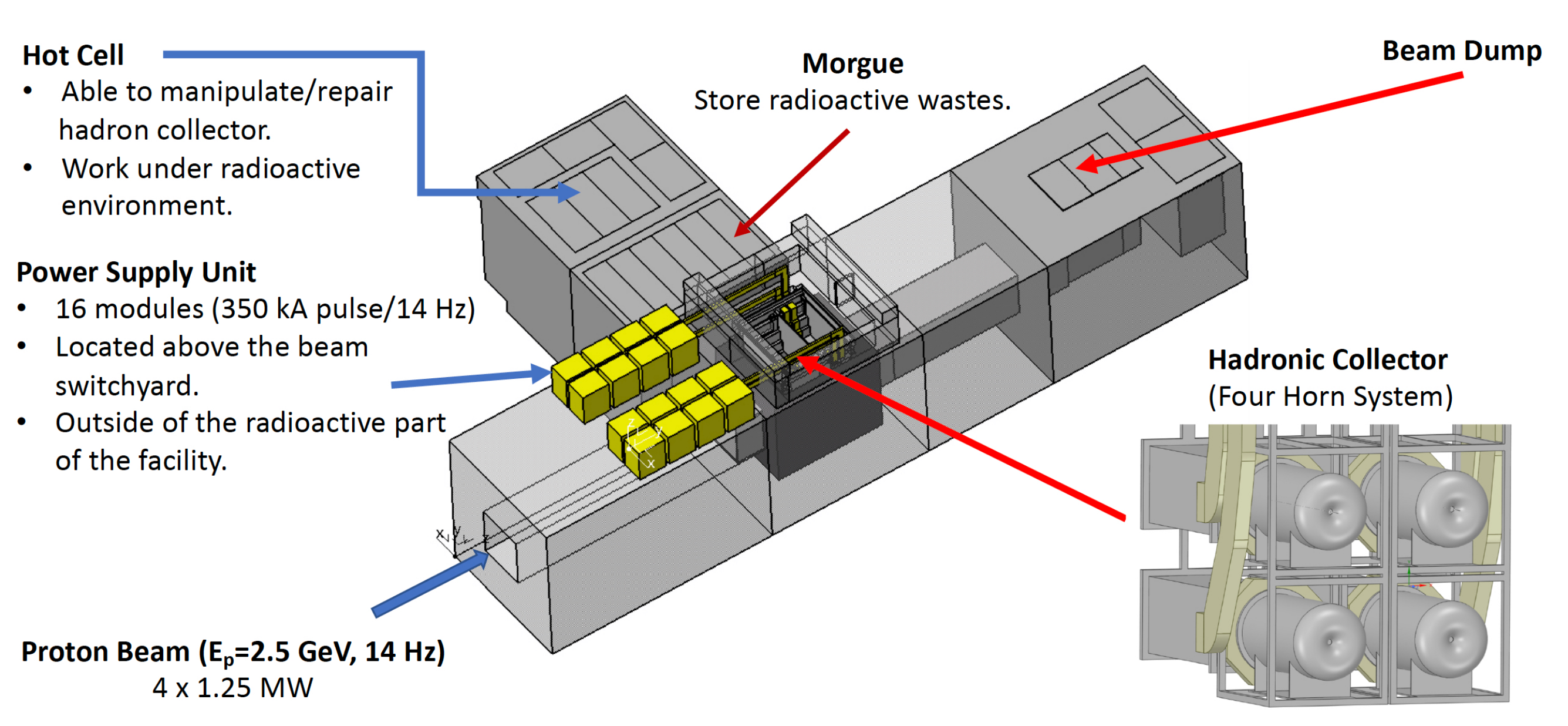}
\caption{Overview of the target station complex with a zoomed view of the 4-horn system.}
\label{fig:TargetStationOverview}
\end{center}
\end{figure}

\begin{figure}[!htbp]
    \centering
           \includegraphics[width=0.4\linewidth]{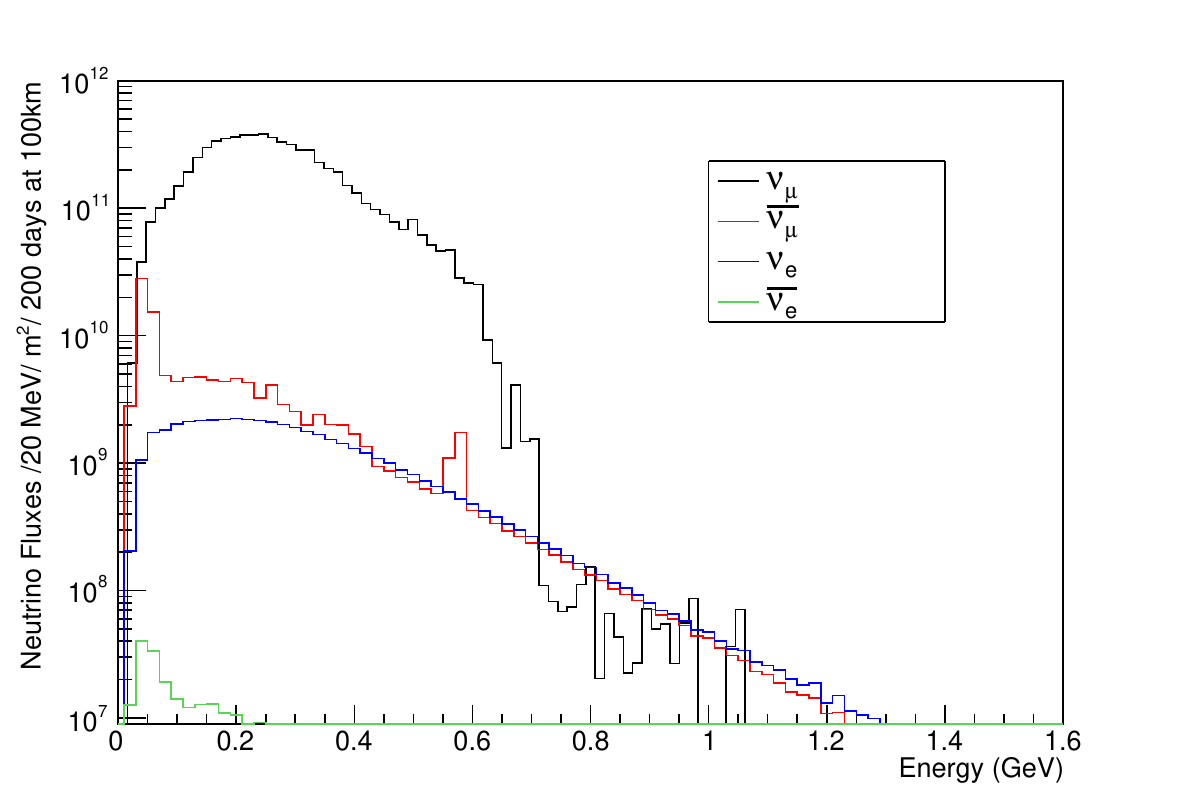}
       \includegraphics[width=0.4\linewidth]{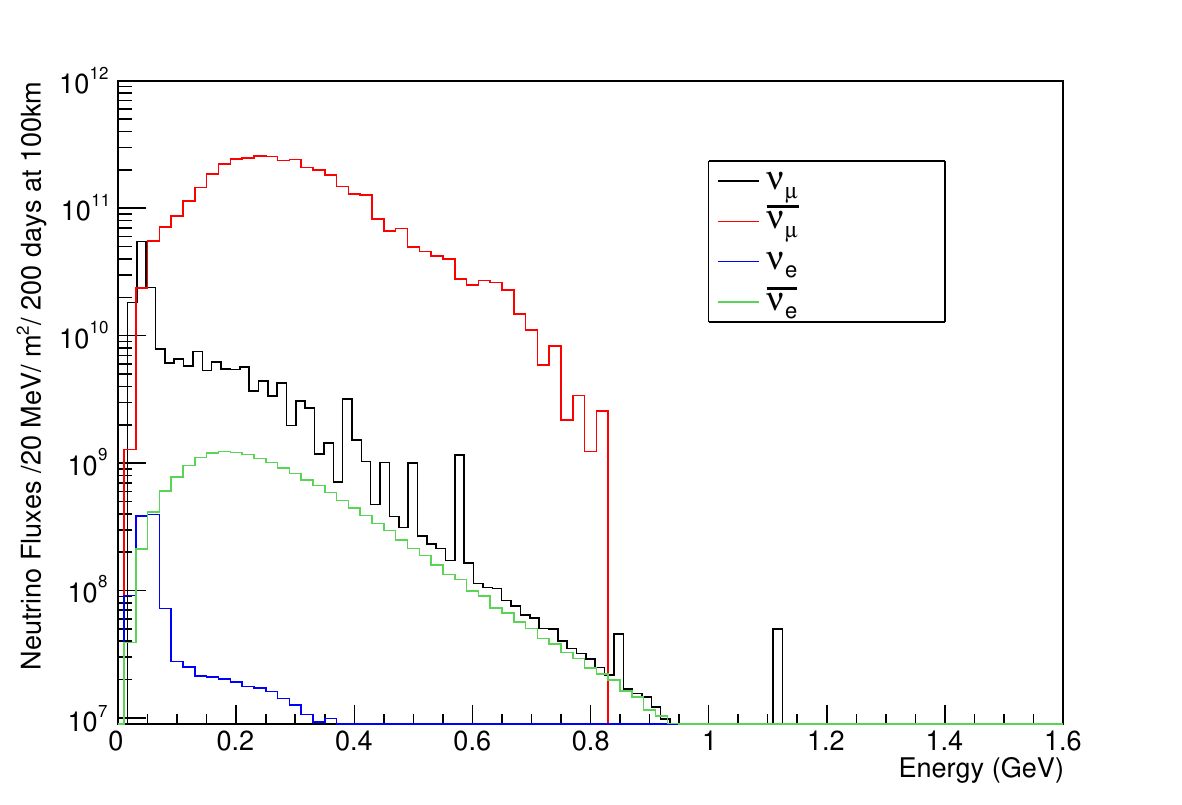} 
    \caption{ESS$\nu$SB neutrino (left) and antineutrino (right) energy spectrum at \SI{100}{\kilo\meter} from the neutrino source.}
    \label{fig:NuFLux1}
\end{figure}

Four identical separate targets will be operated in parallel in order to reduce the beam power that a single target will have to sustain, i.e., from \SI{5}{\mega\watt}/target to \SI{1.25}{\mega\watt}/target. The target design is based on a tube-shaped canister, of ca \SI{3}{\cm} diameter and ca \SI{78}{\cm} length, filled with \SI{3}{\milli\meter} diameter titanium spheres, and cooled using a forced transverse flow of helium gas, pressurised at 10~bar. The primary advantage of such design, in comparison to the monolithic targets, lies in the possibility of making the cooling medium flow directly through the target and by doing so to allow for a better heat removal from the target areas of highest power deposition. Each target is surrounded by a pulsed magnetic horn (mini-Boone type \cite{osti_900360}) providing a strong toroidal field, with a value of the magnetic field strength of $B_{max} = \SI{1.97}{\tesla}$ at peak current. This is required for the focusing of the charged pion beam, produced from the interaction between the impinging protons on the titanium target in the forward direction into a \SI{50}{\meter} long decay tunnel filled with He gas to reduce interactions. The charge sign of the pions being focused will be changed by reverting the direction of the current in the horn. Each horn will have a separate power supply-unit capable of providing a \SI{350}{\kilo\ampere} semi-sinusodial current pulse of order \SI{100}{\micro\second} length, which at a peak current provides a sufficiently constant current within \SI{1.3}{\micro\second}. These current pulses will be sent to the horn through strip lines. Several horn designs were investigated in this study, of which the Van der Meer horn structure \cite{vanderMeer:278088} showed the best performance. Figure~\ref{fig:TargetStationOverview} shows a 3D CAD drawing of the target station complex with a zoom-in on the 4-horn system. The geometry of the horn has been designed and optimised using the so-called genetic algorithm to provide an optimal signal efficiency in the Far Detector. Figure~\ref{fig:NuFLux1} (right) left shows the (anti) neutrino flux distribution at a \SI{100}{\kilo\meter} distance from the neutrino source, resulting from the horn and decay tunnel geometry optimisations.

The magnetic field in each horn is produced by a Power Supply Unit (PSU), which delivers the current pulses to the four horns, synchronised with the proton beam pulses coming from the switchyard. The PSU unit consists of 16 modules connected in parallel and capable of delivering \SI{350}{\kilo\ampere} with \SI{100}{\micro\second} time width pulse at \SI{14}{\hertz} to each horn. Each proton pulse will deliver $2.23 \times 10^{14}$ protons per target and will have a total energy of \SI{89}{\kilo\joule}. Due to the high power and short pulse duration of the proton beam, the ESS$\nu$SB target will be operating under severe conditions. The energy deposition and the resulting displacement per atom (DPA) in the target have been studied using Fluka. The simulations show a total energy deposition of \SI{0.276}{\giga\electronvolt}/pot/pulse, corresponding to \SI{138}{\kilo\watt}, in the target body. The results of the DPA analysis show that ca 8~DPA/yr are produced along the target. If we consider 1~DPA as a reference value, to estimate the lifetime of the target, the previous value of the DPA corresponds to $ 3.024 \times 10^{7} $ pulses.

At the end of the decay tunnel there will be a water-cooled beam-dump (BD) that will absorb the non-interacted primary proton beam and the undecayed muons. The BD core structure will consists of four independent graphite blocks, each block facing one of the four horns. The four segments are supported on a cross-like structure, made of a Copper-Chromium-Zirconium (CuCr1Zr-UNS C18150 \cite{CuAlloy}) alloy, similar to that used for the new PSB \cite{PSBBeamDump} and ESS beam dumps. Each segment is in turn constructed from two zigzag blocks with \SI{1}{\cm} spacing (play) between them, to allow for the thermal expansion of the individual blocks. The \SI{30}{\cm} support plates are used as heat sinks with water channels drilled inside the plate body.



\subsubsection{Detectors}
\label{detectors}
\noindent\textbf{Near Detector}

\begin{figure}[H]
    \centering
       \includegraphics[width=0.45\linewidth]{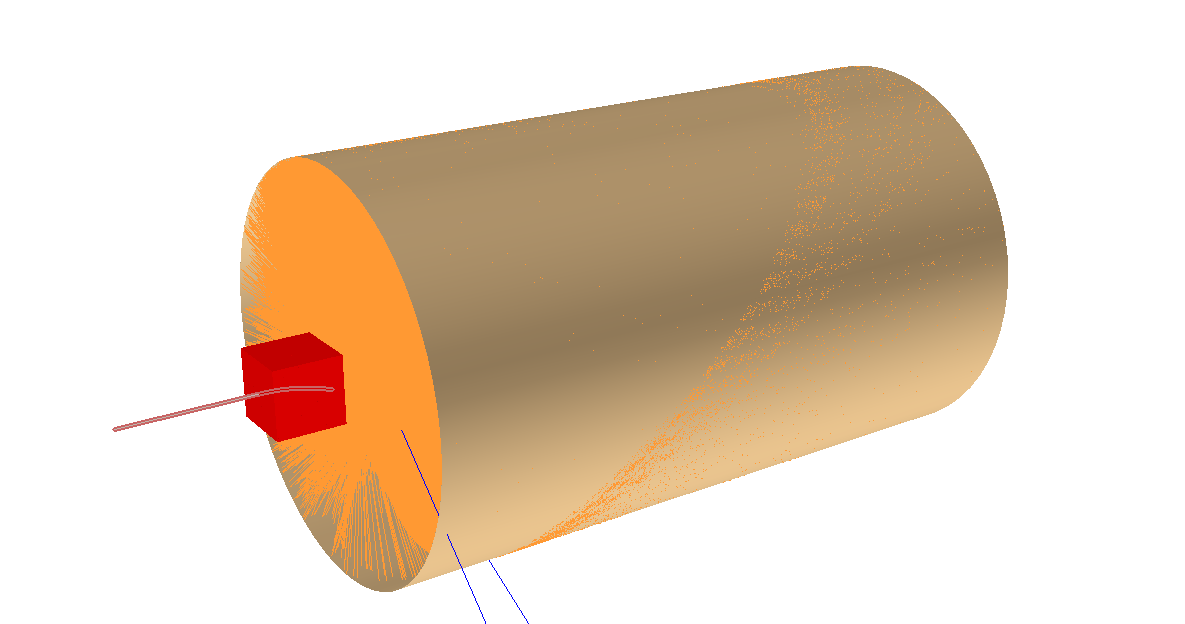}
       \includegraphics[width=0.45\linewidth]{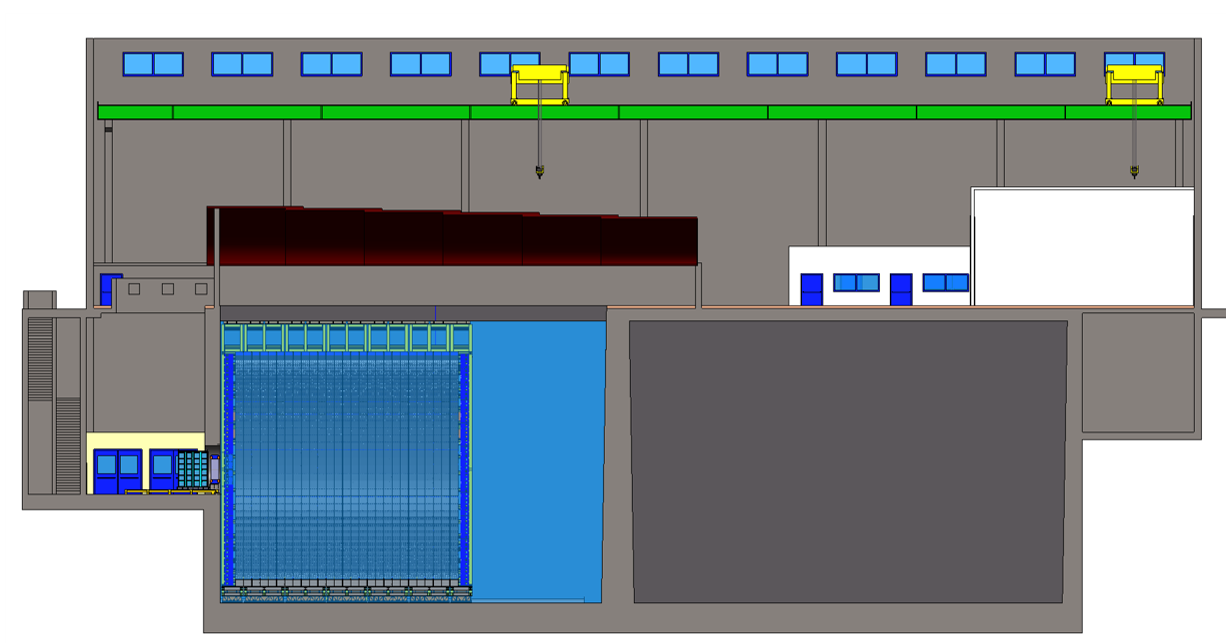} 
    \caption{The ESS$\nu$SB near detector layout. (Left) An artistic view of the detector, with an event in the sFGD. The trajectory of a positive muon (red) in the sFGD (bended by the magnetic field) and the two neutrinos (blue) from its decay in the WatCh. Emitted Cherenkov photons in the WatCh are shown in orange. (Right), an engineering design of the detector and the cavern.}
    \label{fig:detectors:NDcomplex}
\end{figure}

The purpose of the near detector is to monitor the neutrino beam intensity and to measure the muon- and electron-neutrino cross sections, in particular their ratio, which is important for minimizing the systematic uncertainties in the experiment. The near detector will be located underground within the ESS site ca \SI{250}{\meter} from the target station. It will be composed of three coupled detectors: A kiloton mass \textbf{Water Cherenkov detector (WatCh)}, which will be used for event rate measurement, flux normalisation and for event reconstruction comparison with the far detector, a \textbf{magnetised super Fine-Grained Detector (sFGD)} \cite{Sgalaberna:2017}, located inside a \SI{1}{\tesla} dipole for measurements of the poorly known neutrino cross sections in the energy region below \SI{600}{\mega\electronvolt}, and placed upstream of, and adjacent to the water volume, and \textbf{an emulsion detector} of similar type as in the NINJA experiment \cite{NINJA_2020}. Figure~\ref{fig:detectors:NDcomplex} (right) shows the layout of the near detector complex.

\begin{figure}[!htbp]
    \centering
     \includegraphics[width=0.55\linewidth]{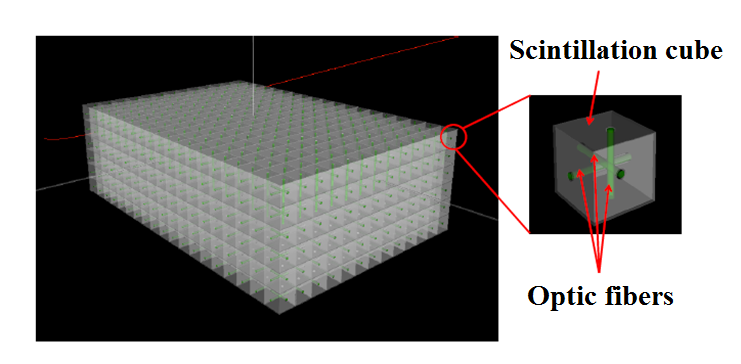}
     \includegraphics[width=0.33\linewidth]{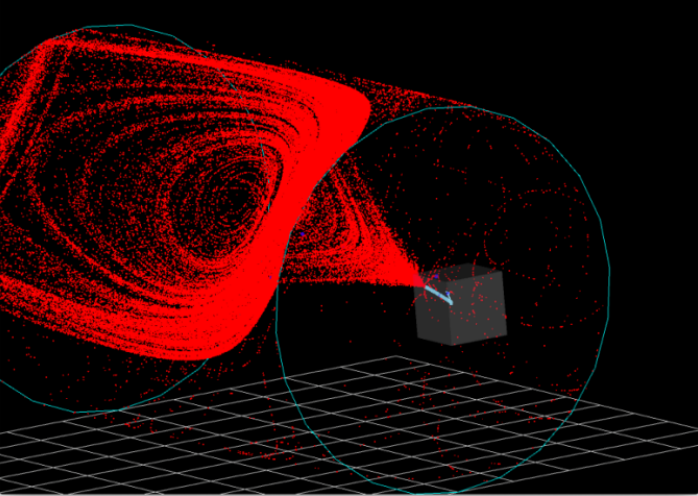}
    \caption{(Left) design of the Super Fine-Grained Detector with three-dimensional read-out. (Right) An $\nu_{\mu}$ interaction event in the SFGD cube with a secondary muon producing a Cherenkov light in the near water Cherenkov detector.}
    \label{fig:detectors:SFGD-1}
\end{figure}

The WatCh water-tank will be a horizontal cylinder of $\sim$~\SI{11}{\meter} length and \SI{4.72}{\meter} radius (total volume $767 m^3$) with the inner walls having a 30\% coverage of 3.5~inch Hamamatsu R14689 photomultiplier tubes (PMTs). The WatCh detector consists of modules, each of them housing 16 PMTs. There are in total 1258 modules altogether containing 20128 PMTs. The 1.4 $\times$ 1.4 $\times$ 0.5~$m^{3}$ sFGD, located in front of the WatCh detector, will be composed of $10^{6}$ \SI{1}{\cm}$^{3}$ plastic scintillator cubes read out by three-dimensional pattern of wave-length-shifting optical fibres (Figure~\ref{fig:detectors:SFGD-1} (left)). The thickness of \SI{0.5}{\meter} along the beam axis is chosen in order to have a sufficient number of charged leptons that continue into the WatCh and thus to allow the combination of the information from the two  detectors. A magnetic field of up to \SI{1}{\tesla} and perpendicular to the beam is applied in the tracker by a dipole magnet. In front of the tracker, a NINJA type emulsion detector is situated that will be used for cross section measurements. The NINJA detector is an emulsion-based detector with a water target, currently under operation at J-PARC with the T2K near detectors. Its primary purpose is to measure precisely the neutrino interaction topology, double differential cross sections, and search for sterile neutrino \cite{NINJA_2020,NINJA_2017,Oshima_2022}. Members of the NINJA Collaboration are prepared to add a similar detector to the ESS$\nu$SB suite of the near detector.

\noindent\textit{\textbf{sFGD and WatCh combined analysis}}

Figure~\ref{fig:detectors:SFGD-1} (right) shows an example of what is here called a cross-over event in which a muon neutrino interaction in the sFGD and the secondary muon continues into the WatCh and produces Cherenkov light. Around 12\% (20\%) of positive (negative) muons produced in the sFGD will continue into the WatCh and be detected there. For electron neutrino interactions such events represent about 6\% of the sample. For the crossover events we aim at a good purity of the electron neutrino event sample by efficiently rejecting events originated in the sFGD that have muons continue into the WatCh by exploiting the sFGD and WatCh ID capabilities together, which will be used for electron neutrino cross section measurement.

 \begin{figure}[!htbp]
    \centering
     \includegraphics[width=0.45\linewidth]{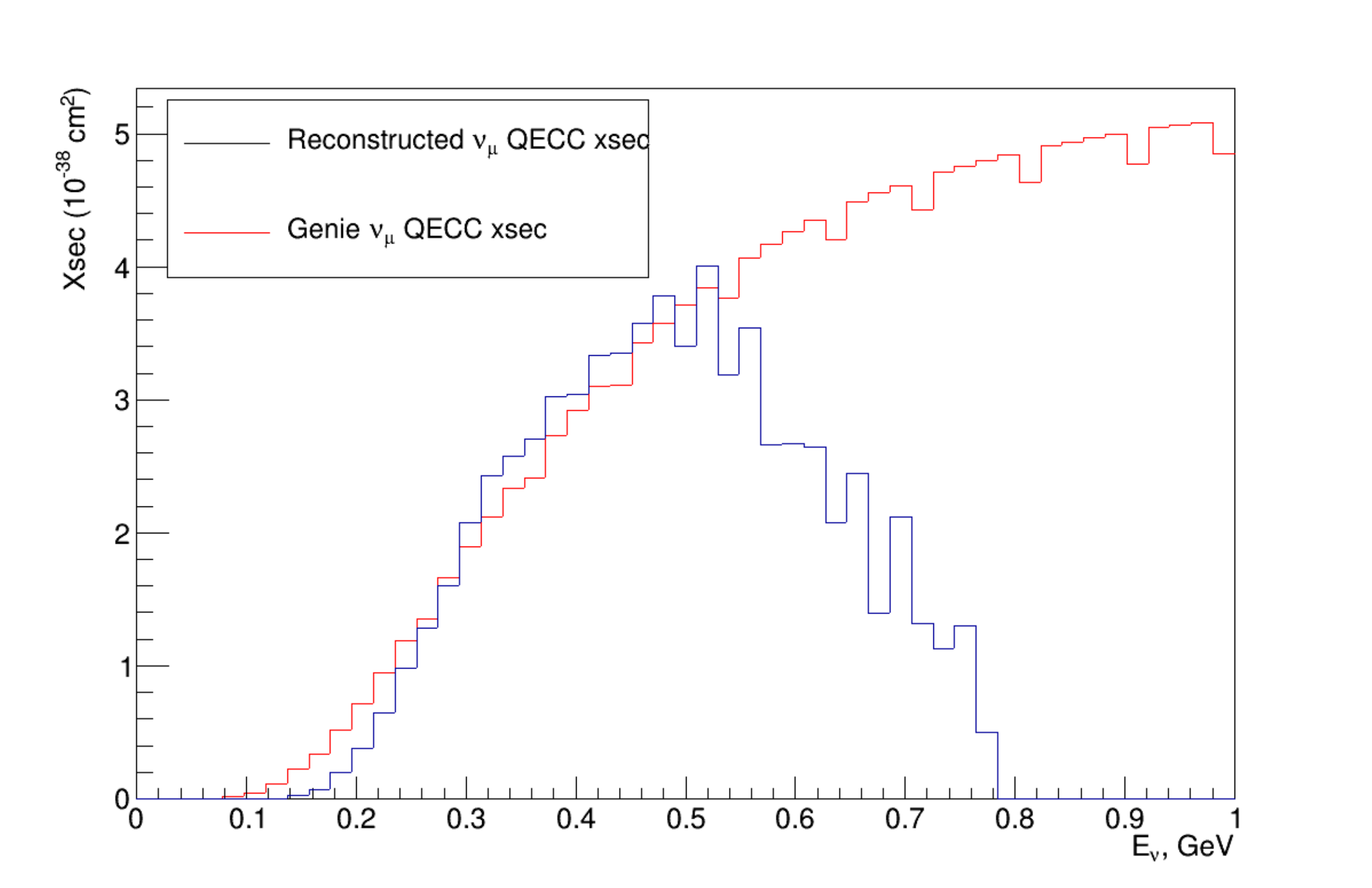} 
     \includegraphics[width=0.45\linewidth]{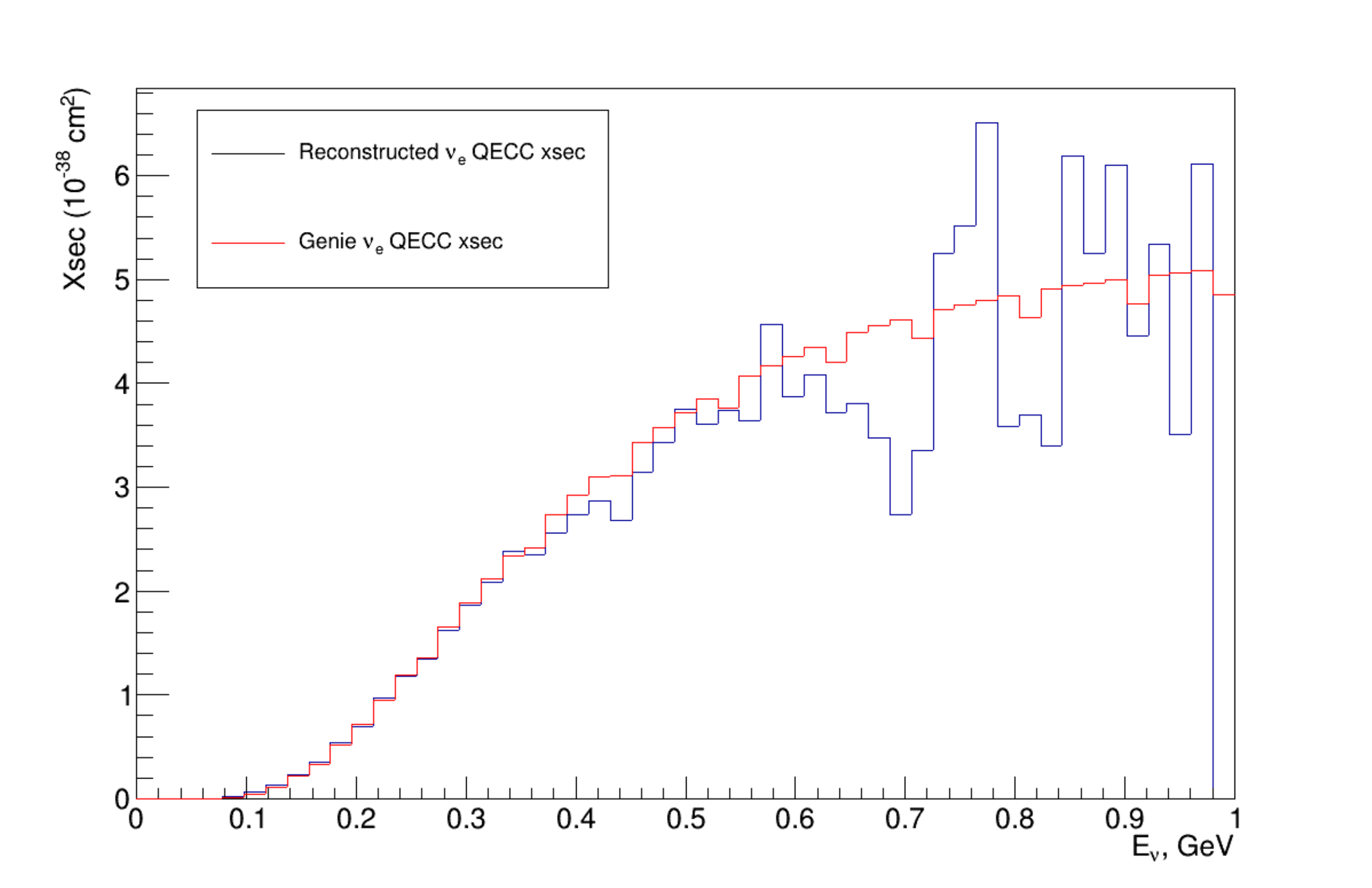} 
     \caption{Left: ``Measured'' $\nu_\mu$ cross section (blue) compared to the cross section used by GENIE (red). Right: ``Measured'' $\nu_e$ cross section  (blue) compared to the cross section used by GENIE (red). }
    \label{fig:detectors:SFGD_xsec}
\end{figure}

\noindent\textit{\textbf{Neutrino cross section measurements at near detector}}

\begin{figure}[!htbp]
    \centering
     \includegraphics[width=0.5\linewidth]{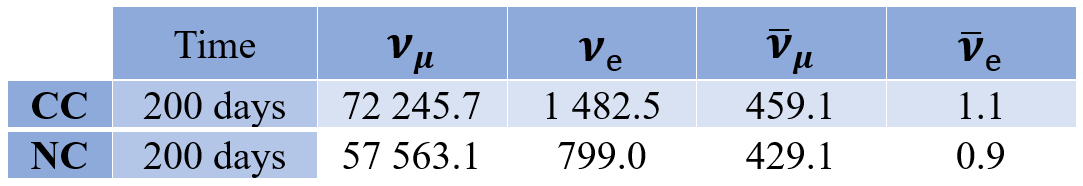}
     \includegraphics[width=0.5\linewidth]{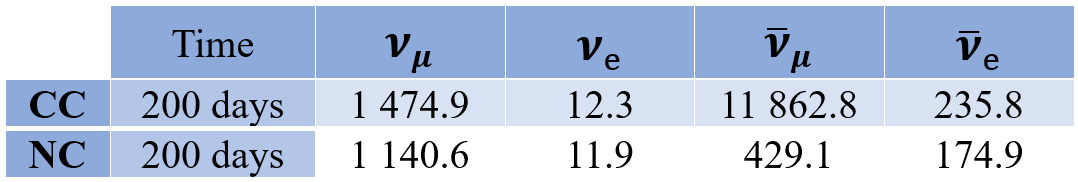}
    \caption{Number of expected interactions in the sFGD: (top) positive, i.e. $\nu$, horn polarity. (Bottom) negative, $\overline{\nu}$, horn polarity.}
    \label{fig:detectors:SFGD_rate}
\end{figure}

As pointed out above, one of the tasks of the sFGD is to measure neutrino cross sections. In Figure~\ref{fig:detectors:SFGD_xsec}, ``measured'' $\nu_\mu$ and $\nu_e$ cross sections are compared with the ones used for simulation of neutrino events. Good agreement up to $E_{\nu} \sim 500$ MeV is observed in both cases. Above that, the discrepancy is big, especially in the muon case. The reason is related to the limited statistics due to the limited number of neutrinos above this momentum in our low-energy neutrino beam. The number of expected (anti)neutrino interactions in the sFGD, obtained by using GENIE neutrino event generator \cite{GENIE:2010, GENIE:2015, GENIE:2021}, is given in the tables in  Figure~\ref{fig:detectors:SFGD_rate} (bottom) top.\\

\noindent\textbf{The Far Detector}

The water Cherenkov Far Detector, which will detect the rate and energy distributions of the muon- and electron-neutrinos, respectively, will be composed of two vertical cylinders of ca \SI{74}{\meter} in height and ca \SI{74}{\meter} in diameter, installed in caverns ca \SI{1000}{\meter} under the ground level to protect them from the cosmic radiation background. Two locations for the far detector have been under consideration in the design study, both near the position of the second $\nu_\mu-\nu_e$ oscillation maximum, thereby resulting in a majority of the events being collected at the second oscillation maximum. One location is in the Zinkgruvan mine, \SI{360}{\kilo\meter} from the ESS, and the other is in the Garpenberg mine, \SI{540}{\kilo\meter} from the ESS. Both mines are active, presenting the advantage of local services like access shafts and declines, ventilation, drainage and other services that are in operation. The locations of the detector caverns are planned to be outside the region of exploitable ore, but still accessible from the mine drifts and at a sufficient distance (a few hundred meters) from the mining activity areas to avoid any kind of mutual disturbance. After careful investigation of the physics performance of the two Far Detector positions, Zinkgruvan has be selected as having the preferred position as discussed in Section~\ref{physicspotential}. The detailed design of the cavern locations in the mine must be preceded by core drilling to enable measurements of the rock strength and pressure in the planned underground regions.The photomultiplier tubes coverage of the ca \SI{25000}{\meter}$^{2}$ inner detector walls will be 40\% (requiring, e.g., ca 50,000 PMTs of 20 inch diameter). The water in the detector tanks will be purified using an industrial-sized water-cleaning plant in order to achieve about \SI{100}{\meter} absorption length for the light wavelength that the PMTs are sensitive to. The ability to dissolve gadolinium salt for increased neutron detection efficiency, and thereby of electron-antineutrinos, will be included.

\subsection{Optimizing the neutrino baseline}
\label{physicspotential}

\begin{figure}[H]
\centering
\includegraphics[width=4.cm]{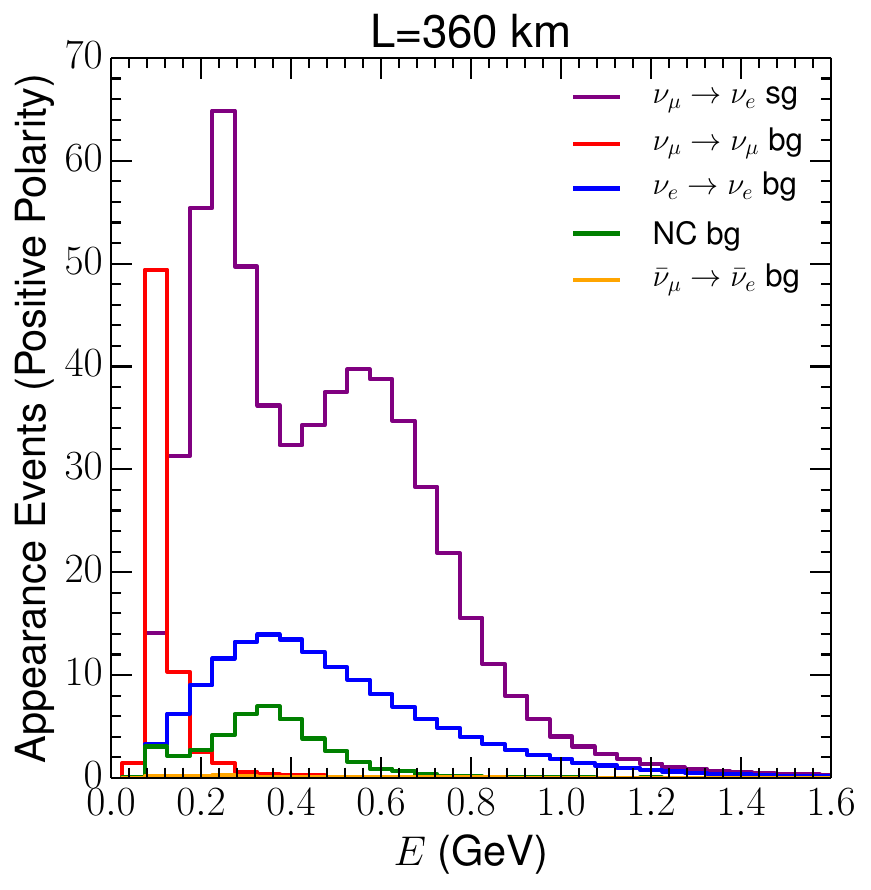}
\includegraphics[width=4.cm]{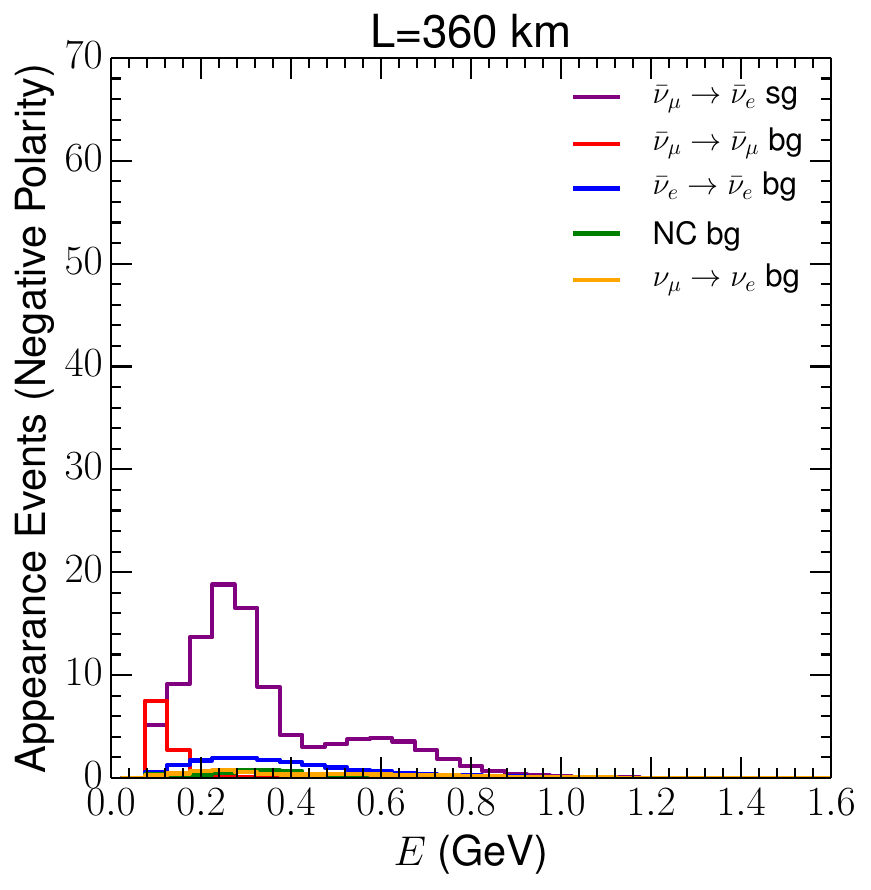}

\includegraphics[width=4.cm]{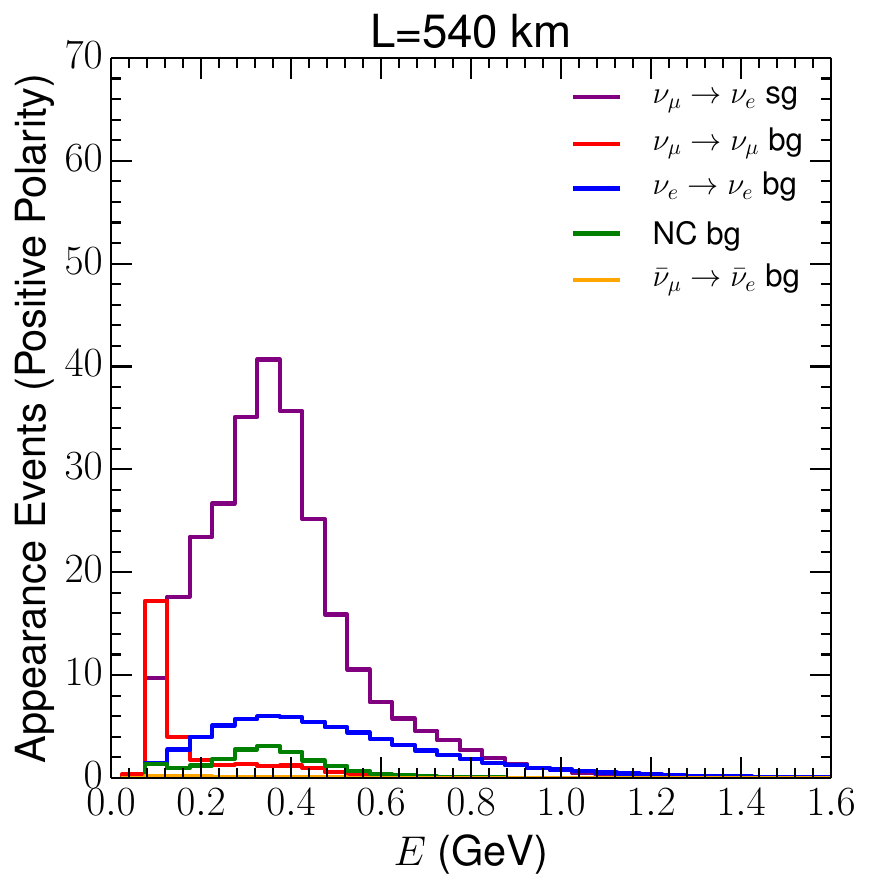}
\includegraphics[width=4.cm]{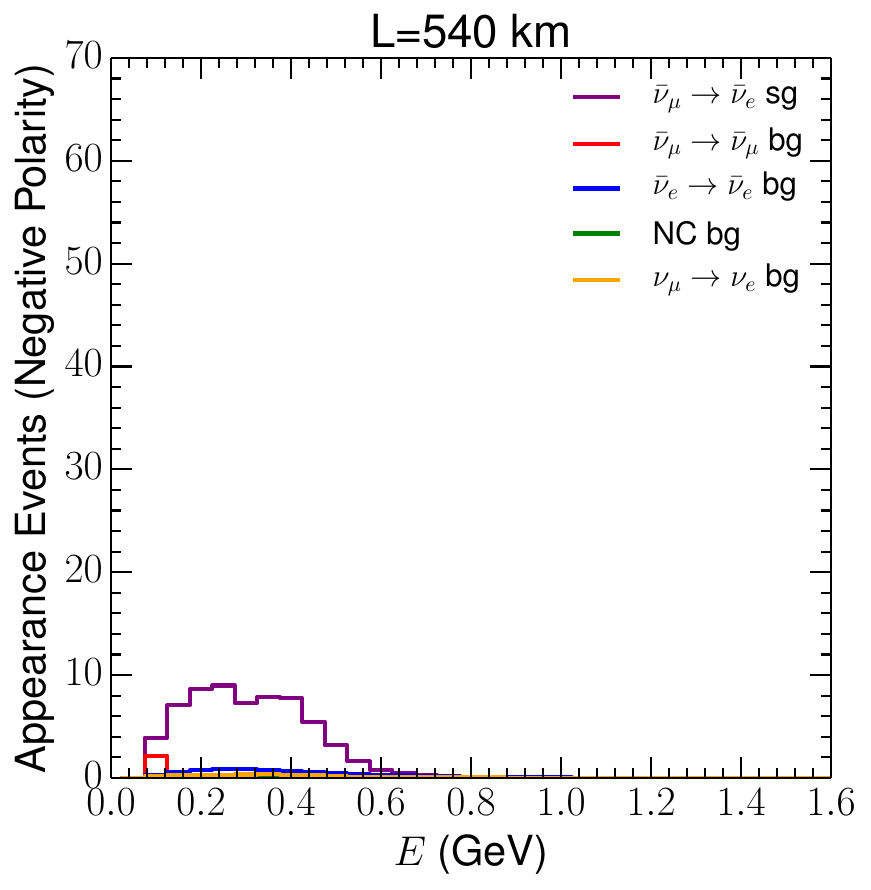}
\caption{Event rates for the different signal and background components for the e-like sample with positive (left panels) and negative focusing (right panels) and for the Zinkgruvan \SI{360}{\kilo\meter} (upper panels) and Garpenberg \SI{540}{\kilo\meter} baselines.}
\label{fig:rates}
\end{figure}

The preliminary optimisation studies presented in Ref.~\cite{Baussan:2013zcy}, as well as the follow-up studies~\cite{Agarwalla:2014tpa,Chakraborty:2017ccm,Chakraborty:2019jlv,Ghosh:2019sfi,Blennow:2019bvl,ESSnuSB:2021azq}, of the physics reach of the ESS$\nu$SB facility for leptonic CP-violation, demonstrate that the best baseline at which to study the neutrino beam would be between \SI{350}{\kilo\meter} and \SI{550}{\kilo\meter}. This makes the ESS$\nu$SB design unique, as the neutrino flux observed by the detector mainly corresponds to the second maximum of the $\nu_\mu \to \nu_e$ oscillation probability, with a more marginal contribution of events at the first oscillation peak. However, as discussed before, there is a price to pay in order to observe the oscillation probability at its second maximum. Even though it is the optimal choice to maximise the dependence of the oscillation probability on the $\delta_{CP}$, the ratio of the oscillation baseline to the neutrino energy ($L/E$) needs to be a factor 3 larger compared to the first maximum. This means that the statistics will be about an order of magnitude smaller than if the detector had been located at the first oscillation peak. Furthermore, the neutrino cross section decreases and beam spread increases for smaller neutrino energies. The \SI{360}{\kilo\meter} baseline option, corresponding to a point between the first and the second oscillation maxima as seen in Figure.~\ref{fig:probs}, represents a compromise between the two choices. The neutrino flux would be 2.25 times larger than that at the \SI{540}{\kilo\meter} option and roughly the same number of events belonging to the second oscillation peak would be observed at either site. At the higher energy end of the spectrum, events from the first oscillation maximum would also be observed for the shorter baseline. This can be seen in the event rates expected at each of the two detector locations depicted in Figure~\ref{fig:rates}. In any case, for the ESS$\nu$SB the choice close to the second oscillation maximum was shown to be optimal for its increased dependence on $\delta_{CP}$ and despite the reduced event rate.

\subsection{Physics reach of the ESS$\nu$SB experiment}
\label{ESSnuSBPhysics}

\begin{figure}[H]
\centering
\includegraphics[width=5.5cm]{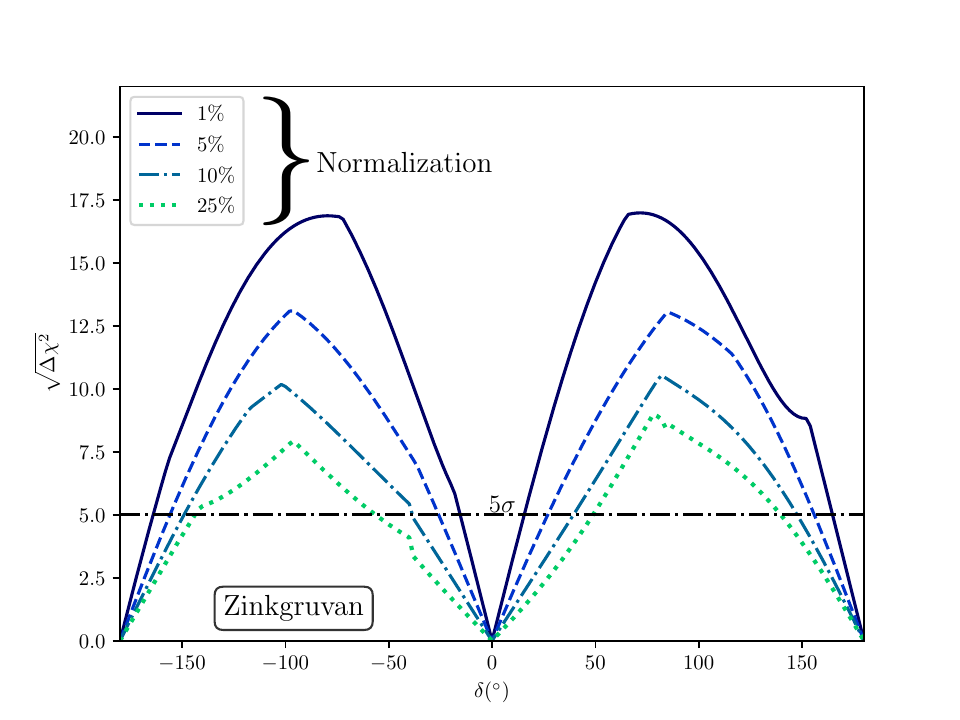}
\includegraphics[width=5.5cm]{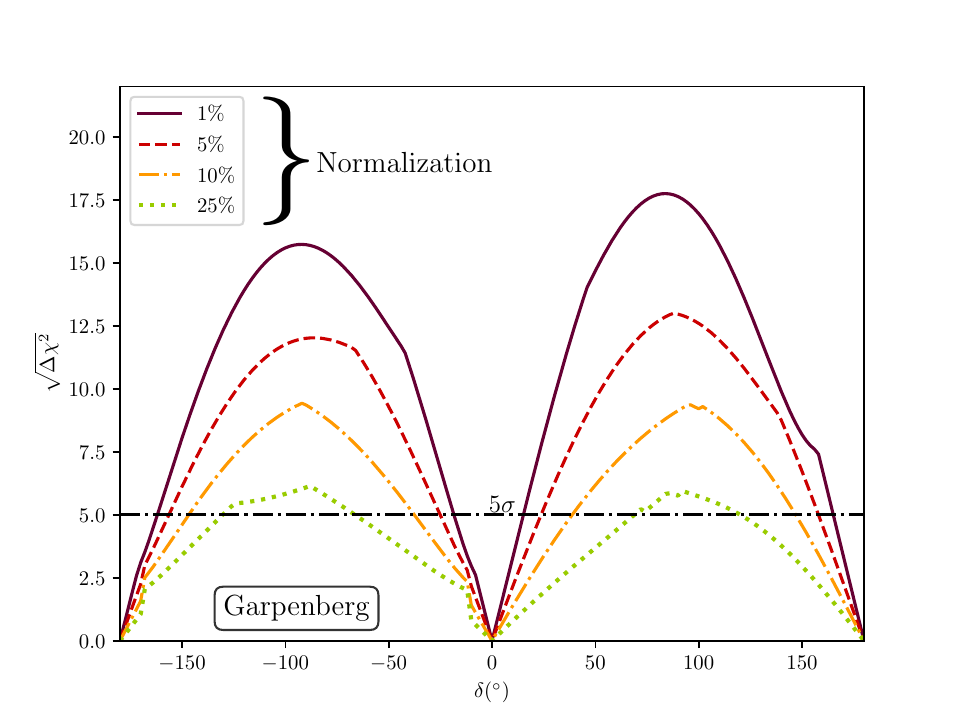}
\includegraphics[width=5.5cm]{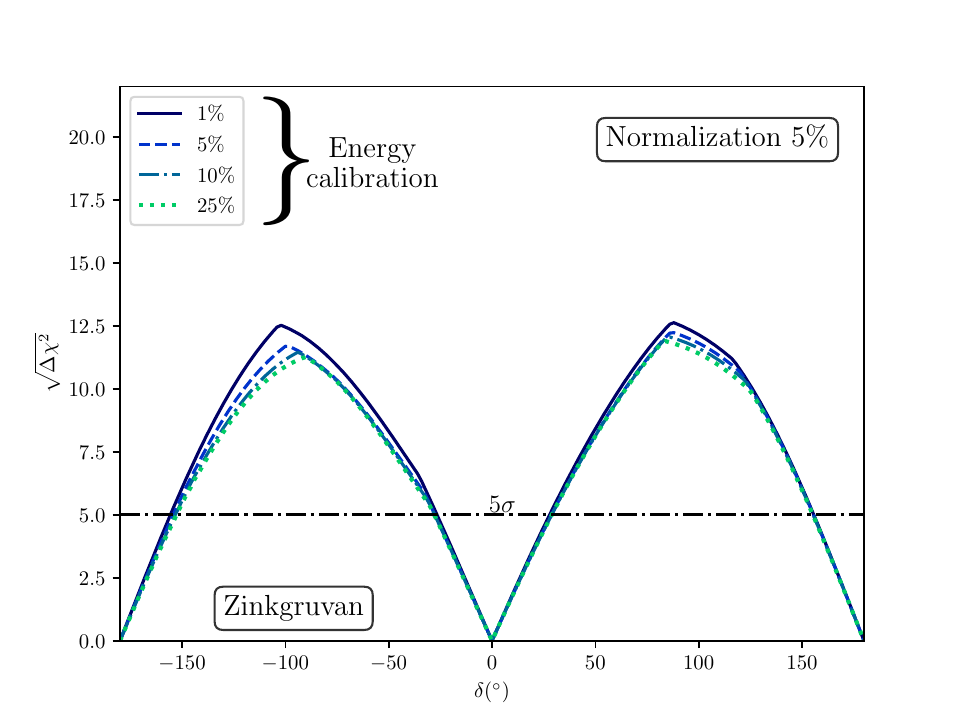}
\includegraphics[width=5.5cm]{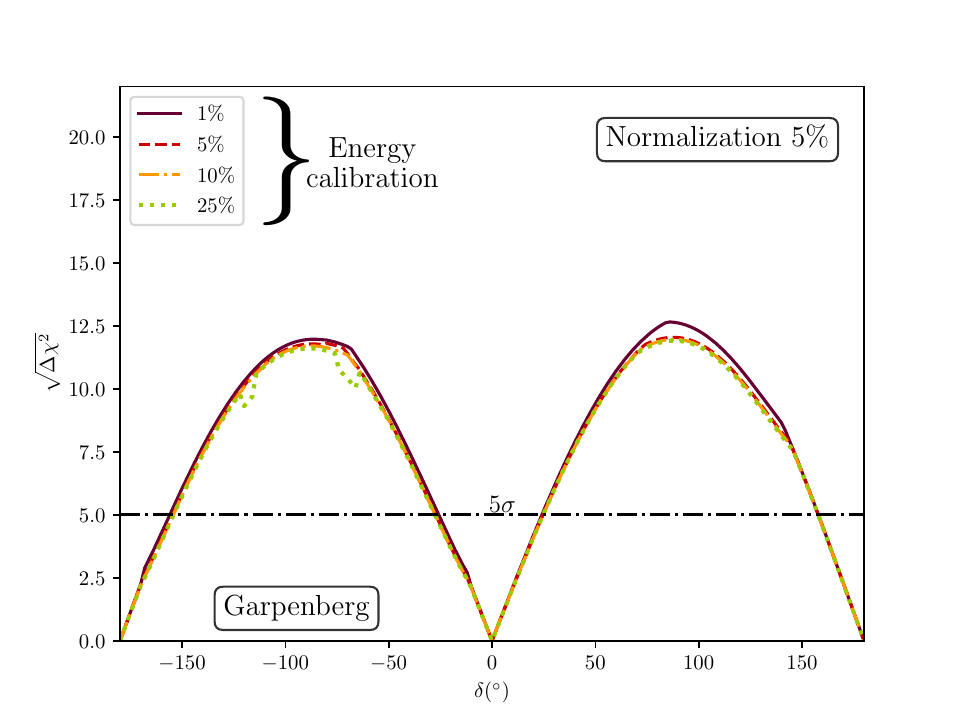}
\includegraphics[width=5.5cm]{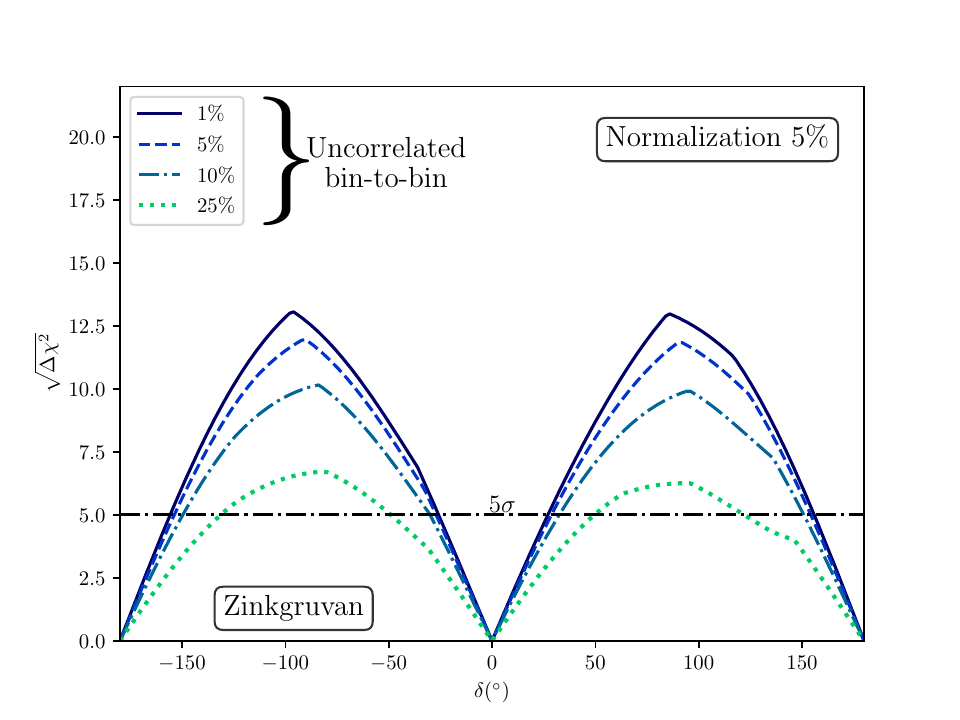}
\includegraphics[width=5.5cm]{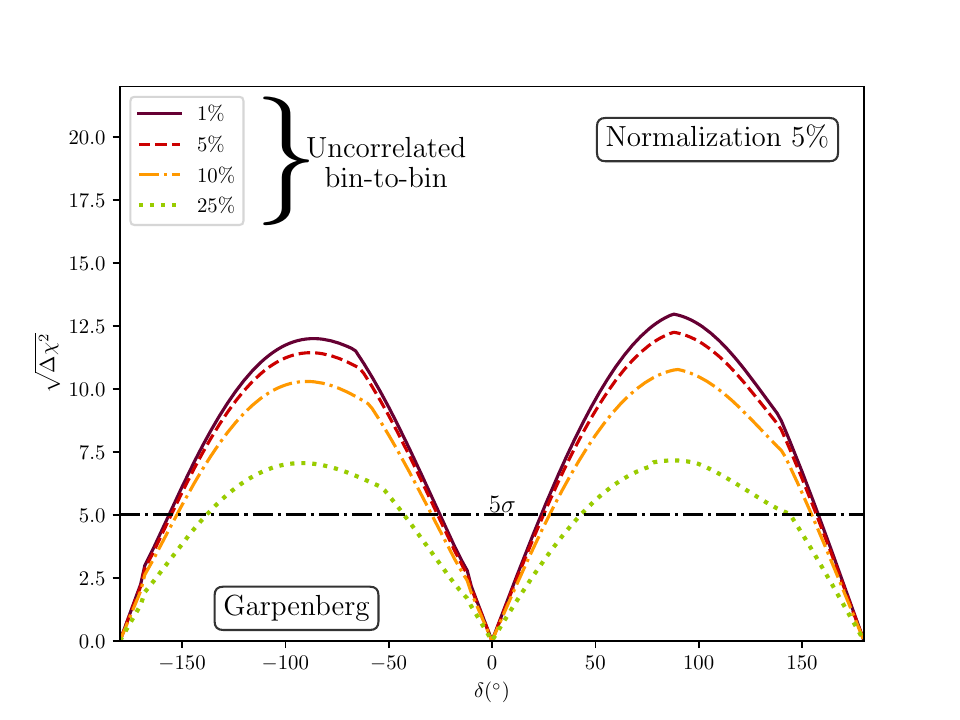} 
\caption{CP discovery potential of the ESS$\nu$SB. Significance in number of standard deviations on the vertical axis and value in degrees of the CP phase angle on the horizontal axis. Left (right) panels for the Zinkgruvan \SI{360}{\kilo\meter} (Garpenberg \SI{540}{\kilo\meter}) baseline.}
\label{fig:cpsens}
\end{figure}

Figure~\ref{fig:cpsens} shows the CP discovery potential obtained for the two baselines for different assumptions about the size of the systematic uncertainties. Generally, we find that the performance of both baselines is very similar, with only slightly different areas covered above the 5~$\sigma$ mark, depending on the systematic uncertainties considered. In the upper panels, the impact of an overall normalization uncertainty, uncorrelated among the different signal and background samples of $1\%$, $5\%$, $10\%$ and $25\%$ uncertainties, is shown.

It is remarkable that, even for an extremely large uncertainty value of $25\%$, a significant portion of the values of $\delta_{CP}$ would still allow a discovery of CP violation above the 5~$\sigma$ level. In the simulation, this uncertainty is uncorrelated between the neutrino and antineutrino samples and therefore is able to mimic the effect of CP violation. Nevertheless, a discovery would be possible even in this scenario. This can be understood from Figure~\ref{fig:probs} where, close to the second oscillation peak, changes in $\delta_{CP}$ can lead to changes in the probability, even above the $25\%$. The dependence of the shape of the oscillation probability on the value of $\delta_{CP}$ may also contribute to the sensitivity. In the middle and lower panels of Figure~\ref{fig:cpsens}, we also explore how robust the results obtained are against other systematic uncertainties that might affect the shape of the measured spectrum. In particular, in the middle panels, the impact of a $1\%$, $5\%$, $10\%$ and $25\%$ uncertainty in the energy calibration is shown. In the lower panels, a more general bin-to-bin uncorrelated set of nuisance parameters has been considered. In both cases, a $5\%$ normalization uncertainty has been added to allow the possible interplay between the different sets of systematic uncertainties that may be relevant. We find that the energy calibration uncertainty has a rather minor impact in the CP discovery potential for both baselines under study. Conversely, the more general implementation of uncorrelated uncertainties in each bin can have a more significant impact, but still a smaller one than the overall normalization considered in the upper panels. The results demonstrate that the ESS$\nu$SB setup has remarkable CP discovery potential even for very conservative assumptions on the systematic uncertainties that could affect the far detector. Indeed, the measurements of the near detector will keep these uncertainties at the few $\%$ level, for the present generation of neutrino oscillation experiments. In particular, assuming a $5\%$ normalization uncertainty in line with assumptions made for similar facilities, we find that CP violation could be established for a $71\%$ ($73\%$) of the values of $\delta_{CP}$ for the \SI{360}{\kilo\meter} (\SI{540}{\kilo\meter}) baseline.

\begin{figure}[H]
\centering
\includegraphics[width=5.cm]{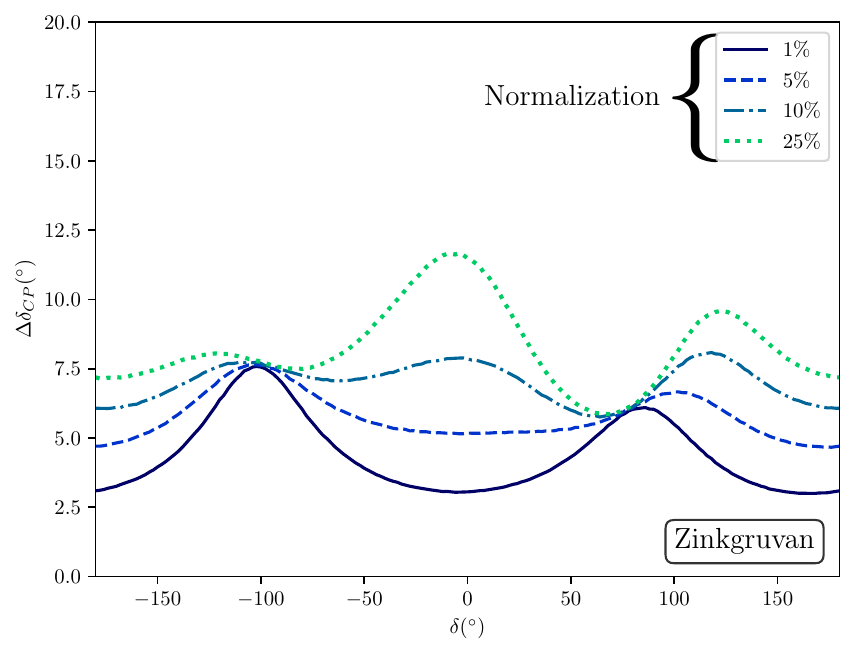}
\includegraphics[width=5.cm]{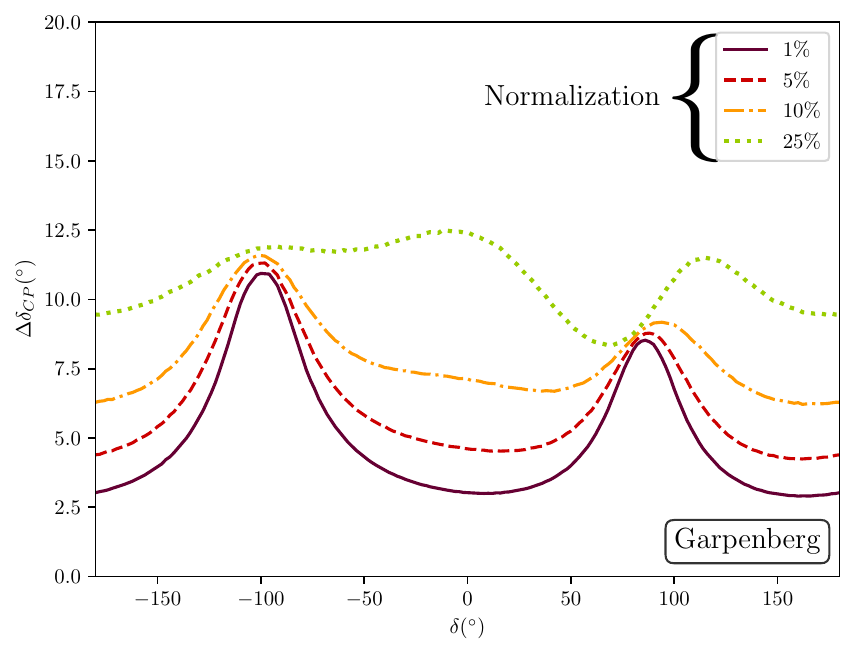}

\includegraphics[width=5.cm]{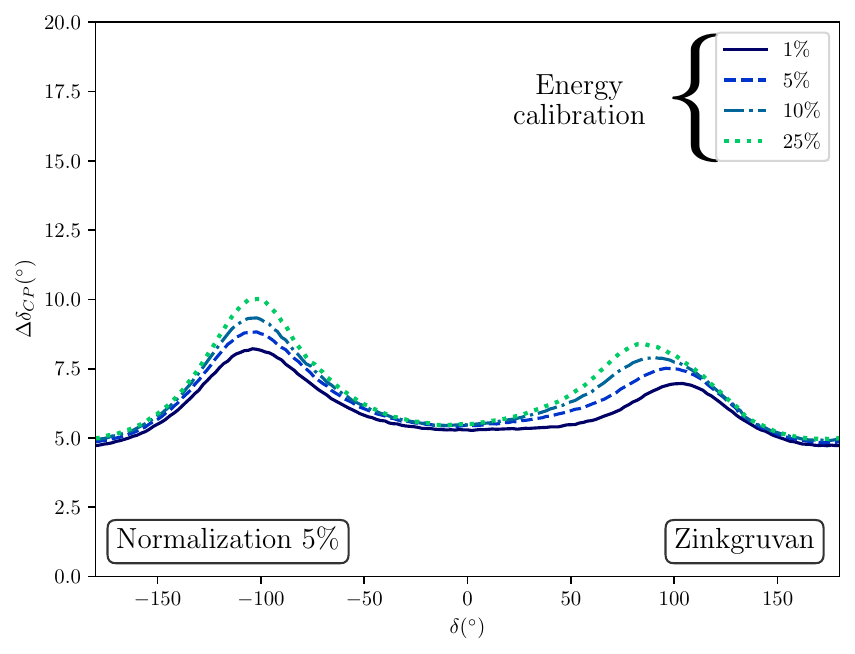}
\includegraphics[width=5.cm]{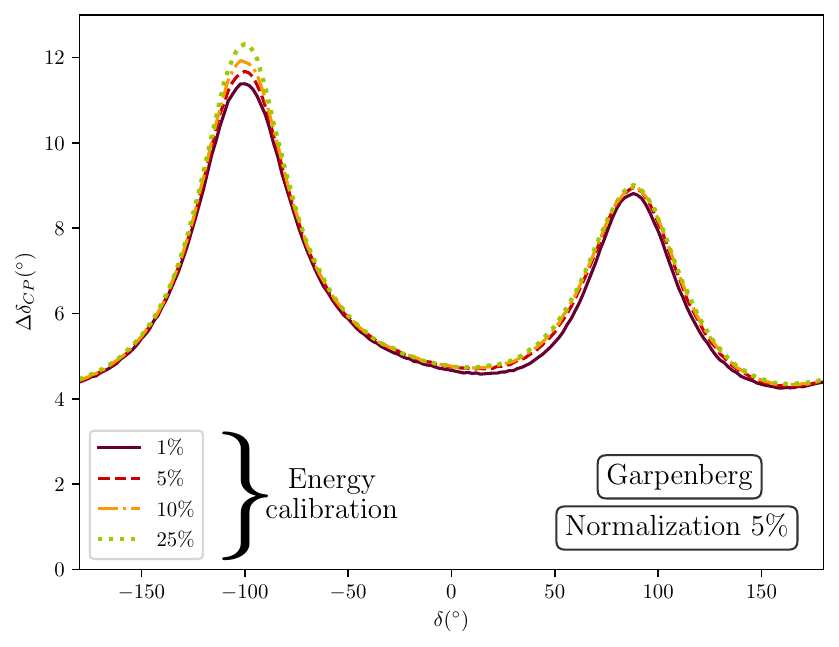}

\includegraphics[width=5.cm]{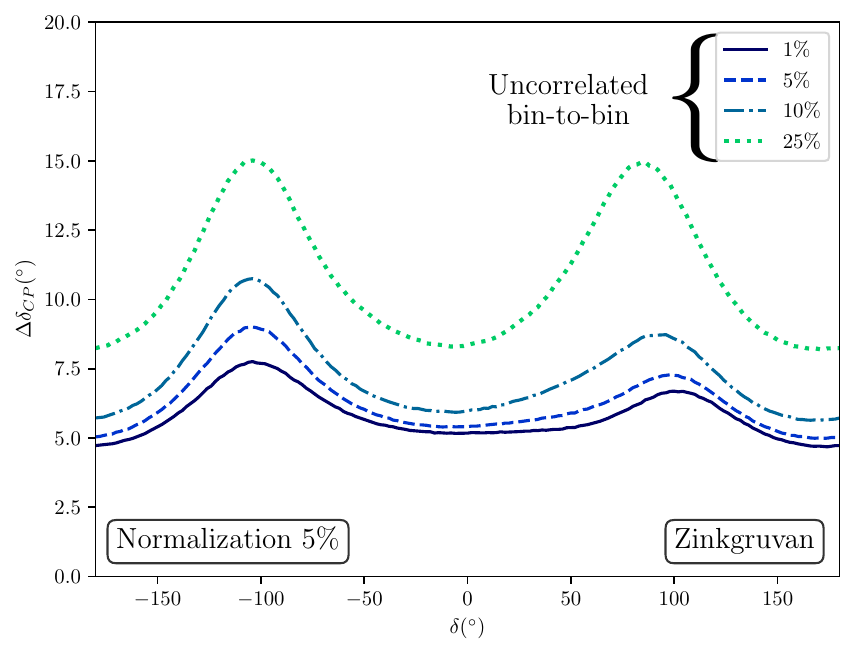}
\includegraphics[width=5.cm]{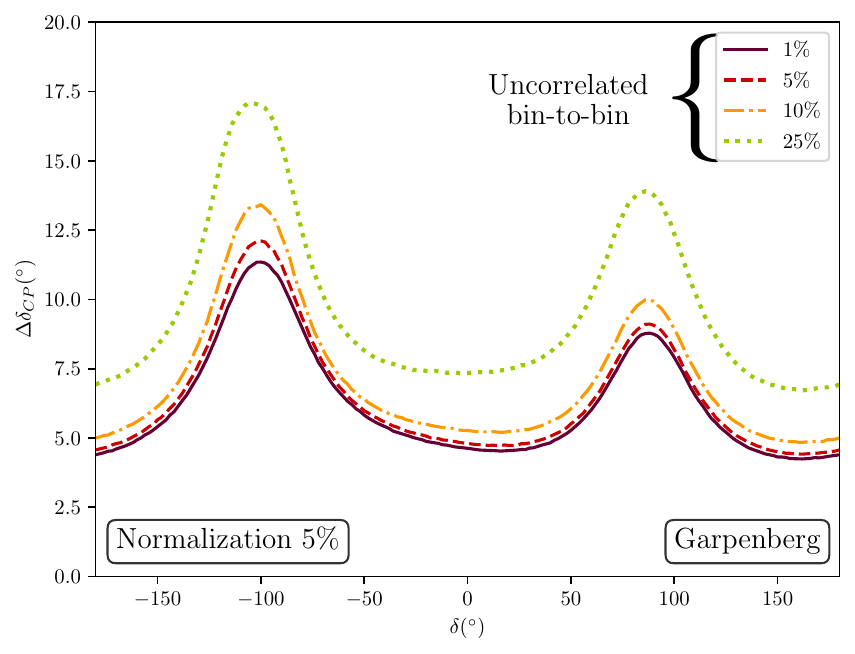}
\caption{ESS$\nu$SB precision measurement of $\delta_{CP}$. Standard error in degrees in the CP phase angle on the vertical axis and value in degrees of the CP phase angle on the horizontal axis. Left (right) panels for the Zinkgruvan \SI{360}{\kilo\meter} (Garpenberg \SI{540}{\kilo\meter}) baseline.}
\label{fig:prec}
\end{figure}

In Figure~\ref{fig:prec}, we estimate the precision with which the ESS$\nu$SB will be able to measure the CP-violating phase $\delta_{CP}$. For each given possible value of $\delta_{CP}$, Figure~\ref{fig:prec} shows the standard error with which $\delta_{CP}$ is determined marginalizing over all nuisance parameters and oscillation parameters other than $\delta_{CP}$. We again observe a similar behaviour to that of Figure~\ref{fig:cpsens}. In particular, we find that the energy calibration uncertainty has a very minor impact, while the other two studied sources of uncertainties have a more important effect. Interestingly, the uncertainty on the overall normalisation is most important for values of $\delta_{CP} \sim 0$. Conversely, the bin-to-bin uncorrelated systematics that can also affect the shape of the recovered spectrum are more relevant close to maximally CP violating values, that is $\delta_{CP} \sim \pm \pi/2$.

\begin{figure}[ht]
\centering
\includegraphics[width=5.3cm]{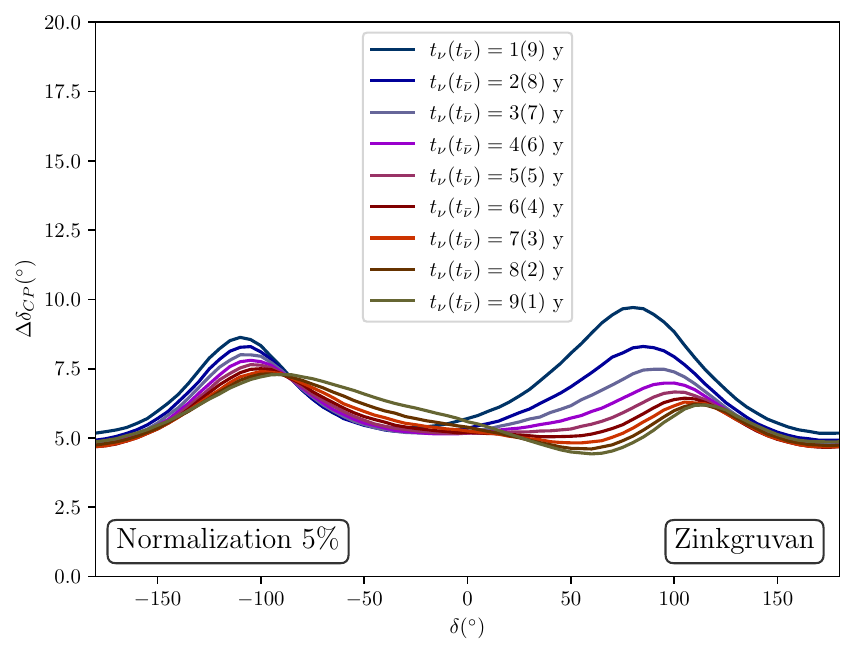}
\includegraphics[width=5.3cm]{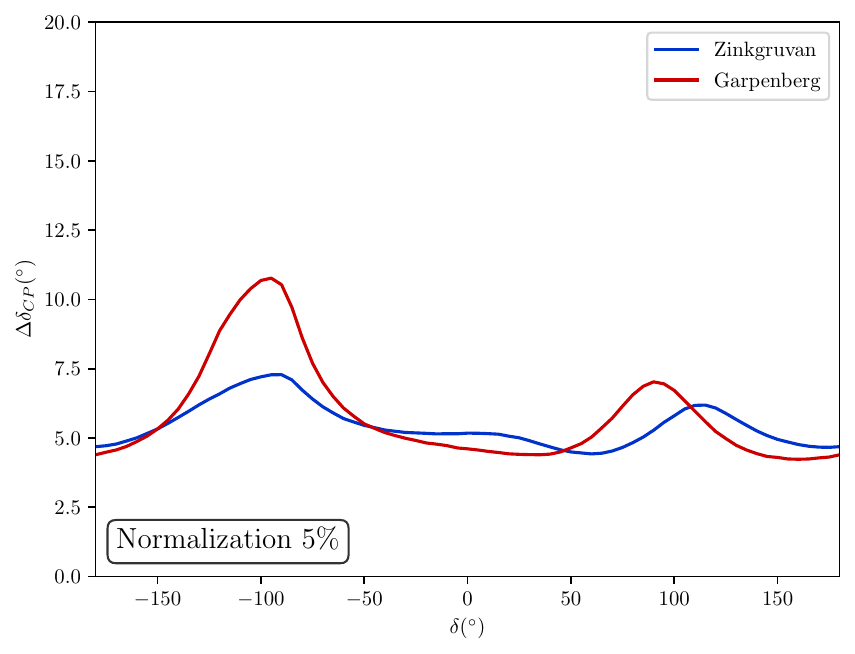}
\caption{Left, precision on the ESS$\nu$SB measurement of $\delta_{CP}$ for different splittings of the running time between neutrino and antineutrino modes at the Zinkgruvan baseline. Standard error in degrees in the CP phase angle on the vertical axis and value in degrees of the CP phase angle on the horizontal axis. The lines span from 9 (1) years to 1 (9) in (anti)neutrino, the total running time is always 10 years. Right, the precision on the measurement of $\delta_{CP}$ when the running time is optimised as in the left panel comparing the Zinkgruvan and Garpenberg options.}
\label{fig:optprec}
\end{figure}

Finally, in the left panel of Figure~\ref{fig:optprec} the dependence of the precision with which $\delta_{CP}$ would be measured is studied as a function of the splitting of the total running time between positive focusing (neutrino mode) and negative focusing (antineutrino mode). As an example, the Zinkgruvan (the \SI{360}{\kilo\meter}) option is shown, but the behaviour is very similar for Garpenberg (the \SI{540}{\kilo\meter}). As can be seen, the optimal splitting depends in the actual value of $\delta_{CP}$. Given the larger fluxes and cross sections, it is easier to accumulate statistics in neutrino mode and thus the best precision would be obtained by devoting longer periods of data taking to positive focusing. Conversely, around $\delta_{CP}=0$ or $\pi$ the complementary between the neutrino and antineutrino samples pays off and more even splits of the running time provide better sensitivity. The measurement strategy of the ESS$\nu$SB can profit from previous hints by preceding oscillation experiments and adapt the splitting between neutrino and antineutrino modes according to the left panel of Figure~\ref{fig:optprec}, depending on the value of $\delta_{CP}$ that the data point to. Following such a strategy, if the best splitting between neutrino and antineutrino modes is adopted for each value of $\delta_{CP}$, the precision that could be obtained is presented in the right panel of Figure~\ref{fig:optprec}. While around CP conserving values the precision achievable in the measurement of $\delta_{CP}$ is around $5^\circ$ for both the Garpenberg and Zinkgruvan options, Zinkgruvan outperforms Garpenberg around $\delta_{CP} = \pm \pi/2$, with the former providing a sensitivity better than $7^\circ$ for any possible value of $\delta_{CP}$. The conclusion of this study is thus that the ESS$\nu$SB experiment with its far detector located at \SI{360}{\kilo\meter}, in the Zinkgruvan site, can provide  unprecedented precision on the measurement of $\delta_{CP}$ ranging between $5^\circ$ and $7^\circ$ depending on its value. The same setup could deliver a $5 \sigma$ discovery of CP violation for a $71 \%$ of all possible values of $\delta_{CP}$.

\subsection{ESS$\nu$SB Synergies with Future Projects}
\label{essnusb_syn}

\subsubsection{Low Energy nuSTORM}
\label{lenustorm}

The basic idea of nuSTORM is to store muons, generated from pion decays, in a racetrack storage ring and use the muon- and electron-neutrinos that are created, from the muon decays, in one of the two straight sections to form a beam that can be used to measure neutrino cross sections as well as search for sterile neutrinos. Contrary to a neutrino beam generated from pion decay, which contains nearly muon neutrinos only, like that of the ESS$\nu$SB, the nuSTORM neutrino beam will contain equal amounts of muon- and electron-neutrinos, thus making high statistics measurements of both electron- and muon-neutrino cross sections possible. In particular, precise electron-neutrino nuclear cross sections are needed for the interpretation of the electron-neutrino spectrum detected by the ESS$\nu$SB far detector.

So far, nuSTORM design studies have been made for the Fermilab and CERN accelerators with proton energies in the order of \SI{100}{\giga\electronvolt} \cite{adey2013nustorm,Ahdida:2654649}. Pions produced from protons of such energies have an average energy of ca \SI{5}{\giga\electronvolt} and will decay to muons with an average ca \SI{4}{\giga\electronvolt} energy, which will in turn decay to neutrinos of average ca \SI{3}{\giga\electronvolt} energy. The length of the racetrack straight sections was chosen to be about \SI{180}{\meter}. Even if the neutrino momentum distribution so produced is broad, it will hardly cover the low neutrino momentum region of ESS$\nu$SB project, which has a neutrino beam of ca \SI{0.4}{\giga\electronvolt} average energy, nor that of the Hyper-K project, which will use a water Cherenkov detector located at the first neutrino oscillation maximum and for which the neutrino beam average energy is ca \SI{0.6}{\giga\electronvolt}.

The ESS-based Low Energy nuSTORM (LEnuSTORM) would use the pulses of protons of \SI{2.5}{\giga\electronvolt} extracted from the ESS$\nu$SB accumulator ring from which a neutrino average energy of ca \SI{0.4}{\giga\electronvolt} will be obtained. It would be difficult to generate a sufficiently powerful beam from the CERN PS, \SI{1.4}{\giga\electronvolt}, Booster or the CERN, \SI{26}{\giga\electronvolt}, PS to cover the low neutrino energies of ESS$\nu$SB and Hyper-K with sufficient statistics. In view of this, it is proposed to carry out a design study of an LEnuSTORM using the ESS$\nu$SB-upgraded ESS LINAC. The target to produce the muons could be either the already designed targets used for producing the neutrino Super Beam with the \SI{1.3}{\micro\second} proton pulses or a new dedicated target. The length of the racetrack straight sections will be much shorter than in the Fermilab and CERN designs and the required strength of the large aperture magnets in the racetrack arcs will be much lower. The lattice design could be of the Fixed-Field Alternating-Gradient accelerator or FODO type or a mixture of both. Figure~\ref{fig:nuSTORM} shows a lay-out of ESS$\nu$SB on the ESS site with proposed positions of the LEnuSTORM racetrack ring and, in this case, a dedicated target station. The LEnuSTORM straight section is directed such that the neutrino beam produced will first hit a LEnuSTORM near detector and then the ESS$\nu$SB near detector,  which is not visible in this figure but located to the right and just above the figure, that would be used as the far detector for the LEnuSTORM beam.

\subsubsection{A Proton Complex Test Facility for a Muon Collider at ESS}
\label{muoncollider}

It is proposed to make, as part of the International Muon Collider design study project \cite{muoncolider}, a design study of a Muon Collider Proton Complex Test Facility which will be based on the use of the ESS LINAC, on the already designed ESS$\nu$SB accumulator ring and on a new compressor/buncher ring. If CERN should decide around 2030 to go forward with the construction of a high energy Muon Collider with a collision energy of 3, 10 or \SI{14}{\tera\electronvolt}, the construction of such a Proton Complex Test Facility at ESS could be started around the same time for the purpose of demonstrating that \SI{2}{\nano\second} proton pulses of 10$^{14}$-10$^{15}$ protons at a rate of \SI{14}{\hertz} can actually be produced in practice. Such a test facility at ESS would require substantial funding but even more funding would be required to start the construction at CERN of such a test  facility, as the build-up of a \SI{5}{\mega\watt} proton accelerator would have to be initiated and this well before 2030 to be in time. Moreover, with such a Proton Complex in operation at ESS and with muon cooling at low intensity being demonstrated in practice in a test facility at CERN, this would open the way - in a longer perspective - for the construction at ESS of a \SI{125}{\giga\electronvolt} Higgs Factory Muon Collider with a unique potential for measurements of the Higgs self-coupling, extremely rare decays and the width of the Higgs boson \cite{rubbia2019searches} and a 3, 10 or \SI{15}{\tera\electronvolt} Muon Collider at CERN for Energy Frontier experiments.

The design study of a Muon Collider Proton Complex at ESS would be based on, inter alia, a faster chopping scheme for the LINAC, a new operation scheme for the accumulator ring, a new design of a compressor/bunch rotation ring and, in a second phase, a separate target station with a target and capture system (horn or solenoid) that could withstand the \SI{2}{\nano\second} short bunches of 10$^{15}$ protons. The basic principle for the generation of the \SI{2}{\nano\second} long pulses from the \SI{2.86}{\milli\second} 10$^{15}$ proton LINAC pulses is illustrated in Figure~\ref{fig:accumulator_LEnuSTORM}. The LINAC H$^{-}$ pulse is chopped into many short pulses that are injected into the accumulator ring from which the proton pulses are  extracted into the compressor/buncher ring where they are phase-rotated to ca \SI{2}{\nano\second} length (\SI{1.5}{\nano\second} in the Figure). This calls for the development of a high frequency chopper acting at the level of the LINAC H$^{-}$ source and an adaptation of the accumulator ring acceptance, RF system, timing and optics. As to the design of the accumulator and the compressor/buncher rings, there has been a design based on the use of the \SI{5}{\giga\electronvolt} \SI{4}{\mega\watt} SPL proton LINAC, that was planned for construction at CERN \cite{protonRutherford} as well as a design based on the use of the \SI{8}{\giga\electronvolt} high power Project-X proton LINAC, that was planned at Fermilab \cite{Flanagan:2010zza}. These designs will be used as starting points for the design and simulation of a compressor/buncher ring based on the use of the ESS LINAC. 

\begin{figure}[H]
\centering
    \includegraphics[width=0.45\linewidth]{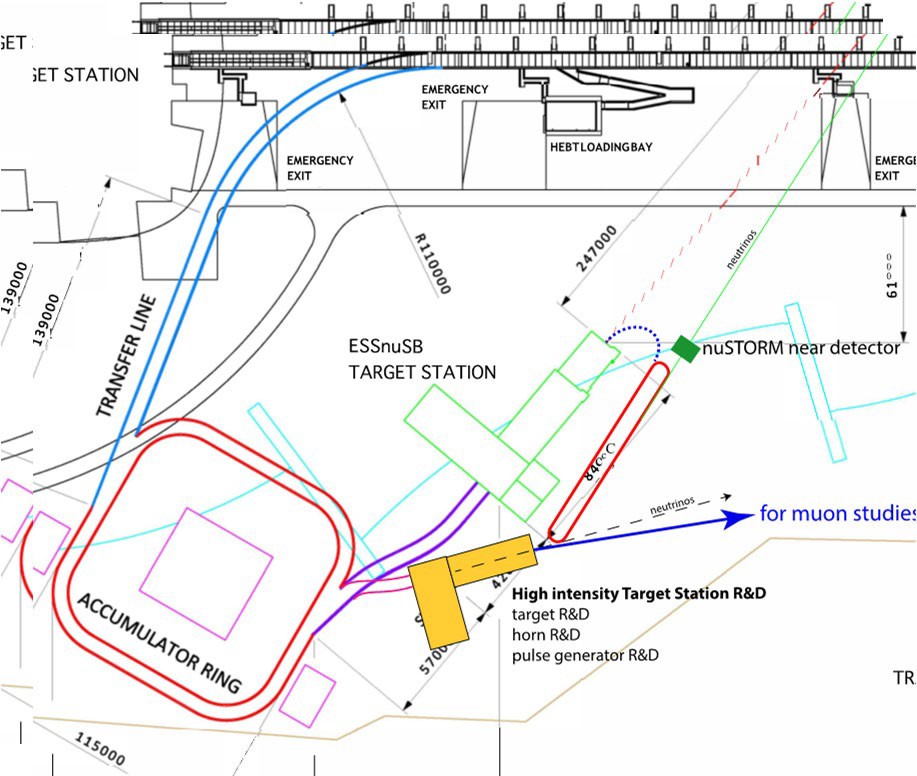}
    \caption{Layout of the ESS$\nu$SB with the proposed LEnuSTORM elongated racetrack ring (in red) and a dedicated target station (in yellow).}
\label{fig:nuSTORM}
\end{figure}

In Figure~\ref{fig:nuSTORM}, there is an indication of the direction of the ejected \SI{2}{\nano\second} pulsed muon beam towards an area at ESS, where there is free space for a second phase project to use the beam so-produced to build and test a target station and cooling front-end set-up there. The compressor ring is tentatively assumed to be located in the same tunnel as the accumulator (the red ring in Figure~\ref{fig:nuSTORM}).

\begin{figure}[hbt]
  \centering
    \includegraphics[width=0.4\linewidth]{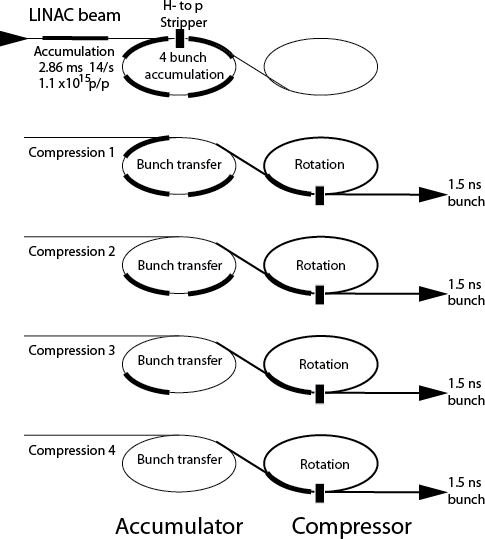}
\caption{The proposed accumulator ring layout dedicated for the Muon Collider Complex at ESS. The transport of the four 1.5 ns pulses will be done through lines of different lengths, such that the four pulses reach the target at the same time}
    \label{fig:accumulator_LEnuSTORM}
\end{figure}

\subsection{Sensitivity with respect to other experiments}
\begin{figure}[H]
\centering
\includegraphics[width=5.cm]{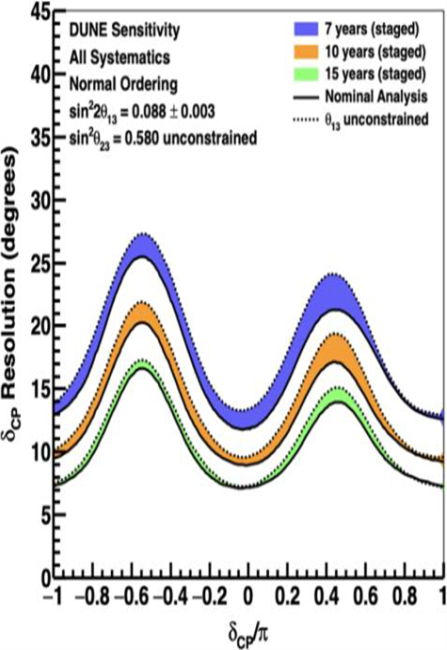}
\includegraphics[width=7.5cm]{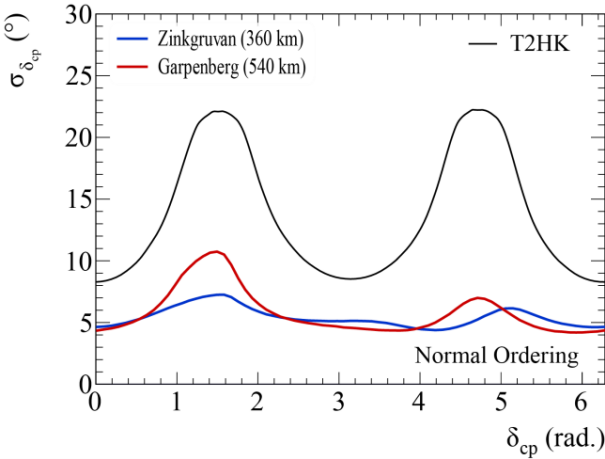}
\caption{The standard error $\Delta \delta_{CP}$ as function of $\delta_{CP}$ for DUNE (left pane, intermediate orange/yellow curve, 10-years operation) and that of T2HK (right pane, upmost black line) compared to that of ESS$\nu$SB (right pane, lower blue line).}
\label{fig:dune-t2hk}
\end{figure}

The higher precision with which $\delta_{CP}$ can be measured as compared to the other two long-baseline neutrino experiments now under construction, DUNE and T2HK, is illustrated in Figure~\ref{fig:dune-t2hk}. This figure shows the standard error $\Delta\delta_{CP}$ with which $\delta_{CP}$ can be measured, by DUNE~\cite{DUNE:2020jqi} in the left pane and by T2HK~\cite{Hyper-Kamiokande:2016srs} in the right pane (the figure in the left pane has been expanded linearly in the vertical direction such that the scales on the vertical axes in the left and the right panes are the same). In the right pane the ESS$\nu$SB standard error has been inserted to enable comparison with the other two experiments clearly showing the superior precision of ESS$\nu$SB. In addition, for both the DUNE and T2HK, the standard errors $\Delta\delta_{CP}$ were obtained by assuming significantly lower systematic uncertainties in the calculations than what was assumed for ESS$\mu$SB (5\% for the signal and 10\% for the background). There are a number of different theoretical models that can be used to derive from the observed amount of baryon asymmetry in the universe a value of $\delta_{CP}$, different for each model. The most precise measurement of $\delta_{CP}$ will provide the sharpest discrimination between these models and select the one that most likely provides a correct description of the origin of the baryon asymmetry, thereby shedding light on the processes occurring right after the Big Bang. A precise measurement of $\delta_{CP}$ will also help to distinguish between different flavour models.

\subsection{Proposed Project Time Line} \label{timeplan}

The proposed time line of the ESS$\nu$SB experiment is the following:
\begin{itemize}
  \item 2022: End of ESS$\nu$SB Conceptual Design Study, CDR, and preliminary costing
  \item 2023-2026: Civil engineering, technical build-up schedule and LEnuSTORM Conceptual Design Study
  \item 2026-2027: TDR, Preparatory Phase 
  \item 2027-2030: Preconstruction Phase, International Agreement 
  \item 2030-2037: Construction of the neutrino beam facility and the detectors, including commissioning
  \item 2037- : Start of data taking
\end{itemize}

The details of the construction schedule 2027-2037 will be studied in 2023-2026 period.

\clearpage 
\section{Conclusion}

The ESS will open up new intensity frontiers for particle physics.  At
design performance, it will be the world’s brightest neutron source and
provide intense neutrino pulses. The flexible design of the target area
allows for the installation of an intensity-optimised moderator and the
extraction of the world’s most intense neutron beams. Its world’s most
powerful proton accelerator offers the possibility to be upgraded and
complemented by a compressor ring to produce the world's most intense
microsecond-pulsed neutrino beams.

These world-leading capabilities enable an ambitious programme of
particle physics experiments at the neutron and neutrino intensity
frontiers. Among the projects presented in this paper are precision
measurements of neutron decay, neutrino-nucleus scattering, the neutron
electric dipole moment and neutrino flavour oscillations at the second
oscillation maximum, and searches for and measurements of
neutron-antineutron oscillations, leptonic CP violation, sterile
neutrons and sterile neutrinos. These measurements and searches tackle
several outstanding questions in our understanding of nature, such as
the structure of fundamental interactions, the origin of the
matter-antimatter asymmetry in the Universe, or the nature of dark
matter or dark energy.

This paper presents concepts and expected performances of possible
experiments. Some of the proposed experiments can be realised within the
mission of the ESS as a user facility and require the installation of
state-of-the-art instrumentation in line with the usual investment at
such facilities, as for example a neutron beam line for particle
physics. Some neutrino-nucleus scattering experiments are already under
construction to be ready for early ESS operation. Other projects require
additional design studies and further investment in the ESS and
experiment infrastructure. Such investment includes both the provision
of a high-intensity neutron moderator as well as funding for large scale
experiments.

The proposed experimental program of particle physics illustrates the
unique scientific potential of the ESS. The realisation of this
programme would place the ESS at the forefront of particle physics with
neutrons and neutrinos.

\section*{Acknowledgements}
J.B. acknowledges support from the Swedish research council grants 2016-05996 and 2019-03779. %
L.B. and F.G. acknowledge support from from the U.S. Department of Energy (DOE), Office of Science, Office of Nuclear Physics. Grant No. DE-AC05-00OR22725. %
The work of Z.B. was supported in part by the MIUR research grant "The Dark Universe: A Synergic Multimessenger Approach" PRIN 2017 No. 2017X7X85K, and in part by the SRNSF. %
J.I.C. acknowledges support by the European Research Council under Grant Agreement No.101055120-ESSCEvNS ERC-2021-ADG. %
V.C. acknowledges support from the U.S. Department of Energy Grant No. DE-FG02-00ER41132. %
M.G.-A. acknowledges support from Generalitat Valenciana (Spain) through the plan GenT program (CIDEGENT/2018/014), and MCIN/AEI/10.13039/501100011033 Grant No. PID2020-114473GB-I00. %
L.L. is supported by the predoctoral training program of non-doctoral research personnel of the Department of Education of the Basque Government. L.L. and F.M. are also supported by the European Research Council (ERC) under Grant Agreement No. 101039048-GanESS and the Severo Ochoa Program grant CEX2018-000867-S. %
V.N. acknowledges support from L’Agence nationale de la recherche, France, ANR-20-CE08-0034. %
D. M. gratefully acknowledges support from the Swedish Research Council and the Swedish Research Council’s Council for Research infrastructure. %
T.O. acknowledges support from the Swedish Research Council (Vetenskapsrådet), Contract No. 2017-03934. %
A.S.P. acknowledges support from the Marie Skłodowska-Curie grant agreement No. 101026628. %
V.S. acknowledges support from the HighNESS project at the European Spallation Source. HighNESS is funded by the European Framework for Research and Innovation Horizon 2020, under grant agreement 951782. %







\newpage
\end{document}